%% file: main.tex
\documentclass[b5paper]{book}

\usepackage{multirow}
\newcommand{\algname}[1] {{\fontfamily{cmtt}\selectfont {#1}}}
\usepackage{mathtools}
\DeclarePairedDelimiter{\ceil}{\lceil}{\rceil}
\usepackage{amsmath}
\usepackage{algorithm}
\usepackage{rotating}
\usepackage{longtable,tabu}

\input{metadata}

\input{config}

\makeindex

\begin{document}

\frontmatter

\include{FrontBackMatter/Titlepage}
\cleardoublepage
\include{FrontBackMatter/Titleback}
\cleardoublepage
\include{FrontBackMatter/TitlepageOfficial}
\cleardoublepage
\include{FrontBackMatter/Abstract}

\tableofcontents

\listoffigures
\listoftables

%
%
\mainmatter

\input{./Chapters/01_Intro}

\input{./Chapters/02_Background}

\input{./Chapters/03_Simulation}

\input{./Chapters/04_ExposureBias}

\input{./Chapters/05_Methodology}

\input{./Chapters/06_Solution1}
\input{./Chapters/07_Solution2}

\input{./Chapters/08_Conclusion}

\tikzsetfigurename{main_}

%
%

%
%
\appendix

\include{./FrontBackMatter/Bibliography}

\cleardoublepage
\printindex

\backmatter

\cleardoublepage
\include{FrontBackMatter/Summaries}

\cleardoublepage
\include{./FrontBackMatter/Acknowledgments}

\include{./FrontBackMatter/curriculumVitae}

\include{./FrontBackMatter/SIKSdissertations}
\end{document}

%% file: metadata.tex
\newcommand{\myTitle}{Understanding and Mitigating Multi-sided Exposure Bias in Recommender Systems\xspace}
\title{\myTitle}

\newcommand{\myName}{M. Mansoury\xspace}
\author{\myName}

\newcommand{\myFaculty}{Mathematics and Computer Science\xspace}

\newcommand{\myUni}{Eindhoven University of Technology\xspace}

\date{} 

\newcommand{\myKeywords}{Recommender systems, Exposure bias, Popularity bias, Multi-sided fairness, Feedback loop\xspace}

%% file: config.tex
\newcommand{\bd}[1]{\begin{definition}(#1)\\}
\newcommand{\ed}{\end{definition}}

\usepackage[dvipsnames,table]{xcolor} 
\definecolor{darkred}{rgb}{0.5,0,0}
\definecolor{darkgreen}{rgb}{0,0.5,0}
\definecolor{darkblue}{rgb}{0,0,0.5}
\definecolor{halfgray}{gray}{0.55} 
\definecolor{webgreen}{rgb}{0,.5,0}
\definecolor{webbrown}{rgb}{.6,0,0}

\providecommand{\mLyX}{L\kern-.1667em\lower.25em\hbox{Y}\kern-.125emX\@}

\makeatletter
  \providecommand*\setfloatlocations[2]{\@namedef{fps@#1}{#2}}
\makeatother
\setfloatlocations{figure}{b!}
\setfloatlocations{table}{t!}

\makeatletter
\def\blfootnote{\xdef\@thefnmark{}\@footnotetext}
\makeatother

\newtheorem{definition}{Definition}[chapter]  

\usepackage{algpseudocode}   

\usepackage{amssymb}        

\usepackage[dutch,english]{babel} 
\usepackage{booktabs}       
\usepackage{bm}             
\usepackage{bibentry}       
\nobibliography*

\usepackage{calc}           

\usepackage{caption}        
\usepackage{cite}           
\usepackage{comment}        

\usepackage{datetime}   
\yyyymmdddate
\usepackage{dsfont}         

\usepackage{emptypage}      

\usepackage{fancyhdr}
\usepackage[Lenny]{fncychap} 
\usepackage[T1]{fontenc}
\usepackage{fourier}        

\usepackage{graphicx}       
\graphicspath{{./Figures/}} 
\usepackage[xindy,toc]{glossaries}  

\usepackage{hhline}         
\usepackage{hyphenat}
\usepackage[utf8]{inputenc}  

\usepackage{lastpage}       
\usepackage{lipsum}         
\usepackage{longtable}      

\usepackage{makeidx}        
\usepackage{mathtools}      
\usepackage{mparhack}       
\usepackage{multirow}       

\usepackage{paralist}       
\usepackage{pdflscape}      
\usepackage{pifont}         
\usepackage{placeins}       
\usepackage[draft]{prelim2e}       

\usepackage[nocites,norefs]{refcheck}       
\usepackage{rotating}       

\usepackage{standalone}     

\usepackage{tabularx}       
\usepackage[nottoc]{tocbibind}      

\input{config-TikZ}

\def\graffito\emph{\marginpar}
\usepackage[colorinlistoftodos,backgroundcolor=white,linecolor=gray]{todonotes}  

\makeatletter
\renewcommand{\todo}[2][]{\tikzexternaldisable\@todo[noline,#1]{#2}\tikzexternalenable}
\makeatother

\usepackage{wasysym}        

\usepackage{xspace}         

\usepackage{fixltx2e} 
\usepackage{textcomp} 

\usepackage{listings}
\PassOptionsToPackage{hidelinks,pdftex,hyperfootnotes=false,pdfpagelabels,ocgcolorlinks}{hyperref}
\RequirePackage{hyperref}

\newcommand{\backrefnotcitedstring}{\relax}
\newcommand{\backrefcitedsinglestring}[1]{(Cited on page~#1.)}
\newcommand{\backrefcitedmultistring}[1]{(Cited on pages~#1.)}
		\PassOptionsToPackage{hyperpageref}{backref}
\usepackage[hyperpageref]{backref} 
   \renewcommand*{\backref}[1]{}  
   \renewcommand*{\backrefalt}[4]{
      \ifcase #1 %
         \backrefnotcitedstring%
      \or%
         \backrefcitedsinglestring{#2}%
      \else%
         \backrefcitedmultistring{#2}%
      \fi}%

\hypersetup{%
    colorlinks=false,
    hidelinks=true,
    linktocpage=true, pdfstartpage=3, pdfstartview=FitV,%
    linktoc=all, 
    breaklinks=true, pdfpagemode=UseNone, pageanchor=true, pdfpagemode=UseOutlines,%
    plainpages=false, bookmarksnumbered, bookmarksopen=true, bookmarksopenlevel=1,%
    hypertexnames=true, pdfhighlight=/O,
    urlcolor=webbrown, linkcolor=RoyalBlue, citecolor=webgreen, 
    pdftitle={\myTitle},%
    pdfauthor={\textcopyright\ \myName, \myUni, \myFaculty},%
    pdfsubject={},%
    pdfkeywords={\myKeywords},%
    pdfcreator={pdfLaTeX},%
    pdfproducer={LaTeX with hyperref}
    ocgcolorlinks%
}

\makeatletter
\addto\extrasenglish{
}
\makeatother

\fancyheadoffset[LE,RO]{\marginparsep+\marginparwidth}

\fancypagestyle{plain}{%
 \fancyhead{} 
}
\makeatletter

\ChNumVar{\fontsize{36}{80}\usefont{OT1}{pag}{m}{n}\selectfont}
\renewcommand{\DOCH}{%
    \settowidth{\px}{\CNV\FmN{\@chapapp}}
    \addtolength{\px}{2pt}

    \setlength{\py}{0pt}
    \setlength{\mylen}{0pt}

    \settowidth{\pxx}{\CNoV\thechapter}
    \addtolength{\pxx}{-1pt}
    \par
    \parbox[b]{\textwidth}{%
    \raggedright%
    \color{gray}
    \CNoV\FmN{\@chapapp}
    \hskip1pt%
    \thechapter%
    \hskip1pt%
    \color{black}
    \rule{\RW}{\pyy}\par\nobreak%
    \vskip -\baselineskip%
    \vskip -\pyy%
    \hskip \mylen%
    \vskip \pyy}%
    \vskip 20\p@}

\renewcommand{\DOTI}[1]{%
    \raggedright
    \CTV\FmTi{#1}\par\nobreak
    \vskip 40\p@}

  \renewcommand{\DOTIS}[1]{%
    \raggedright
    \CTV\FmTi{#1}\par\nobreak
    \vskip 40\p@}

\makeatother

\makeglossaries


%% file: config-TikZ.tex
\usepackage{subcaption}     

\usepackage{pgfplots}       
\usepgfplotslibrary{colormaps,groupplots,patchplots,statistics} 
\pgfplotsset{
 every axis/.append style={
      legend image post style={mark=none},
      cycle list name = linestyles*},
      yticklabel style={/pgf/number format/fixed,/pgf/number format/fixed zerofill,/pgf/number format/precision=3},
      every axis plot /.append style={mark=none},
      every colorbar/.append style={
        colorbar sampled line,
        colorbar horizontal,
  },
  colormap/autumn,
}
\pgfplotsset{compat=1.10}
\pgfplotscreateplotcyclelist{customLineList}{solid,dashed,dashdotted,dashdotdotted,dotted}

\usepackage{tikz}           
\usepackage{tikz-qtree}     
\usepackage[external]{forest}         

\usetikzlibrary{arrows,automata,backgrounds,calc,decorations.pathmorphing,decorations.text,decorations.markings,fit,petri,positioning,scopes,shadows,shapes,spy}

\usetikzlibrary{external}
\tikzset{small/.style={level distance=15pt,sibling distance=-2pt,outer ysep=-3pt, label distance=-8pt}}
\forestset{.style={for tree={l-=2pt, l sep-=2pt,s sep=2pt}}} 
\tikzset{smaller/.style={scale=.5}}
\tikzset{configuration/.style={rectangle callout, callout pointer arc=30,callout absolute pointer={#1},draw=gray!50,fill=gray!15,font=\tiny}}
\tikzstyle{markedTreeNode} = [circle,fill=gray!30]
\tikzstyle{circledTreeNode} = [circle,dashed,draw]

\tikzset{node distance=6mm} 
\tikzset{PNsmall/.append style={node distance=3mm}}
\tikzset{place/.append style={circle,draw=black,thick,inner sep=0pt,minimum size=6mm,label position=below}} 
\tikzset{transition/.append style={rectangle,draw=black,thick,inner sep=0pt,minimum size=6mm}}
\tikzset{tau/.style={transition,fill=black,minimum size=4mm}}
\tikzset{point/.style={circle,inner sep=0pt,draw=black,minimum size=1mm,fill=black}}

\tikzstyle{BPMN_task} = [transition, rounded corners,align=center,minimum width=4em, thin]
\tikzstyle{BPMN_event} = [place]
\tikzstyle{BPMN_end event} = [BPMN_event,ultra thick]
\tikzstyle{BPMN_gateway_xor} = [diamond, thin, draw, label=center:$\boldsymbol\times$]
\tikzstyle{BPMN_gateway_or}  = [diamond, thin, draw, label=center:$\boldsymbol\ocircle$]
\tikzstyle{BPMN_gateway_and} = [diamond, thin, draw, label=center:\textbf{+}]

\tikzstyle{EPC_function} = [transition, rounded corners, fill=green!50, thick, draw=black, minimum width=3em]
\tikzstyle{EPC_event} = [chamfered rectangle, chamfered rectangle angle=15, thick, draw=black, fill=red!50, minimum width=3em]

\tikzstyle{flw_block} = [rectangle, draw, fill=blue!20, text centered, rounded corners] 
\tikzstyle{flw_decision} = [diamond, draw, fill=blue!20,  text badly centered, inner sep=0pt] 
\tikzstyle{flw_line} = [draw, -latex']
\tikzstyle{flw_cloud} = [draw, cloud, cloud ignores aspect,fill=red!20] 

\tikzstyle{eventlog} = [cylinder, cylinder uses custom fill, cylinder end fill=blue!50, cylinder body fill=blue!25, draw=blue, minimum width=2em, minimum height = 3em, shape border rotate = 90,  label=center:EL]
\tikzstyle{ETM} = [signal, signal to = east, signal from = west, fill=green!50, minimum width=4em, minimum height=2em]
\tikzstyle{processtree} = [dart, shape border rotate = 90, draw, fill=brown!75, minimum height=4em, label=center:PT]
\tikzstyle{processtreeConfigurable} = [dart, shape border rotate = 90, draw, fill=brown!75, minimum height=4em, label=center:PT, label=above:C]
\tikzstyle{processtreeConfigurations} = [rectangle split, rectangle split part fill={red!50, green!50, blue!50, yellow!50},draw,minimum width=.5cm]

\pgfplotsset{
    every axis/.append style={
        scale only axis,          
    },
    label style={font=\footnotesize},
    legend style={font=\footnotesize}
}

\makeatletter
\def\customrevertcolormap#1{%
    \pgfplotsarraycopy{pgfpl@cm@#1}\to{custom@COPY}%
    \c@pgf@counta=0
    \c@pgf@countb=\pgfplotsarraysizeof{custom@COPY}\relax
    \c@pgf@countd=\c@pgf@countb
    \advance\c@pgf@countd by-1 %
    \pgfutil@loop
    \ifnum\c@pgf@counta<\c@pgf@countb
        \pgfplotsarrayselect{\c@pgf@counta}\of{custom@COPY}\to\pgfplots@loc@TMPa
        \pgfplotsarrayletentry\c@pgf@countd\of{pgfpl@cm@#1}=\pgfplots@loc@TMPa
        \advance\c@pgf@counta by1 %
        \advance\c@pgf@countd by-1 %
    \pgfutil@repeat
}%
\makeatother


%% file: FrontBackMatter/Titlepage.tex

\title{Understanding and Mitigating Multi-Sided Exposure Bias in Recommender Systems}
\author{Masoud Mansoury}


\maketitle 

%% file: FrontBackMatter/Titleback.tex

\thispagestyle{empty}
\noindent
\begin{tabular}{p{109mm}}

%
\myTitle by Masoud Mansoury.\\
Eindhoven: Technische Universiteit Eindhoven, 2021.\ Proefschrift.\\[6mm]


\noindent
A catalogue record is available from the Eindhoven University of Technology Library\\[6mm]
ISBN: 978-90-386-5376-1\\[6mm]

\end{tabular}

\vfill

\noindent
\begin{tabular}{p{104mm}}
\\
\includegraphics[scale=0.06]{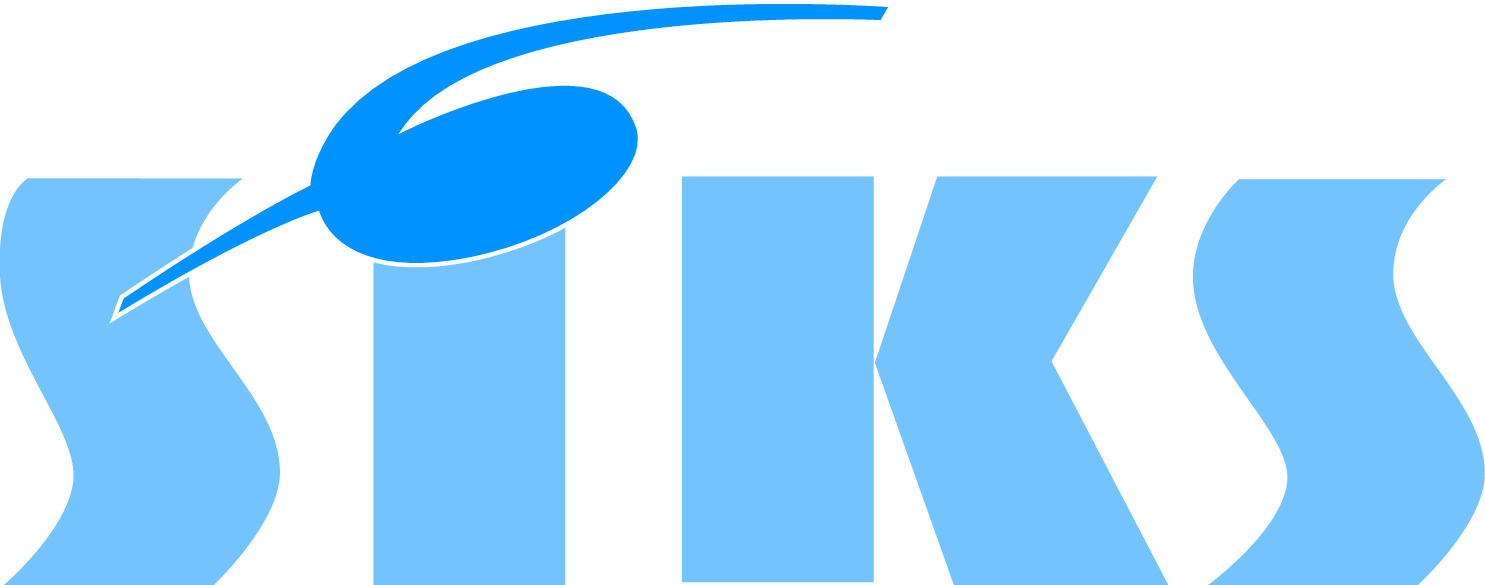}

SIKS Dissertation Series No. 2021-20
\\
The research reported in this thesis has been carried out under the auspices of SIKS, the Dutch Research School for Information and Knowledge Systems.
\\[6mm]
\end{tabular}

%% file: FrontBackMatter/TitlepageOfficial.tex

%
%

\thispagestyle{empty}
\noindent

\begin{center}

\huge 


\myTitle

\vspace{3em}

\Large
PROEFSCHRIFT

\vspace{3em}

ter verkrijging van de graad van doctor aan de Technische Universiteit Eindhoven, op gezag van de rector magnificus prof.dr.ir. F.P.T. Baaijens,
voor een commissie aangewezen door het College voor Promoties, in het openbaar te verdedigen op \\donderdag 14 oktober 2021 om 16:00 uur

\vspace{3em}

door

\vspace{3em}

Masoud Mansoury

\vspace{2em}

geboren te Noor, Iran

\end{center}

\newpage
\thispagestyle{empty}
\normalsize

\noindent
Dit proefschrift is goedgekeurd door de promotoren en de samenstelling van de promotiecommissie is als volgt:

\vspace{1em}
\noindent
\begin{tabular}{lp{9cm}}
1e promotor:       & prof. dr. Mykola Pechenizkiy \\
1e co-promotor:    & prof. dr. Bamshad Mobasher (DePaul University) \\
2e co-promotor:    & prof. dr. Robin Burke (University of Colorado Boulder) \\
leden:             & prof. dr. Nava Tintarev  (Maastricht University) \\
                   & prof. dr. Martha Larson (Radboud University) \\
                   & dr. Martijn Willemsen \\
adviseur(s):       & dr. Maryam Tavakol \\
\end{tabular}

\vfill

\noindent Het onderzoek of ontwerp dat in dit proefschrift wordt beschreven is uitgevoerd in overeenstemming met de TU/e Gedragscode Wetenschapsbeoefening.

%% file: FrontBackMatter/Abstract.tex


\chapter*{Abstract}
\addcontentsline{toc}{chapter}{Abstract}
\thispagestyle{empty}
\emph{Fairness is a critical system-level objective in recommender systems that has been the subject of extensive recent research. It is especially important in multi-sided recommendation platforms where it may be crucial to optimize utilities not just for the end user, but also for other actors such as item sellers or producers who desire a fair representation of their items. Existing solutions do not properly address various aspects of multi-sided fairness in recommendations as they may either solely have one-sided view (i.e. improving the fairness only for one side), or do not appropriately measure the fairness for each actor involved in the system. In this thesis, I aim at first investigating the impact of unfair recommendations on the system and how these unfair recommendations can negatively affect major actors in the system. Then, I seek to propose solutions to tackle the unfairness of recommendations. I propose a rating transformation technique that works as a pre-processing step before building the recommendation model to alleviate the inherent popularity bias in the input data and consequently to mitigate the exposure unfairness for items and suppliers in the recommendation lists. Also, as another solution, I propose a general graph-based solution that works as a post-processing approach after recommendation generation for mitigating the multi-sided exposure bias in the recommendation results. For evaluation, I introduce several metrics for measuring the exposure fairness for items and suppliers, and show that these metrics better capture the fairness properties in the recommendation results. I perform extensive experiments to evaluate the effectiveness of the proposed solutions. The experiments on different publicly-available datasets and comparison with various baselines confirm the superiority of the proposed solutions in improving the exposure fairness for items and suppliers.}


%% file: Chapters/01_Intro.tex
\chapter{Introduction}
\label{chap:intro}
\tikzsetfigurename{intro_}

\section{Motivation}
Recommender systems are tools that act as decision guides, helping users to find their desired items by predicting their preferences and suggesting the preferred items to them. These systems use historical data on interactions between users and items to generate personalized recommendations for the users. Recommender systems are used in a variety of different applications including movies, music, e-commerce, online dating, and many other areas where the number of options from which the user needs to choose can be overwhelming. Examples of real-world recommendation systems are movie recommendation in Netflix, music recommendation in Spotify, and product recommendation in Amazon. There are various recommendation models and approaches for generating recommendations for the users. These approaches will be discussed in detail in Chapter~\ref{chap:back}.

For a long time, the main concern in research on recommender systems was improving the accuracy of the recommendations. In those works, the researchers tried to design new recommendation algorithms or enhance the existing recommendation algorithms to generate recommendations to the users that are better matched with their preferences. However, new challenges have recently emerged in recommender systems research domain such as novelty, diversity, serendipity, and fairness of recommendations. The main focus of this dissertation is improving the fairness of recommender systems (or addressing algorithmic bias or unfairness in recommender systems).

The topic of fairness in recommender systems is concerned with the fair treatment of all entities in the system when generating recommendations. This means that the recommendation algorithm should serve all users (minority and majority) or items/suppliers (popular and non-popular) equally by satisfying their expectations and utilities. For example, when the system gives more exposure or visibility to certain items or suppliers, it raises the issue of discrimination or unfairness. The goal of this dissertation is understanding these issues and proposing solutions for addressing them. 

It has been shown that recommendations generated by recommender systems generally suffer from bias against certain groups of users or items \cite{ekstrand2018,mansoury2019bias,yao2017,yin2012challenging,patro2020fairrec}. The problem with biased recommendation is that it raises the issue of unfairness in recommendation results as certain groups of users or items may receive more benefit from the recommender systems than others. Therefore, tackling this bias for equalizing the benefits from recommender systems between different groups of users and items is the objective of fairness-aware recommendation systems. 

Bias in recommendation output can originate from different sources: 1) it may stem from the underlying biases in the input data used for training \cite{burke2017,virginia2018}: some groups of users may represent the majority in terms of number of individual users or number of ratings provided by these users, or some items may receive large proportion of ratings while other items may not receive much attention from the users, or 2) it may be due to the algorithmic bias where recommendation algorithms propagate the existing bias in data \cite{kamishima2011,yao2017,zembel2013,dietmar2013,abdollahpouri2020multi} and, in some cases, intensify it by recommending these popular items even to the users who are not interested in popular items \cite{mansoury2020feedback,abdollahpouri2019managing}.

Addressing these biased and unfair recommendations is a challenging task and requires careful design of algorithms as improving the fairness of recommendations leads to loss in recommendation accuracy \cite{mehrotra2018towards}. Thus, an algorithmic solution for tackling unfairness in recommender systems should also take into account the trade-off with accuracy of recommendations as the most important goal of recommender systems. In addition, addressing fairness of recommender systems can be even more challenging when multiple \textit{stakeholders}, \textit{entities}, \textit{sides}, or \textit{sides} are involved in the system. In this situation, the fairness of each side should be properly addressed when generating recommendation lists. 

Recommender systems often operate in a \textit{multi-sided} environment where different sides or actors are involved in the system: users, items, suppliers \cite{himan2019a,burke2017b,burke2016}. Examples of multi-sided recommendation platforms are song recommendations on Spotify in which the sides in this platform are users who listen the songs, items are the songs, and artists are the suppliers. Or another example of multi-sided recommendation platform is GooglePlay in which the sides are the users who download and install apps on their smart phones, apps are the items, and apps developers are the suppliers. Finally, on Netflix, users are the movie watchers, items are the movies, and suppliers are the movie producers.

\begin{figure}[t]
    \centering
    \includegraphics[width=0.95\textwidth]{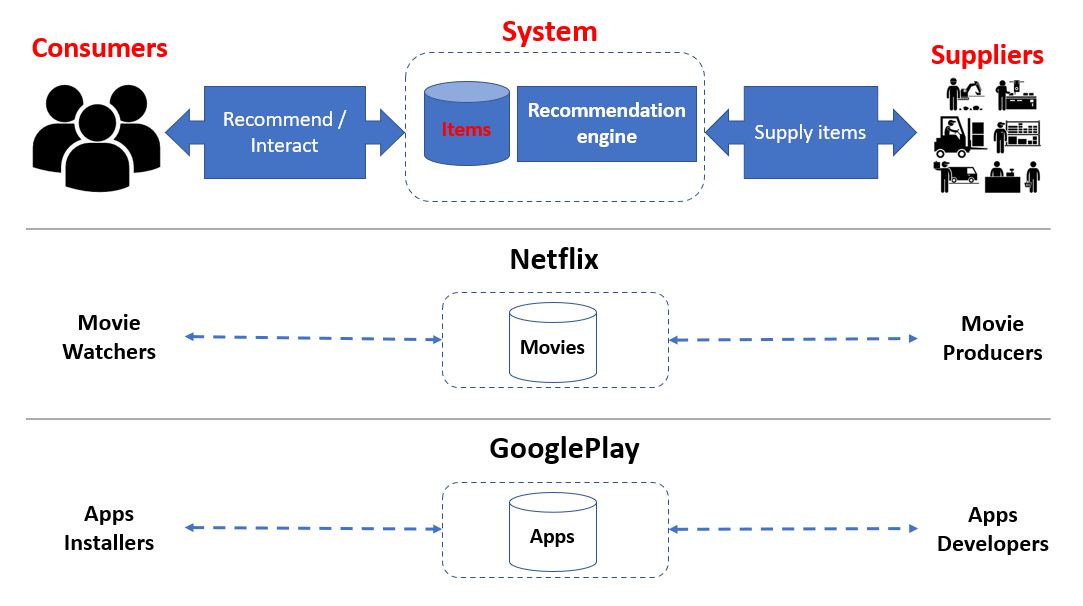}
\caption{Multi-sided nature of recommender systems. The first row shows the sides in a typical recommendation platform. Major sides are shown in red color. The second and third rows illustrate the major sides in two real-world recommender applications.}\label{multisided}
\end{figure}

Figure~\ref{multisided} shows the major sides in recommender systems. Consumers (or users) are the side who interact with the items in the system and provide their feedback about those items. The system also processes the feedback from consumers and generates recommendation lists for them. On the other hand, the suppliers are the sides who feed the system with contents and items. Finally, item is another side in the system that works as a bridge between consumers and suppliers: \textit{items supplied by suppliers are consumed by consumers}. All these sides are closely connected and influence each other when interact with the system or are interacted by the system. For instance, when consumers frequently interact with certain items in the system, it adds bias to the rating data collected by the system. It can also cause recommendation algorithms focus on those certain items when generating recommendations that would cause unfair representation of items and suppliers in the recommendation lists. Therefore, an algorithmic solution to address the fairness of all sides is needed.

In multi-sided recommendation platforms, addressing the needs and utilities of each side is critical. For users, utility can be achieved by delivering accurate recommendations that are matched with their preferences. For items and suppliers, utility can be defined as providing equal chance for each item or supplier to be shown in the recommendation lists. This means that each item or supplier must have fair exposure or visibility in the recommendation lists. In this dissertation, I aim at addressing exposure bias for items and suppliers in the recommendation results.


Generally, item-side or supplier-side exposure refers to the fact that how much an item or a supplier is represented in the recommendation lists to the users \cite{abdollahpouri2020multi}. For example, when an item is recommended to two users out of 100 users in the system, we say that this item received exposure of 2\%. In this definition, the position of the item in the recommendation lists is not taken into account and the exposure of the items or suppliers at any position in the list is computed similarly. Singh and Joachims in \cite{singh2018fairness} defined exposure of items and suppliers according to their position in the recommendation list. This definition not only considers the number of times an item is recommended, but also it takes into account the position that the item is exposed in the recommendation list. In this definition, when an item is recommended at the first position in the recommendation list, it signifies that the item has higher exposure than the items in the lower position. 

In this dissertation, I use the first definition where the position of the item in the recommendation list is not taken into account.\ Given this definition for exposure of items or suppliers, exposure bias in recommender systems refers to the fact that some items and suppliers are over-represented in recommendation results, while other items and suppliers are not adequately represented \cite{abdollahpouri2020multi,chen2020bias,singh2018fairness}. Due to the interactive nature of recommendation systems, this bias can be even amplified over time as users interact with the recommended items at each time and their interactions would be used as input for recommendation algorithm at the next time \cite{mansoury2020feedback}. Therefore, addressing this bias is critical for achieving fair treatment of items and suppliers in the system.


\section{Research objectives and approaches}

\begin{itemize}
    \item This dissertation aims at studying the problem of unfairness in recommender systems for different sides in the system. In particular, it investigates the exposure unfairness for items and suppliers by analyzing the recommendation outputs in terms of how fairly items and suppliers are appeared or represented in the recommendation lists delivered to the users. 
    \item To understand the negative impacts of exposure unfairness on each side in the system and the possible sources/reasons for this issue, this thesis seeks to perform a simulation study on interaction between users and recommender systems over time. Using a recommendation algorithm, in this simulation, a set of items will be recommended to the user at each time point and user feedback/click on the recommended items will be recorded. Users' feedback on recommended items will be added to their profile and will be used as input for recommendation algorithm in the next time point. Various bias analysis will be conducted on input data and recommendation lists at each time point. 
    \item As part of these analysis, the objective is also to study the existing evaluation metrics for measuring exposure bias in recommender systems, and how well they are able to reveal the exposure bias of a recommendation algorithm. This analysis helps us to better understand the limitations and strengths of the existing metrics. Then, this thesis seeks to introduce new metrics or modify the existing metrics to properly evaluate the exposure bias of recommendation algorithms. 
    \item After understanding the issues and their consequences on each side in the system, this dissertation aims at proposing solutions to mitigate the unfair exposure of items and suppliers in recommendation results. The objective is to address the issue by processing either the input data or the recommendation results.
    \item In the first solution, a pre-processing approach is considered to mitigate the inherent bias (e.g.\ popularity bias) in the input data. This is done by proposing a rating transformation technique that compensates for the influence of the popular items (suppliers) in the learning process and provides chance for other items (i.e.\ non-popular items) to appear in the recommendation lists.
    \item In the second solution, a post-processing approach is considered that processes a longer recommendation list and generates the final/shorter recommendation list for each user. In this approach, the goal is to increase the exposure or visibility of under-represented items or suppliers in the final recommendation lists. 
    \item To show the effectiveness of the proposed solutions, a comprehensive set of experiments are performed on several datasets and the outputs are compared with state-of-the-art mitigation techniques (baselines). 
\end{itemize}

\section{Thesis outline}
\begin{itemize}
    \item I study the unfairness problem in recommender systems on the perspective of different sides in the system. For this purpose, I investigate the performance of existing recommendation algorithms on generating fair results for each side in the system. This investigation includes fairness of recommendations for users who receive recommendations and exposure fairness of items and suppliers in the recommendation lists (Chapter~\ref{chap:back}) \cite{mansoury2019bias,mansoury2020investigating,mansoury2019relationship,mansoury2020feedback}.
    \item I simulate the recommendation process over time by iteratively generating recommendations at each time point and getting users' feedback on delivered recommendations. In this simulation, I investigate the negative impacts of unfair recommendations on the system in a long-run (Chapter~\ref{chap:back}, Section~\ref{fairness_impact}) \cite{mansoury2020feedback}.
    \item I simulate a multi-sided matching and recommendation problem where different sides are involved in the system. I use an educational system as an example of a multi-sided environment and simulate the problem over that platform. The goal of this study is simply showing the importance of having multi-sided view when building a recommendation model. In this simulation, I optimize the system in different situations including optimization based on the utility of each side and optimization based on the utility of all sides (multi-sided view). The analysis of this simulation emphasizes the importance of multi-sided view for optimizing the system (Chapter~\ref{chap:simulation}).
    \item I particularly study the multi-sided exposure bias in recommender systems and its impact on different sides in the system. I perform sets of experiments using existing recommendation algorithms on different datasets and investigate the fairness and equality of exposure for items and suppliers (Chapter~\ref{chap:expo}) \cite{masoud2021wsdm,mansoury2021tois}. 
    \item To mitigate the multi-sided exposure bias in recommender system, I propose a pre-processing approach that transforms the rating data into percentile values. Through extensive experiments, I show that the proposed percentile transformation is able to improve the exposure fairness for both items and suppliers by compensating the high rating values of popular items in the input data. The experimental results show that using the percentile values as input for recommendation algorithms can significantly improve the accuracy of the recommendations compared to other input values and transformation techniques (Chapter~\ref{chap:solution1}) \cite{mansoury2021flatter}.
    \item As another solution for addressing the multi-sided exposure bias in recommender systems, I propose a graph-based approach, \textit{FairMatch} algorithm, to mitigate the exposure bias in recommendation lists. The proposed technique works as a post-processing recommendation and is able to mitigate exposure bias for items and suppliers with negligible loss in recommendation accuracy. 
    A comprehensive set of experiments on different datasets and comparison with various state-of-the-art baselines show the effectiveness of FairMatch algorithm on improving exposure fairness for items and suppliers in recommender systems (Chapter~\ref{chap:solution2}) \cite{mansoury2021tois,mansoury2020fairmatch}.
    \item I review the existing metrics for measuring the exposure bias and discuss the limitations of those metrics. I show that existing metrics cannot properly measure the exposure bias and hide important aspects of measuring exposure bias. I propose several metrics that can better measure the effectiveness of an algorithm in mitigating the exposure bias for items and suppliers (Chapter~\ref{chap:methodology}) \cite{mansoury2021tois}.
\end{itemize}

%% file: Chapters/02_Background.tex
\part{Understanding Multi-Sided Exposure Bias in Recommender Systems}
\chapter{Fairness in Recommender Systems}
\label{chap:back}
\tikzsetfigurename{back_}

In this chapter, I review the literature in recommender systems and introduce the related contributions. In Section~\ref{back_rs}, I review well-known recommendation approaches, evaluation methods, and various challenges introduced in the literature. One of these challenges is bias and unfairness in recommendation results which is the main focus of this dissertation. In Section~\ref{fairness}, I review the literature on fairness in recommender systems and in particular, I review the existing definitions for fairness, metrics for evaluating fairness of recommendation results, and techniques for improving fairness of recommendations. As part of the contributions in this dissertation, in Section~\ref{fairness_definition}, I introduce a metric for measuring the unfairness of recommender systems which is published in \cite{mansoury2019bias}. Also, in Section~\ref{fairness_factors}, I study different factors leading to unfairness in recommendation and show the relationship between each factor and unfairness in recommender systems. These contributions are published in \cite{mansoury2020investigating,mansoury2019relationship,himan2019c}. Moreover, in Section~\ref{fairness_impact}, I study the impact of unfair recommendations on each actor. In this investigation, I simulate the recommendation process over time and show the negative impacts of algorithmic bias in the system. These contributions are published in \cite{mansoury2020feedback}.    

\section{Recommender systems}\label{back_rs}

Development of e-commerce has led to behavioral changes in traditional businesses where users increasingly tend to buy products via the internet. However, the proliferation of information by the internet companies has caused information overload that leads to a decline in customer satisfaction. One way to deal with this problem is to create \textit{Recommender Systems}\cite{resnick1997recommender,lu2012recommender,melville2010recommender,jannach2010recommender} (RS) that extract information about products which are desired by each customer. 


Recommender systems use historical data on interactions between users and items to generate personalized recommendations for the users. These systems are used in a variety of different applications including movies, music, e-commerce, online dating, and many other areas where the number of options from which the user needs to choose can be overwhelming. Examples of these applications are the recommendation of
books in Amazon \cite{brynjolfsson2003,smith2017two,linden2003amazon}, photo groups in Flickr \cite{zheng2010photo}, videos in YouTube \cite{baluja2008video,davidson2010youtube,zhou2010impact} and results in the Web search \cite{jeh2003scaling,liu2004personalized,sun2005cubesvd}.


\subsection{Types of recommender systems}

There are various types of recommendation approaches including content-based, collaborative filtering, demographic, utility-based, knowledge-based, and hybrid recommender systems \cite{burke2002hybrid,burke2007hybrid}. Three main classes of these systems are content-based \cite{pazzani2007content}, collaborative filtering, and hybrid models.

\subsubsection{Content-based recommendation}

Content-based recommender systems \cite{pazzani2007content,balabanovic1997fab} use item content or item features (e.g.\ name, genre, location, description, etc) to extract users' preferences. Extracting users' preferences builds an interest profile for each user that shows the overall taste of the user toward different items. Based on the user's interactions with the items, the recommendation algorithm builds user's profile toward types of contents that user likes. For example, in a movie recommendation system, when a user shows interest toward Action movies, then the system builds the profile for that user accordingly as she is more interested in Action movies. In recommendation generation process, each user's profile is compared with the items' content and the items that match a user's interest would be recommended to that user.

There are several advantages and disadvantages with the content-based recommender systems \cite{lops2011content}. Regarding the advantages, first, content-based filtering works only based on the interaction made by the target user with different items/content and does not use the interactions of other users. Thus, it can work well when the data in the system is sparse. Second, this approach can work well when a new item is added to the system and even recommends those new items to the users. This is because of the assumption that items contents are always available and based on the profiles of users, if new item matches a user's profile, then it would be recommended. Third, the recommendations generated by content-based recommender systems are explainable and transparent. By explicitly listing the content features or descriptions of the recommended items, the user understands the reasons why those items are recommended which increases trust to the recommendation system.    

With respect to disadvantages, content-based filtering does not allow exploration and learning new preferences. This means that users will always receive recommendations based on their past interests and will not experience new contents. For example, if a user showed interest towards Drama movies, then the system only recommends Drama movies to her which raises two concerns: first, the recommendations delivered to this user is not good in terms of diversity (only one type of contents is recommended) and second, the user will not have a chance to see other contents and possibly shows interest towards those contents as well. Finally, another drawback of content-based filtering is the assumption that item content is always available. Although some platforms automatically collect item content, it may not be always available. 

\subsubsection{Collaborative filtering}

Collaborative Filtering (CF) \cite{Resnick:1994a,Koren:2008a} is a well-known recommendation technique that most of the researches have been conducted on it. CF works on the basis of rating behavior of similar users to the target user. There are two type of CF: 1) memory-based CF, 2) model-based CF.

Memory-based CF utilizes \textit{k-nearest neighbor (kNN)} algorithm \cite{fix1951discriminatory} for predicting the rating that target user will give to target item. There are two different approaches of memory-based CF: 1) \textit{user-based collaborative filtering (\algname{UserKNN})} \cite{Resnick:1994a}, and 2) \textit{item-based collaborative filtering (\algname{ItemKNN})} \cite{sarwar2001}. 

In \algname{UserKNN}, rating prediction for a target user is done using the opinions/ratings given by similar users to the target user on target item. On the other hand, in \algname{ItemKNN}, the ratings assigned to the similar items to what target user rated in the past is used to predict the rating of target item. The similarity between users (in \algname{UserKNN}) or items (in \algname{ItemKNN}) is calculated using \textit{Pearson Correlation Coefficient} or \textit{Cosine similarity}. Then, in \algname{UserKNN}, the rating that target user $u$ will give to target item $i$ is calculated as:

\begin{equation}\label{userknn}
    \hat{r}_{ui}=\Bar{r}_u + \frac{\sum_{v \in S_u}{(r_{vi}-\Bar{r}_v).Sim(u,v)}}{\sum_{v \in S_u}{Sim(u,v)}}
\end{equation}

\noindent where $\Bar{r}_u$ is the average rating given by user $u$ to different items, $Sim(u,v)$ is the similarity values between $u$ and $v$, and $S_u$ is the set of similar users to $u$ (i.e.\ $\{\forall v \in U, Sim(u,v)>0\}$). The same calculation process can be used for \algname{ItemKNN} where similar items to the target item are considered instead of $S_u$. Also, it is worth noting that the aforementioned process is used for rating prediction task. Deshpande and Karypis in \cite{mukund2004} adapted memory-based CF for ranking task where only similarity between the target user and other users are considered as the exact predicted rating value does not matter in ranking task, instead an ordered list of items based on predicted scores are used.

In model-based CF \cite{Koren:2008a,sarwar2000application}, a model is built based on interactions between users and items by learning the latent factors of users and items. For this purpose, optimization techniques are used to learn the latent factors of users and items by minimizing an objective function that fits the model to the observed user-item interactions. Given latent factor for user $u$ as $p_u$ and latent factor for item $i$ as $q_i$, the rating that $u$ will give to $i$ can be calculated by dot product between vectors $p_u$ and $q_i$ as:

\begin{equation}\label{modelbased}
    \hat{r}_{ui}=p_u^T.q_i
\end{equation}

The advantage of the CF approach is that it does not require any additional information about users and items (only uses rating/interaction data) and is powerful method to accurately predict the ratings. However, a disadvantage of the CF approach is that it needs some degree of density in interaction data to work well. The data sparsity problem is a well-known issue in CF methods \cite{grvcar2005data}. In \algname{UserKNN}, for instance, the algorithm needs a sufficient number of similar users (i.e.\ neighbors in kNN algorithm) to accurately predict the unknown ratings. In sparse datasets, neighbors for a target user are often rare as the target user may not have enough commonly rated items with other users.   
 
\subsubsection{Hybrid recommendation model}

Hybrid recommendation models \cite{burke2002hybrid,burke2007hybrid} combine two or more recommendation algorithms for generating recommendations to users. This is analogous to ensemble learning techniques \cite{opitz1999popular} in machine learning where several classifiers are combined to fulfill a prediction task. Hybrid recommendation models overcome the limitations of existing recommendation models by utilizing the benefits of both models.

Burke in \cite{burke2007hybrid} identified seven strategies for combining recommendation models and building hybrid recommendation models: weighted, switching, mixed, feature combination, feature augmentation, cascade, and meta-level. 

The first three strategies (i.e.\ weighted, switching, and mixed) use multiple recommendation models to separately generate the recommendations and finally, each strategy uses its own criterion to combine the recommendations generated by the models. The weighted strategy combines the output of each recommendation model using a linear weighting scheme. In this strategy, the scores predicted by each recommendation model on candidate items are considered and a linear combination of those scores determines which candidate items are high-quality to be recommended. Switching strategy examines the outputs of each recommendation model and chooses the one that has the highest confidence and reliability in generating recommendations. The idea behind the switching strategy is that recommendation models may not have consistent performance for all types of users. The mixed strategy simply merges the outputs of multiple recommendation models and show them to the users.

Feature combination and augmentation strategies simultaneously employ automatic feature engineering techniques and recommendation generation. Unlike the first three strategies (weighted, switching, and mixed) where the outputs from different recommendation models were combined, in these strategies, knowledge from different sources/models are involved in recommendation generation process. In the feature combination strategy, features derived from different recommendation models are combined and are passed to a single recommendation model to generate the final recommendations to the users, while in feature augmentation strategy, one recommendation model is used to extract the features and then those features are used as input to another recommendation model for recommendation generation.

Finally, in last two strategies (i.e.\ cascade and meta-level), the recommendation models are sequentially connected and the output from one model is used as input to the next model. In the cascade strategy, a priority level is assigned to each recommendation model and the model with the lowest priority is used as a tie-breaker for the output generated by the model with the highest priority. In meta-level strategy, the model built by one recommendation model is used as input for the next recommendation model.  

\subsubsection{Sequential recommendation}

Sequential recommendation models \cite{kang2018self,chen2018sequential,tang2018personalized} process the sequences of users' interaction data and produce an ordered list of recommendations. In these models, the order in which a user interacts with items over time plays an important role and the recommendation algorithm learns this order when modelling the user's preferences. The output of the sequential recommendation models are similar to those in traditional recommendation approaches, but in some scenarios, the order of the recommended items also matters. For example, in a music recommender system, the recommended songs can be delivered to the users in an ordered list in which users may click or ignore the recommended items. In this situation, either user plays a song or clicks on next button, user's interaction data would be recorded and used for later recommendation generation \cite{hariri2012context}.

In sequence-aware recommender system, users' preferences are usually defined as long-term and short-term preferences \cite{xiang2010temporal}. The former refers to the whole user's historical interaction data, while the latter refers to the user's recent interaction data or user's session data. Although users' short-term preferences are more relevant to their current interests, the long-terms preferences also play an important role in modelling users' general preferences. Therefore, there are a line of research that attempt to properly trade-off between these two types of preferences when building recommendation models and generating recommendations \cite{yu2019adaptive}.  

Sequence-aware recommender systems are used to address different problems \cite{quadrana2018sequence}. These systems can be used to adapt the recommendations according to the users' contextual information. Context adaptation refers to the fact that the relevance of the sets of the recommendable items, while considering the user's general preferences, depends on the situation and context that user resides. Examples of such contextual situations are users' geographical position or the time of the day. In each situation, the user may have specific preferences out of his/her general preferences. Sequence-aware recommender systems learn such patterns by modelling the sequence of users' interaction data and their related contexts \cite{liu2016context}.

Modelling repeated behavior or consumption is another research topic that sequence-aware recommender systems help to address. In some recommendation domains, providing repeated recommendations is beneficial for the users. Example of such scenario can be food recommendation in groceries. In these scenarios, items are consumable and the users need to repeatedly purchase those items. Sequence-aware recommender systems model users' repeated consumption behaviors and attempt to recommend previously interacted items at the right time \cite{wang2013opportunity}.

\subsection{Evaluation}

Evaluation of recommender systems answers these questions: how to assess the performance of a recommendation algorithm? If we have multiple recommendation algorithms, how to choose one? Evaluation of recommender systems consists of choosing appropriate metrics, performing a set of experiments to generate the recommendation lists, and measuring the specified metrics on recommendation lists. Three main types of evaluation techniques are online evaluation, user studies, and offline evaluation \cite{shani2011evaluating}.  

Online evaluation is performed on a real-world platform with a steady stream of data. In this evaluation, users' feedback\footnote{Depending on the type of recommendation algorithm, users' feedback can be ratings provided on items or clicking of the items.} on recommendation lists generated by a recommendation algorithm is collected and then the performance of recommendation algorithm is calculated using evaluation metrics. Based on the users' feedback and calculated performance, the system designer can refine the recommendation algorithm for improving the performance of recommendation system. The advantage of online evaluation is that users real interests and tastes can be captured over time and used to refine the recommendation algorithm accordingly. However, performing online evaluation requires access to a real-world platform, which is not always available. Even when access to a real-world platform is possible, there is a risk of degrading users' satisfaction by delivering low-quality recommendations to them. This is because experimentation and algorithm development are tested on a platform that users are interacting and it is possible that the algorithm does not work well. 

To overcome the aforementioned issues for online evaluation, user studies can be an alternative. In user studies, instead of performing experiments and evaluation on a live platform, a set of users are asked to interact with the system and provide their opinions about the items within the system. Then, based on the data collected from the subjects, the recommendation algorithm would be evaluated and refined. Although user studies helps to mimic the real-world evaluation without taking the risk of online evaluation, they have some limitations. User studies are expensive and complicated. They are expensive because they require a fair number of subjects to be participated in the study. Also, they require a specific design regarding the selection of subjects, generating recommendations to them, and collecting feedback. 

Finally, offline evaluation is the widely-used evaluation method in research on recommender systems. In this evaluation, previously collected data is used to perform experiments and evaluate the performance of recommendation algorithm. In this method, the collected data is divided into training and test sets. The recommendation model is built on training set and using this model, the recommendation lists for all users are generated. Then, by comparing the generated recommendation lists and users' true preferences in the test set, the performance of recommendation algorithm is evaluated. Due to the simplicity and possibility of performing this method, it is the commonly used method among researchers. A disadvantage of offline evaluation is that it may not necessarily reveal the real performance of a recommendation model as it may be observed on a real-world platform or online evaluation. Also, this method sometimes requires access to a dataset with specific attributes such as users' gender or items' genre information which is not always available.  

\subsubsection{Recommendation tasks and metrics}

Besides the taxonomy mentioned above, there are generally two different tasks in recommendations: rating prediction and ranking. The rating prediction task only seeks to predict the rating that a target user may assign to a target item (unseen item). On the other hand, the ranking task aims to generate a ranked list of items for a target user that he/she might like to receive. Each of these tasks has specific evaluation design and methodology which would be discussed in this Section.

\textbf{Metrics for rating prediction tasks.}
The well-known metrics for evaluating the performance of recommender systems in rating prediction tasks are \textit{Mean Absolute Error (MAE)} \cite{hyndman2006another} and \textit{Root Mean Square Error (RMSE)} \cite{willmott2005advantages}. 

Given $r_{ui}$ as the rating given by user $u$ to item $i$ and $\hat{r}_{ui}$ as the predicted rating by recommendation model that user $u$ will give to item $i$, MAE computes the average deviation of the predicted rating with the true rating for all user-item pairs in test set as follows:

\begin{equation}\label{mae}
    MAE=\frac{\sum_{(u,i)\in R_{test}}{|r_{ui}-\hat{r}_{ui}|}}{|R_{test}|}
\end{equation}

\noindent where $R_{test}$ is the test set and $|R_{test}|$ is the size of test set. On the other hand, RMSE is computed as follows:

\begin{equation}\label{rmse}
    RMSE=\sqrt{\frac{\sum_{(u,i)\in R_{test}}{(r_{ui}-\hat{r}_{ui})^2}}{|R_{test}|}}
\end{equation}

In contrast to MAE, the square of deviation between true and predicted ratings in RMSE emphasises on larger errors and if there is large deviation between true rating and predicted rating, it will result in higher error value.

There are some other variations of error estimation derived from MAE and RMSE. For example, in \cite{Massa:2007a,massa2004trust}, \textit{User Mean Absolute Error (UMAE)} has been proposed that computes the MAE separately for each user and then takes the average over the MAE values for all users. The purpose of this metric is to account for user-level errors. Assume that, in the test set, there are 100 users with small profile (e.g.\ one rating) and one user with large profile (e.g.\ hundred ratings). If in this situation, recommendation model predicts the ratings for users who have smaller profile with high error and predicts the ratings for users who have larger profile with lower error, then MAE measures the performance of this model as good enough. But, the fact is that in this situation, there are 100 unhappy users and the system needs to capture this. On the other hand, UMAE properly captures this situation and computes the performance of the model separately for each user. 

\textbf{Metrics for ranking tasks.} In ranking tasks, instead of predicting the rating value for a target item by a target user, the goal is to predict a list of items as recommendations for a target user with which she might like to interact \cite{Cremonesi:2010a}. To measure how good the generated list for a target user is, \textit{accuracy} or \textit{ranking quality} metrics are considered. Accuracy metrics measure how well the recommendation list is matched with the user's profile in the test set, while ranking quality metrics measure how well the recommended items are ordered in the list according to the true preferences in the test set.

The well-established accuracy metrics in machine learning that are also used in recommender systems are \textit{precision} and \textit{recall}. Before presenting the equations for precision and recall, I introduce some notations. Let $L_u$ be the recommendation list generated for user $u$ and $P_u^{test}$ be the $u$'s profile in test set. For each item $i$ in $L_u$, if $i$ exists in $P_u^{test}$, then we have a \textit{hit}, otherwise we have a \textit{miss}.
 
Precision computes the average ratio of hit in recommendation lists generated for all user and can be calculated as:

\begin{equation}\label{precision}
    precision=\frac{1}{|U|}.\sum_{u \in U}{\frac{\sum_{i \in L_u}{\mathds{1}(i \in P_u^{test})}}{|L_u|}}
\end{equation}

\noindent where $\mathds{1}(.)$ is the indicator function returning zero when its argument is False and 1 otherwise. $U$ is the whole users in the system. A disadvantage of precision is that it does not take into account the number of relevant items in user's profile in test set. For example, the precision value of one hit in recommendation list of size 10 for a user with 100 item in her profile in test set is as same as that for a user with 1 item in his profile in test set (both 0.1). Thus, it does not consider the number of items in user's profile in test set.  On the other hand, recall computes the ratio of the user's profile in test set that are appeared in the user's recommendation list and can be calculated as:

\begin{equation}\label{precision}
    recall=\frac{1}{|U|}.\sum_{u \in U}{\frac{\sum_{i \in L_u}{\mathds{1}(i \in P_u^{test})}}{|P_u^{test}|}}
\end{equation}

Although recall overcome the issue mentioned for precision, it does not properly measure the quality of recommendations in some situations. One may achieve higher recall value by increasing the size of the recommendation lists. In extreme case, we may get a perfect recall by recommending all unseen items to a user (though very low precision). Therefore, a better way for measuring the quality of the recommendations is considering both metrics. 

Precision and recall do not consider the rank of the items in the recommendation lists. These metrics do not distinguish whether the relevant items are placed on the bottom of the list or on top of the list. To address this issue, a ranking-based metric is needed. For measuring the ranking quality of the recommendation lists, \textit{Normalised
Discount Cumulative Gain (nDCG)} is a well-known metric in information retrieval for measuring the quality of search results. nDCG can be calculated as:

\begin{equation}\label{ndcg}
    nDCG=\frac{1}{|U|}.\sum_{u \in U}{\frac{1}{iDCG_u}\sum_{i \in L_u}\frac{2^{\mathds{1}(i \in P_u^{test})}-1}{\log (K(i)+1) }}
\end{equation}

\noindent where $K(i)$ returns the position of item $i$ in the list and $uDCG_u$ is the normalisation factor and can be calculated as:

\begin{equation}\label{idcg}
    iDCG_u=\sum_{j=1}^{min(|L_u|,|P_u^{test}|)}{\frac{1}{\log (j+1)}}
\end{equation}
 
\subsection{Challenges}

For a long time, the main challenge and research question in recommender system was increasing the accuracy of the recommendations to the users. Various studies have been conducted and different recommendation algorithms have been developed to further improve the accuracy of recommendations \cite{dubois2009improving,gedikli2013improving,ma2014improving,zhao2017improving,lee2016improving}. However, other research questions have emerged. 

One of these challenges was \textit{cold-start users} problem \cite{mansoury2016improving,camacho2018social,gogna2015comprehensive,safoury2013exploiting,lika2014facing,lam2008addressing}. In this problem, the question is: how to accurately generate the recommendation list or predict the rating of an item for a new user who recently joined the system or there is not sufficient information about her in the system? The problem with these users is that the system does not have enough information about them and cannot properly learn their preferences. The same issue can be considered for items as well. Thus, the question is: how to accurately recommend a new item to the users or predict the rating that a user might assign to a new item? Addressing cold-start users and items in recommender systems are important considerations as new users/items often appear in the system in the real-world. This can also be related to the problem of \textit{data sparsity} \cite{natarajan2020resolving,sarwar2001sparsity} which there are many cold-start users or items in the input data. In this situation, the number of users and items is large, but the number of the ratings is low which means the user-item matrix has too many empty cells.

Scalability of recommender system is also a challenge \cite{sarwat2013lars,ghazanfar2010scalable}. In real-world platforms, there are many users and items in the system which makes the recommendation generation process difficult. First, the system has to generate the recommendations for a large number of users. Second, the system has to manage the large pool of candidate items for generating the recommendation list for each user. Handling this issue requires designing an efficient recommendation algorithm.  

Another challenge in recommender system is diversity and novelty of the recommended items for the users. Diversity refers to the fact that recommended items should be from different sets of categories \cite{kaminskas2016,castells2015novelty,vargas2014}. In movie recommendation, for example, recommending movies that are from different genres is one way to achieve diversified recommendation. Novelty, on the other hand, refers to recommending novel items to a user such that it makes her surprised when seeing those recommended items \cite{gravino2019towards,zhang2013definition,mendoza2020evaluating}. In this regard, improving diversity and novelty always brings the cost of losing accuracy. Therefore, the trade-off between diversity/novelty and accuracy is another research problem in this field \cite{javari2015probabilistic,isufi58accuracy}.

Addressing algorithmic bias is another area of challenge in recommender systems. Algorithmic bias refers to the fact that recommendation algorithms tend not to treat different users and items in the system equally. From the users' perspective, the recommendation algorithm may not deliver the recommendations with the same level of quality to every user in the system \cite{ekstrand2018,mansoury2020investigating}. This means that some groups of users may represent the majority in the system and dominates the preferences of the minority groups in the system. In this situation, the recommendation algorithm only learns the preference of the majorities and the delivered recommendations better match the preferences of the majorities than the minorities. Another types of bias can happen on the item side where some items may frequently appear in the interaction data, while some other items may rarely be interacted by the users. This is known as popularity bias \cite{powell2017love,abdollahpouri2020popularity}. This bias leads to unfair exposure of items in the recommendation lists due to the algorithmic bias as recommendation algorithms frequently recommend popular items, while rarely recommend non-popular ones \cite{abdollahpouri2020multi}. Considering the recommender system as a multi-sided or multi-stakeholder environment \cite{himan2019a,himan2019a}, popularity bias can also negatively affect suppliers (i.e.\ content providers) of the system \cite{abdollahpouri2020unfair,mehrotra2018towards}. 

The main focus of this dissertation is on mitigating algorithmic bias in recommender systems, in particular when the system is operating in a multi-sided platform. This topic will be further discussed in Section~\ref{fairness}.

\section{Fairness in predictive modelling} \label{fairness}

Research on fairness in decision making and machine learning can be traced back to 2008-2010 \cite{pedreshi2008discrimination,calders2009building,calders2010three}. Recently, the topic of fairness has been extended to recommender systems and received extensive attention from researchers \cite{beutel2019fairness,xiao2017fairness,burke2017b,ekstrand2019fairness}. This Section focuses on various aspects of fairness-aware recommender systems including the definitions of fairness, evaluation metrics, factors leading to unfair recommendations, impact of unfair recommendations, and existing techniques for tackling unfairness in recommender systems. 

Fairness in Machine Learning is mainly concerned with the fair treatment of individuals based on human-aspect criteria and attempts to treat all individuals equally regardless of their sensitive attributes (e.g.\ gender, ethnicity, sexual orientation, disability, etc.). Given a dataset $D=<A,X,Y>$ where $A$ is the whole attributes of $D$, $X\subseteq A$ is the sensitive attributes, and $Y$ is the label of each instance in $D$ as the ground truth, the goal of predictive models is to predict the target variable in a way that it does not use any information about $X$ in $D$ \cite{gajane2017formalizing}. For example, in the recidivism domain, if certain race group showed higher risk in reoffending, then this should not cause a predictive model to assign a higher risk to an individual belonging to that race group. In fact, the predictive model should use other information about the individuals for making decision, not the sensitive attributes. In this situation, the predictive model either should not use the sensitive attributes as input data, or should utilize bias mitigation techniques to not make a biased decisions.

On the other hand, fairness in recommender systems is mainly concerned with delivering accurate and high-quality recommendations to all users such that the recommended items are matched with the users' preferences. There is also another concern in recommender systems with respects to the suppliers in the systems. In this regard, the system attempts to provide fair exposure to all suppliers at least at the same level of their merit in the input data (i.e.\ representation in input data such as popularity of the item). Note that in majority of recommendation algorithms, no sensitive attributes are used as input for building the recommendation model, but the fairness evaluation using the sensitive attributes is done on the recommendation outputs.

\subsection{Fairness definitions}\label{fairness_definition}

Various definitions have been proposed for defining fairness in Machine Learning. Examples include Error Parity \cite{buolamwini2018gender}, False Discovery or Omission rates \cite{kleinberg2016inherent}, Envy-freeness \cite{kyropoulou2020almost}, Demographic Parity \cite{dwork2012}, Equality of False Positive or False Negative rates \cite{hardt2016equality}.

Equality of False Positive or False Negative rates requires the percentage of users falsely predicted to be positive or negative to be the same across true negative or positive individuals belonging to each group. Envy-freeness requires that each individual prefer his allocation to anyone else’s allocation. Demographic parity aims at equalizing the percentage of users who are predicted to be positive across different groups. False discovery or Omission rates aims at equalizing the percentage of false positive or negative predictions among individuals predicted to be positive or negative in each group. 

Some of the definitions mentioned above are adapted in recommender systems to measure the fairness of recommendation results. In the following, I review the ones that are more relevant to the topic in this dissertation.

Fairness-aware recommender systems aims to provide fair treatment to each entity in the system. Depending on the domain that the system is operating and the goals defined by the system owner/designer, fair treatment can have certain meaning and definition. Since recommender systems are operating in a multi-sided platforms \cite{abdollahpouri2020addressing,zheng2017multi,burke2016,zheng2018utility,himan2019a,himan2019a} where different actors are involved in the system, addressing the fairness for each actor may require specific definition. In this Section, I review the fairness definitions provided for each actor in the literature.

\subsubsection{User-side fairness}

\textbf{Definition 1.} On the user side, the main fairness goal is delivering recommendations with the same level of quality to each user or user group based on their interest \cite{ekstrand2018,mansoury2019relationship,mansoury2020investigating}. This definition concerns about the accuracy of recommendations for individual users or users' groups. For example, if user \textit{A} receives recommendations that are 60\% matched with her preferences, while the recommendations for user \textit{B} only matches 10\% of his interests, then the recommendation algorithm would be called unfair against user \textit{B}. Thus, the recommendation algorithm needs to be refined to better learn the interests of all users and generates recommendations that are matched with the interests of each user.

\noindent \textbf{Definition 2.} User-side fairness can also be defined with respect to user groups' interest toward item categories (e.g.\ movie genres). In other words, the fairness is defined as the degree to which a group’s preferences on various item categories is reflected in the recommendations they receive \cite{virginia2018,mansoury2019bias,steck2018calibrated,abdollahpouri2020connection,abdollahpouri2019impact}. In some cases, biases in the original data may be amplified or reversed by the underlying recommendation algorithm. This happens when the preferences of one user group is dominant in the input data and a biased recommendation algorithm only learns these preferences, which causes the generated recommendations fail to represent the preferences of other user groups. In a movie recommendation system, for example, if the profile of one user's group consists of 40\% Action movies and 60\% Comedy movies, then the recommendation sets for this group should also consist of the same ratio from Action and Comedy movies. Thus, this definition concerns the ability of a recommendation algorithm to properly capture user groups preferences toward item categories.

\noindent \textbf{Definition 3.} User-side fairness can be defined as the interests of users toward popular or non-popular items. Some users may have niche tastes, which make them more interested in non-popular items, while some other users may have blockbuster tastes which make them more interested in popular items. Due to the nature of recommendation algorithms that are more biased toward popular items \cite{ciampaglia2018algorithmic,harald2011,powell2017love,abdollahpouri2017controlling}, the generated recommendations  usually fail to represent the preferences of niche-taste users \cite{himan2019c,kowald2020unfairness}. Therefore, it is important that a recommendation algorithm properly takes into account the interest of users toward popularity of items when generating recommendations.  

\subsubsection{Item-side and supplier-side fairness}

The fairness definitions for item and supplier sides concentrate on fair exposure for items and suppliers. This means that all items and suppliers have equal chance to appear in the recommendation lists. To achieve fair exposure for items and suppliers, items and suppliers should be recommended to the equal number of users as much as possible, which leads to a uniform distribution for recommended items in the recommendation lists.


\subsection{Metrics for evaluating fairness in recommendations}

Based on the first definition for user-side fairness in Section~\ref{fairness_definition}, various metrics are introduced to measure the disparity and difference in the quality of recommendation delivered to different groups of users. Given $protected$ and $unprotected$ as the group of users who belong to the minority and majority groups, respectively, based on a sensitive attribute (e.g.\ gender, race, age, ethnicity, etc) and $hit$ as the number of times that the recommended items to a user matched the items in her profile in test set, \textit{Statistical parity difference} \cite{dwork2012} measures the deviations from statistical parity as follows:  
    \begin{equation}\label{spd}
        SPD=Pr(hit|unprotected)-Pr(hit|protected)
    \end{equation}
\noindent where $Pr(hit|unprotected)$ is the probability of correct recommendations (based on users' preferences) for $unprotected$ group. 
$hit$ can be interpreted as the precision of recommendation results. Other accuracy metrics such as recall, f, nDCG, and calibration can also be considered. Lower $SPD$ means that the quality of recommendation for both groups are close and signifies a fairer recommendation system. Another similar metric is \textit{disparate impact} \cite{chouldechova2017fair} which replaces the difference in this Equation with a ratio. \textit{Equal opportunity difference} is a relaxed version of equality of opportunity \cite{hardt2016equality,zafar2017fairness}, which returns the difference in recall scores (True Positive Rate, TPR) between the unprotected and protected groups. A value of 0 indicates equality of opportunity.

With respect to the second definition provided in Section~\ref{fairness_definition}, the degree to which a group’s preferences on various item categories is reflected in the recommendations they receive, various metric have been developed. \textit{Bias disparity} \cite{virginia2018} measures how much an individual's recommendation list deviates from his or her original preferences in the training set. Given a group of users, $G$, and an item category, $C$, bias disparity is defined as follow:

\begin{equation} \label{eq:bd}
BD(G,C)=\frac{B_R(G,C)-B_T(G,C)}{B_T(G,C)}
\end{equation}
where $B_T$ ($B_R$) is the \textit{bias} value of group $G$ on category $C$ in training data (recommendation list). $B_T$ is defined by:

\begin{equation} \label{eq:b}
B_T(G,C)=\frac{PR_T(G,C)}{P(C)}
\end{equation}

where $P(C)$ is the fraction of item category $C$ in the dataset defined as $|C|/|m|$, $m$ is the size of the dataset. $PR_T$ is the preference ratio of group $G$ on category $C$ calculated as:

\begin{equation} \label{eq:pr}
PR_T(G,C)=\frac{\sum_{u \in G} \sum_{i \in C} T(u,i)}{\sum_{u \in G} \sum_{i \in I} T(u,i)}
\end{equation}

where $T$ is the binarized user-item matrix. If user $u$ has rated item $i$, then $T(u,i)=1$, otherwise $T(u,i)=0$. The \textit{bias} value of group $G$ on category $C$ in the recommendation list, $B_R$, is defined similarly. 

Bias disparity separately measures the deviation of each group's interests toward item categories from the represented interest in recommendation sets and does not give an overall view about the fairness of the system. To overcome this issue, I introduced \textit{average disparity} \cite{mansoury2019bias} that measures how much preference disparity between the training data and the recommendation lists for one group of users (e.g.\ unprotected groups) is different from that for another group of users (e.g.\ protected group). Inspired by \textit{value unfairness} metric proposed by Yao and Huang \cite{yao2017}, I introduce the average disparity as:

\begin{equation} \label{eq:fairness}
\begin{aligned}
\overline{disparity}=\frac{1}{|C|} \sum_{i=0}^{|C|}|(N_R(G_U,C_i)-N_T(G_U,C_i)) \\
-(N_R(G_P,C_i)-N_T(G_P,C_i))|
\end{aligned}
\end{equation}

\noindent where $G_U$ and $G_P$ are unprotected and protected groups, respectively. $N_R(G,C)$ and $N_T(G,C)$ return number of items from category $C$ in recommendation lists and training data, respectively, that are rated by users in group $G$. 

\begin{figure*}[t]
    \centering
    \includegraphics[width=0.99\textwidth]{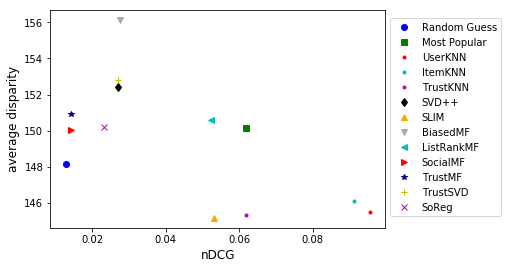}
    \caption{Comparison of recommendation algorithms by ranking quality and average disparity.}\label{fig:avg_disparity}
\end{figure*}

Figure~\ref{fig:avg_disparity} compares the performance of thirteen recommendation algorithms with respect to how accurately recommendation algorithms generate recommendations with low disparity for unprotected and protected groups on Yelp dataset\footnote{\url{https://github.com/masoudmansoury/yelp_core40}} \cite{mansoury2019bias}. Gender is used to define protected and unprotected groups where male users represent unprotected group and female users represent protected group. In this figure, horizontal axis is the ranking quality of recommendations (nDCG) and vertical axis is average disparity calculated by Equation~\ref{eq:fairness}. Details about the recommendation algorithms, dataset, and experimental results are discussed in Chapter~\ref{chap:methodology}.

The results in Figure~\ref{fig:avg_disparity} show that neighborhood models generate recommendations with the highest ranking quality and lowest average disparity. Also, the results show that side information like trust information can even generate better results in terms of average disparity compared to other recommendation algorithms.

Analogous to the average disparity, the calibration of the recommendation list(s) for an individual user or groups of users can be used to measure the degree to which a group’s preferences on various item categories is reflected in the recommendations they receive. Given the distribution of item categories in user's profile and recommendation list for that user as $p$ and $q$, respectively, miscalibration \cite{steck2018calibrated} can be calculated as the distance between $p$ and $q$. For this purpose, Kullback-Leibler divergence (KLD) \cite{kullback1997information} can be used to calculate the distance between $p$ and $q$ as follows:

\begin{equation} \label{eq:kld}
\begin{aligned}
KLD(p|\widetilde{q})=\sum_{c \in C}{p_{c}  log\frac{p_{c}}{\widetilde{q}_{c}}}
\end{aligned}
\end{equation}

\noindent where $C$ is item categories (e.g.\ genres in movie recommendations) and $\widetilde{q}$ is approximately similar to $q$ calculated as:

\begin{equation} \label{eq:q}
\begin{aligned}
\widetilde{q}_{c}=(1-\alpha).q_{c} + \alpha.p_{c}
\end{aligned}
\end{equation}

The purpose of $\widetilde{q}$ is to overcome the issue of zero values for some categories in $q$. Small value for $\alpha>0$ guarantees $\widetilde{q} \approx q$. 

With respect to the third fairness definition provided in Section~\ref{fairness_definition}, in \cite{abdollahpouri2020multi}, we introduced \textit{User Propensity Deviation (UPD)} metric that calculates the deviation between the ratio of item popularity groups in user profile and her recommendations. Given popular items as \textit{head}, non-popular items as \textit{tail}, and the rest of the items as \textit{mid}, this metric measures the distance between the ratio of each of these item groups in users' profile and their recommendation lists. Lower distance indicates that the recommendation list is generated based on user's interest toward (un)popular items. For example, when a user profile consists of 20\% head items, 30\% mid items, and 50\% tail items, then it is expected to see the same ratio in recommendation list for that user. Similar to calibration metric mentioned above, given $p$ and $q$ as the distribution of item popularity groups (head, mid, and tail) in user profile and recommendation lists, respectively, UPD can be calculated as the KLD between $p$ and $q$. 

Finally, with respect to the item and supplier side fairness, various metrics have been developed to measure the exposure of items and suppliers in recommendation lists. \textit{Aggregate diversity} \cite{adomavicius2011improving,javari2015probabilistic,shi2013trading,adomavicius2011maximizing,isufi58accuracy,eskandanianUMAP2020} is a widely used metric for measuring the coverage of items in recommendation lists. It measures the fraction of items in catalog that appear at least once in recommendation lists. The limitation of aggregate diversity is that it does not take into account the whether or not the recommended items are popular or non-popular. A recommendation system may achieve high aggregate diversity by recommending many popular items which would not be considered as fair as it does not give opportunity to non-popular items to be seen by the users. Another issue with aggregate diversity is that it does not take into account the frequency of recommended items. For example, suppose a recommendation system recommends popular items frequently (recommending those items to many users), but recommends non-popular items to few users (e.g.\ only once). Although this system achieves high aggregate diversity, it is not fair as it does not give enough exposure to all items.

To address the first issue, \textit{long-tail coverage} metric is introduced \cite{park2008long,yin2012challenging,li2017}. Long-tail coverage measures the fraction of long-tail items that appear in the recommendation lists. Long-tail items are the non-popular items that received less attention (i.e.\ few interactions) from users. Also, to overcome the second issue, \textit{Gini Index} and \textit{Entropy} are used to measure the fairness of distribution of recommended items \cite{vargas2014improving,mansoury2020fairmatch}. Both of these metrics measure the uniformity of distribution of recommended items, with uniform distribution indicating a fair recommendation as it provides equal exposure to all items. Gini Index is the measure of fair distribution of recommended items. It takes into account how uniformly items appear in recommendation lists. Uniform distribution will have Gini index equal to zero which signifies equal exposure for the items or suppliers in recommendation lists (lower Gini index is better). Given all the recommendation lists for users, $L$, and $p(i_{k}|L)$ as the probability of the $k$-th least recommended item being drawn from $L$ calculated as \cite{vargas2014improving}:
    
\begin{equation}
    p(i|L)=\frac{\sum_{u \in U}^{} \mathds{1}_{i \in L_{u}}}{\sum_{u \in U}^{} \sum_{j \in I}^{} \mathds{1}_{j \in L_{u}}} 
\end{equation}
    
\noindent where $L_{u}$ is the recommendation list for user $u$. Now, Gini index of $L$ can be computed as:
    
\begin{equation}
    Gini(L)=\frac{1}{|I|-1} \sum_{k=1}^{|I|} (2k-|I|-1)p(i_{k}|L) 
\end{equation}

Also, given the distribution of recommended items, entropy measures the uniformity of that distribution. Uniform distribution has the highest entropy or information gain, thus higher entropy is more desired when the goal is increasing diversity. 
    
\begin{equation}
    Entropy(L)=- \sum_{i \in I}^{} p(i|L) \log p(i|L) 
\end{equation}
    
\noindent where $p(i|L)$ is the observed probability value of item $i$ in recommendation lists $L$.

In terms of supplier fairness, I adapted the aforementioned metrics for measuring the fairness of suppliers in recommendation results. Those metrics will be discussed in Chapter~\ref{chap:methodology}.

\subsection{Factors leading to unfair recommendations}\label{fairness_factors}

There are various factors that may affect the fairness of recommendations. One of the contributions of this dissertation is investigation of factors leading to unfairness in recommender systems. These contributions are published in \cite{mansoury2020investigating,mansoury2019relationship}. In my investigation, I explored the relationship between several characteristics of users' profile with the quality of recommendations delivered to them. For this purpose, I used the definition provided in Equation~\ref{spd} with precision and (mis)calibration of recommendations as the measures of quality of recommendations. The factors that I investigated are as follows:

\begin{itemize}
    \item \textbf{Profile anomaly ($\mathcal{A}$):} One factor that could impact recommendation performance is the degree of anomalous rating behavior relative to other users. The authors in this paper showed that users whose rating behavior is more consistent with other users in the system as a whole receive better recommendations than those who have more anomalous ratings. This happens because users who rate more in line with typical users are likely to find more matching items or users. We measure the degree of profile anomaly based on how similarly a user rates items compared to the majority of other users who have rated that item. Since collaborative filtering approaches use opinions of other users (e.g.\ similar users) for generating recommendations for a target user, it is highly possible that users with anomalous ratings receive less accurate recommendations. Given a target user, $u$, and $I_{u}$ as all items rated by $u$, profile anomaly of $u$ can be calculated as:
    \begin{equation} \label{eq:consistency}
        \begin{aligned}
            \mathcal{A}_{u}=\frac{\sum_{i \in I_{u}}|r_{u,i}-\overline{r_{i}}|}{N_{u}}
        \end{aligned}
    \end{equation}
    \noindent where $r_{u,i}$ is the rating given by $u$ to item $i$, $\overline{r_{i}}$ is  the average rating assigned to item $i$, and $N_{u}$ is the number of items rated by $u$ (i.e.\ the profile size of $u$).
    \item \textbf{Profile entropy ($\mathcal{E}$):}
    Another possible factor that could impact recommendation performance is how informative a user's profile is. The more diverse a user's ratings are, the higher their entropy is. For example, has the user only given high (or low) ratings to different items? Or are there a wide range of different ratings given by the user? We measure the entropy of user $u$'s profile as follows:
    \begin{equation}
        \mathcal{E}_{u}=-\sum_{v \in V}D_{u}(v)\log D_{u}(v)
    \end{equation}
    \noindent where $V$ is the set of discrete rating values (for example, 1,2,3,4,5) and $D_{u}$ is the observed probability distribution over those values in $u$'s profile. 
    \item \textbf{Profile size ($\mathcal{S}$):} The last factor I investigate in this paper is the profile size of each user. I believe users who are more active in the system (and have rated a larger number of items) receive better recommendations compared to those with shorter profiles. 
\end{itemize}

For this investigation, I performed a set of experiments on MovieLens1M dataset \cite{Harper:2016} using four recommendation algorithms: user-based collaborative filtering (\algname{UserKNN}) \cite{Resnick:1994a}, item-based collaborative filtering (\algname{ItemKNN}) \cite{sarwar2001}, singular value decomposition (\algname{SVD++}) \cite{Koren:2008a}, and list-wise matrix factorization (\algname{ListRankMF}) \cite{shi2010}. The MovieLens1M dataset has 6,040 users provided around 1M ratings (4,331 males provided 753,769 ratings and 1,709 females provided 246,440 ratings) on 3,706 movies. The ratings are in the range of 1-5 and the density of the dataset is 4.468\%. Also, each movie is assigned either a single genre or a combination of several genres. Overall, there are 18 unique genres in this dataset. Details of the dataset and algorithms are explained in Chapter~\ref{chap:methodology}. 

\input{./Tables/2.2.3.factors1}

Table~\ref{tab:stat} shows the specification of MovieLens1M dataset for male and female users. As shown in this table, there are more male users in the dataset than female users. Moreover, on average, male users have larger profiles, and their profile entropy is also higher than female users. In addition, the average anomaly of male users' profiles is slightly lower than female users. 

I divided the dataset into training and test sets in an 80\% - 20\% ratio, respectively. The training set is then used to build the model. After training different recommendation algorithms, recommendation lists of size 10 are generated for each user in the test set.

I created 20 user groups separately for males and females by measuring different factors: degree of anomaly, entropy, and profile size. Specifically, I sort users based on each factor and then split them into 20 buckets in an ascending order. Users that fall within each bucket represent one group. In order to calculate the anomaly, entropy, profile size, precision, and miscalibration for each group, I average the corresponding measure over all the users in the group. All recommendation models are optimized using gridsearch over hyperparameters and the configuration with the highest precision is selected. The precision values for \algname{UserKNN}, \algname{ItemKNN}, \algname{SVD++}, and \algname{ListRankMF} are 0.214, 0.223, 0.122, and 0.148, respectively.

\input{./Tables/2.2.3.factors2}

Table~\ref{tab:accuracy} shows the performance of recommendation algorithms for male and female users. In terms of precision, male users consistently receive more accurate recommendations than females and in terms of miscalibration, except for \algname{SVD++}, male users receive less miscalibrated (i.e.\ more calibrated) recommendations than females. Lower miscalibration for females than males on \algname{SVD++} shows an interesting result in the experiments that needs further investigation.

\begin{figure*}[htp]
  \centering
  \begin{subfigure}[b]{1\textwidth}
        \includegraphics[width=0.32\textwidth]{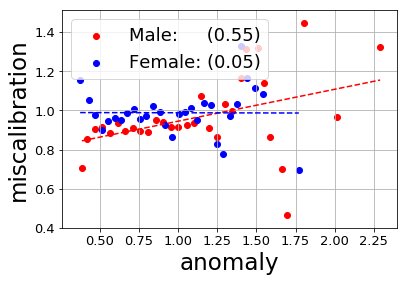}
        \includegraphics[width=0.32\textwidth]{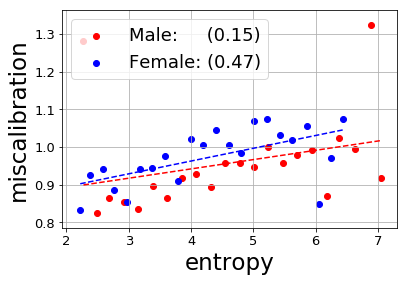}
        \includegraphics[width=0.32\textwidth]{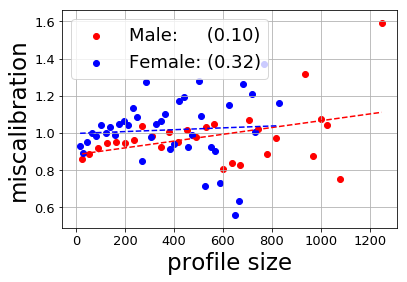}
        \caption{UserKNN}
  \end{subfigure}
  \begin{subfigure}[b]{1\textwidth}
        \includegraphics[width=0.32\textwidth]{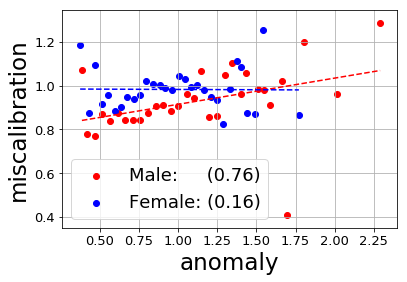}
        \includegraphics[width=0.32\textwidth]{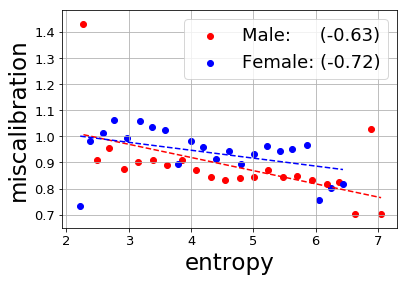}
        \includegraphics[width=0.32\textwidth]{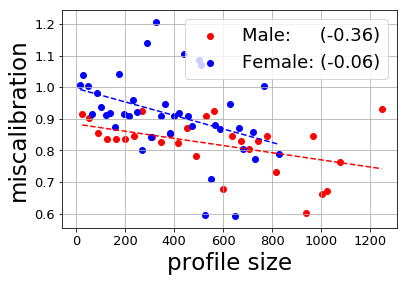}
        \caption{ItemKNN}
  \end{subfigure}
  \begin{subfigure}[b]{1\textwidth}
        \includegraphics[width=0.32\textwidth]{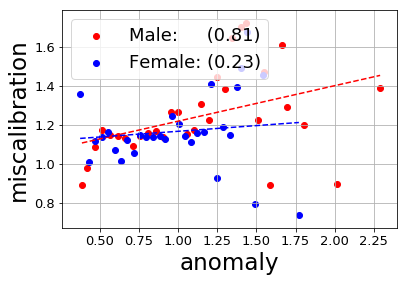}
        \includegraphics[width=0.32\textwidth]{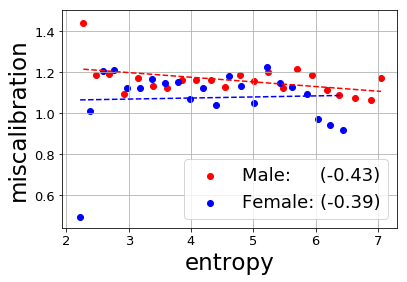}
        \includegraphics[width=0.32\textwidth]{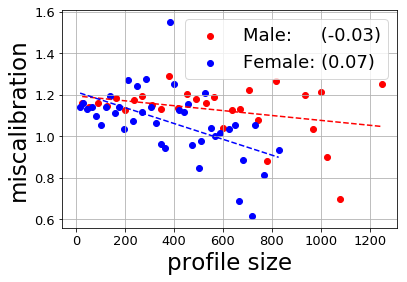}
        \caption{SVD++}
  \end{subfigure}
  \begin{subfigure}[b]{1\textwidth}
        \includegraphics[width=0.32\textwidth]{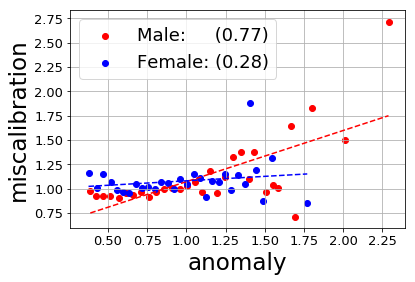}
        \includegraphics[width=0.32\textwidth]{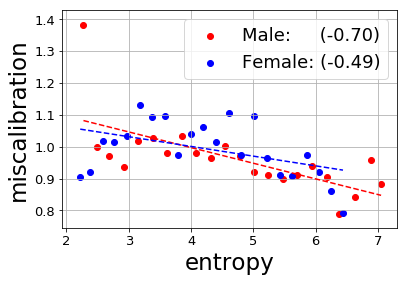}
        \includegraphics[width=0.32\textwidth]{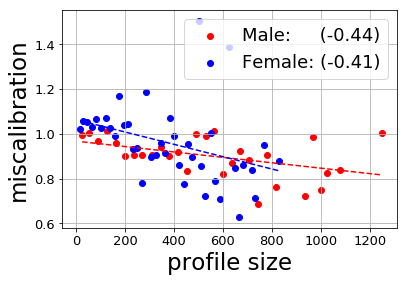}
        \caption{ListRankMF}
  \end{subfigure}
\caption{The Correlation between anomaly, entropy, and size of the users' profiles and  miscalibration of the recommendations generated for them. Numbers next to the legends in the plots show the correlation coefficient for each user group.} \label{fig:corr_cons}
\end{figure*}

Figure~\ref{fig:corr_cons} shows the relationship between the degree of anomaly, entropy, and profile size for 20 user groups for both male and female users and the miscalibration of the recommendations they received. As it can be seen in the first column (anomaly vs miscalibration), in all algorithms except for \algname{SVD++}, the recommendations given to the female users have higher miscalibration (they are less calibrated) regardless of the anomaly of their ratings compared to the male user groups. Also, it can be seen that the positive correlation between profile anomaly and recommendation miscalibration discussed in \cite{mansoury2019relationship} can only be seen on male users. 

The second column of Figure~\ref{fig:corr_cons} shows the relationship between the entropy of the ratings and the miscalibration of their recommendations. Again, it can be seen that except for \algname{SVD++}, for all other algorithms, female user groups have higher miscalibration in their recommendations regardless of the amount of entropy of their ratings. 

Finally, the last column of Figure~\ref{fig:corr_cons} shows the correlation between the average profile size of different user groups and the miscalibration of their recommendations. Looking at this plot, we can see that there is no significant correlation between these two factors, indicating the profile size of the users does not affect the miscalibration of their recommendations. However, except for \algname{SVD++}, again all algorithms have higher miscalibration for female user groups regardless of their profile size. It seems that \algname{SVD++} is indeed the \textit{fairest} algorithm among the four as it gives a comparable performance for both male and female users. It can also be seen that there is no data point for female groups when the value of the $x$ axis is larger than 400, meaning the largest average profile size for female groups is 400 while there are some male user groups with an average profile size of around 700.

\begin{figure*}[htp]
  \centering
  \begin{subfigure}[b]{1\textwidth}
        \includegraphics[width=0.32\textwidth]{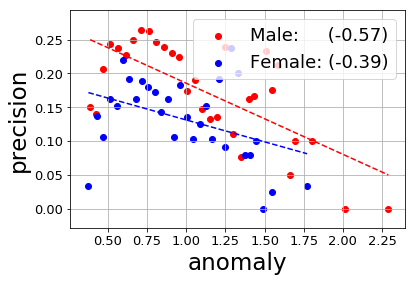}
        \includegraphics[width=0.32\textwidth]{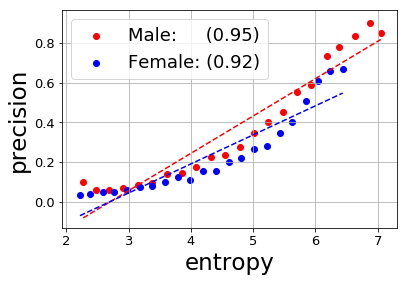}
        \includegraphics[width=0.32\textwidth]{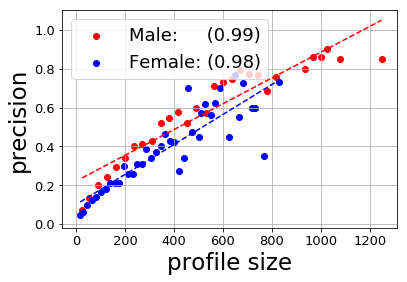}
        \caption{UserKNN}
  \end{subfigure}
  \begin{subfigure}[b]{1\textwidth}
        \includegraphics[width=0.32\textwidth]{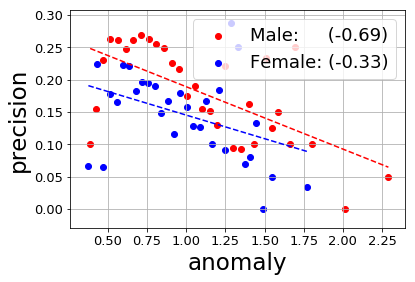}
        \includegraphics[width=0.32\textwidth]{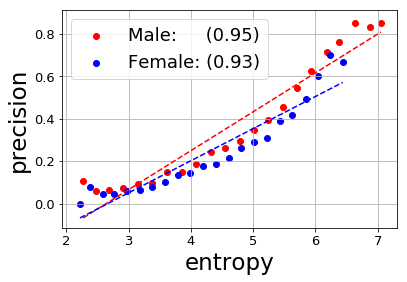}
        \includegraphics[width=0.32\textwidth]{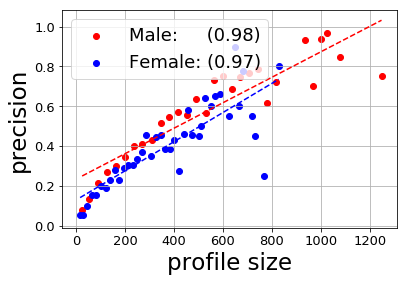}
        \caption{ItemKNN}
  \end{subfigure}
  \begin{subfigure}[b]{1\textwidth}
        \includegraphics[width=0.32\textwidth]{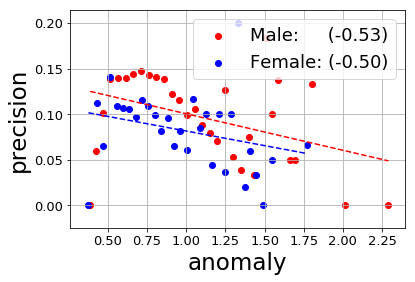}
        \includegraphics[width=0.32\textwidth]{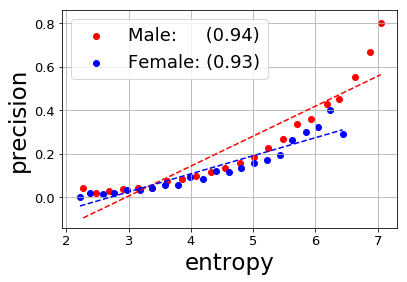}
        \includegraphics[width=0.32\textwidth]{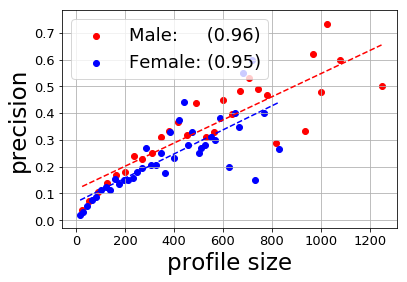}
        \caption{SVD++}
  \end{subfigure}
  \begin{subfigure}[b]{1\textwidth}
        \includegraphics[width=0.32\textwidth]{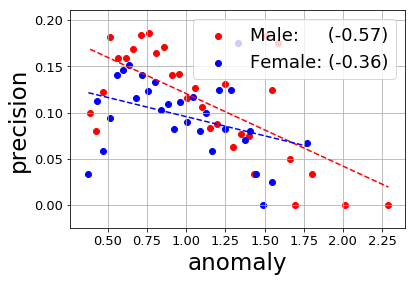}
        \includegraphics[width=0.32\textwidth]{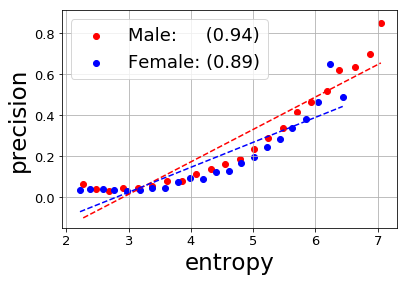}
        \includegraphics[width=0.32\textwidth]{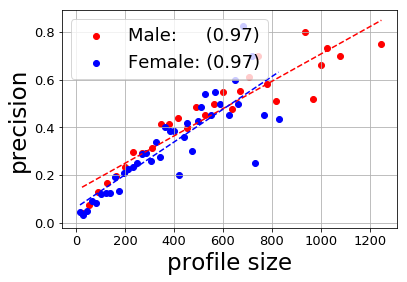}
        \caption{ListRankMF}
  \end{subfigure}
\caption{The Correlation between anomaly, entropy, and size of the users' profiles and precision of the recommendations generated for them. Numbers next to the legends in the plots show the correlation coefficient for each user group.} \label{fig:corr_precs}
\end{figure*}

Figure~\ref{fig:corr_precs} shows the correlation between the aforementioned factors for different user groups and the precision of their recommendations. Unlike miscalibration, it seems the correlations of these three factors with precision are much stronger. For example, the first column of this figure shows that the higher the inconsistency of the ratings, the lower the precision is, which is what we expected. 

The second column shows a strong correlation (correlation coefficient $\approx$ 0.9) indicating that user groups with higher entropy (more information gain) in their ratings receive more accurate (higher precision) recommendations. Also, from the same figure, we can see for the lower values of entropy, the algorithms behave more fairly, but, the larger the entropy gets, the discrimination between female and male user groups becomes more apparent (higher precision for male user groups). 

The relationship between the average profile size and precision is also shown in the last column of Figure~\ref{fig:corr_precs}. As expected, user groups with larger profiles benefit from more accurate recommendations for both males and females. However, the discrimination can still be seen for some algorithms such as \algname{UserKNN} where female users with the same profile size still receive recommendations with lower precision compared to the male users. 

The factors described above mainly come from the input data where users' interactions with the system are recorded. However, there are some other factors that come from the algorithm. It means that the algorithm even amplify the bias when generating the recommendation lists. One of the well-known algorithmic bias is \textit{popularity bias} \cite{ciampaglia2018algorithmic,powell2017love,harald2011}: the tendency of recommender system to frequently recommend popular items and rarely recommend non-popular items. Recommending popular items may not bring much benefit for users and they might be already known to the users. Also, recommending popular items hinders the system ability to better learn users' preferences through exploration and deliver more accurate recommendations.

In addition, in \cite{himan2019c}, we showed that popularity bias can cause unfairness on the user side. A recommendation algorithm that is biased toward popular items will not properly capture the interests of users who are interested in non-popular items. This means that a biased recommendation system only learns the interests of blockbuster users (users who are more interested in popular items), while does not serve niche-taste users well. Also, in \cite{abdollahpouri2020unfair}, we investigated the supplier side unfairness of popularity bias in music recommendation domains. This analysis showed that due to the algorithmic popularity bias, unpopular artists may not have a chance to appear in recommendation lists to users. 

In the next Section, I describe how algorithmic bias, in particular popularity bias, can negatively affect the performance of the recommender system and different actors in the system over time.

\subsection{Impacts of unfair recommendations}\label{fairness_impact}

Unfair recommendation can adversely impact the satisfaction of different actors in the system:

\begin{itemize}
    \item In item-side, unfair distribution of recommended items can amplify the \textit{popularity bias} \cite{ciampaglia2018algorithmic,abdollahpouri2020connection}. In this situation, popular items would be over-recommended even more that their merit\footnote{Depending on the domain, item merit can be defined in various ways, but on simple approach is considering the popularity of the item.} in the rating data and tail items would be under-recommended. This skew in the frequency of recommended items in recommendation lists not only adversely impacts the whole ecosystem, but also negatively affects the experience of both suppliers (some suppliers may not receive enough attention) \cite{abdollahpouri2020unfair} and users (those who are not interested in popular items still receive popular items in their recommendations) \cite{abdollahpouri2019managing}.
    \item Unfair treatment of suppliers may cause the items belonging to some suppliers to appear frequently in the recommendation lists, while the items belong to the other suppliers do not receive proportionate or deserved attention, leading to skew in the appearance of suppliers in the recommendations \cite{abdollahpouri2020multi}. This skewness in the appearance of suppliers in recommendation lists may stimulate under-recommended suppliers to leave the system, which will have negative impact on the whole system in long run.
    \item Finally, user-side unfairness not only perpetuates undesirable social dynamics, it can also degrade the satisfaction of certain groups of users \cite{edizel2020}. Measuring user-side fairness is a challenging task and requires online evaluation of recommendation lists on a real-world platform with steady stream of data, which is not always available. But, in an offline setting, it can be simply defined as any disparity in the quality of recommendation generated for the users. For instance, if group $A$ receives recommendations which are better matched with their preferences than those of group $B$, it shows that the recommendation algorithm mainly learned the preferences of group $A$, showing algorithmic bias against group $B$. 
\end{itemize}

The algorithmic bias could be intensified over time as users interact with the given recommendations that are biased towards popular items and this interaction is added to the data. Users receiving recommendation lists may select (e.g.\ by rating or clicking) some of the recommended items and the system will add those items to their profiles as part of their interaction history. In this way, recommendations and user profiles form a feedback loop \cite{damour2020,chaney2018}; the users and the system are in a process of mutual dynamic evolution where user profiles get updated over time via recommendations generated by the recommender system and the effectiveness of the recommender system is also affected by the profile of users.

The study on feedback loop in machine learning and particularly recommender systems has recently received more attention from researchers \cite{damour2020,chaney2018,sinha2016,sun2019debiasing}. D'Amour et al. \cite{damour2020} analyzed the long-term fairness of machine learning based decision-making systems in three different domains through simulation studies: bank loans, allocation of attention, and college admission in an agent-based environment. Their analysis showed that common single-step analysis does not show the dynamic behavior of the system and the need for exploring the long-term effect of the decision-making systems. In another work which is also based on a simulation using synthetic data, Chaney et al. \cite{chaney2018} showed that feedback loop causes homogenization of the user experience and shift in item consumption. Homogenization in their study was measured as the ratio of commonly rated items in a target user's profile and her nearest neighbor's profile, and showed that homogenization leads to lower utility for the users. 

In this section, I study the negative impacts of unfair recommendation on the system, in particular the effect of feedback loop on amplifying bias in recommender systems. This contribution is published in \cite{mansoury2020feedback}. I investigate popularity bias amplification and the impact of this bias on other aspects of a recommender system including declining aggregate diversity, shifting the representation of the users' taste, and also homogenization of the users. In particular, I show that the impact of feedback loop is generally stronger for the users who belong to the minority group. For the experiments, I simulate the users interaction with recommender systems over time in an offline setting. The concept of time here is not chronological but rather consecutive interactions of users with the recommendations in different iterations. That is, in each iteration, users' profile is updated by adding selected items from the recommendation lists generated at previous iteration to their profile. 

\subsubsection{Feedback loop simulation}\label{simulation}

The ideal scenario for investigating the effect of feedback loop on amplifying bias in recommender systems is to perform online testing on a real-world platform with steady stream of data. 
However, due to the lack of access to real-world platforms for experimentation, I simulated the recommender system process in an offline setting. To do so, I simulated the recommendation process over time by iteratively generating recommendation lists to the users and updating their profile by adding the selected items from those recommendation lists based on an acceptance probability. 

Given the rating data $D$ as an $m \times n$ matrix formed by ratings provided by the users $U=\{u_1,...,u_m\}$ on different items $I=\{i_1,...,i_n\}$, the mechanism for simulating feedback loop is to generate recommendation lists for the users in each iteration $t \in \{1,...,T\}$ and update their profile based on the delivered recommendations in each iteration. The following steps show this mechanism: 

\begin{itemize}
    \item[\textbf{1)}] Given $D^t$ as the rating data in iteration $t$, we split $D^t$ into training and test sets as 80\% for $train^t$ and 20\% for $test^t$. 
    \item[\textbf{2)}] We build the recommendation model on $train^t$ to generate the recommendation lists $R^t$ to all users. 
    \item[\textbf{3)}]  For each user $u$ and recommendation list $R_{u}^{t}$ generated for $u$, we follow the \textit{acceptance probability} concept proposed in \cite{himan2019a} to decide which item from the recommendation list the user might select. The acceptance probability assigns a probability value to each item in $R_{u}^{t}$ where more relevant items (higher ranked) are assigned higher probability to be selected. Formally, for each item $i$ in $R_{u}^{t}$, the acceptance probability can be calculated as follows:
    
    \begin{equation}\label{acceptanceprob}
        prob(i|R_{u}^{t})=e^{\alpha \times rank_i}
    \end{equation}
    
    \noindent where $\alpha$ is a negative value ($\alpha < 0$) for controlling the probability assigned to each recommended item and $rank_i$ is the rank of the item $i$ in $R_{u}^{t}$. 
    Equation~\ref{acceptanceprob} is only a selection probability and does not assign a potential rating a user might give to the selected item. This is particularly important if we want to also include rating-based algorithms such as \algname{UserKNN} in our simulation as we have done in this paper. To estimate the rating a user might give to the selected item, we follow the \textit{Item Response Theory} used in \cite{sinha2016,ho2008}. More formally,
    
     \begin{equation}\label{irt1}
        \omega=\overline{s}_u + (sd(s_u) \times \overline{s}_i) + \eta_{u,i}
    \end{equation}

    \noindent where $\overline{s}_u$ is the average of the ratings in $u$'s profile, $sd(s_u)$ is the standard deviation of the ratings in $u$'s profile, $\overline{s}_i$ is the average of ratings assigned to $i$, and $\eta_{u,i}$ is a noise term derived from a Gaussian distribution (i.e.\ $\eta_{u,i} \sim N(0,1)$). In order to estimate an integer rating value in the range of $[a,b]$ where $a$ and $b$ are the minimum and maximum rating values, respectively, we use the Equation $\hat{s}_{u,i}=max(min(round(\omega),b),a)$ as proposed in \cite{sinha2016}.
    After estimating $\hat{s}_{u,i}$, we add $(i,\hat{s}_{u,i})$ to $u$'s profile if $i$ is not already in $u$'s profile and we repeat this process for all users to form $D^{t+1}$. 
    
\end{itemize}

The steps 1 through 3 are repeated in each iteration.

\subsubsection{Modeling Bias Amplification}\label{biasmodel}

In this section, we formally model the propagation of this bias due to the feedback loop phenomenon. Let $\overline{P}_{D^t}$ and $\overline{P}_{R^t}$ be the average popularity (i.e.\ the expected values) of the items in the rating data and the recommended items in iteration $t$, respectively. 

\begin{equation}\label{modeling}
    \overline{P}_{R^t} \propto \overline{P}_{D^t}+\theta^t 
\end{equation}

\noindent where $\theta^t$ is the percent increase of the popularity of the recommendations compared to that of rating data in iteration $t$. Now, assuming, out of all the recommendations given to the users, we add $K$ interactions ($K>=0$) to the profiles of the users, the size of the rating data in the next iteration would be $|D^t|+K$ and its average popularity will be

\begin{equation*}
    \overline{P}_{D^{t+1}} \approx \frac{|D^t|\times \overline{P}_{D^t} + K \times (\overline{P}_{D^t}+\theta^t) }{|D^t|+K}
\end{equation*}

\noindent which can be simplified as

\begin{equation*}
    \frac{(|D^1|+K) \times \overline{P}_{D^t} +K \times \theta^t}{|D^t|+K}=\overline{P}_{D^t}+\frac{K \times \theta^t}{|D^t|+K}
\end{equation*}

\noindent which means the average popularity of the items in the rating data is now increased by

\begin{equation*}
    \frac{K \times \theta^t}{|D^t|+K}
\end{equation*}


Based on Equation~\ref{modeling}, by definition, the average popularity of the recommended items in each iteration is proportional to the average popularity of the rating data in the same iteration plus a positive value and since $\overline{P}_{D^{t+1}}$ has increased compared to $\overline{P}_{D^t}$, $\overline{P}_{R^{t+1}}$ will be also higher than $\overline{P}_{R^t}$ due to transitivity. In other words, in each iteration $t$, $\overline{P}_{R^{t+1}}>\overline{P}_{R^{t}}$ indicating the popularity propagation/intensification from one iteration to the next one.  

\subsubsection{Experiments}

In investigation of the effect of feedback loop on amplifying bias, I performed a set of experiments on well-known MovieLens1M dataset \cite{Harper:2016} using three different recommendation algorithms: user-based collaborative filtering (\algname{UserKNN}) \cite{Resnick:1994a}, Bayesian Personalized Ranking (\algname{BPR}) \cite{rendle2009bpr}, and \algname{MostPopular}. \algname{MostPopular} recommends the most popular items to everyone (the popular items that a user has not seen yet). We set the number of factors in \algname{BPR} and the number of neighbors in \algname{UserKNN} to 50 to achieve the best performance in terms of precision. For our simulation, we performed the steps 1-3 in Section~\ref{simulation} for 20 iterations ($T=20$). Details of the dataset and algorithms are explained in Chapter~\ref{chap:methodology}.

\subsubsection{Popularity bias amplification}

\begin{figure*}[t]
    \centering
    \includegraphics[width=0.99\textwidth]{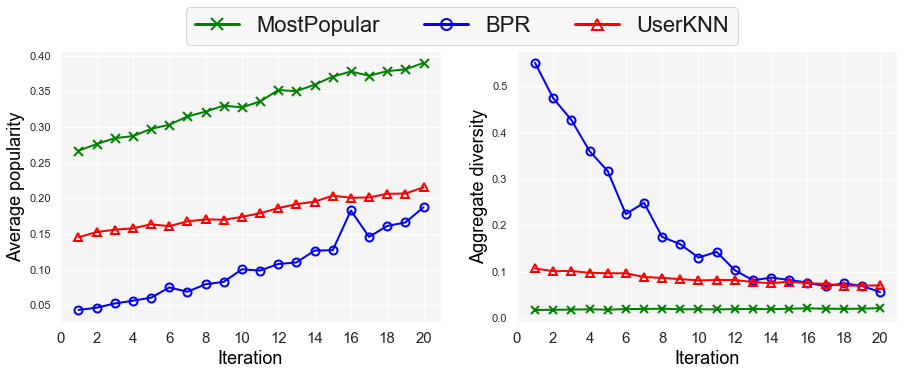}
    \caption{Average popularity (left) and aggregate diversity (right) of the recommendations.}\label{fig:popularity}
\end{figure*}

As I formally showed in previous section (entitled Modeling Bias Amplification), recommendation models can intensify the popularity bias in input data over time due to the feedback loop. Figure~\ref{fig:popularity} (left) shows the effect of this loop on the average popularity of recommendation lists over time (i.e.\ in different iterations). As shown in this plot, even though these algorithms start with different average popularity values due to their inherent nature, they all show an ascending pattern in terms of the average popularity over different iterations. The curve for \algname{BPR} seems to have a steeper slope compared to the other algorithms indicating a stronger bias propagation of this algorithm. The exact reason for these performance differences across different algorithms needs further investigation and I leave it for future work. 

Figure~\ref{fig:popularity} (right) shows the aggregate diversity (aka catalog coverage) of recommendation algorithms: the percentage of items that appear at least once in the recommendation lists across all users. As a recommender system concentrates more on popular items, it will necessarily cover fewer items in its recommendations and that effect is clear here, especially for \algname{BPR}, which starts out with a relatively high aggregate diversity. 

This bias amplification over different iterations could lead to two problems: 1) shifting the representation of the user's taste over time, and 2) the domination of the preferences of one group of users (the majority group) over another (the minority group) which eventually could diminish the differences between the groups and create homogenization.

\begin{figure*}[t]
    \centering
    \includegraphics[width=0.99\textwidth]{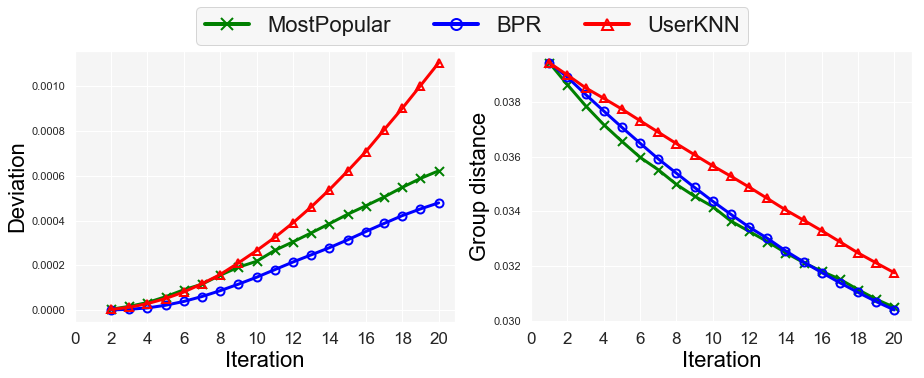}
    \caption{Deviation from the initial preferences (left) and the distance between the representation of genre preferences of males and females in different iterations (right).}\label{fig:homogenity}
\end{figure*}

\subsubsection{Shifting users' taste representation}

One consequence of the feedback loop is shifting the representation of the users' taste revealed in user profiles. I define the users interest toward various movie genres based on the rated items in their profile which creates a genre distribution over rating data. This genre distribution is calculated as the ratio of the movies associated with each genre over different genres in the users' profiles. In the MovieLens1M dataset, some movies are assigned multiple genres hence, in those case, I assign equal probability to each genre. For example, if an item has genres $a$ and $b$, the probability of either of $a$ and $b$ is 0.5. 

Given genre distribution in iteration $t=1$ as initial preferences represented in the system, I am interested in investigating how initial users' taste representation changes over time due to the feedback loop. For this purpose, in each iteration $t>1$, I calculate the Kullback-Leibler divergence ($KLD$) between the initial genre distribution and the genre distribution in iteration $t$ for each user. Higher $KLD$ value indicates higher deviation from the initial preference.  

Figure~\ref{fig:homogenity} (left) shows the deviation of users taste from their initial preferences. In all recommendation algorithms, we observe that the deviation of users' profiles from their initial preferences increases over time. It is worth noting that the change in users preferences shown in this figure is the change in the representation of users' preferences in the system, not the change in users' intrinsic preferences. One consequence of this change is that recommendation models may not be able to capture the users' true preferences when generating recommendations for the users.

\subsubsection{Homogenization}

A shift in the users' taste representation could happen in two situations: when the recommendations given to the users are more diverse than what the users are interested in (i.e.\ exploration), or when the recommendations are over-concentrated on few items when the users' profiles are more diverse. In the latter case, since all users are exposed to a limited number of items over time, their profiles all converge towards a common range of preferences.

Figure~\ref{fig:homogenity} (right) shows the distance between the representation of males (majority group) and females (minority group) preferences over time. In each iteration $t$, given the genre distribution separately extracted from males and females ratings as $G_M$ and $G_F$, respectively, I calculate the $KLD$ of $G_M$ and $G_F$, $KLD(G_M||G_F)$, which measures the distance between the preferences of males and females. As shown in the plot, the $KLD$ value dramatically decreases over time in all algorithms showing the strong homogenization of users' preferences.

Now, an interesting question is that the preferences of which user group is dominating the other. To answer this question I separately compare genre preferences of males and females with the preferences of the whole population. Given $G$ as the initial genre preferences of all users, I calculate $KLD(G||G_F)$ and $KLD(G||G_M)$ in each iteration $t$. 

\begin{figure*}[t]
    \centering
    \includegraphics[width=0.99\textwidth]{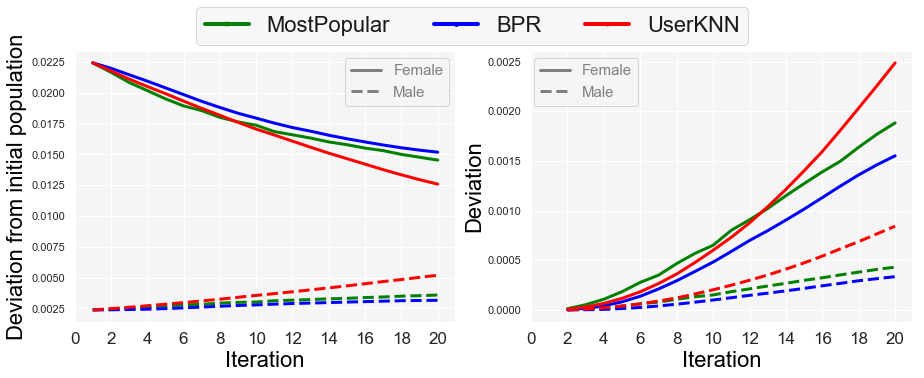}
    \caption{Deviation of the representation of male and female preferences from the representation of the population initial preferences (left) and deviation of the representation of male and female preferences from their initial preferences (right).}\label{fig:population}
\end{figure*}

Figure~\ref{fig:population} (left) separately shows $KLD(G||G_F)$ and $KLD(G||G_M)$ in different iterations. We can see that, for all algorithms, the representation of females preferences are approaching toward the representation of initial preferences of the population. However, this value is slightly increasing for males showing that they become distant from the preferences of the initial population. I believe the reason is that male users are taking up the majority of the ratings in the data and hence, initially, the population is closer to the male profiles. Over time, since the recommended items are more likely to be those rated by males (as males have rated more items), when added to the users' profiles, causes the female profiles to get closer to the initial population, which was dominated by the male users. 

Figure~\ref{fig:population} (right) shows the deviation from the representation of initial preferences of each user in the system separately for males and females. In all algorithms, the deviation for females is significantly higher than males, demonstrating the severity of the impact of the feedback loop on the minority group.

\subsubsection{Limitations}

Simulation studies are often designed to mimic the real-world scenarios as much as possible, but due to the complexity of these scenarios, it is not always possible. Therefore, the complexities are usually relaxed by making certain assumptions on the simulation process. I also made some assumptions in the proposed simulation study which limit the ability of perfectly mimicking the real-world scenario. 

First, the selection technique in Equation~\ref{acceptanceprob} I used in this simulation leverages the ranking position of the items in the list in order to define whether it would be selected by the user or not. It assigns higher probability to the items on top of the list to be selected than the items on bottom of the list. However, selection based on the ranking of the recommended items is not the only factor that a real user may consider when selecting an item. A real user may sometimes find the lowered ranked items more desired than the items on top of the list based on his/her actual preference. Therefore, this limitation can be addressed by modeling user behavior on selecting an item from the list or considering other selection policies such as random selection.

Second, in some recommendation domains such as
music, it is very common for a user to listen to the same song repeatedly. Therefore, the restriction I imposed on the selection algorithm in this simulation regrading the items that were already in
the users’ profile (those items were not added to the users’ profiles in the next iteration) can be lifted and, instead, the rating for that item would be updated in each iteration.

\subsection{Techniques for addressing unfair recommendations}

The problem of unfair recommendations and the challenges it creates for the recommender system has been well studied by other researchers. The solutions for tackling bias in the machine learning and recommender systems literature can be categorized into three groups: pre-processing, in-processing, and post-processing approaches \cite{kamiran2013techniques}. In pre-processing approaches, the input data is modified (e.g.\ over-sampled or under-sampled) to reduce the inherent bias or skew in independent/dependent variables. In in-processing approaches, the predictive algorithm is modified to mitigate the algorithmic bias. In post-processing approaches, the results of the predictive model is processed to remove any possible bias in the final output. In this Section, I review existing techniques for tackling bias and unfairness in recommender systems.  

Previous research has raised concerns about discrepancies in recommendation accuracy across different users \cite{yao2017,zhu2018,kamishima2011}. For instance, \cite{ekstrand2018} shows that women on average receive less accurate, and consequently, less fair recommendations than men using a movie dataset.

Burke et. al. in \cite{burke2017} have shown that inclusion of a balanced neighborhood regularization term to SLIM algorithm \cite{Ning2011} can improve the fairness of the recommendations for protected and unprotected groups. Based on their definition for protected and unprotected groups, their solution takes into account the group fairness of recommendation outputs. Analogously, Yao and Huang in \cite{yao2017} improved the fairness of recommendation results by adding fairness terms to objective function in model-based recommendation algorithms. They proposed four fairness metrics that capture the degree of unfairness in recommendation outputs and added these metrics to learning objective function to further optimize it for fair results.

Zhu et al. in \cite{zhu2018} proposed a fairness-aware tensor-based recommender systems to improve the equity of recommendations while maintaining the recommendation quality. The idea in their paper is isolating sensitive information from latent factor matrices of the tensor model and then using this information to generate fairness-aware recommendations.

It is well-known that popularity bias leads to unfair exposure of items in recommendation lists to the users \cite{powell2017love,abdollahpouri2020popularity}. Also, research conducted by authors in \cite{himan2019c,kowald2020unfairness} have shown that popularity bias can cause unfairness from the users' perspective as users are not equally treated based on their interests toward popular items. Hence, several studies have been conducted to mitigate popularity bias in recommender systems \cite{anderson2006long,longtailnichesriche,park2008long}. Authors in this work have mainly explored the overall accuracy of the recommendations in the presence of long-tail distribution in rating data. In addition, some other researchers have proposed algorithms that can control this bias and give more chance to long-tail items to be recommended \cite{adomavicius2011improving,DBLP:conf/recsys/KamishimaAAS14,abdollahpouri2019managing}.

Jannach et al., \cite{Jannach2015} conducted a comprehensive set of analysis on popularity bias of several recommendation algorithms. They analyzed recommended items by different recommendation algorithms in terms of their average ratings and their popularity. While it is very dependent to the characteristics of the data sets, they found that some algorithms (e.g.\ \algname{SlopeOne} \cite{lemire2005slope}, KNN techniques \cite{Resnick:1994a,sarwar2001}, and ALS-variant of factorization models \cite{takacs2012alternating}) focus mostly on high-rated items which bias them toward a small sets of items (low coverage). Also, they found that some algorithms (e.g.\ ALS-variants of factorization model) tend to recommend popular items, while some other algorithms (e.g.\ \algname{UserKNN} and \algname{SlopeOne}) tend to recommend less-popular items. Abdollahpouri et al., \cite{abdollahpouri2017controlling} addressed popularity bias in learning-to-rank algorithms by inclusion of fairness-aware regularization term into objective function. They showed that the fairness-aware regularization term controls the recommendations being toward popular items.

Vargas and Castells in \cite{vargas2011} proposed probabilistic models for improving novelty and diversity of recommendations by taking into account both relevance and novelty of target items when generating recommendation lists. In other work \cite{vargas2014}, they proposed the idea of recommending users to items for improving novelty and aggregate diversity. They applied this idea to nearest neighbor models as an inverted neighbor and a factorization model as a probabilistic reformulation that isolates the popularity components. 

Adomavicius and Kwon \cite{adomavicius2011} proposed the idea of diversity maximization using a maximum flow approach. They used a specific setting for the bipartite recommendation graph in a way that the maximum amount of flow that can be sent from a source node to a sink node would be equal to the maximum aggregate diversity for those recommendation lists. In their setting, given the number of users is $m$, the source node can send a flow of up to $m$ to the left nodes, left nodes can send a flow of up to 1 to the right nodes, and right nodes can send a flow of up to 1 to the sink node. Since the capacity of left nodes to right nodes is set to 1, thus the maximum possible amount of flow through that recommendation bipartite graph would be equivalent to the maximum aggregate diversity.

A more recent graph-based approach for improving aggregate diversity was proposed by Antikacioglu and Ravi in \cite{antikacioglu2017}. They generalized the idea proposed in \cite{adomavicius2011} and showed that the minimum-cost network flow method can be efficiently used for finding recommendation subgraphs that optimizes the diversity. In this work, an integer-valued constraint and an objective function are introduced for discrepancy minimization. The constraint defines the maximum number of times that each item should appear in the recommendation lists and the objective function aims to find an optimal subgraph that gives the minimum discrepancy from the constraint. This work shows improvement in aggregate diversity with a smaller accuracy loss compared to the work in \cite{vargas2011} and \cite{vargas2014}. Similar to this work, the proposed FairMatch algorithm in this dissertation (see Chapter~\ref{chap:solution2}) also uses a graph-based approach to improve aggregate diversity. However, unlike the work in \cite{antikacioglu2017} which tries to minimize the discrepancy between the distribution of the recommended items and a target distribution, the FairMatch algorithm has more freedom in promoting high-quality items with low visibility since it does not assume any target distribution of the recommendation frequency.

In addressing exposure bias in domains like job recommendations where job seekers or qualified candidates are recommended, Zehlike et. al. \cite{zehlike2017fa} proposed a re-ranking algorithm to improve the ranked group fairness in recommendations. 
The algorithm creates queues of protected and unprotected items and merges them using normalized scoring such that protected items get more exposure. 

Finally, in addressing supplier-side unfairness, Surer et al. in \cite{surer2018} proposed a multi-stakeholder optimization model that works as a post-processing approach for standard recommendation algorithms. In this model, a set of constraints for providers are considered when generating recommendation lists for end users. Also, Liu and Burke in \cite{liu2018} proposed a fairness-aware re-ranking approach that iteratively balances the ranking quality and provider fairness. In this post-processing approach, users' tolerance for diversity list is also considered to find trade-off between accuracy and provider fairness. 

Mehrotra et al. \cite{mehrotra2018towards} investigated the trade-off between the relevance of recommendations for users and supplier fairness, and their impacts on users' satisfaction. Relevance of the recommended items to a user is determined by the score predicted by a recommendation algorithm. To determine the supplier fairness in recommendation list, first, suppliers are grouped into several bins based on their popularity in rating data and then the supplier fairness of a recommendation list is measured as how diverse the list is in terms of covering different supplier popularity bins.

%% file: Tables/2.2.3.factors1.tex
\captionsetup[table]{skip=4pt}
\begin{table}[t!]
\small
\centering
\captionof{table}{Specification of ML1M for male and female users} \label{tab:stat}
\begin{tabular}{l|cccc}
\toprule
  & $\#users$ & $\overline{\mathcal{A}}$ & $\overline{\mathcal{E}}$ & $\overline{\mathcal{S}}$ \\
 \midrule
 Male & 4,331 & 0.781 & 4.174 & 139.2 \\
 Female & 1,709 & 0.808 & 3.995 & 115.4 \\
 \bottomrule
\end{tabular}
\end{table}

%% file: Tables/2.2.3.factors2.tex
\captionsetup[table]{skip=4pt}
\begin{table}[t!]
\small
\centering
\captionof{table}{Precision and miscalibration of recommendation algorithms for male and female users} \label{tab:accuracy}
\begin{tabular}{llllll}
\toprule
 \multirow{2}{*}{algorithm} & \multicolumn{2}{c}{Precision} & & \multicolumn{2}{c}{Miscalibration} \\ \cline{2-3}\cline{5-6}
 & Male & Female & & Male & Female \\
 \bottomrule
 \algname{UserkNN} & 0.235 & 0.162 & & 0.915 & 0.971 \\
 \algname{ItemkNN}  &  0.242 & 0.175 & & 0.874 & 0.973 \\
 \algname{SVD++} &  0.133 & 0.095 & & 1.156 & 1.130 \\
 \algname{ListRankMF} & 0.160 & 0.118 & & 0.970 & 1.032 \\
 \bottomrule
\end{tabular}
\end{table}

%% file: Chapters/03_Simulation.tex
\chapter{Multi-Sided Matching and Recommendation Problem}
\label{chap:simulation}
\tikzsetfigurename{simulation_}

In a multi-sided platform, various actors are involved in the system. Optimizing an objective in this situation may need specific design and constraints. In this chapter, I study multi-sided matching and recommendation problem through a simulation on educational system.

\section{Simulation study}

In this simulation, I consider the matching between students and supervisors in an educational system where the goal is to assign (match) students to the supervisors under various settings. The number of settings and assumptions can be overwhelming and may significantly complicate the problem. In this simulation, I define the settings in a way to represent the real-world scenario while keeping it simple.

The problem that I simulate in this chapter is the assignment of students to supervisors in universities or any educational institutions. The goal is to find a match between students and supervisors in a way to satisfy the demands and requirements for both and make them happy. In realistic scenario, students are not on the same level of qualification and supervisors also are not on the same level of knowledge and expertise. Thus, I considers these facts when generating synthetic data for students and supervisors. Students always want to work with successful or reputable supervisors, and supervisors prefer to advise high-qualified students. A fair match is the one that assigns students to supervisors according the level of qualification and expertise of both. For simplicity, I do not consider the information about the topics of interest for supervisors and students in this simulation.    

\input{./Tables/tbl_simulation_notation}

\subsection{Notations and variables}

There are $n$ students $P=\{p_1,p_2,...,p_n\}$ and $m$ supervisors $Q=\{q_1,q_2,...,q_m\}$ and the goal is to match them by assigning students to each supervisor. Students expressed their preferred ranked list of supervisors that shows the students' preference toward supervisors. The preferred ranked list of student $p$ is shown by $R_p$ which signifies an ordered list of supervisors that $p$ wishes to work with. For example, $R_p=\{q_3,q_1,q_2\}$ shows that student $p$ is interested in being assigned to supervisor $q_3$, but if for some reasons this assignment is impossible, she is interested in working with supervisor $q_1$, and finally if this assignment is not also possible, she would like to be assigned to supervisor $q_2$. In this simulation, I assume that each student expressed her preferred ranked list on all supervisors.

Additionally, each student has certain degree of qualification based on her background and past achievements such as course grades and GPA. Thus, for each student $p$, $X_p$ shows the qualification of $p$. Indeed, $X_p$ is a multivariate or column vector variable as student qualification is measured based on multiple elements. However, for simplicity, I assume that all those elements of student qualification are aggregates and eventually ends up to a real number for each student as her qualification. The reason for this assumption is that since this study is on synthetically generated data, generating synthetic multivariate data as student qualification may add noise to the experiments. 

The notation introduced above are summarized in Table~\ref{tab:simulation_notations}. 

\begin{figure*}[btp]
    \centering
    \begin{subfigure}[b]{0.7\textwidth}
        \includegraphics[width=\textwidth]{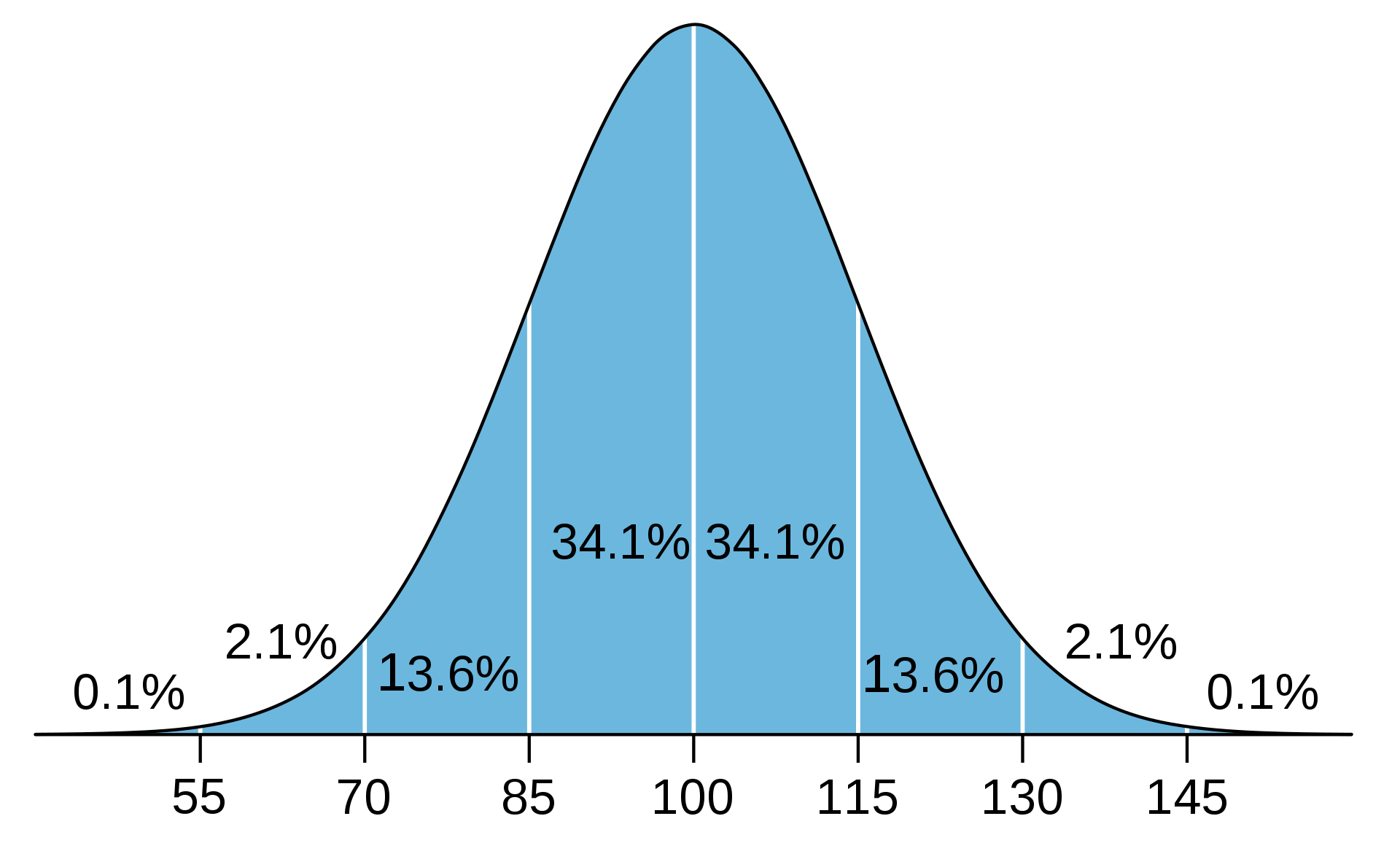}
    \end{subfigure}
\caption{Normalized IQ distribution with mean 100 and standard deviation 15 \cite{IQ}.} \label{iq_dist}
\end{figure*}

\subsection{Synthetic data generation} 

Variables $P$ and $Q$ as students and supervisors, respectively, are generated by successive integer numbers where each number represents a student (or supervisor): $P=\{1,2,3,...,n\}$ and $Q=\{1,2,3,...,m\}$. In this simulation, the number of students is set to 200 ($n=200$) and the number of supervisors is set to 20 ($m=20$). 

The data for student qualification variable, $X_p, p \in P$, can be generated equivalent to the distribution of Intelligence Quotient (IQ) test\footnote{\url{https://en.wikipedia.org/wiki/Intelligence_quotient}}. It is well-known that the distribution of IQ scores for a population follows Gaussian distribution or, informally, bell curve. Figure~\ref{iq_dist} shows this distribution. According to this figure:

\begin{itemize}
    \item Approximately 95\% of the population has IQ scores between 70 and 130. 
    \item Approximately 99.7\% of the population has IQ scores between 55 and 145.
    \item Only approximately 0.3\% of the population has IQ scores outside of this
interval (less than 55 or higher than 145).
\end{itemize}

\begin{figure*}[btp]
    \centering
    \begin{subfigure}[b]{0.6\textwidth}
        \includegraphics[width=\textwidth]{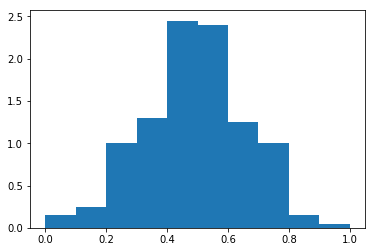}
    \end{subfigure}
\caption{Distribution of student qualification.} \label{qualification_dist}
\end{figure*}

Therefore, in this simulation, $X \sim \mathcal{N}(\mu,\,\sigma^{2})$, where $\mu=100$ and $\sigma^{2}=15$. After generating this distribution, the values of $X$ are mapped to $[0,1]$ using min-max normalization technique where 0 means lowest qualification and 1 means highest qualification. Note that $|X|=n$ and each value in $X$ represents the qualification of a students. For example, the first value of $X$ shows the qualification of student $1$, the second value of $X$ shows the qualification of student $2$, and so on. Figure~\ref{qualification_dist}, shows the distribution of student qualification, variable $X$, as explained above.

\begin{figure*}[t]
  \centering
  \begin{subfigure}[b]{1\textwidth}
        \includegraphics[width=0.45\textwidth]{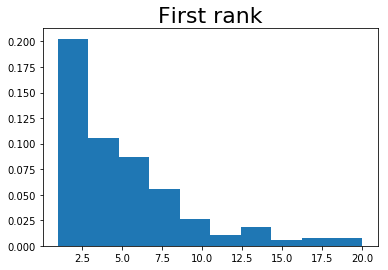}
        \includegraphics[width=0.45\textwidth]{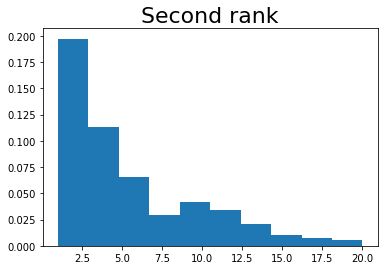}
  \end{subfigure}
  \begin{subfigure}[b]{1\textwidth}
        \includegraphics[width=0.45\textwidth]{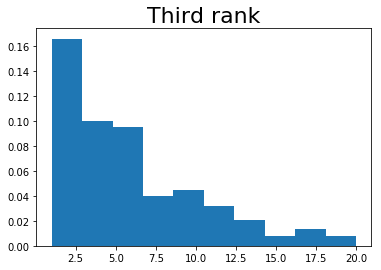}
        \includegraphics[width=0.45\textwidth]{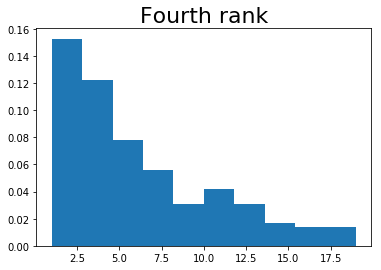}
  \end{subfigure}
\caption{Supervisors distribution in students preferred ranking list in first, second, third, and fourth ranks.} \label{supervisor_dist}
\end{figure*}

The data for variable $R$, preferred ranked list of supervisors for each student, is generated in a way that interests toward supervisors follow a \textit{Normal distribution}. This means that some supervisors receive more attentions from students than other supervisors and this attention forms a normal distribution. This is what we usually see in real-world where some supervisors in a university have better reputation than others. Therefore, when generating the ranked preferred list of supervisors for each student, the distribution of selected supervisors at each rank should eventually have normal shape. Since variable $R$ is a matrix (or $2D$ array) with dimensions $n \times m$ ($n$ students as rows and $m$ supervisors as columns), the distribution of selected supervisors by students at each rank (first column, second column, ..., $m$-th column) will have normal shape, meaning that some supervisors are selected more than other ones by students. Figures~\ref{supervisor_dist} shows the distribution of selected supervisors for first four ranks of $R$.

Finally, each supervisor has certain capacity for supervising a specific number of students. For this purpose, this capacity for supervisor $q$ is defined as $[q_a,q_b]$ which means that $q$ can supervise at least $q_a$ and at most $q_b$ students. 

\section{Fairness and utility metrics}\label{utility_metric}

In this section, I introduce the utility and fairness metrics for students and supervisors. In the literature, the term \textit{utility}
is used to represent the desired outcome for each side. In the present simulation, it is meaningful to also interpret it as \textit{happiness} of students and supervisors. To be consistent with the notations in the literature, I also use the term utility for defining the desired outcome for each side.

\subsection{Modeling utility for both sides}

For students, the desired outcome can be achieved when the preferred supervisor for each student $p$ according to $R_p$ is assigned to her. For example, if student $p$ stated her preferred ranked list of supervisors as $R_p=\{q_3,q_1,q_2\}$, then assigning $q_3$ to $p$ will result in the highest utility for $p$. Therefore, for each student $p \in P$, the utility of $p$, $U_p^{student}$, can be defined as:

\begin{equation}\label{stu_utility}
    U_p^{student}=\frac{|R_p|-Rank(A(p),R_p)}{|R_p|}
\end{equation}

\noindent where $R_p$ is the ranked list of preferred supervisors specified by student $p$, $A(p)$ returns the supervisor assigned to $p$, and $Rank(A(p),R_p)$ returns rank of $A(p)$ (assigned supervisor to $p$). Higher value for $U_p^{student}$ means higher utility for $p$ and lower value for $U_p^{student}$ means lower utility.

On the other hand, the desired outcome for supervisors can be achieved when students with high qualification are assigned to each supervisors. Since students have various qualification levels, the matching model should maximize the average qualification of students assigned to a supervisor in a way that each supervisor receives certain degree of qualified students. Therefore, the utility of supervisor $q$, can be defined as:

\begin{equation}\label{sup_utility}
    U_q^{supervisor}=\frac{\sum_{p \in A'(q)}X_p}{|A'(q)|}
\end{equation}

\noindent where $A'(q)$ returns all students assigned to supervisor $q$ and $X_p$ is the qualification of student $p$. Higher $U_q^{supervisor}$ means higher utility for supervisor $q$ and lower $U_q^{supervisor}$ means lower utility for $q$.

It is also worth noting that Equation~\ref{sup_utility} does not take into account the merit of supervisors, meaning that all supervisors are treated equally. However, another definition for supervisor utility is proportional equality to supervisor's merit based on students preferences. According to this definition, a supervisor with higher merit may be assigned students with higher qualification. Merit can be measured as the number of students who selected the supervisor as the lower rank (more preferred) of her ranked preferred list of supervisors. Hence, proportional equality can be defined as:

\begin{equation}\label{sup_utility_prop}
    {PU}_q^{supervisor}=\frac{U_q^{supervisor}}{\sum_{p \in P}Rank(q,R_p)}
\end{equation}

\noindent where ${PU}_q^{supervisor}$ measures the utility of supervisor $q$ (i.e.\ Equation~\ref{sup_utility}) proportional to the merit of $q$ based on students preferences. The denominator is sum of the ranking position of a supervisor in students' ranked preferred list of supervisors ($R$). This means that higher $\sum_{p \in P}Rank(q,R_p)$ means that students ranked supervisor $q$ in a higher rank (lower preference) of their ranked list of preferred supervisors, while lower $\sum_{p \in P}Rank(q,R_p)$ means that students ranked supervisor $q$ in lower rank (higher preference) of their ranked list of preferred supervisors. Therefore, higher value for denominator (lower merit) results in lower utility (i.e.\ low value for ${PU}_q^{supervisor}$) and lower value for denominator (higher merit) results in higher utility (i.e.\ high value for ${PU}_q^{supervisor}$) 

\subsection{Modeling fairness for both sides}

Fairness for students can be defined as how equally students are treated on satisfying their utility. More precisely, it can be defined as how equally similar students are treated on assigning their preferred supervisors. For example, if two students are similar in terms of their qualifications, it is expected that they receive their preferred supervisors by matching model. Therefore, student fairness can be defined as:

\begin{equation}\label{stu_fairness}
    \displaystyle equality^{student}=\frac{\sum_{i=1}^{|P|}{\sum_{\substack{j>i\\Sim(i,j)>\alpha}}^{|P|}{|U_i-U_j|}}}{\sum_{i=1}^{|P|}{\sum_{j>i}^{|P|}{\mathds{1}(Sim(i,j)>\alpha)}}}
\end{equation}

\noindent where $\mathds{1}(.)$ is the indicator function returning zero when its argument is False and 1 otherwise. $U_i$ is the utility value for student $i$ calculated as Equation~\ref{stu_utility}, $Sim(i,j)$ calculates the similarity between students $i$ and $j$ based on their qualification and can be calculated as $Sim(i,j)=1-|X_i-X_j|$, and $\alpha$ is a threshold for determining whether two students are similar or not. This Equation measures how similar students are treated equally by the matching algorithm on assigning their preferred supervisors to them. 

Analogously, fairness for supervisors can be defined as equal treatment of matched students. In other words, the average qualification of all students assigned to each supervisor should be equal. As an unfair situation, for instance, when one supervisor is assigned students with the highest qualification (i.e.\ high utility), while another supervisor is assigned students with the lowest qualification (i.e.\ low utility), this will raise the issue of unfairness which needs to be addressed. Therefore, based on the utility defined for supervisors in Equation~\ref{sup_utility}, one may seek to equalize the utility for all supervisors to achieve the supervisor fairness. Considering the utility values of all supervisors as distribution over these utilities, we need to calculate the uniformity of this distribution (how close the values are).

There are various ways of measuring the uniformity of a distribution such as standard deviation, Entropy, and Gini Index. Here, I use entropy to measure how close the utility of supervisors is and can be calculated as:

\begin{equation}\label{ent}
        Entropy=- \sum_{q=1}^{|Q|} {U_q^{supervisor} }\log U_q^{supervisor}
\end{equation}

\noindent where $U_q^{supervisor}$ is the utility of supervisor $q$ calculated by Equation~\ref{sup_utility}.

In this formulation, equal treatment of supervisors in terms of their utility calculated by Equation~\ref{sup_utility} is considered. In other words, no matter what the merit of supervisors is, it defines fairness as equalizing their utility. However, another way of defining fairness is equal treatment of supervisors by considering their merit. To do so, entropy will be computed over utilities defined by Equation~\ref{ent}: instead of $U_q^{supervisor}$, $PU_q^{supervisor}$ is used for calculating entropy. The supervisor equality using $PU_q^{supervisor}$ can be calculated as follows:

\begin{equation}\label{ent_prop}
        Entropy^{prop}=- \sum_{q=1}^{|Q|} {PU_q^{supervisor} }\log PU_q^{supervisor}
\end{equation}

\section{Matching models}

In this section, various matching models will be discussed. The matching models include optimizing only for student utility, optimizing only for supervisor utility, and optimizing for both students and supervisors utility. I implemented all data processing infrastructure and algorithms
in Python using the Python interface of the Gurobi
Software\footnote{http://www.gurobi.com} for solving the optimizations.

\subsection{Optimizing for student utility}\label{opt_only_student}

This optimization seeks to maximize students' utility and can be computed as:

\begin{equation}
    \begin{aligned}
        \max_{p\in P} \quad & U_p^{student}
\end{aligned}    
\end{equation}

\noindent where $U_p^{student}$ is the utility of student $p$ calculated by Equation~\ref{stu_utility}. In other words, this Equation aims to match student to the supervisor based on student preference as much as possible. More precisely, this Equation can be written as:

\begin{equation}\label{stu_opt}
    \begin{aligned}
        \min_{p \in P} \quad & Rank(A(p),R_p)\\
        \textrm{s.t.} \quad & |A(p)|=1 \\ &\forall q, q_a \leq |A'(q)| \geq q_b
\end{aligned}    
\end{equation}

\noindent where $Rank(A(p),R_p)$ returns the rank of assigned supervisor to student $p$ (i.e.\ $A(p)$) in ranked preferred list of supervisors specified by $p$. There are two constraints: $|A(p)|=1$ which specifies that only one supervisor can be assigned to each student, and $\forall q, q_a \leq |A'(q)| \geq q_b$ specifies that the number of students assigned to each supervisor should follow the supervisors' capacity.

\subsection{Optimizing for supervisor utility}\label{opt_only_supervisor}

This optimization seeks to maximize supervisors' utility and can be computed as:

\begin{equation}\label{sup_opt}
    \begin{aligned}
        \max_{q\in Q} \quad & {U^{supervisor}_q}\\
        \textrm{s.t.} \quad & |A(p)|=1 \\ &\forall q, q_a \leq |A'(q)| \geq q_b
\end{aligned}    
\end{equation}

\noindent where $U^{supervisor}_q$ is the utility of supervisor $q$ and can be computed as Equation~\ref{sup_utility}.

\subsection{Optimizing for both sides}\label{opt_only_both}

This optimization seeks to maximize both students and supervisors' utility by simultaneously taking into account the rank of matched supervisor to a student and qualification of students, and can be computed as:

\begin{equation}\label{supstu_opt}
    \begin{aligned}
        \max_{p \in P, q \in Q} \quad & \lambda \times \frac{|R_p|-Rank(q,R_p)}{|R_p|} + (1-\lambda)(1-X_p)\\
        \textrm{s.t.} \quad & |A(p)|=1 \\ &\forall q, q_a \leq |A'(q)| \geq q_b
\end{aligned}    
\end{equation}

\noindent where $\frac{|R_p|-Rank(q,R_p)}{|R_p|}$ is the utility for student $p$ similar to Equation~\ref{stu_utility}, $X_p$ is the qualification of student $p$, and $\lambda$ is a hyperparameter for controlling the trade-off between maximizing the utility of students and supervisors. The constraints have the same definition as Equation~\ref{stu_opt}.

\input{./Tables/tbl_simulation_results}

\section{Experiments}

Experiments are performed using the optimization models for students, supervisors, and both introduced in subsection \ref{opt_only_student}. Also, the results for each optimization models are evaluated using utility and fairness metrics introduced in subsection \ref{utility_metric} for students and supervisors. The experimental results are reported in Table~\ref{tab:simulation_results}. 

\begin{figure*}[t]
    \centering
    \begin{subfigure}[b]{0.98\textwidth}
        \includegraphics[width=\textwidth]{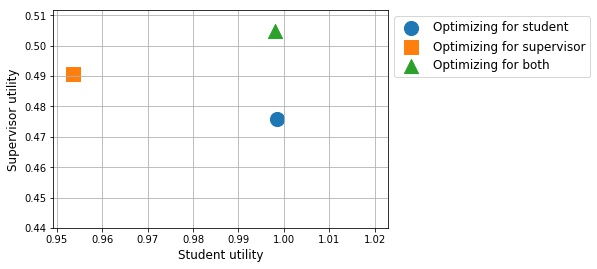}
    \end{subfigure}
\caption{Trade-off between student utility (Equation~\ref{stu_utility}) and supervisors utility (Equation~\ref{sup_utility}) with various optimization models.} \label{utility_stu_sup}
\end{figure*}

\begin{figure*}[t]
    \centering
    \begin{subfigure}[b]{0.98\textwidth}
        \includegraphics[width=\textwidth]{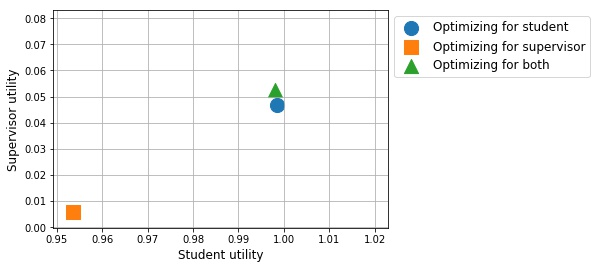}
    \end{subfigure}
\caption{Trade-off between student utility (Equation~\ref{stu_utility}) and supervisors proportional utility (Equation~\ref{sup_utility_prop}) with various optimization models.} \label{utility_stu_sup_prop}
\end{figure*}

\begin{figure*}[t]
    \centering
    \begin{subfigure}[b]{0.98\textwidth}
        \includegraphics[width=\textwidth]{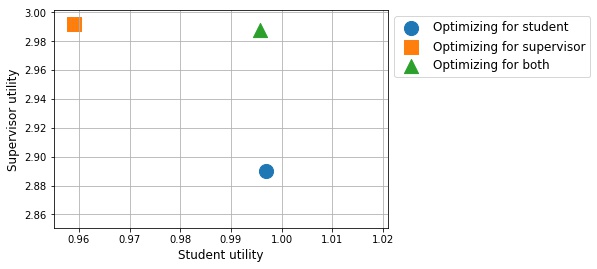}
    \end{subfigure}
\caption{Trade-off between student fairness (Equation~\ref{stu_fairness}) and supervisors fairness (Equation~\ref{ent}) with various optimization models.} \label{fairness_stu_sup}
\end{figure*}

\begin{figure*}[t]
    \centering
    \begin{subfigure}[b]{0.98\textwidth}
        \includegraphics[width=\textwidth]{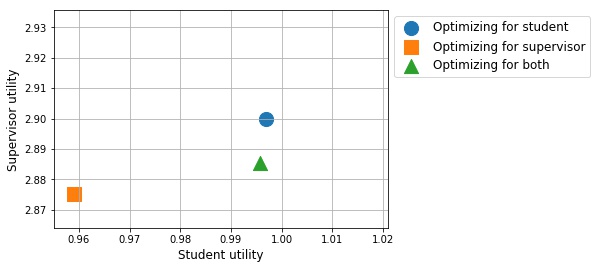}
    \end{subfigure}
\caption{Trade-off between student fairness (Equation~\ref{stu_fairness}) and supervisors proportional fairness (Equation~\ref{ent_prop}) with various optimization models.} \label{fairness_stu_sup_prop}
\end{figure*}

The results show that although optimizing only for one side (only student or supervisor) improves the utility and fairness for that side, it will negatively affect the utility and fairness of the other side. As shown in Table~\ref{tab:simulation_results}, optimizing only for students yields the highest utility for students (i.e.\ 0.9985) and the lowest utility for supervisors (i.e.\ 0.4759). Also, optimizing only for students yields the same results based on fairness: the highest fairness for students (i.e.\ 0.9969) and the lowest fairness for supervisors (i.e.\ 2.890). The same patterns can also be observed in Table~\ref{tab:simulation_results} when only optimizing for supervisors. However, optimizing for both sides yields better results for both sides. This means that optimizing for both sides balances the outcome for both students and supervisors by not greedily only taking into account improving the utilities of one side and losing utility for the other side.    
Figures~\ref{utility_stu_sup}, \ref{utility_stu_sup_prop}, \ref{fairness_stu_sup}, and \ref{fairness_stu_sup_prop} show the trade-off between the utilities and fairness for students and supervisors using various optimization models. The horizontal axis shows the utility/fairness for students and vertical axis shows the utility/fairness for supervisors. For all metrics the higher values are desired (upper right in the plots has better results, higher utility/fairness for both sides). 

In all these plots, it can be observed that optimizing for students (blue circle) achieves the highest utility/fairness for students and lowest utility/fairness for supervisors. Also, optimizing for supervisors (orange square) achieves the highest utility/fairness for supervisors and lowest utility/fairness for students (except for supervisor utility in Figures \ref{utility_stu_sup_prop} and \ref{fairness_stu_sup_prop} that is due to the definition of proportional utility). However, when simultaneously optimizing for both sides (green triangle) yields fairly high utility/fairness for both sides. 

The results shown in this simulation show the importance of having multi-sided perspective for optimizing the utility and fairness of actors in a multi-sided platform.

\section{Discussion}

This simulation showed the importance of multi-sided view when optimizing an objective for multiple stakeholders in the system. In this simulation, I studied the assignment of students and supervisors in universities. To simplify the simulation and avoid any possible noise, I made several assumptions. However, there are several other settings that can be considered to further improve the simulation and to make it closer to real-world scenarios.

Incorporating the topic of interest to the simulation for students and supervisors  can reveal interesting patterns. Supervisors usually have certain expertise and their knowledge falls into a specific topics in a field, and also students have certain research interests and prefer to work with a supervisor on those topics. Thus, a constraint for topics of interest for supervisors and students needs to be added to the optimization objective. This constraint controls the topic of interest for students and supervisors when forming an assignment and assigns students to supervisors that are interested in the same topics.

Incorporation of research interests for students and supervisors can also add new challenges to the simulation, specially to the synthetic data generation process. For example, a student can have an excellent fit for topic $A$, less so for topic $B$, and not at all for topic $C$. Then, the challenge in data generation phase is how to generate such topic interests for students that also represents the real-world scenario for the whole population. Also, defining utility and fairness for students and supervisors in this situation can be challenging. I leave these challenges as a future work for this dissertation. 

%% file: Tables/tbl_simulation_notation.tex
\captionsetup[table]{skip=4pt}
\begin{table}[t!]
\small
\centering
\captionof{table}{Summary of notation used in the simulation} \label{tab:simulation_notations}
\begin{tabular}{p{2cm}p{5cm}}
    \toprule
     Notation & Description \\
     \midrule
     $P$ & The set of all students \\
     \midrule
     $Q$ & The set of all supervisors \\
     \midrule
     $R$ & The ranked list of preferred supervisors specified by students. $R_p$ is the list of preferred supervisors for student $p$ \\
    \midrule
    $X$ & Students qualification in the range of $[0,1]$. $X_p$ is the qualification of student $p$ \\
    \midrule
    A(p) & Returns the supervisor assigned to student $p$ \\
    \midrule
    ${A'}(q)$ & returns the students assigned to supervisor $q$ \\
    \midrule
    $Rank(q,R_p)$ & Returns the rank of supervisor $q$ in ranked list of preferred supervisors by student $p$ \\
    \midrule
    $q_a,q_b$ & The capacity of supervisor $q$ for supervising students, $q_a$ as the minimum number and $q_b$ as the maximum number of students that $q$ can supervise \\
    \bottomrule
\end{tabular}
\end{table}

%% file: Tables/tbl_simulation_results.tex
\captionsetup[table]{skip=4pt}
\begin{table}[t!]
\small
\centering
\captionof{table}{Utility and fairness of student-supervisor matching problem.} \label{tab:simulation_results}
\begin{tabular}{p{3cm}p{2cm}p{2cm}p{2.2cm}}
    \toprule
    \multirow{2}{*}{Metrics} & \multicolumn{3}{c}{Optimization models} \\\cline{2-4}
    
      & Student (equation \ref{stu_opt}) & Supervisor (equation \ref{sup_opt}) & Both sides   (equation \ref{supstu_opt})  \\
     \midrule
     $U^{student}$ & 0.9985 & 0.9536 & 0.9980 \\
     $equality$ & 0.9969 & 0.9590 & 0.9956 \\
     $U^{supervisor}$ & 0.4759 & 0.4906 & 0.5047 \\
     $PU^{supervisor}$ & 0.0467 & 0.0059 & 0.0525 \\
     $Entropy$ & 2.890 & 2.992 & 2.987 \\
     $Entropy^{prop}$ & 2.899 & 2.875 & 2.885 \\
    \bottomrule
\end{tabular}

\end{table}

%% file: Chapters/04_ExposureBias.tex
\chapter{Multi-Sided Exposure Bias in Recommendation}
\label{chap:expo}
\tikzsetfigurename{expo_}

There are certain types of bias in recommender systems that adversely impact the performance of these systems and distort the recommendation process. One type of these biases is \textit{Exposure Bias} that causes skew in the representation/distribution of recommended items and suppliers in recommendation lists. Exposure bias refers to the fact that some items or suppliers appear frequently in the recommendation lists and some other items or suppliers have appeared rarely. In other words, frequently recommended items (items belong to those suppliers) will be exposed to many users even if those items are not matched with the users preferences. This way, exposure bias can impact various actors in the recommendation systems. 

In this chapter, I discuss exposure bias in recommender systems with multi-sided perspective and its impact on various actors in the system. I also empirically show how existing recommendation algorithm are affected by exposure bias. Analysis and contributions in this chapter are published in \cite{masoud2021wsdm} and \cite{mansoury2021tois}\footnote{Accepted in ACM Transactions on Information Systems (TOIS)}.

\section{Types of bias in recommender systems}

Various types of bias are recognized in recommender systems \cite{chen2020bias}. In this Section, I briefly review well-known biases in recommender systems.

\begin{itemize}
    \item \textbf{Selection bias: }Selection bias refers to the fact that users only observe and rate part of the whole items in the catalog and this partial observation is not a proper representation of all ratings. In fact, the system only has information about the rated items and there is no information about the rest of the items. It is not clear unobserved items are either positive or negative samples for a user. This type of bias comes from the input data where users' interaction with the system are collected \cite{hernandez2014probabilistic,marlin2012collaborative,steck2013evaluation}.
    \item \textbf{Conformity bias: }Conformity bias refers to the fact that the ratings provided by users do not always represents users' true preferences as users usually tend to follow the opinions of the majority when rating an item. This means that if users may rate an item similar to other users even if they find it against their true preferences. This type of bias comes from the input data and can end up with inaccurate recommendations for users as ratings are not perfectly matched with users' preferences \cite{wang2014amazon,liu2016you,krishnan2014methodology}.
    \item \textbf{Exposure bias: }Exposure bias refers to the fact that only few items or suppliers have chance to appear in the recommendation lists to the users and other items and suppliers may not receive proportionate attention. This skew in representation of items and suppliers raises the issue of unfair treatment of items and suppliers. Although this is an issue in item and supplier sides, it can also negatively affect users as users are only exposed to specific sets of items or suppliers which may not properly captures the interests of them. This type of bias usually originates from the data due to the inherent popularity bias in the input data and is even intensified by the recommendation algorithm.  \cite{abdollahpouri2020multi,abdollahpouri2020unfair,zheng2020disentangling}. 
    \item \textbf{Position bias: }Position bias refers to the fact that users tend to interact with the few items on top of the recommendation lists and may ignore the rest of the items in the lists. Those ignored items are usually considered as low-quality ones by the recommendation algorithms as user did not show an interest to them, but the fact is that users usually do not even examine those items (i.e.\ neither like nor dislike). Although this is a result from low effort from user side that does not examine the whole list, recommendation algorithms can be designed to avoid this bias \cite{collins2018study,o2006modeling,klockner2004depth}. As a solution, for instance, recommendation algorithms can mitigate this bias by providing equal chance to different items to appear on top of the recommendations lists delivered to the users. As another solution used in \textit{Cascading Bandit} algorithms \cite{zong2016cascading,hiranandani2020cascading,li2020cascading,mansoury2021unbiased}, the recommendation algorithms can be designed to consider the items on the bottom of the list as unobserved, not as disliked.  
    \item \textbf{Popularity bias: }Popularity bias refers to the fact that few popular items are frequently recommended and majority of unpopular items are rarely recommended. This is a serious issue as popular items are recommended even more that what their popularity in rating data warrants. This over-recommendation of popular items and under-recommendation of unpopular items would also intensify the inherent popularity bias in input data over time as users interact with the system and add the recommended items to their profile. This type of bias originates from the input data as users interact with the popular items more than unpopular items and recommendation algorithm also intensifies this bias by over-recommending popular items to the users \cite{harald2011,abdollahpouri2020multi,himan2019c}.
\end{itemize}

In this thesis, I mainly focus on exposure bias and popularity bias to address the unfairness in recommendation results by mitigating those types of biases.

\section{Exposure bias in recommendations}

It is well-known that recommendation algorithms favor popular items which leads to an unfair exposure of other items that might not be as popular \cite{harald2011,abdollahpouri2017controlling}. This bias towards popular items can negatively affect the less popular items, items that are new to the system (aka cold start items), and even the supplier of the items \cite{himan2019a,patro2020fairrec}. In this Section, I illustrate the exposure bias of several recommendation algorithms from both the items and suppliers perspective.

\begin{figure*}[btp]
    \centering
    \begin{subfigure}[b]{0.33\textwidth}
        \includegraphics[width=\textwidth]{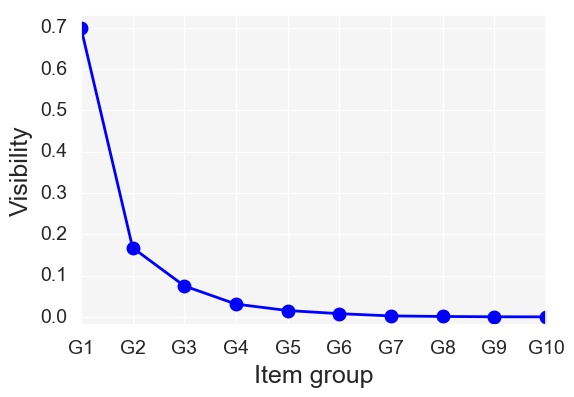}
        \caption{BPR} \label{dist_ml_50_item_bpr}
    \end{subfigure}
    \begin{subfigure}[b]{0.33\textwidth}
        \includegraphics[width=\textwidth]{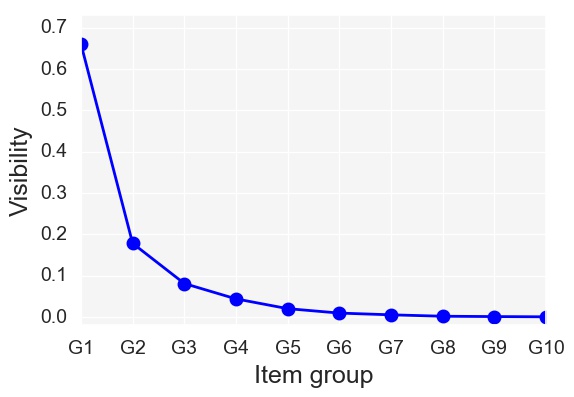}
        \caption{NCF} \label{dist_ml_50_item_ncf}
    \end{subfigure}%
    \begin{subfigure}[b]{0.33\textwidth}
        \includegraphics[width=\textwidth]{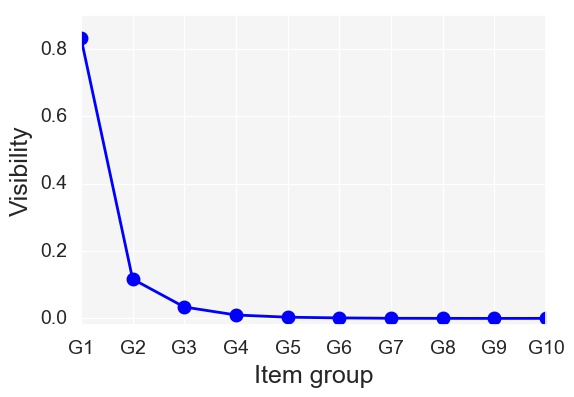}
        \caption{UserKNN} \label{dist_ml_50_item_userknn}
    \end{subfigure}%
\caption{Visibility of recommended items for different recommendation algorithms on ML1M dataset.} \label{dist_ml_50_item}
\end{figure*}

\subsection{Bias in item exposure}

An exposure for an item is the percentage of the times it has appeared in the recommendations \cite{khenissi2020modeling,singh2018fairness}. Recommendation algorithms are often biased towards more popular items giving them more exposure than many other items. Figure~\ref{dist_ml_50_item} shows the visibility of different items in the recommendations produced by three recommendation algorithms \algname{NCF}, \algname{UserKNN}, and \algname{BPR}. Items are binned into ten groups based on their visibility in recommendation lists. We can see that in all three algorithms, there is a long-tail shape for the visibility of the items indicating few popular item groups are recommended much more frequently than the others creating an item exposure bias in the recommendations. Not every algorithm has the same level of exposure bias for different items. For instance, we can see that \algname{UserKNN} has recommended items from group $G_1$ to roughly 80\% of the users while this number is near 70\% and 65\% for \algname{BPR} and \algname{NCF}, respectively. On the other hand, $G_2$ has received less exposure in \algname{UserKNN} (10\%) compared to \algname{BPR} and \algname{NCF} which have given 17\% and 19\% visibility to items in this group, respectively. 

\subsection{Bias in supplier exposure}

\begin{figure*}[btp]
    \centering
    \begin{subfigure}[b]{0.33\textwidth}
        \includegraphics[width=\textwidth]{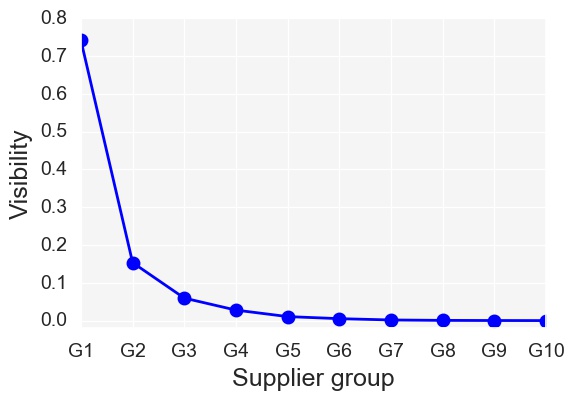}
        \caption{BPR} \label{dist_ml_50_sup_bpr}
    \end{subfigure}
    \begin{subfigure}[b]{0.33\textwidth}
        \includegraphics[width=\textwidth]{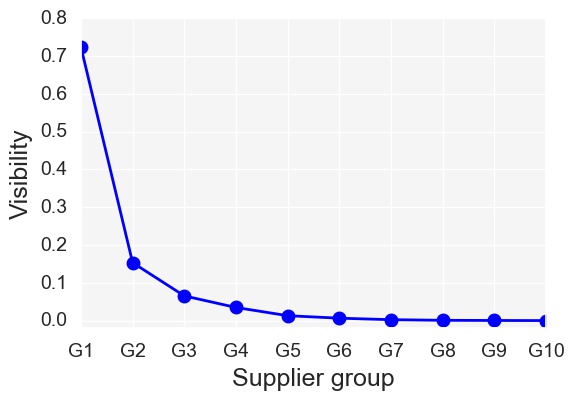}
        \caption{NCF} \label{dist_ml_50_sup_ncf}
    \end{subfigure}%
    \begin{subfigure}[b]{0.33\textwidth}
        \includegraphics[width=\textwidth]{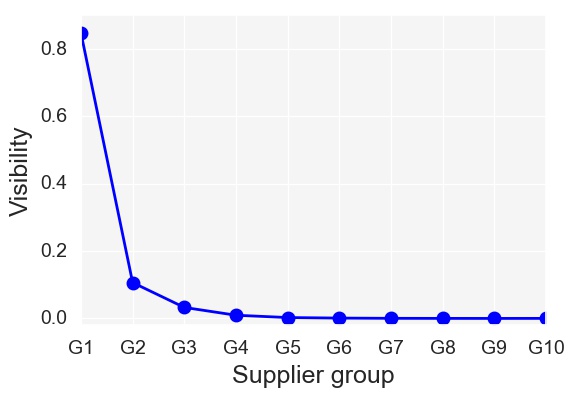}
        \caption{UserKNN} \label{dist_ml_50_sup_userknn}
    \end{subfigure}%
\caption{Visibility of suppliers for different recommendation algorithms on ML1M dataset.} \label{dist_ml_50_sup}
\end{figure*}

The unfair exposure does not only affect the items in a recommender system. We know that in many recommendation platforms the items that are candidates to be recommended are provided by different suppliers. Therefore, the dynamic of how recommendation algorithms can impact the experience of the suppliers is also crucial. Authors in \cite{abdollahpouri2020multi} empirically show that recommendation algorithms often over-promote items from popular suppliers while suppressing the less popular ones. Figure~\ref{dist_ml_50_sup} shows a similar plot to Figure~\ref{dist_ml_50_item} but for the suppliers of the items. Similar to items, suppliers are binned into ten groups based on their visibility in recommendation lists. The same problem that we observed in Figure~\ref{dist_ml_50_item} also exists here: in all three algorithms, there is a long-tail shape for the visibility of the suppliers indicating few supplier groups are recommended much more frequently than the others. 

There are many existing works for improving the visibility of different items in the recommendations and reducing the exposure bias in items. However, the same cannot be said about the suppliers and there has not been much attention to improving the supplier visibility/exposure. Although, improving the item visibility can, indirectly, help suppliers as well as it was demonstrated in \cite{abdollahpouri2020addressing}, a more explicit incorporation of suppliers in the recommendation process can yield fairer outcomes for different suppliers in terms of visibility. 

This dissertation aims to address this problem by directly incorporating the suppliers in the recommendation process to mitigate the exposure bias from the suppliers perspective.

\section{Exposure bias mitigation techniques}

The concept of popularity bias has been studied by many researchers often under different names such as long-tail recommendation \cite{abdollahpouri2017controlling,yin2012challenging}, Matthew effect \cite{moller2018not}, and aggregate diversity \cite{liu2015trust,adomavicius2011improving} all of which refer to the fact that the recommender system should recommend a wider variety of items across all users. 



Authors in \cite{abdollahpouri2017controlling} proposed a regularization term to control the popularity of recommended items that could be added to an existing objective function of a learning-to-rank algorithm \cite{karatzoglou2013learning} to improve the aggregate diversity of the recommendations. In another work, Vargas and Castells in \cite{vargas2011rank} proposed probabilistic models for improving novelty and diversity of recommendations by taking into account both relevance and novelty of target items when generating recommendation lists. Moreover, authors in  \cite{vargas2014improving}, proposed the idea of recommending users to items for improving novelty and aggregate diversity. They applied this idea to nearest neighbor models as an inverted neighbor and a factorization model as a probabilistic reformulation that isolates the popularity components. 


Adomavicius and Kwon \cite{adomavicius2011maximizing} proposed the idea of diversity maximization using a maximum flow approach. They used a specific setting for the bipartite recommendation graph in a way that the maximum amount of flow that can be sent from a source node to a sink node would be equal to the maximum aggregate diversity for those recommendation lists. In their setting, given the number of users is $m$, the source node can send a flow of up to $m$ to the left nodes, left nodes can send a flow of up to 1 to the right nodes, and right nodes can send a flow of up to 1 to the sink node. Since the capacity of left nodes to right nodes is set to 1, thus the maximum possible amount of flow through that recommendation bipartite graph would be equivalent to the maximum aggregate diversity. 

A more recent graph-based approach for improving aggregate diversity which also falls into the reranking category was proposed by Antikacioglu and Ravi in \cite{antikacioglu2017}. They generalized the idea proposed in \cite{adomavicius2011maximizing} and showed that the minimum-cost network flow method can be efficiently used for finding recommendation subgraphs that optimizes the diversity. In this work, an integer-valued constraint and an objective function are introduced for discrepancy minimization. The constraint defines the maximum number of times that each item should appear in the recommendation lists and the objective function aims to find an optimal subgraph that gives the minimum discrepancy from the constraint. This work shows improvement in aggregate diversity of the items with a smaller accuracy loss compared to the work in \cite{vargas2011rank} and \cite{vargas2014improving}. Our algorithm is also a graph-based approach that not only is it able to improve aggregate diversity and the exposure fairness of items, it also gives the suppliers of the recommended items a fairer chance to be seen by different users. Moreover, unlike the work in \cite{antikacioglu2017} which tries to minimize the discrepancy between the distribution of the recommended items and a target distribution, our FairMatch algorithm has more freedom in promoting high-quality items or suppliers with low visibility since it does not assume any target distribution of the recommendation frequency.

Another work that also uses a re-ranking approach is by Abdollahpouri et al. \cite{abdollahpouri2019managing} where authors proposed a diversification method for improving aggregate diversity and long-tail coverage in recommender systems. Their method was based on \textit{eXplicit Query Aspect Diversification} (xQuAD) algorithm \cite{santos2010exploiting} that was designed for diversifying the query result such that it covers different aspects related to a given query. In \cite{abdollahpouri2019managing}, the authors used xQuAD algorithm for balancing the ratio of popular and less popular (long-tail) items in final recommendation lists. 

In addition, in \cite{mansoury2020fairmatch}, I proposed a graph-based algorithm that finds high quality items that have low visibility in the recommendation lists by iteratively solving the maximum flow problem on recommendation bipartite graph. An extended version of this algorithm that also takes into account the supplier side exposure is presented in Chapter~\ref{chap:solution2}. 

In addressing exposure bias in domains like job recommendations where job seekers or qualified candidates are recommended, Zehlike et. al. \cite{zehlike2017fa} proposed a re-ranking algorithm to improve the ranked group fairness in recommendations. 
The algorithm creates queues of protected and unprotected items and merges them using normalized scoring such that protected items get more exposure. In their setting, a recommendation set of candidates satisfies the ranked group fairness criterion if it fairly represents candidates belong to the protected group, contains the most qualified candidates, and orders the candidates based on their qualification (more qualified candidates should appear on top of the recommendation set). The fair representation of protected candidates in a recommendation list is determined by comparing the number of protected candidates in the ranked list and the expected number of candidates if they were selected at random. The algorithm optimizes to re-rank the candidates to achieve the fair representation of protected candidates in recommendation lists while considering the qualification of the candidates.


Most of the existing works in the literature for improving aggregate diversity and exposure fairness have only concentrated on the items and ignored the fact that in many recommendation domains the recommended items are often provided by different suppliers and hence their utility should also be investigated. To the best of our knowledge, there are only few prior works that have addressed this issue such as \cite{abdollahpouri2020addressing} and \cite{mehrotra2018towards}. In \cite{abdollahpouri2020addressing}, authors illustrated how popularity bias is a multistakeholder problem and hence they evaluated their solution for mitigating this bias from the perspective of different stakeholders. Mehrotra et al. \cite{mehrotra2018towards} investigated the trade-off between the relevance of recommendations for users and supplier fairness, and their impacts on users' satisfaction. Relevance of the recommended items to a user is determined by the score predicted by a recommendation algorithm. To determine the supplier fairness in recommendation list, first, suppliers are grouped into several bins based on their popularity in rating data and then the supplier fairness of a recommendation list is measured as how diverse the list is in terms of covering different supplier popularity bins. 

The work in this dissertation also observes the importance of evaluating algorithms from the perspective of multiple stakeholders and I propose algorithms that can directly improve the visibility of the suppliers without losing much accuracy from the users' perspective.

\subsection{Limitations of existing bias mitigation techniques}

Various algorithmic solutions have been proposed to tackle the unfairness in recommender systems. However, there are several limitations in those solutions. In this Section, I discuss the limitations of existing solutions for addressing unfairness in recommender systems.

First, a large body of existing solutions has one-sided view on addressing the unfairness of recommendation systems, meaning that they optimize to improve the fairness of one actor, overlooking the fairness of other actors in the system. For example, research works in \cite{burke2017,edizel2020,zehlike2017fa,ekstrand2012,yao2017,ekstrand2018} optimize for user-side fairness or research works in \cite{abdollahpouri2019managing,wang2016multi,abdollahpouri2017controlling,antikacioglu2017,vargas2014improving,adomavicius2011maximizing,vargas2011} only optimize for item-side fairness. Although these works showed improvement on one side, it is not clear how they are affecting other sides, most likely those approaches adversely affected the other sides of the system . It has been shown that only improving the fairness of one side will hurt the fairness of the other sides \cite{mehrotra2018towards}.   

Second, existing solutions with multi-sided view usually optimize to generate a list of recommendations for each user to contain items from different suppliers while maintaining the relevance of items for each user \cite{mehrotra2018towards,suhr2019two}. In these approaches, the proposed models optimize to locally improve the supplier fairness for the list generated for one user in the hope that improving supplier fairness separately for a user's list results in global optimum for the whole recommendation lists. However, I argue that optimizing for supplier fairness requires a holistic view over the whole recommendation lists, not the recommendation list for a user. For instance, consider a recommendation algorithm that recommended 10 items from 10 different suppliers to each user. In this situation, the recommendation list for each user will have items from different suppliers, while the overall recommendation lists only have items from 10 suppliers which is unfair against other suppliers. Thus, it is important to keep track of supplier fairness over the whole recommendation lists. 

Third, some of the existing solutions are designed to specifically address the fairness in a domain and cannot be generalized to fairness definitions of other domains \cite{mehrotra2018towards,yao2017,steck2018calibrated}. 

For example, Mehrota et al., in \cite{mehrotra2018towards} used equality of attention as the utility of supplier and the relevance score of target item predicted by the base recommender for target user as the utility of users. As another example, S{\"u}hr et. al., in \cite{suhr2019two} addressed multi-sided fairness in ride-hailing platforms as a case study for their research. They designed a multi-sided objective function to address the fairness for both passengers and drivers by specific definition for each side. However, fairness is a general concept and may have different definitions depending on the application and the domain under study. Therefore, a flexible algorithmic solution with capability of being generalized for different notion of fairness is needed.

%% file: Chapters/05_Methodology.tex
\part{Mitigating Multi-Sided Exposure Bias in Recommender Systems}
\chapter{Experimental Methodology}
\label{chap:methodology}
\tikzsetfigurename{methodology_}

In this chapter, I introduce the datasets, recommendation algorithms, and re-ranking techniques that are used for performing the experiments. The details about the experiments such as the choice of model, hyperparameter tuning, and comparison would be discussed. To prepare the datesets for my experimental design, I used online APIs from different datasets to collect necessary information. As part of my contributions in this dissertation, in section \ref{metrics}, I discuss the limitations of existing metrics for measuring the exposure bias in recommender systems and introduce appropriate metrics for evaluating the exposure bias of recommendations results. These contributions are published in \cite{mansoury2021tois}\footnote{Accepted in ACM Transactions on Information Systems (TOIS)}. Finally, I introduce a recommendation tool, \textit{librec-auto}\footnote{The source code can be found in \url{https://github.com/that-recsys-lab/librec-auto} and the documentation can be found in \url{librec-auto.readthedocs.io}.}, that I worked on during my PhD program. I used this tool for my experimentation in this dissertation. My contributions on this tool are published in \cite{mansoury2019algorithm,masoud2018,sonboli2020fairness,burke2020experimentation}.

\section{Data}

Experiments are performed on four publicly available datasets: Last.fm\footnote{http://www.cp.jku.at/datasets/LFM-1b/} \cite{schedl2016lfm}, two versions of MovieLens \cite{Harper:2016}, 
and Goodreads\footnote{https://www.kaggle.com/bahramjannesarr/goodreads-book-datasets-10m}. The characteristics of the datasets are summarized in Table~\ref{tab:dataset}.

Last.fm dataset contains user interactions with songs (and the corresponding albums). I used the same methodology in \cite{kowald2020unfairness} to turn the interaction data into rating data using the frequency of the interactions with each item (more interactions with an item will result in higher rating). In addition, I used albums as the items to reduce the size and sparsity of the item dimension, therefore the recommendation task is to recommend albums to users. I considered the artists associated with each album as the supplier of that album. In pre-processing step, I removed users with less than 50 ratings and items less than 200 ratings to create a denser dataset and then, I randomly sampled 2,000 users from the data. 

\input{./Tables/tbl_data}

The MovieLens dataset is a movie rating data and was collected by the GroupLens\footnote{https://grouplens.org/datasets/movielens/} research group. I considered the movie-maker associated with each movie as the supplier of that movie. Since this dataset does not originally contain information about the movie-makers, I used the API provided by OMDB (not to be confused with IMDB) website\footnote{http://www.omdbapi.com/} to extract the information about movie-makers associated with different movies. I created a subset of MovieLens dataset by filtering out the movies that information about their movie-makers was not found. Therefore, there are two different versions of MovieLens dataset in this dissertation which I distinguish them by MovieLens1M which refers to the original one without movie-makers information and ML1M which refers to the sampled data with movie-makers information.


Finally, Goodreads dataset contains users' feedback on books. In this dataset, each user has rated at least 10 books and each book is rated by at least 10 users. Also, publishers of the books are considered as the suppliers.

\begin{figure*}[btp]
    \centering
    \begin{subfigure}[b]{0.7\textwidth}
        \includegraphics[width=\textwidth]{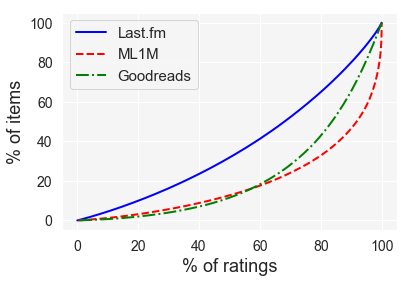}
    \end{subfigure}
\caption{Distribution of item popularity for Last.fm, ML1M, Goodreads, and Bookcrossing datasets. Items are ordered according to popularity (most popular at the bottom).} \label{train_dist}
\end{figure*}

These datasets are from different domains, have different levels of sparsity, and are different in terms of popularity distribution of different items. Figure~\ref{train_dist} shows the distribution of item popularity for all datasets. 
Vertical axis shows the percentage of items in the datasets ordered according to their popularity, most popular at the bottom. Horizontal axis shows the percentage of ratings that are assigned to items. It can be observed that these datasets have different characteristics in terms of popularity. In Goodreads dataset, only 4.7\% of popular items have collectively taken 33\% of the ratings in the datasets. 
The distribution in ML1M dataset is slightly less long-tailed such that 6.5\% of the items have collectively taken 33\% of the ratings.
Finally, Last.fm dataset shows even lesser long-tail properties in which 18.3\% of items have collectively taken 33\% of the ratings.  

\begin{figure*}[btp]
    \centering
    \begin{subfigure}[b]{0.33\textwidth}
        \includegraphics[width=\textwidth]{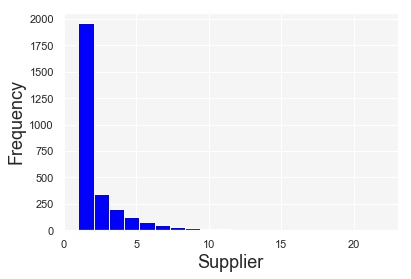}
        \caption{Last.fm} \label{lf-hist}
    \end{subfigure}
    \begin{subfigure}[b]{0.33\textwidth}
        \includegraphics[width=\textwidth]{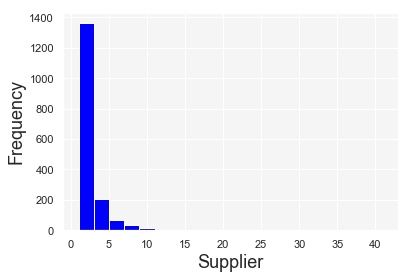}
        \caption{ML1M} \label{ml-hist}
    \end{subfigure}%
    \begin{subfigure}[b]{0.33\textwidth}
        \includegraphics[width=\textwidth]{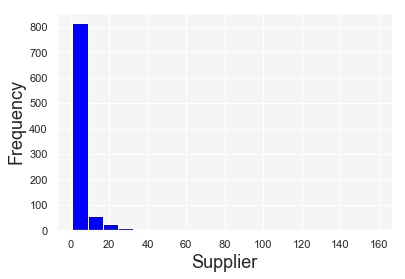}
        \caption{Goodreads} \label{lf-hist}
    \end{subfigure}
\caption{Histogram of suppliers inventory (number of items each supplier owns).} \label{hist}
\end{figure*}

Also it is worth noting that different suppliers do not own the same number of items as we can see in Figure~\ref{hist} where the majority of suppliers have only one item. Because of this, it will be seen that both versions of the proposed FairMatch algorithm (Chapter~\ref{chap:solution2}) perform relatively similar in some cases since improving item visibility for those items that belong to suppliers with only one item is indeed equivalent to improving the visibility of the corresponding supplier.


\section{Setup}

For experiments, I used 80\% of each dataset as the training set and the other 20\% for the test. The training set was used for building a recommendation model and generating recommendation lists, and the test set was used for evaluating the performance of generated recommendations. As mentioned earlier, in this dissertation, two solutions for mitigating exposure bias in recommender systems are proposed: pre-processing and post-processing solutions. 

In pre-processing solution, the pre-processed training set is used as input for a recommendation algorithm, while in post-processing solution, longer recommendation lists generated by a recommendation algorithm is processed to generate the final shorter recommendation lists. In other words, I generated recommendation lists of size $t=50$ (longer recommendation lists) for each user using each recommendation algorithm. I then extract the final recommendation lists of size $n=10$ using the proposed and each reranking method by processing the recommendation lists of size 50. 
I used \textit{librec-auto} and LibRec for running the experiments \cite{masoud2018,mansoury2019algorithm,Guo2015}. 

The recommendation algorithms for generating the recommendation lists of size 10 in pre-processing solution are Biased Matrix Factorization (\algname{BiasedMF}) \cite{Koren:2009a}, Singular Value Decomposition (\algname{SVD++}) \cite{Koren:2008a}, and List-wise Matrix Factorization (\algname{ListRankMF}) \cite{shi2010list}. The recommendation algorithms for generating the longer recommendation lists of size 50 in post-processing solution are Bayesian Personalized Ranking (\algname{BPR}) \cite{rendle2009bpr}, Neural Collaborative Filtering (\algname{NCF}) \cite{he2017neural}, User-based Collaborative Filtering (\algname{UserKNN}) \cite{Resnick:1994a}. I chose these algorithms to cover different approaches in recommender systems: matrix factorization, neural networks, and neighborhood models. 

The reason why different recommendation algorithms are used for each solution is due to the limitations of the proposed pre-processing approach. The pre-processing solution is a rating transformation technique that only works on recommendation algorithms that use rating values for their internal processing/optimization. For instance, implicit feedback algorithms that use unary data, such as \algname{BPR}, would be inappropriate to use with the proposed pre-processing approach because they use binary interaction information and ignore rating values. Therefore, only recommendation algorithms that use rating values are used for experiments on pre-processing approach. On the other hand, the proposed post-processing approach has more flexibility on the choice of recommendation algorithm and for this reason, I chose more advanced and diverse sets of algorithms for experiments\footnote{Improvement over other recommendation algorithms are also observed, though not reported. For instance, in \cite{mansoury2020fairmatch}, I showed that the proposed post-processing technique yields superior performance compared to other baselines using \algname{ListRankMF}.}.


Each recommendation algorithm involves several hyperparameters. To identify the best-performing sets of hyperparameters for each algorithms, I performed gridsearch on hyperparameters space and selected the results with the highest precision for the next analysis. Table~\ref{tab:conf} shows the hyperparameter values that gridsearch was performed.

\input{./Tables/4.2.tbl_model_conf}




\section{Evaluation metrics}\label{metrics}

For evaluation, we use the following metrics to measure different aspects of the effectiveness of each method:

\begin{enumerate}
    \item \textbf{Precision ($P$)}: The fraction of the recommended items shown to the users that are part of the users' profile in the test set.
    \item \textbf{Item Visibility Shift ($IVS$)}: The percentage of increase or decrease in the visibility of item groups in final recommendation lists generated by a reranking algorithm compared to their visibility in recommendation lists generated by a base recommender. Given long recommendation lists of size $t$, $L'$, generated by a base recommender and the visibility of each item $i$ computed as the fraction of times that it appears in the recommendation lists of different users 
    , I create 10 groups of items based on their visibility in $L'$. To do so, first, I sort the recommended items based on their visibility in $L'$ in descending order, and then I group the recommended items into 10 equal-sized bins where the first group represents the items with the highest visibility and 10th group represents the items with the lowest visibility in $L'$. 
    Item Visibility ($IV$) of each item $i$ in final recommendation lists can be calculated as:
    
    
    \begin{equation}\label{v_i}
        IV(i)=\frac{\sum_{j \in L}\mathds{1}(j=i)}{|L|}
    \end{equation}
    
    where $\mathds{1}(.)$ is the indicator function returning zero when its argument is False and 1 otherwise. Item Group Visibility ($IGV$) for each item group $\tau$ can be calculated as:
    
    \begin{equation}
        IGV({\tau})=\frac{\sum_{i \in \tau}{IV(i)}}{|\tau|}
    \end{equation}
    
    Therefore, Item Visibility Shift ($IVS$) of group $\tau$ can be calculated as:
    
    \begin{equation}\label{IVS_g}
        IVS(\tau)= \frac{IGV(\tau)^{Reranker}-IGV(\tau)^{Base}}{IGV(\tau)^{Base}}    
    \end{equation}
    
    where $IGV(\tau)^{Reranker}$ and $IGV(\tau)^{Base}$ are the visibility of item group $\tau$ in recommendation lists of size $n$ generated by reranking algorithm and the base algorithm, respectively.
    
    \item \textbf{Supplier Visibility Shift ($SVS$)}: The percentage of increase or decrease in the visibility of supplier groups in final recommendation lists generated by a reranking algorithm compared to their visibility in recommendation lists generated by a base recommender. $SVS$ can be calculated similar to $IVS$, but instead of calculating the percentage change over item groups, I calculate it over supplier groups in $SVS$. Thus, given long recommendation lists $L'$ generated by a base recommender and the visibility of each supplier $s$ computed as the fraction of times the items belonging to that supplier appear in the recommendation lists of different users 
    , analogous to $IVS$, I create 10 groups of suppliers based on their visibility in $L'$. 
    Supplier Visibility ($SV$) of each supplier $s$ in final recommendation lists $L$ can be calculated as:
    
    
    \begin{equation}\label{SV(s)}
        SV(s)=\sum_{s \in g}{\sum_{i \in A(s)}{IV(i)}}
    \end{equation}
    
    where $A(s)$ returns the items belonging to supplier $s$. Supplier Group Visibility ($SGV$) for each supplier group $g$ can be calculated as:
    
    \begin{equation}
    SGV(g)=\frac{SV(s)}{|g|}
    \end{equation}
    
    Therefore, Supplier Visibility Shift ($SVS$) of group $g$ can be calculated as:
    
    \begin{equation}\label{IVS_g}
        SVS(g)= \frac{SGV(g)^{Reranker}-SGV(g)^{Base}}{SGV(g)^{Base}}    
    \end{equation}
    
    where $SGV(g)^{Reranker}$ and $SGV(g)^{Base}$ are the visibility of item group $g$ in recommendation lists of size $n$ generated by reranking algorithm and the base algorithm, respectively.

    \item \textbf{Item Aggregate Diversity ($\alpha\mbox{-}IA$)}: I propose $\alpha\mbox{-}IA$ as the fraction of items which appear at least $\alpha$ times in the recommendation lists and can be calculated as:
    
    \begin{equation}
        \alpha\mbox{-}IA=\frac{\sum_{i \in I}\mathds{1}{(\sum_{j \in L}\mathds{1}(j=i) \geq \alpha)}}{|I|} ,\quad (\alpha \in \mathbb{N})
    \end{equation}

    This metric is a generalization of standard aggregate diversity as it is used in \cite{vargas2011rank,adomavicius2011maximizing} where $\alpha=1$.
    
    
    \item \textbf{Long-tail Coverage ($LT$):} The fraction of the long-tail items covered in the recommendation lists. To determine the long-tail items, I separated the top items which cumulatively make up 20\% of the ratings in train data as short-head and the rest of the items are considered as long-tail items. Given these long-tail items, I calculated $LT$ as the fraction of these items appeared in recommendation lists.
    
    \item \textbf{Supplier Aggregate Diversity ($\alpha\mbox{-}SA$)}: I propose $\alpha\mbox{-}SA$ as the fraction of suppliers which appear at least $\alpha$ times in the recommendation lists and can be calculated as:
    
    \begin{equation}
        \alpha\mbox{-}SA=\frac{\sum_{s \in S}\mathds{1}{(\sum_{i \in A(s)}\sum_{j \in L}\mathds{1}(j=i) \geq \alpha)}}{|S|} ,\quad (\alpha \in \mathbb{N})
    \end{equation}
    
    where $A(s)$ returns all the items belonging to supplier $s$ and $S$ is the set of all suppliers. 
    
    \item \textbf{Item Gini Index ($IG$)}: The measure of fair distribution of recommended items. It takes into account how uniformly items appear in recommendation lists. Uniform distribution will have Gini index equal to zero which is the ideal case (lower Gini index is better). $IG$ is calculated as follows over all the recommended items across all users:

    
    \begin{equation}\label{IG(L)}
        IG=\frac{1}{|I|-1} \sum_{k=1}^{|I|} (2k-|I|-1)IV(i_{k}) 
    \end{equation}
    
    where $IV(i_k)$ is the visibility of the $k$-th least recommended item being drawn from $L$ and is calculated using Equation \ref{v_i}.
    
    \item \textbf{Supplier Gini Index ($SG$)}: The measure of fair distribution of suppliers in recommendation lists. This metric can be calculated similar to $IG$, but instead of considering the distribution of recommended items, I consider the distribution of recommended suppliers and it can be calculated as:
    
    
    
     \begin{equation}\label{IG(L)}
        SG=\frac{1}{|S|-1} \sum_{k=1}^{|S|} (2k-|S|-1)SV(s_{k}) 
    \end{equation}
    
    where $SV(s_{k})$ is the visibility of the $k$-th least recommended supplier being drawn from $L$ and is calculated using Equation \ref{SV(s)}. 
    
    \item \textbf{Item Entropy ($SE$)}: Given the distribution of recommended items, entropy measures the uniformity of that distribution. Uniform distribution has the highest entropy or information gain, thus higher entropy is more desired when the goal is increasing diversity. 
    
    \begin{equation}
        IE=- \sum_{i \in I}^{} {IV(i) }\log IV(i)
    \end{equation}
    
    
    \item \textbf{Supplier Entropy ($SE$)}: The measure of uniformity of the distribution of suppliers in the recommendation lists. Similar to Gini where we had both $IG$ and $SG$, we can also measure the entropy for suppliers as follows:
    
    \begin{equation}
        SE=- \sum_{s \in S} SV(s) \log SV(s)
    \end{equation}
\end{enumerate}

\input{./Tables/tbl_metrics}

%% file: Tables/tbl_data.tex
\captionsetup[table]{skip=4pt}
\begin{table}[t!]
\footnotesize
\centering
\captionof{table}{Statistical properties of datasets} \label{tab:dataset}
\begin{tabular}{lrrrrrrr}
    \toprule
     Dataset & \#users & \#items & \#ratings & range & density & supplier & \#suppliers \\
     \midrule
     Last.fm & 2,000 & 6,817 & 218,985 & [1,5] & 1.61\% & artist & 2,856 \\
     MovieLens1M & 6,040 & 3,706 & 1,000,209 & [1,5] & 4.47\% & - & - \\
     ML1M & 6,040 & 3,079 & 928,739 & [1,5] & 4.99\% & movie-maker & 1,699 \\
     Goodreads & 2,225 & 3,423 & 137,045 & [1,5] & 1.8\% & publisher & 912 \\
    \bottomrule
\end{tabular}
\end{table}

%% file: Tables/4.2.tbl_model_conf.tex
\captionsetup[table]{skip=4pt}
\begin{table}[t!]
\small
\centering
\captionof{table}{Hyperparameter configuration for recommendation algorithms.} \label{tab:conf}
\begin{tabular}{|l|l|l|}
\hline
 model & hyperparameter & values \\ 
 \hline
 \multirow{4}{*}{Factorization models} & regularizers & $\{0.0001,0.001,0.01\}$ \\
 \cline{2-3}
                                        & number of iterations & $\{30,50,100,200,300\}$ \\
 \cline{2-3}                                        
                                        & number of factors & $\{50,100,200,300,400\}$ \\
 \cline{2-3}
                                        & learning rate & $\{0.0001,0.001,0.005,0.01\}$ \\
\cline{1-3}
 \multirow{3}{*}{Neural models} & epochs & $\{10,20\}$ \\
 \cline{2-3}
                                & number of factors & $\{8,15,30\}$ \\
 \cline{2-3}
                                & learning rate & $\{0.0001,0.001\}$ \\
\cline{1-3}
 \multirow{2}{*}{Neighborhood models} & number of neighbors & $\{10,30,50,100,200,300\}$ \\
 \cline{2-3}
                                      & shrinkage & $\{10,30,50,100,200,300\}$ \\
\hline
\end{tabular}
\end{table}

%% file: Tables/tbl_metrics.tex
\captionsetup[table]{skip=4pt}
\begin{table}[t!]
\small
\centering
\captionof{table}{Summary of evaluation metrics} \label{tab:metrics}
\begin{tabular}{p{4cm}p{1cm}p{5cm}}
    \toprule
     Metric & Abb. & Description \\
     \midrule
     Precision & Precision & The fraction of correctly recommended items. \\
     \midrule
     Item Visibility Shift & $IVS$ & The percentage of increase or decrease in the visibility of item groups in final recommendation lists generated by a reranking algorithm compared to their visibility in recommendation lists generated by a base recommender \\
     \midrule
     Supplier Visibility Shift & $SVS$ & The percentage of increase or decrease in the visibility of supplier groups in final recommendation lists generated by a reranking algorithm compared to their visibility in recommendation lists generated by a base recommender. \\
     \midrule
     Item Aggregate Diversity & $\alpha\mbox{-}IA$ & The fraction of items which appear at least $\alpha$ times in the recommendation lists.\\
    \midrule
    Long-tail Coverage & $LT$ & The fraction of the long-tail items covered in the recommendation lists.\\
    \midrule
    Supplier Aggregate Diversity & $\alpha\mbox{-}SA$ & The fraction of suppliers which appear at least $\alpha$ times in the recommendation lists.\\
    \midrule
    Item Gini Index & $IG$ & The measure of fair distribution of recommended items.\\
    \midrule
    Supplier Gini Index & $SG$ & The measure of fair distribution of suppliers in recommendation lists.\\
    \midrule
    Item Entropy & $SE$ & Given the distribution of recommended items,  entropy measures the uniformity of that distribution.\\
    \midrule
    Supplier Entropy & $SE$ & The measure of uniformity of the distribution of suppliers in the recommendation lists.\\
    \bottomrule
\end{tabular}
\end{table}

%% file: Chapters/06_Solution1.tex
\chapter{Solution 1: A Pre-processing Approach for Mitigating Multi-sided Exposure Bias}
\label{chap:solution1}
\tikzsetfigurename{solution1_}

In this chapter, I introduce a pre-processing technique that transforms the item ratings before recommendation generation. The proposed technique transforms the ratings provided by users on different items into percentile values and then those percentile values are used as input for recommendation algorithm. I originally proposed this technique in \cite{mansoury2021flatter} and showed its superiority on improving the ranking quality of recommender systems. In this chapter, I adapt this technique for tackling multi-sided exposure bias in recommender systems. The experimental results show that the proposed technique is able to mitigate exposure bias by outperforming other pre-processing techniques on different datasets. I am currently working on a paper with all the contributions in this chapter to submit to a relevant venue. 

\section{Introduction}

Recommender systems use information from user profiles to generate personalized recommendations. User profiles are either implicitly inferred by the system through user interaction, or explicitly provided by users \cite{Adomavicius:2005,Adomavicius:2015}. In the latter case, users are asked to rate different items based on their preferences and may have individual differences in how they use explicit rating scales: some users may tend to rate higher, while some users may tend to rate lower; some users may use the full extent of the rating scale, while others might use just a small subset. \cite{Herlocker:1999a}. 

\begin{figure*}[btp]
    \centering
    \begin{subfigure}[b]{0.45\textwidth}
        \includegraphics[width=\textwidth]{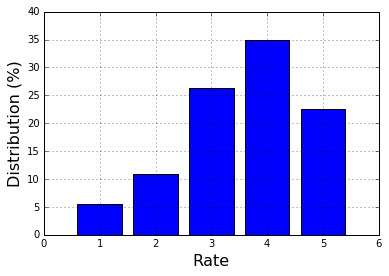}
        \caption{ML1M} \label{ml_rate_dist}
    \end{subfigure}
    \begin{subfigure}[b]{0.45\textwidth}
        \includegraphics[width=\textwidth]{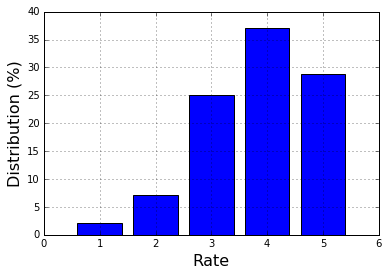}
        \caption{Goodreads} \label{gr_rate_dist}
    \end{subfigure}%
\caption{Rating distribution of ML1M and Goodreads datasets.} \label{rate_dist}
\end{figure*}

When a user concentrates his or her ratings in only a small subset of the rating scale, this often results in ratings distributions that are skewed -- most often towards the high end of the scale. This is because items are not rated at random, but rather preferred items are more likely to be selected and therefore rated due to selection bias~\cite{marlin2007collaborative}. Figure~\ref{rate_dist} shows the overall rating distribution of two datasets that exhibit typically right-skewed distributions. Users in the ML1M dataset in Figure~\ref{ml_rate_dist}, for example, have assigned less than 17\% of the ratings to ratings 1 and 2 and some 57\% of ratings are values 4 and 5. As another example, users in the Goodreads dataset in \ref{gr_rate_dist} have assigned less than 10\% of the ratings to ratings 1 and 2 and more than 65\% of ratings are values 4 and 5. We can assume, in ML1M dataset for instance, this is not because there are so many more good movies than bad, but rather than users are selecting movies to view that they are likely to enjoy and the ratings are concentrated among those selections. A drawback of this skew to the distribution is that we have more information about preferred items and less information about items that are not liked as well. It also means that a given rating value may be ambiguous in meaning.

As an example, assume that Alice and Bob both purchase an item \textit{X} and rate it. Alice is a user who tends to rate lower and tends to use the whole rating scale, while Bob is a user who tends to rate higher and never uses ratings at the bottom of the scale. Their profiles, sorted by rating value, are shown in Table~\ref{tab:uexample}. After using item \textit{X}, Alice is fully satisfied with it, but Bob is only partially satisfied. As a result, both rate the item \textit{X} as 4 out of 5 although they have different levels of satisfaction toward that item. These ratings, while identical, do not carry the same meaning. A transformation based on percentiles, shown in the bottom rows of the Table, captures this distinction well: a rating of 4 for Alice is percentile 80; whereas for Bob, the same score has a score of 50. In addition, unlike the original profiles, where the users' ratings are distributed over different ranges, these profiles span the same numerical range from 20 to 90. 

In \cite{mansoury2021flatter}, we showed that percentile transformation on users' profile (as illustrated in Table~\ref{tab:uexample}) can improve the accuracy of recommendations. In percentile transformation on users' profile, each value associated with an item in the users' profile reflects its rank among all of the items that the user has rated. Thus, the percentile captures an item's position within a user's profile better than the raw rating value and compensates for differences in users' overall rating behavior. Also, the percentile, by definition, will span the whole range of rating values and gives rise to a more uniform rating distribution. These two properties of the percentile transformation on users' profile -- its ability to compensate for individual user biases and its ability to create a more uniform rating distribution -- lead to enhanced recommender system performance.

\addtolength{\parskip}{-0.5mm}
\begin{table}[t]
\centering
\captionof{table}{\label{tab:uexample} User rating profiles with percentile transformation}
\begin{tabular}{ l | l | l}
\hline
Alice & rating & $\langle 1,1,2,2,3,3,3,4,5 \rangle$ \\
Bob & rating & $\langle 3,3,4,4,4,5,5,5,5 \rangle$ \\
\hline
Alice & percentile & $\langle 20,20,40,40,70,70,70,80,90 \rangle$ \\
Bob & percentile & $\langle 20,20,50,50,50,90,90,90,90 \rangle$\\
\hline
\end{tabular}
\end{table}

\begin{figure*}[btp]
    \centering
    \begin{subfigure}[b]{0.45\textwidth}
        \includegraphics[width=\textwidth]{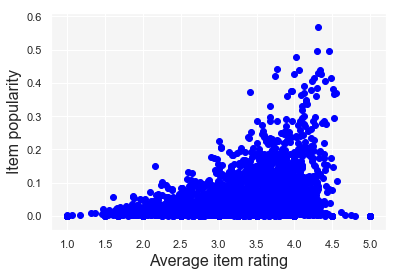}
        \caption{ML1M} \label{ml_avgrating_pop}
    \end{subfigure}
    \begin{subfigure}[b]{0.45\textwidth}
        \includegraphics[width=\textwidth]{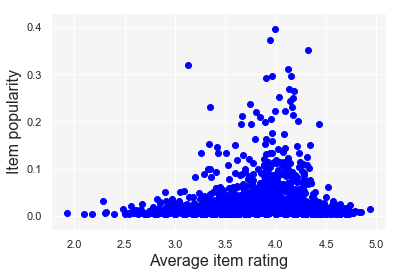}
        \caption{Goodreads} \label{gr_avgrating_pop}
    \end{subfigure}%
\caption{Average rating assigned to items versus popularity of items based on the number of interactions made on those items.} \label{avgrating_pop}
\end{figure*}

In this chapter, I formalize a rating transformation model as above that converts the ratings assigned to items into percentile values as a pre-processing step before recommendation generation. This transformation would be applied on items' profiles and is able to mitigate multi-sided exposure bias in recommender systems. As discussed in previous chapters and shown in \cite{abdollahpouri2017controlling,abdollahpouri2019popularity}, popularity bias in input data is the major source of exposure bias in recommendation results where popular items frequently appear in recommendation lists, while non-popular items rarely appear in recommendation lists. 
Popular items are the ones that not only, by definition \cite{abdollahpouri2017controlling}, received many interactions and ratings from different users, but also are assigned high rating values. Figure~\ref{avgrating_pop} shows the average ratings assigned to different items with different degree of popularity on ML1M and Goodreads datasets. In both datasets, it can be seen that popular items received high ratings. This pattern is even stronger in ML1M dataset as popular items are mainly assigned rating 4 on average. These two properties of popular items -- large number of ratings and high rating values -- cause the recommendation algorithm to mainly focus on them and recommend them to many users. 

\begin{figure*}[btp]
    \centering
    \begin{subfigure}[b]{0.45\textwidth}
        \includegraphics[width=\textwidth]{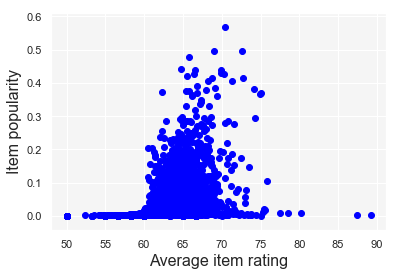}
        \caption{ML1M} \label{ml_avgrating_pop_last}
    \end{subfigure}
    \begin{subfigure}[b]{0.45\textwidth}
        \includegraphics[width=\textwidth]{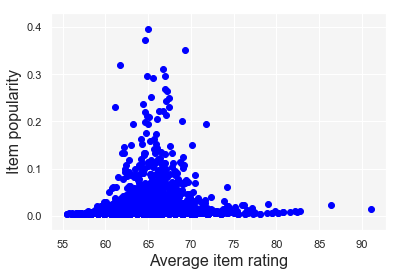}
        \caption{Goodreads} \label{gr_avgrating_pop_last}
    \end{subfigure}%
\caption{Average percentile assigned to items versus popularity of items based on the number of interactions made on those items.} \label{avgrating_pop_last}
\end{figure*}

Percentile transformation is able to compensate for the high rating values assigned to popular items and alleviate the existing bias in input data. Figure~\ref{avgrating_pop_last} shows the average percentile values assigned to different items with different degree of popularity on ML1M and Goodreads datasets. Unlike the ratings values in Figure~\ref{avgrating_pop} that popular items were assigned high ratings, after transforming the ratings into percentile values in Figure~\ref{avgrating_pop_last}, the percentile values assigned to popular items are shifted to almost neutral values (i.e.\ around percentile value 65 in the range of 1 to 100). I hypothesize that this compensation on high rating values assigned to the popular items will mitigate exposure bias in recommendation lists. 

As an example, Table~\ref{tab:iexample} shows the ratings for two items. Assume that item $B$ is a popular items that received high ratings from 9 users, while item $A$ is not that popular and received various rating values from 4 users. The corresponding percentile values for each item show that the percentile transformation assigned lower percentile values to the many high rating values assigned to the popular items. The average of ratings assigned to items $B$ is 4 (in the scale of 1 to 5), while the average of percentile values is 63.3 (in the scale of 1 to 100). On the other hand, the average of ratings assigned to item $A$ is 2.75 (a bit lower than the neural rating 3), while the average of percentile values is 55 (a bit higher than the neutral percentile 50). This signifies that the percentile transformation is able to alleviate the emphasis or weight assigned to the popular items in the rating data, while slightly promote the non-popular items. This can help to mitigate the over-recommendation of the popular items.

To show the effectiveness of the proposed percentile technique on items' profiles, extensive experiments are performed using three different recommendation algorithms on two publicly-available datasets and the results are evaluated in terms of various metrics including recommendation accuracy, item aggregate diversity, supplier aggregate diversity, and fair distribution of recommended items and suppliers. Also, comparison with original rating values and z-score transformation show the superiority of percentile transformation on improving multi-sided exposure fairness in recommendation results.  

\addtolength{\parskip}{-0.5mm}
\begin{table}[t]
\centering
\captionof{table}{\label{tab:iexample} Item rating profiles with percentile transformation}
\begin{tabular}{ l | l | l}
\hline
A & rating & $\langle 1,3,3,4 \rangle$ \\
B & rating & $\langle 3,3,4,4,4,4,4,5,5 \rangle$ \\
\hline
A & percentile & $\langle 20,60,60,80 \rangle$ \\
B & percentile & $\langle 20,20,70,70,70,70,70,90,90 \rangle$\\
\hline
\end{tabular}
\end{table}



\section{Percentile transformation}

In statistics, given a series of measurements, percentile (or quantile) methods are used to estimate the value corresponding to a certain percentile. Given the \textit{P}\textsuperscript{th} percentile, these methods attempt to put \textit{P}\% of the data set below and (100-\textit{P})\% of the data set above. There are a number of different definitions in the literature for computing percentiles \cite{Hyndman:1996,Langford:2006}. Although they are apparently different, the answers produced by these methods are very similar and the slight differences are negligible \cite{Langford:2006}. In this paper, we use a definition from \cite{Hyndman:1996}.\\
\indent The percentile value, \(p\), corresponding to a measurement, \(x\), in a series of measurements, \(M\), is computed with regard to the position of \(x\) in the ordered list \(M\), \(o(M)\), as follows: 

\begin{equation} \label{eq:percentile}
p(x,M)=\frac{100\times position(x,o(M))}{|M|+1}
\end{equation}

\noindent where \(position(x, o(M))\) returns the index of occurrence of \(x\) in \(o(M)\), or the position in the order where $x$ would appear if it is not present, and \(|M|\) is the number of measurements in \(M\). For more details see \cite{Hyndman:1996}.\\
\indent This transformation assumes that values are distinct and there is no repetition in the series. However, with rating data, we often have a different situation. User profiles usually contain many repetitive ratings, and it is unclear how to specify the position of a rating. For example, in a series of ratings $v=\langle 2, 3, 3, 3, 3, 3, 5, 5, 5 \rangle$, it is not clear what the position of rating 3 should be. We could take the first occurrence, position 2, or the last occurrence 6, or something in between.\\
\indent I explored the performance of the proposed percentile technique by taking the index of the first and the last occurrence of repeated ratings in the ordered vector. But, the experiments showed that last index percentile transformation better compensate for the high rating values in input data and consequently yielded consistent results in terms of mitigating multi-sided exposure bias. Therefore, for the rest of this chapter, the percentile transformation is performed using the last index occurrence of the repetitive ratings in items' profile\footnote{See https://github.com/masoudmansoury/percentile for the code for computing these and other transformations described in this thesis.}.\\
\indent Even in contexts where ratings are gathered implicitly, they are often converted into numeric scores representing user preference or relevance. For example, time spent on a page is often considered a measure of user interest \cite{yi2014beyond} or number of seconds watched of a video \cite{zhao2019recommending}. Profiles generated in these ways can also be normalized using the percentile transform as well, although they are less likely to have repeated entries.\\
\indent For the purposes of this dissertation, the entire set of ratings provided on an item \(i\) is considered as a rating vector for \(i\), denoted by \(R_{i}\) with an individual rating given by a user $u$, denoted as $r_{ui}$. Let $p(v,\ell)$ be the percentile mapping in Equation~\ref{eq:percentile} from a rating value $v$ in a list of values $\ell$, using the first and last index methods. Then, the percentile value of a rating \(r\) provided by user \(u\) on an item \(i\) is computed by taking the rating $r_{ui}$ and calculating its percentile value within the whole profile of the item. For example, based on the first index rule, for the item $A$ from Table~\ref{tab:iexample}, rating 3 would have percentile value $100*2/(4+1) = 40$. We define the percentile function, $Per$, as follows:
\begin{equation} \label{eq:percu}
\mbox{Per}(u,i)= p(r_{ui}, R_i)
\end{equation}

\section{Experiments}

I evaluated the performance of percentile transformation on ML1M and Goodreads datasets. The characteristics of the datasets are summarized in Chapter~\ref{chap:methodology}. These datasets are from various domains and have different degrees of sparsity.\\
\indent Extensive experiments are performed to evaluate the effect of the percentile transformation on mitigating the exposure bias of a number of recommendation algorithms. Due to the nature of the proposed percentile technique, the experiments are only performed with algorithms that make use of rating magnitude. 
Therefore, the experiments include biased matrix factorization (\algname{BiasedMF}) \cite{Koren:2009a}, singular value decomposition (\algname{SVD++}) \cite{Koren:2008a}, and list-wise ranking matrix factorization (\algname{ListRankMF}) \cite{shi2010list}.\\ 
\indent The results produced by percentile values as input for the recommendation algorithms are compared with the original rating values (no transformation) and z-score values. Z-score values are computed using well-known z-score transformation \cite{larsen2005introduction}. In statistics, z-score transformation is used to standardize the raw scores and measures how a value deviates from the population mean. Given $\Bar{R}_i$ and $sd_{R_i}$ as the average and standard deviation of ratings assigned to item $i$, respectively, the z-score value corresponding to rating $r_{ui}$ is computed as follows:

\begin{equation}
    zscore(r_{ui},\Bar{R}_i,sd_{R_i})=\frac{r_{ui}-\Bar{R}_i}{sd_{R_i}}
\end{equation}

The results are reported for nine experimental conditions. Three recommendation algorithms evaluated over three different inputs: the original ratings, the results of the percentile transformation, and the results of the z-score transformation.

\subsection{Best-performing results}

For each algorithm used for experiments, I optimized the recommendation algorithms using gridsearch over hyperparameters (details are explained in Chapter~\ref{chap:methodology}) to achieve the highest possible precision. What I am interested in by reporting these results is to see which input value is able to generate more accurate recommendations to users. Since there is always a trade-off between accuracy (i.e.\ precision) and non-accuracy (i.e.\ aggregate diversity, gini index, and entropy) metrics, improving one group of metrics would cause performance loss in another group of metrics. Thus, it is expected that an experimental condition that yields the high accuracy will results in poor performance in terms of non-accuracy metrics.\\
\input{Tables/6.3.2.best_result}
\indent Table~\ref{tab:bestpercentileresult} shows the best-performing results in terms of precision for each experimental condition on ML1M and Goodreads datasets. Using \algname{BiasedMF}, percentile transformation yielded the highest precision by 0.112 and 0.062 on ML1M and Goodreads datasets, respectively. However, the highest precision that could be achieved by original ratings is 0.097 and 0.030, and by z-score transformation is 0.085 and 0.027 on ML1M and Goodreads, respectively. Although z-score transformation achieved the highest performance in terms of aggregate diversity and fair distribution of recommended items and suppliers compared to original ratings and percentile transformation, this is not a reliable improvement as the precision value for z-score is significantly lower than other input values. This shows that percentile transformation provides more meaningful and informative input for recommender systems and enables the algorithm to achieve higher precision than other input values. In the next Section, the results with the same precision value would be further discussed.\\
\indent Using \algname{ListRankMF}, the best-performing precision for all input values on both datasets are at the same level. These results show an interesting pattern. With the same level of precision for all input values, percentile transformation significantly outperformed other input values in terms of mitigating exposure bias. This result indicates the ability of percentile transformation in improving multi-sided exposure fairness compared to other input values.\\
\indent Using \algname{SVD++} on Goodreads datset, again, percentile transformation yielded the highest precision by 0.066 compared to 0.034 and 0.028 for original ratings and z-score transformation, respectively. On ML1M, original ratings and z-score transformation achieved the highest and lowest precision, respectively. Although percentile transformation was outperformed by original ratings, it significantly achieved outperformed original ratings in terms of improving exposure fairness. For instance, with 4.8\% loss in precision compared to original ratings, percentile transformation achieved 0.288, 0.197, 0.306, 0.222, 0.924, 5.76, 0.924, 5.10 compared to 0.024, 0.019, 0.032, 0.028, 0.994, 3.18, 0.991, 2.87 for original ratings in terms of $1\mbox{-}IA$, $5\mbox{-}IA$, $1\mbox{-}SA$, $5\mbox{-}SA$, $IG$, $IE$, $SG$, and $SE$.\\
\indent For a fair comparison, results with the same precision values for each input value and experimental condition is reported in Table~\ref{tab:percentileresult}. Those results would be discussed in the following Sections. 

\input{Tables/6.3.2.results}

\subsection{Item aggregate diversity}

Table~\ref{tab:percentileresult} shows the $1\mbox{-}IA$ and $5\mbox{-}IA$ for the same precision value for each input value on both datasets. Using \algname{BiasedMF} on Goodreads dataset, percentile transformation yielded significantly higher item aggregate diversity compared to rating and z-score transformation\footnote{We can think of the original ratings as a null transformation.}. On ML1M dataset, although in terms of $1\mbox{-}IA$, percentile transformation is outperformed by z-score transformation, in terms of $5\mbox{-}IA$, it outperformed both original ratings and z-score transformation. This result can also be observed on Goodreads dataset using \algname{SVD++}. These results show how regular item aggregate diversity ($1\mbox{-}IA$) can be misleading on measuring the exposure bias. When many items are recommended to only few users, even though it achieves high item aggregate diversity, the exposure for each item would not be fair because many items are rarely appeared in the recommendation lists, while few items are frequently appeared in the recommendation lists.\\
\begin{figure*}[t]
    \centering
    \begin{subfigure}[b]{0.99\textwidth}
        \includegraphics[width=\textwidth]{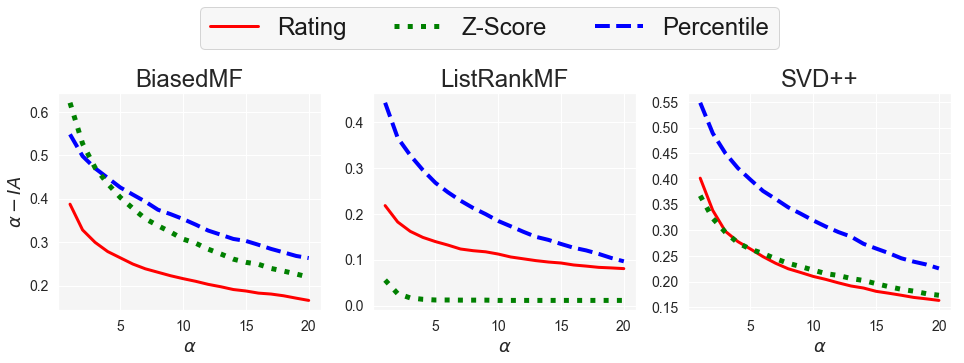}
        \caption{ML1M} \label{ml_item_agg_k}
    \end{subfigure}
    \begin{subfigure}[b]{0.99\textwidth}
        \includegraphics[width=\textwidth]{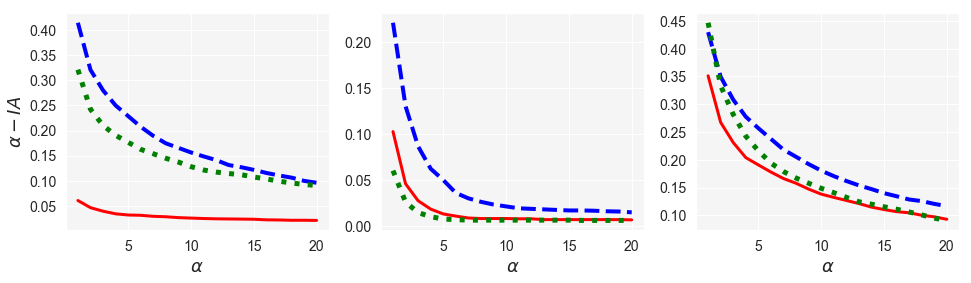}
        \caption{Goodreads} \label{gr_item_agg_k}
    \end{subfigure}%
\caption{Comparison of recommendation algorithms with different input values in terms of item aggregate diversity ($\alpha\mbox{-}IA$) with varying $\alpha$ on ML1M and Goodreads datasets.} \label{item_agg_k}
\end{figure*}
\indent To clarify that percentile transformation really outperforms other input values using \algname{BiasedMF}, Figure~\ref{item_agg_k} shows the performance of each input values in terms of $\alpha\mbox{-}IA$ with varying $\alpha$ from 1 to 20. As shown in this Figure, on Goodreads dataset, percentile transformation clearly outperforms original ratings and z-score transformation, but on ML1M dataset, although for $\alpha \in \{1,2,3,4\}$ the percentile transformation is outperformed by z-score transformation, for $\alpha > 4$, percentile transformation outperformed the z-score and original ratings.\\ 
\indent Using \algname{ListRankMF} on both datasets, percentile transformation outperformed original ratings and z-score transformation in terms of both $1\mbox{-}IA$ and $5\mbox{-}IA$. The improvement on both datasets is significant and substantial. For example, on ML1M dataset, $1\mbox{-}IA$ and $5\mbox{-}IA$ for percentile values are 0.443 and 0.268, for original ratings are 0.218 and 0.140, and for z-score values are 0.056 and 0.012, respectively. Also, Figure~\ref{item_agg_k} shows that using \algname{ListRankMF} on both datasets, percentile transformation significantly outperforms other transformation techniques for all $\alpha$ values.\\
\indent Finally, using \algname{SVD++} on ML1M, percentile transformation outperformed other transformations in terms of $1\mbox{-}IA$ and $5\mbox{-}IA$ with even higher precision value. On Goodreads dataset, for $5\mbox{-}IA$, percentile transformation outperformed both original ratings and z-score transformation. Also, Figure~\ref{item_agg_k} confirms the improvement by percentile transformation where on ML1M the improvement is significant for all $\alpha$ values and on Goodreads, for $\alpha>4$ it outperformed other transformations.

\begin{figure*}[t]
    \centering
    \begin{subfigure}[b]{0.99\textwidth}
        \includegraphics[width=\textwidth]{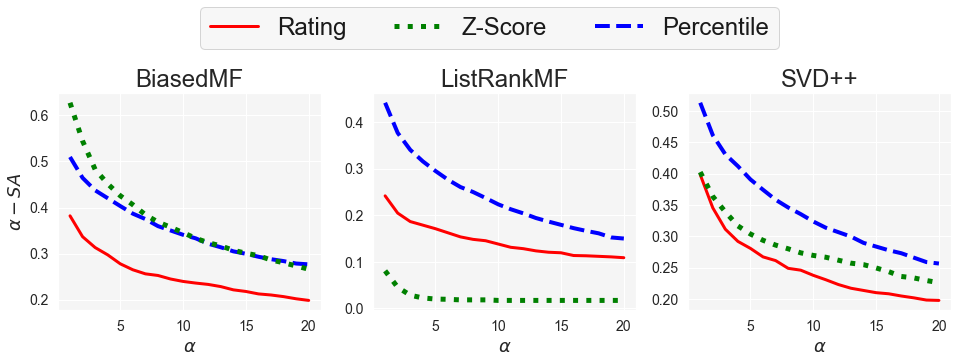}
        \caption{ML1M} \label{ml_supp_agg_k}
    \end{subfigure}
    \begin{subfigure}[b]{0.99\textwidth}
        \includegraphics[width=\textwidth]{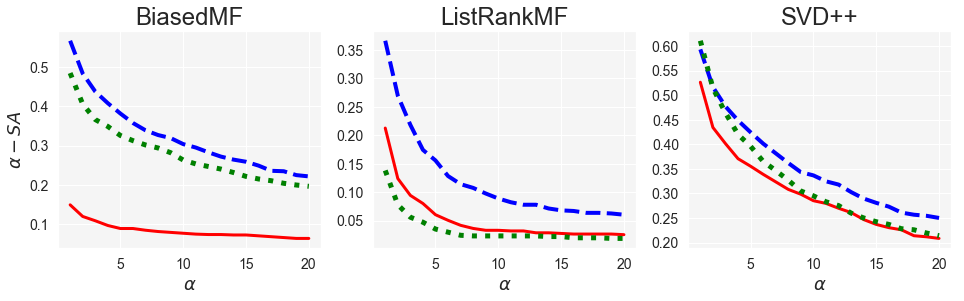}
        \caption{Goodreads} \label{gr_supp_agg_k}
    \end{subfigure}%
\caption{Comparison of recommendation algorithms with different input values in terms of supplier aggregate diversity ($\alpha\mbox{-}SA$) with varying $\alpha$ on ML1M and Goodreads datasets.} \label{supp_agg_k}
\end{figure*}

\subsection{Supplier aggregate diversity}

Looking at Table~\ref{tab:percentileresult} reveals that percentile transformation outperformed other transformations except for \algname{BiasedMF} on ML1M dataset and for \algname{SVD++} on Goodreads only in terms of $1\mbox{-}SA$. The improvement by percentile transformation on \algname{ListRankMF} is even more significant as it increased supplier aggregate diversity by 82.6\% and 71.8\% in terms of $1\mbox{-}SA$ on ML1M and Goodreads datasets, respectively, and by 72.5\% and 160\% in terms of $5\mbox{-}SA$ on ML1M and Goodreads, respectively. The same results can also be observed for \algname{BiasedMF} on Goodreads and \algname{SVD++} on ML1M.\\
\indent Also, Figure~\ref{supp_agg_k} shows that percentile transformation significantly improved supplier aggregate diversity ($\alpha\mbox{-}SA$) compared to other transformations for different values of $\alpha$ except for \algname{BiasedMF} on ML1M dataset. For \algname{BiasedMF} on ML1M dataset, although percentile transformation is outperformed for $\alpha \leq 10$, it yielded the same supplier aggregate diversity with z-score transformation for $\alpha > 10$. In other cases, percentile transformation outperformed other transformations for $\alpha$ values.

\subsection{Long-tail analysis}

Long-tail coverage ($LT$) is also another metric for showing how much an algorithm give chance to non-popular items to be shown in recommendation lists. Hence, $LT$ measures the fraction of non-popular items that appeared in the recommendation lists. According to Table~\ref{tab:percentileresult}, percentile transformation resulted in improved long-tail coverage in all experimental conditions except for \algname{BiasedMF} on ML1M and \algname{SVD++} on Goodreads.

\subsection{Fair distribution of recommended items}

Fair distribution of recommended items refers to the fact that how equally each item is represented in recommendation lists. If distribution of recommended items represents the number of times each item appeared in the recommendation lists, a uniform distribution signifies that all items are equally appeared in the lists. Thus, uniform distribution for the recommended items would be an ideal distribution for achieving a fair exposure for recommended items. For this purpose, Gini index and Entropy (discussed in \ref{chap:methodology}) are used to measure how fair the distribution of recommended items is.\\
\indent Results reported in Table~\ref{tab:percentileresult} shows that percentile transformation yielded a fairer distribution of recommended items compared to rating and z-score transformations. The results are consistent over all recommendation algorithms, datasets, and evaluation metrics except for \algname{ListRankMF} on ML1M dataset in terms of item Entropy. Even for \algname{ListRankMF} on ML1M dataset, although percentile transformation is outperformed by rating values in terms of item Entropy, it outperformed rating and z-score transformations in terms of item Gini index.

\subsection{Fair distribution of suppliers in recommendation lists}

Fair distribution of suppliers refers to the fact that how items belong to different suppliers are recommended such that those suppliers have equal representation in the recommendation lists. Analogous to fair distribution of recommended items, the recommendation lists that give uniform distribution for suppliers is an ideal situation for achieving fair exposure in suppliers perspective. This can be measured using Gini index and Entropy on distribution of suppliers.\\
\indent According to Table~\ref{tab:percentileresult}, percentile transformation consistently gives fairer exposure to suppliers compared to other transformations on Goodreads dataset for all recommendation algorithms and both $SG$ and $SE$ metrics. On ML1M dataset, percentile transformation shows superior on achieving fair exposure suppliers only based on one of the evaluation metrics.

\section{Discussion and Limitations}

Experimental results showed the superiority of percentile transformation on improving the multi-sided exposure fairness in recommendation results. This includes recommending more items from the catalog (high aggregate diversity) and providing equal representation or exposure to recommended items or suppliers. In addition, the experimental results showed that input data derived from percentile transformation can lead to more accurate recommendation results. This means that percentile values provide meaningful and informative input for recommendation algorithms in which enable those algorithms to optimize to achieve the highest precision in recommendation results.\\
\indent There are several limitations associated with the proposed percentile technique. First, the percentile transformation does not work on binary data. According to the definition for percentile transformation, a range of more than two (non-binary) rating members is needed to find the position of a rating in the vector, otherwise, finding the position of a rating in the vector may be impossible. However, the proposed transformation can be applied on explicit rating data (as shown in this chapter) and implicit feedback data. Example of implicit feedback data can be listening history of songs by users: number of times that a user listened to a song in his/her profile can be converted into percentile value to show the degree of interest toward that song.\\
\indent Second, the proposed percentile technique only makes sense to be used as input in the recommendation algorithms that utilize the rating values as part of its optimization or model learning. When a recommendation algorithm does not utilize the rating values, using the percentile values as input for that recommendation algorithm will not make any improvement in the recommendation performance.\\
\indent Finally, the proposed percentile technique may not work well on sparse datasets. In a sparse dataset, there can be many items with few ratings assigned to them which makes the percentile transformation inaccurate. For example, for an item with only one rating, it is not clear what would be the position of that rating in the item's profile. As another example, for items with several identical ratings, calculating the percentile values would not be accurate for the same reason that accurately finding the position of rating in the profile would not be possible.\\
\indent These limitations can be considered as possible future works and further improving the performance of the proposed percentile transformation. For example, the third limitation, weakness on sparse data, may be lifted by considering \textit{smoothed percentile transformation} \cite{mansoury2021flatter}. Although my initial experimental results did not show the effectiveness of this method on overcoming data sparsity issue, I plan to further investigate it in the future.

%% file: Tables/6.3.2.best_result.tex
\captionsetup[table]{skip=4pt}
\begin{table}[t!]
\footnotesize
\centering
\captionof{table}{Best-performing performance of recommendation algorithms in terms of precision at top-n=10. The bolded entries show the best values and the underlined entries show the statistically significant change from the second-best result with $p<0.05$.} \label{tab:bestpercentileresult}
\begin{tabular}{llrrrrrrr}
\toprule
 \multirow{2}{*}{algorithm} & \multirow{2}{*}{metric} & \multicolumn{3}{c}{\textbf{ML1M}} & & \multicolumn{3}{c}{\textbf{Goodreads}} \\ \cline{3-5} \cline{7-9}
 & & rating & z-score & percentile & & rating & z-score & percentile \\
 \midrule

 \multirow{10}{*}{\algname{BiasedMF}} & precision & 0.097 & 0.085 & \underline{\textbf{0.112}} & & 0.030 & 0.027 & \underline{\textbf{0.062}} \\
                           & $1\mbox{-}IA$ & 0.245 & \underline{\textbf{0.500}} & 0.368 & & 0.060 & \underline{\textbf{0.321}} & 0.107 \\
                           & $5\mbox{-}IA$ & 0.161 & \underline{\textbf{0.359}} & 0.258 & & 0.032 & \underline{\textbf{0.176}} & 0.064 \\
                           & $LT$ & 0.194 & \underline{\textbf{0.468}} & 0.325 & & 0.043 & \underline{\textbf{0.273}} & 0.051 \\
                           & $1\mbox{-}SA$ & 0.261 & \underline{\textbf{0.469}} & 0.359 & & 0.149 & \underline{\textbf{0.485}} & 0.206 \\
                           & $5\mbox{-}SA$ & 0.187 & \underline{\textbf{0.348}} & 0.267 & & 0.089 & \underline{\textbf{0.326}} & 0.146 \\
                           & $IG$ & 0.950 & \underline{\textbf{0.827}} & 0.902 & & 0.990 & \underline{\textbf{0.900}} & 0.978 \\
                           & $IE$ & 5.32 & \underline{\textbf{6.62}} & 6.00 & & 3.74 & \underline{\textbf{6.12}} & 4.59 \\
                           & $SG$ & 0.949 & \underline{\textbf{0.867}} & 0.914 & & 0.969 & \underline{\textbf{0.877}} & 0.949 \\
                           & $SE$ & 4.62 & \underline{\textbf{5.66}} & 5.19 & & 3.60 & \underline{\textbf{4.93}} & 4.12 \\
\cline{1-9}

 \multirow{10}{*}{\algname{ListRankMF}} & precision & 0.143 & \textbf{0.151} & 0.150 & & 0.067 & 0.078 & \textbf{0.079} \\
                           & $1\mbox{-}IA$ & 0.103 & 0.028 & \underline{\textbf{0.151}} & & 0.052 & 0.020 & \underline{\textbf{0.111}} \\
                           & $5\mbox{-}IA$ & 0.026 & 0.015 & \underline{\textbf{0.047}} & & 0.010 & 0.009 & \underline{\textbf{0.024}} \\
                           & $LT$ & 0.065 & 0.009 & \underline{\textbf{0.095}} & & 0.030 & 0.007 & \underline{\textbf{0.070}} \\
                           & $1\mbox{-}SA$ & 0.140 & 0.041 & \underline{\textbf{0.174}} & & 0.133 & 0.058 & \underline{\textbf{0.232}} \\
                           & $5\mbox{-}SA$ & 0.041 & 0.021 & \underline{\textbf{0.070}} & & 0.037 & 0.030 & \underline{\textbf{0.078}} \\
                           & $IG$ & 0.993 & 0.995 & \textbf{0.992} & & 0.995 & 0.996 & \textbf{0.993} \\
                           & $IE$ & 3.23 & 3.07 & \textbf{3.24} & & 2.85 & 2.83 & \textbf{3.13} \\
                           & $SG$ & \textbf{0.990} & 0.993 & \textbf{0.990} & & 0.984 & 0.988 & \textbf{0.982} \\
                           & $SE$ & \textbf{2.96} & 2.65 & 2.87 & & 2.81 & 2.55 & \textbf{2.86} \\
 \cline{1-9}

 \multirow{10}{*}{\algname{SVD++}} & precision & \textbf{0.145} & 0.086 & 0.138 & & 0.034 & 0.028 & \underline{\textbf{0.066}} \\
                           & $1\mbox{-}IA$ & 0.024 & \underline{\textbf{0.367}} & 0.288 & & 0.035 & \underline{\textbf{0.447}} & 0.019 \\
                           & $5\mbox{-}IA$ & 0.019 & \underline{\textbf{0.263}} & 0.197 & & 0.023 & \underline{\textbf{0.216}} & 0.015 \\
                           & $LT$ & 0.006 & \underline{\textbf{0.329}} & 0.240 & & 0.025 & \underline{\textbf{0.407}} & 0.004 \\
                           & $1\mbox{-}SA$ & 0.032 & \underline{\textbf{0.403}} & 0.306 & & 0.099 & \underline{\textbf{0.611}} & 0.057 \\
                           & $5\mbox{-}SA$ & 0.028 & \underline{\textbf{0.304}} & 0.222 & & 0.068 & \underline{\textbf{0.396}} & 0.045 \\
                           & $IG$ & 0.994 & \underline{\textbf{0.882}} & 0.924 & & 0.993 & \underline{\textbf{0.862}} & 0.993 \\
                           & $IE$ & 3.18 & \underline{\textbf{6.21}} & 5.76 & & 3.34 & \underline{\textbf{6.40}} & 3.35 \\
                           & $SG$ & 0.991 & \underline{\textbf{0.879}} & 0.924 & & 0.979 & \underline{\textbf{0.849}} & 0.979 \\
                           & $SE$ & 2.87 & \underline{\textbf{5.59}} & 5.10 & & 3.17 & \underline{\textbf{5.11}} & 3.17 \\

\bottomrule
\end{tabular}
\end{table}

%% file: Tables/6.3.2.results.tex
\captionsetup[table]{skip=4pt}
\begin{table}[t!]
\footnotesize
\centering
\captionof{table}{Performance of recommendation algorithms with almost the same precision at top-n=10. The bolded entries show the best values and the underlined entries show the statistically significant change from the second-best result with $p<0.05$.} \label{tab:percentileresult}
\begin{tabular}{llrrrrrrr}
\toprule
 \multirow{2}{*}{algorithm} & \multirow{2}{*}{metric} & \multicolumn{3}{c}{\textbf{ML1M}} & & \multicolumn{3}{c}{\textbf{Goodreads}} \\ \cline{3-5} \cline{7-9}
 & & rating & z-score & percentile & & rating & z-score & percentile \\
 \midrule
 
 \multirow{10}{*}{\algname{BiasedMF}} & precision & 0.072 & 0.072 & 0.072 & & 0.030 & 0.027 & 0.032  \\
                           & $1\mbox{-}IA$ & 0.387 & \textbf{0.621} & 0.548 & & 0.060 & 0.321 & \underline{\textbf{0.415}} \\
                           & $5\mbox{-}IA$ & 0.264 & 0.403 & \textbf{0.426} & & 0.032 & 0.176 & \underline{\textbf{0.229}} \\
                           & $LT$ & 0.346 & \textbf{0.598} & 0.519 & & 0.043 & 0.273 & \underline{\textbf{0.373}} \\
                           & $1\mbox{-}SA$ & 0.382 & \textbf{0.627} & 0.510 & & 0.149 & 0.485 & \textbf{0.567} \\
                           & $5\mbox{-}SA$ & 0.278 & \textbf{0.426} & 0.403 & & 0.089 & 0.326 & \underline{\textbf{0.382}} \\
                           & $IG$ & 0.888 & 0.808 & \underline{\textbf{0.779}} & & 0.990 & 0.900 & \underline{\textbf{0.861}} \\
                           & $IE$ & 6.18 & 6.69 & \textbf{6.82} & & 3.74 & 6.12 & \underline{\textbf{6.43}} \\
                           & $SG$ & 0.902 & \textbf{0.840} & 0.842 & & 0.969 & 0.877 & \underline{\textbf{0.847}} \\
                           & $SE$ & 5.35 & \textbf{5.81} & \textbf{5.81} & & 3.60 & 4.93 & \underline{\textbf{5.17}} \\
\cline{1-9}

 \multirow{10}{*}{\algname{ListRankMF}} & precision & 0.125 & 0.125 & 0.125 & & 0.059 & \textbf{0.066} & 0.059 \\
                           & $1\mbox{-}IA$ & 0.218 & 0.056 & \underline{\textbf{0.443}} & & 0.103 & 0.060 & \underline{\textbf{0.221}} \\
                           & $5\mbox{-}IA$ & 0.140 & 0.012 & \underline{\textbf{0.268}} & & 0.013 & 0.008 & \underline{\textbf{0.050}} \\
                           & $LT$ & 0.164 & 0.024 & \underline{\textbf{0.406}} & & 0.059 & 0.025 & \underline{\textbf{0.173}} \\
                           & $1\mbox{-}SA$ & 0.242 & 0.081 & \underline{\textbf{0.442}} & & 0.213 & 0.138 & \underline{\textbf{0.366}} \\
                           & $5\mbox{-}SA$ & 0.171 & 0.020 & \underline{\textbf{0.295}} & & 0.060 & 0.035 & \underline{\textbf{0.156}} \\
                           & $IG$ & 0.964 & 0.995 & \underline{\textbf{0.936}} & & 0.994 & 0.995 & \underline{\textbf{0.983}} \\
                           & $IE$ & \textbf{4.93} & 2.99 & 4.68 & & 2.91 & 2.82 & \underline{\textbf{3.63}} \\
                           & $SG$ & 0.962 & 0.991 & \underline{\textbf{0.947}} & & 0.982 & 0.985 & \underline{\textbf{0.967}} \\
                           & $SE$ & \textbf{4.31} & 2.85 & 4.12 & & 2.85 & 2.69 & \underline{\textbf{3.38}} \\
 \cline{1-9}

 \multirow{10}{*}{\algname{SVD++}} & precision & 0.096 & 0.086 & \textbf{0.100} & & 0.025 & 0.028 & \textbf{0.029} \\
                           & $1\mbox{-}IA$ & 0.402 & 0.367 & \underline{\textbf{0.549}} & & 0.351 & \textbf{0.447} & 0.430 \\
                           & $5\mbox{-}IA$ & 0.264 & 0.263 & \underline{\textbf{0.399}} & & 0.191 & 0.216 & \underline{\textbf{0.257}} \\
                           & $LT$ & 0.362 & 0.329 & \underline{\textbf{0.520}} & & 0.305 & \textbf{0.407} & 0.389 \\
                           & $1\mbox{-}SA$ & 0.398 & 0.403 & \underline{\textbf{0.513}} & & 0.526 & \textbf{0.611} & 0.593 \\
                           & $5\mbox{-}SA$ & 0.281 & 0.304 & \underline{\textbf{0.391}} & & 0.355 & 0.396 & \textbf{0.424} \\
                           & $IG$ & 0.892 & 0.881 & \underline{\textbf{0.815}} & & 0.889 & 0.862 & \underline{\textbf{0.829}} \\
                           & $IE$ & 6.10 & 6.21 & \underline{\textbf{6.63}} & & 6.20 & 6.40 & \textbf{6.66} \\
                           & $SG$ & 0.904 & 0.879 & \underline{\textbf{0.856}} & & 0.862 & 0.849 & \underline{\textbf{0.821}} \\
                           & $SE$ & 5.31 & 5.59 & \underline{\textbf{5.71}} & & 5.06 & 5.11 & \textbf{5.30} \\

\bottomrule
\end{tabular}
\end{table}



%% file: Chapters/07_Solution2.tex
\chapter{Solution 2: A Post-processing Approach for Mitigating Multi-sided Exposure Bias}
\label{chap:solution2}
\tikzsetfigurename{solution2_}

In this chapter, I introduce a graph-based technique for tackling exposure bias in recommender systems. The proposed technique is general and can be used for mitigating exposure bias of both items and suppliers. The experimental results show the superiority of the proposed technique on mitigating exposure bias compared to other baselines on different datasets. I named the proposed technique as FairMatch algorithm. My contributions in this chapter are published in \cite{mansoury2020fairmatch} and also are submitted to ACM Transactions on Information Systems (TOIS). 

\section{Introduction}

One of the main reasons for different items and suppliers not getting a fair exposure in the recommendations is the popularity bias problem where few popular items/suppliers are over-recommended while the majority of other items/suppliers do not get a deserved attention. For example, in a music recommendation system, few popular artists might take up the majority of the streamings leading to under-exposure of less popular artists. This bias, if not mitigated, can negatively affect the experience of different users and items on the platform \cite{mehrotra2018towards,abdollahpouri2020multi}. It could also be perpetuated over time by the interaction of users with biased recommendations and, as a result, using biased interactions for training the model in the subsequent times \cite{damour2020,chaney2018,sinha2016,sun2019debiasing,mansoury2020feedback}.\\
\indent There are numerous methods to tackle exposure bias by either modifying the underlying recommendation algorithms by incorporating the popularity of each item 
\cite{vargas2014improving,sun2019debiasing,abdollahpouri2017controlling,adamopoulos2014over} or as a post-processing re-ranking step to modify an existing, often larger, recommendation list and extract a shorter list that has a better characteristics in terms of fair exposure of different items or suppliers \cite{adomavicius2011maximizing,adomavicius2011improving,antikacioglu2017,abdollahpouri2019managing}. However, most of these algorithms solely concentrated on mitigating the exposure (visibility) bias in an item level. What these algorithms ignore is the complexity of many real world recommender systems where there are different suppliers that provide the recommended items and hence the fairness of exposure in a supplier level need to be also addressed \cite{himan2019a}.\\
\indent One way to improve supplier exposure fairness is to improve the visibility of items hoping it will also lead to giving a more balanced exposure to different suppliers as often there is a positive correlation between the popularity of suppliers and their items. However, only optimizing for item visibility without explicitly taking into account the suppliers in the recommendations does not necessarily make the recommendations fairer for suppliers. This can be observed in the following example. 

\begin{figure}[btp]
    \centering
    \includegraphics[width=0.9\textwidth]{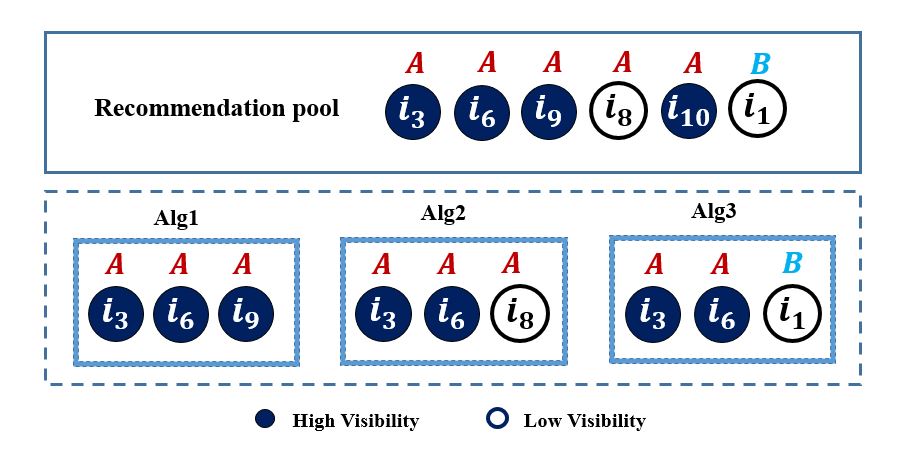}
    \caption{Comparison between a relevance based recommendation algorithm ($Alg1$), item visibility-aware reranker ($Alg2$), and supplier visibility-aware reranker ($Alg3$).}\label{motivation}
\end{figure}

Figure~\ref{motivation} shows a scenario where we have a list of items as candidate pool and the goal is to extract a list of recommendations (in this example the size is 3 for illustration purposes) and recommend it to the user. In addition, items are categorized to either \textit{high visibility} (i.e.\ frequently recommended) and \textit{low visibility} (less frequently recommended). Moreover, each item is also provided by either supplier $A$ or $B$. Three recommendation algorithms (these are just for illustration purposes) are compared in terms of how they extract the final list of three items. The first algorithm \textit{Alg1} extracts the three most relevant items from the top of the list without considering the visibility of items or which supplier they belong to. Obviously, this algorithm performs poorly in terms of fairness of item exposure and supplier exposure since only highly relevant items are recommended and they are all from supplier $A$. In contrast, the second algorithm \textit{Alg2} extracts the final recommendation list by also taking into account the visibility of items. This algorithm could represent many existing approaches to overcome exposure bias in recommendation. However, although the list of recommended items are now more diverse in terms of different type of items (high visibility vs low visibility) it still only contains items from supplier $A$ since the supplier information was not incorporated in the algorithm. The third algorithm \textit{Alg3}, on the other hand, has recommended a diverse list of items not only in terms of items, but also in terms of the suppliers of those items. Therefore, it is important to also optimize for suppliers for achieving fairer recommendation results.

\section{FairMatch algorithm}


FairMatch algorithm is formulated as a post-processing step after the recommendation generation. In other words, first recommendation lists of size larger than what ultimately is desired for each user is generated using any standard recommendation algorithm and then those large recommendation lists are used to build the final recommendation lists. FairMatch works as a batch process, similar to that proposed in \cite{surer2018} where all the recommendation lists are produced at once and re-ranked simultaneously to achieve the objective. In this formulation, a longer recommendation list of size $t$ for each user is produced and then, after identifying candidate items (based on defined utility, more details in Section~\ref{weight_comp}) by iteratively solving the maximum flow problem on recommendation bipartite graph, a shorter recommendation list of size $n$ (where $t>>n$) is generated.\\
\indent Let $G=(I,U,E)$ be a bipartite graph of recommendation lists where $I$ is the set of left nodes, $U$ is the set of right nodes, and $E$ is the set of edges between left and right nodes when recommendation occurred.  
$G$ is initially a uniformly weighted graph, 
 but we will update the weights for edges as part of the algorithm. 
I will discuss the initialization and the weighting method in Section~\ref{weight_comp}.\\
\indent Given a weighted bipartite graph $G$, the goal of our FairMatch algorithm is to improve the exposure fairness of recommendations without a significant loss in accuracy of the recommendations. I define exposure fairness as providing equal chance for items or suppliers to appear in recommendation lists. The FairMatch algorithm does this by identifying items or suppliers with low visibility in recommendation lists and promote them in the final recommendation lists while maintaining the relevance of recommended items for users.

\begin{algorithm}[btp]
\caption{The FairMatch Algorithm}
\small
\begin{algorithmic}
  \Function{FairMatch}{Recommendations $R$, TopN $n$, Suppliers $S$, Coefficient $\lambda$}
    \State Build graph $G=(I,U,E)$ from $R$
    \State Initialize \textit{subgraphs} to empty
    \Repeat
        \State $G$=WeightComputation($G$, $R$, $S$, $\lambda$)
        \State $\mathcal{I}_{C}$ = Push-relabel($G$)
        \State Initialize $subgraph$ to empty
        \For {\textbf{each} $i \in \mathcal{I}_{C}$}
            \If{$label_{i} \geq |I|+|U|+2$}
                \For {\textbf{each} $u \in Neighbors(i)$}
                    \State Append $<i,u,e_{iu}>$ to \textit{subgraph}
                \EndFor
            \EndIf 
        \EndFor
        \If{$subgraph$ is empty}
            \State $break$
        \EndIf
        \State Append $subgraph$ to $subgraphs$
        \State $G$=Remove $subgraph$ from $G$
    \Until($true$)
    \State Reconstruct $R$ of size $n$ based on \textit{subgraphs}
   \EndFunction
\end{algorithmic}

\end{algorithm}

FairMatch algorithm uses an iterative process to identify the subgraphs of $G$ that satisfy the underlying definitions of fairness without a significant loss in accuracy of the recommendation for each user. After identifying a subgraph $\Gamma$ at each iteration, $\Gamma$ will be removed from $G$ and the process of finding subgraphs on the rest of the graph (i.e.\ $G / \Gamma$) will continue. The algorithm keeps track of all the subgraphs as it uses them to generate the final recommendations in the last step. 

Identifying $\Gamma$ at each iteration is done by solving a \textit{Maximum Flow} problem (explained in Section~\ref{fair_max}) on the graph obtained from the previous iteration. Solving the maximum flow problem returns the left nodes connected to the edges with lower weight on the graph. After finding those left nodes, we form subgraph $\Gamma$ by separating identified left nodes and their connected right nodes from $G$. Finally, $<user,item>$ pairs in subgraphs are used to construct the final recommendation lists of size $n$. I will discuss this process in detail in the following Sections.





\begin{figure}[btp]
    \centering
    \includegraphics[width=0.99\textwidth]{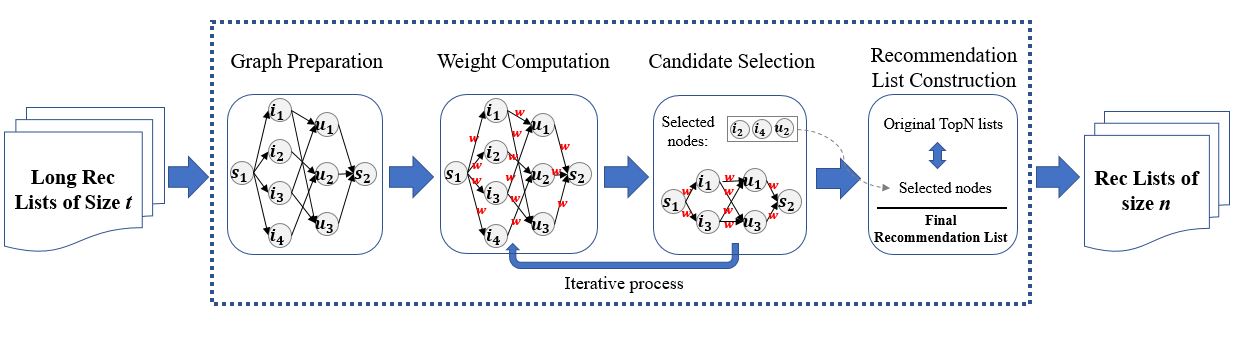}
    \caption{The process of FairMatch algorithm.}\label{fig:process}
\end{figure}

Algorithm 1 shows the pseudocode for FairMatch. Overall, FairMatch algorithm consists of the following four steps: 1) Graph preparation, 2) Weight computation, 3) Candidate selection, and 4) Recommendation list construction. Figure~\ref{fig:process} shows the process of FairMatch algorithm. FairMatch takes the long recommendation lists of size $t$ generated by a base recommendation algorithm as input, and then over four consecutive steps, as mentioned above, it generates the final recommendation lists. The detail about each step in FairMatch algorithm would be discussed in the following Sections. 

\subsection{Graph preparation}

Given long recommendation lists of size $t$ generated by a standard recommendation algorithm, we create a bipartite graph from recommendation lists in which items and users are the nodes and recommendations are expressed as edges.\\
\indent Since FairMatch algorithm is formulated as a maximum flow problem, we also add two nodes, \textit{source} ($s_1$) and \textit{sink} ($s_2$). The purpose of having a source and sink node in the maximum flow problem is to have a start and endpoint for the flow going through the graph. We connect $s_1$ node to all left nodes and also we connect all right nodes to $s_2$. Figure~\ref{fig:ex} shows a sample bipartite graph resulted in this step. 




\subsection{Weight computation}\label{weight_comp}

Weight computation step plays an important role on improving the exposure fairness of recommendations in the proposed model. Depending on the fairness definition that we desire to achieve, weight computation step should be adapted accordingly. In this Section, 
I discuss how weight computation can be adapted for improving the exposure fairness of items or suppliers.\\
\indent Given the bipartite recommendation graph, $G=(I,U,E)$, the task of weight computation is to calculate the weight for edges between the source node and left nodes, left nodes and right nodes, and right nodes and sink node.\\
\begin{figure}[t]
    \centering
    \includegraphics[width=0.5\textwidth]{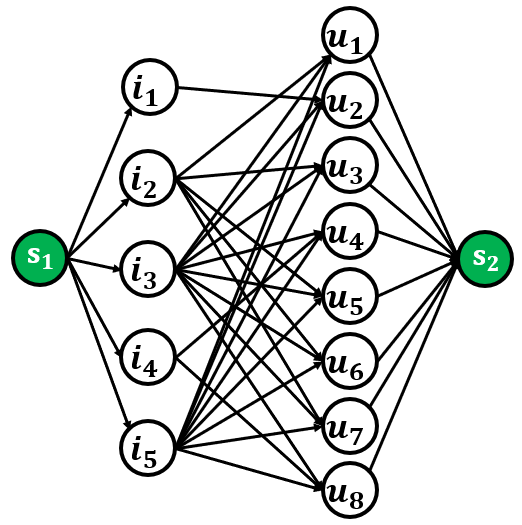}
    \caption{An example of a recommendation bipartite graph of recommendation lists of size 3.}\label{fig:ex}
\end{figure}
\indent For edges between left nodes and right nodes, I define the weights as the weighted sum of user utility and supplier utility (or instead, item utility). The utility of each user is defined as the relevance of recommended items for that user. Given the long recommendation list of size $t$ for user $u$ as $L_u$, in this formulation, I define the relevance of an item $i$ for user $u$ as rank of $i$ in sorted $L_u$ in descending order based on predicted score by the base recommender. This way, items in lower rank will be more relevant to the user (e.g.\ item in the first rank is the most relevant one). 

The utility for each item and supplier is defined as their exposure or visibility in the long recommendation lists. The visibility of each item is defined as the degree of the node corresponding to that item (excluding the edge with the source node). Item degree is the number of edges going out from that node connecting it to the user nodes and that shows how often it is recommended to different users. Analogously, the visibility for each supplier is defined as sum of the degree of all nodes corresponding to the items belonging to that supplier. Therefore, I introduce two separate weight computation schemes, one for item utility and another for supplier utility, which eventually results in two variations of FairMatch algorithm as follows:


\begin{itemize}
    \item \textbf{$FairMatch^{item}$:} For computing the weight for edges between $i \in I$ and $u \in U$, I use the following Equation:

\begin{equation}\label{wi}
    w_{iu}= \lambda \times rank_{iu} + (1 - \lambda) \times degree_i
\end{equation}

\noindent where $rank_{iu}$ is the position of item $i$ in the sorted recommendation list of size $t$ generated for user $u$, $degree_i$ is the number of edges from $i$ to right nodes (i.e.\ $u \in U$), and $\lambda$ is a coefficient to control the trade-off between the relevance of the recommendations and the exposure of items. 

    \item \textbf{$FairMatch^{Sup}$}: For computing the weight for edges between $i \in I$ and $u \in U$, I use the following Equation:
  
  \begin{equation}\label{ws}
    w_{iu}= \lambda \times rank_{iu} + (1 - \lambda) \times       
    \sum_{i \in A(B(i))}{degree_{i}}
\end{equation}

    
    where $B(i)$ returns the supplier of item $i$ and $A(B(i))$ returns all items belonging to the supplier of item $i$. Therefore, the term $\sum_{i \in A(B(i))}{degree_{i}}$ computes the visibility of supplier of item $i$ (i.e.\ sum of visibility of all items that belong to the supplier of item $i$). $rank_{iu}$ and  $\lambda$ have the same definition as Equation~\ref{wi}. 
\end{itemize}




Note that in Equation~\ref{wi} and \ref{ws}, $rank_{iu}$ and visibility for suppliers and items have different ranges. The range for $rank_{iu}$ is from 1 to $t$ (there are $t$ different positions in the original list) and the range of visibility depends on  the frequency
of the item (or its supplier) recommended to the users (the more frequent it is recommended to different users the higher its degree is). Hence, for a meaningful weighted sum, I normalize visibility of items and suppliers to be in the same range as $rank_{iu}$. 

Given weights of the edges between $i \in I$ and $u \in U$, $w_{iu}$, total capacity of $I$ and $U$ would be 
$C_{T}=\sum_{i\in I}^{}\sum_{u \in U}^{}w_{iu}$
which simply shows the sum of the weights of the edges connecting left nodes to the right nodes.\\
\indent For computing the weight for edges connected to the source and sink nodes, first, I equally distribute $C_{T}$ to left and right nodes. Therefore, the capacity of each left node, $C_{eq}(I)$, and right node, $C_{eq}(U)$, would be as follow: 

\begin{equation}
C_{eq}(I)=\ceil[\bigg]{\frac{C_{T}}{|I|}}, \;\;\;\;\;\; C_{eq}(U)=\ceil[\bigg]{\frac{C_{T}}{|U|}}
\end{equation}

\noindent where $\ceil[\big]{a}$ returns the ceil value of $a$. For example, suppose the total capacity, $C_{T}$, is 100. If we have 5 left nodes and 8 right nodes (similar to Figure~\ref{fig:ex}), then the capacity of each left node would be 20 ($\ceil[\big]{100/5}$) and the capacity of each right node would be 13 ($\ceil[\big]{100/8}$).
Then, based on equal capacity assigned to each left and right nodes, we follow the method introduced in \cite{bonchi2018} to compute weights for edges connected to source and sink nodes as follow:

\begin{equation} \label{eq:sl_cap1}
\forall i \in I, w_{s_{1}i}=\ceil[\bigg]{min(\frac{C_{eq}(I)}{gcd(C_{eq}(I),C_{eq}(U))},\frac{C_{eq}(U)}{gcd(C_{eq}(I),C_{eq}(U))})}
\end{equation}


\begin{equation} \label{eq:sl_cap2}
\forall u \in U, w_{us_{2}}=\ceil[\bigg]{\frac{C_{eq}(I)}{gcd(C_{eq}(I),C_{eq}(U))}}
\end{equation}

\noindent where $gcd(C_{eq}(I),C_{eq}(U))$ is the Greatest Common Divisor of the distributed capacity of left and right nodes. 
Assigning the same weight to edges connected to the source and sink nodes guaranties that all nodes in $I$ and $U$ are treated equally and the weights between them play an important role in FairMatch algorithm. In Section~\ref{fair_max}, I further discuss the impact of these weights on effectiveness of FairMatch  algorithm. 

\subsection{Candidate selection}\label{fair_max}

The graph constructed in previous steps is ready to be used for solving the maximum flow problem. In a maximum flow problem, the main goal is to find the maximum amount of feasible flow that can be sent from the source node to the sink node through the flow network. Several algorithms have been proposed for solving a maximum flow problem. Well-known algorithms are Ford--Fulkerson \cite{ford1956}, Push-relabel \cite{goldberg1988}, and Dinic's algorithm \cite{dinic1970}. In FairMatch algorithm, I use Push-relabel algorithm to solve the maximum flow problem on the bipartite recommendation graph as it is one of the efficient algorithms for this matter and also it provides some functionalities that FairMatch algorithm benefits them.\\
\begin{figure*}[t]
    \centering
    \begin{subfigure}[b]{0.32\textwidth}
        \includegraphics[width=\textwidth]{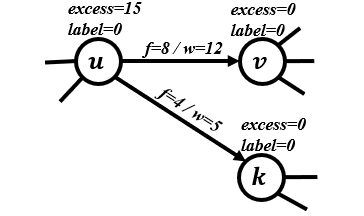}
        \caption{Original graph} \label{fig:push_ex:a}
    \end{subfigure}
    \begin{subfigure}[b]{0.33\textwidth}
        \includegraphics[width=\textwidth]{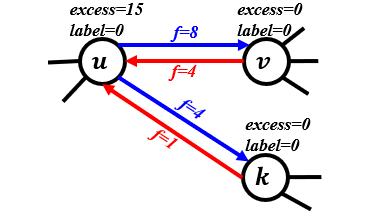}
        \caption{Residual graph} \label{fig:push_ex:b}
    \end{subfigure}%
    \begin{subfigure}[b]{0.33\textwidth}
        \includegraphics[width=\textwidth]{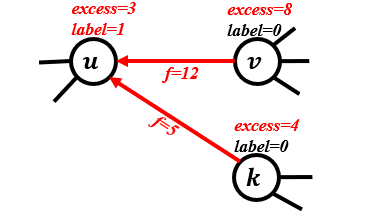}
        \caption{Pushing excess flow of u} \label{fig:push_ex:c}
    \end{subfigure}%
\caption{Example of push and relabel operations.} \label{fig:push_ex}
\end{figure*}
\indent In push-relabel algorithm, each node will be assigned two attributes: \textit{label} and \textit{excess flow}. The label attribute is an integer value that is used to identify the neighbors to which the current node can send flow. A node can only send flow to neighbors that have lower label than the current node. Excess flow is the remaining flow of a node that can still be sent to the neighbors. When all nodes of the graph have excess flow equals to zero, the algorithm will terminate.\\
\indent The push-relabel algorithm combines $push$ operations that send a specific amount of flow to a neighbor, and $relabel$ operations that change the label of a node under a certain condition (when the node has excess flow greater than zero and there is no neighbor with label lower than the label of this node).\\
\indent Here is how the push-relabel algorithm works: Figure~\ref{fig:push_ex} shows a typical graph in the maximum flow problem and an example of push and relabel operations. In Figure~\ref{fig:push_ex:a}, $f$ and $w$ are current flow and weight of the given edge, respectively. In Push-relabel algorithm, a residual graph, $G^{'}$, will be also created from graph $G$. As graph $G$ shows the flow of forward edges, graph $G^{'}$ shows the flow of backward edges 
calculated as $f_{backward}=w-f$. Figure~\ref{fig:push_ex:b} shows residual graph of graph $G$ in Figure~\ref{fig:push_ex:a}. Now, we want to perform a push operation on node $u$ and send its excess flow to its neighbors.\\
\indent Given $x_{u}$ as excess flow of node $u$, $push(u,v)$ operation will send a flow of amount $\Delta=min(x_{u},f_{uv})$ from node $u$ to node $v$ and then will decrease excess flow of $u$ by $\Delta$ (i.e.\ $x_{u}=x_{u}-\Delta$) and will increase excess flow of $v$ by $\Delta$ (i.e.\ $x_{v}=x_{v}+\Delta$). After $push(u,v)$ operation, node $v$ will be put in a queue of active nodes to be considered by the push-relabel algorithm in the next iterations and residual graph would be updated. Figure~\ref{fig:push_ex:c} shows the result of $push(u,v)$ and $push(u,k)$ on the graph shown in Figure~\ref{fig:push_ex:b}. In $push(u,v)$, for instance, since $u$ and all of its neighbors have the same label value, in order to perform push operation, first we need to perform relabel operation on node $u$ to increase the label of $u$ by one unit more than the minimum label of its neighbors to guaranty that there is at least one neighbor with lower label for performing push operation. After that, node $u$ can send flow to its neighbors.\\
\indent Given $x_{u}=15$, $f_{uv}=8$, and $f_{uk}=4$ in Figure~\ref{fig:push_ex:b}, after performing relabel operation, we can only send the flow of amount 8 from $u$ to $v$ and the flow of amount 4 from $u$ to $k$. After these operations, residual graph (backward flow from $v$ and $k$ to $u$) will be updated.\\ 
\indent The push-relabel algorithm starts with a "preflow" operation to initialize the variables and then it iteratively performs push or relabel operations until no active node exists for performing operations. Assuming $\mathcal{L}_v$ as the label of node $v$, in preflow step, we initialize all nodes as follow: $\mathcal{L}_{s_1}=|I|+|U|+2$, $\mathcal{L}_{i \in I}=2$, $\mathcal{L}_{u \in U}=1$, and $\mathcal{L}_{s_2}=0$. This way, we will be able to send the flow from $s_1$ to $s_2$ as the left nodes have higher label than the right nodes. Also, we will push the flow of amount $w_{s_{1}i}$ (where $i \in I$) from $s_1$ to all the left nodes. 

After preflow, all of the left nodes $i \in I$ will be in the queue, $\mathcal{Q}$, as active nodes because all those nodes now have positive excess flow. The main part of the algorithm will now start by dequeuing an active node $v$ from $\mathcal{Q}$ and performing either push or relabel operations on $v$ as explained above. This process will continue until $\mathcal{Q}$ is empty. At the end, each node will have specific label value and the sum of all the coming flows to node $s_2$ would be the maximum flow of graph $G$. For more details see \cite{goldberg1988}

An important question is: \textit{how does the Push-relabel algorithm can find high-quality (more relevant) nodes (items and their suppliers) with low degree (visibility)?} 
I answer this question by referring to the example in Figure~\ref{fig:push_ex:c}. In this figure, assume that $u$ has a backward edge to $s_1$. Since $u$ has excess flow greater than zero, it should send it to its neighbors. However, as you can see in the figure, $u$ does not have any forward edge to $v$ or $k$ nodes. Therefore, it has to send its excess flow back to $s_1$ as $s_1$ is the only reachable neighbor for $u$. Since $s_1$ has the highest label in our setting, in order for $u$ to push all its excess flow back to $s_1$, it should go through a relabel operation so that its label becomes larger than that of $s_1$. Therefore, the label of $u$ will be set to $\mathcal{L}_{s_1}+1$ for an admissible push.

The reason that $u$ receives high label value is the fact that it initially receives high flow from $s_1$ (it is important how to assign weight to edges between $s_1$ and left nodes), but it does not have enough capacity (the sum of weights between $u$ and its neighbors is smaller than its excess flow. i.e.\ 8+4<15) to send all that flow to them. 

In FairMatch, in step 2 (i.e.\ Section \ref{weight_comp}), the same weight is assigned to all edges connected to $s_1$ and $s_2$. This means that the capacity of edges from $s_1$ to all item nodes would be the same and also the capacity of edges from all user nodes to $s_2$ would be the same. However, the weights assigned to edges between item and user nodes depend on the quality and visibility of the recommended items to users in the recommendation lists and play an important role in finding the desired output in FairMatch algorithm. Assume that the weights for edges between $s_1$ and item nodes are $w_{s_1}$ and the weights for edges between user nodes and $s_2$ are $w_{s_2}$.\\
\indent In preflow step, $s_1$ sends flow of amount $w_{s_1}$ to each item node in $I$ and this flow would be recorded in each item nodes as their excess flow. When Push-relabel starts after preflow, the algorithm tries as much as possible to send the excess flow in item nodes to user nodes and then finally to $s_2$. However, the possibility of achieving this objective depends on the capacity of edges between item and user nodes. Items connected to edges with low capacity will not be able to send all their excess flow to their neighbors (user nodes) and will be returned as candidate items in step 3 of FairMatch algorithm.\\
\indent There are two possible reasons for some items to not be able to send all their excess flow to their neighbors: 1) they have few neighbors (user nodes) which signifies that those items are recommended to few users and consequently they have low visibility in recommendation lists, 2) they are relevant to the users' preferences meaning that their rank in the recommendation list (sorted based on the predicted score by a base recommender) for users is low and consequently make those items more relevant to users. Hence, these two reasons--low visibility and high relevance--cause some items to not have sufficient capacity to send their excess flow to their neighbors (user nodes) and have to send it back to $s_1$ similar to what we illustrated above. As a result, sending back the excess flow to $s_1$ means first running relabel operation as $s_1$ has higher label value than item nodes and then push the excess flow to $s_1$. Performing relabel operation will assign the highest label value to those items which makes them to be distinguishable from other nodes after push-relabel algorithm terminated. Therefore, in step 3 (i.e.\ Section \ref{fair_max}), left nodes without sufficient capacity on their edges will be returned as part of the outputs from push-relabel algorithm and are considered for constructing the final recommendation list in step 4 (i.e.\ Section \ref{step4}). FairMatch aims at promoting those high relevance items (or suppliers) with low visibility.

\subsection{Recommendation list construction}\label{step4}

In this step, the goal is to construct a recommendation list of size $n$ by the $<user,item>$ pairs identified in previous step. 
Given a recommendation list of size $n$ for user $u$, $L_{u}$, sorted based on the scores generated by a base recommendation algorithm, candidate items identified by FairMatch connected to $u$ as $\mathcal{I}_{C}$, and visibility of each item $i$ in recommendation lists of size $n$ as $\mathcal{V}_{i}$, I use the following process for generating recommendation list for $u$.\\ 
\indent First, I sort recommended items in $L_{u}$ and $\mathcal{I}_{C}$ based on their $\mathcal{V}_{i}$ in ascending order. Then, I remove $min(\beta \times n,|\mathcal{I}_{C}|)$ from the bottom of sorted $L_{u}$ and add $min(\beta \times n,|\mathcal{I}_{C}|)$ items from $\mathcal{I}_{C}$ to the end of $L_{u}$. $\beta$ is a hyperparameter in $0<\beta \leq 1$ that specifies the fraction of items in the original recommendation lists that we want to replace with the identified items in previous step.\\
\indent This process will ensure that extracted items in the previous step will replace the frequently recommended items meaning that it decreases the visibility of the frequently recommended items/suppliers and increases the visibility of rarely recommended items/suppliers to generate a fairer distribution on recommended items/suppliers.

\section{Experimental results}

In this Section, I analyze the performance of \textit{FairMatch} algorithm in comparison with some of the state-of-the-art re-ranking algorithms using three different standard recommendation algorithms as the base for the re-ranking algorithms on two datasets. The datasets are MovieLens1M and Last.fm which their specifications are described in Chapter~\ref{chap:methodology}. The base recommender algorithms are Bayesian Personalized Ranking (\algname{BPR}) \cite{rendle2009bpr}, Neural Collaborative Filtering (\algname{NCF}) \cite{he2017neural}, User-based Collaborative Filtering (\algname{UserKNN}) \cite{Resnick:1994a}. These base recommendation algorithms are used to generate the long recommendation lists of size 50. These long recommendation lists are used as input for the proposed FairMatch and other re-ranking algorithms to generate the final recommendation lists of size 10.\\ 
\indent The baseline re-ranking algorithms used for comparison with the FairMatch algorithms are as follow. 

\begin{enumerate}
  \item \textbf{FA*IR.} This is the method introduced in \cite{zehlike2017fa} and is originally used for improving the representation of protected group in ranked recommendation lists. However, I adapted this method for improving the visibility of long-tail items in recommendation lists. I defined protected and unprotected groups as long-tail and short-head items, respectively. For separating short-head from long-tail items, I considered those top items which cumulatively make up 20\% of the ratings according the Pareto principle \cite{sanders1987pareto} as the short-head and the rest as long-tail items. Also, I set the other two hyperparameters, proportion of protected candidates in the top $n$ items\footnote{Based on suggestion from the released code, the range should be in $[0.02,0.98]$} and significance level\footnote{Based on suggestion from the released code, the range should be in $[0.01,0.15]$}, to $\{0.2,0.6,0.8\}$ and $\{0.05,0.1\}$, respectively. 
  \item \textbf{xQuAD}. This is the method introduced in \cite{abdollahpouri2019managing}. I specifically included xQuAD method since it attempts to promote less popular items (most likely items with low visibility in recommendation lists) by balancing the ratio of popular and less popular items in recommendation lists. This method involves a hyperparameter to control the trade-off between relevance and long-tail promotion, and I experimented with different values for this hyperparameter in $\{0.2,0.4,0.6,0.8,1\}$. Also, the separation of short-head and long-tail items is done according to Pareto principle as described above.
  \item \textbf{Discrepancy Minimization (DM)}. This is the method introduced in \cite{antikacioglu2017} and was explained in Chapter~\ref{chap:expo}. For hyperparameter tuning, I followed the experimental settings suggested by the original paper for the experiments. I set the target degree distribution to $\{1,5,10\}$ and relative weight of the relevance term to $\{0.01,0.5,1\}$.  
  \item \textbf{ProbPolicy}. This is the method introduced in \cite{mehrotra2018towards} and was mentioned in Chapter~\ref{chap:expo}. I included this method as it was designed for improving supplier fairness and visibility in recommendation lists. This method involves a hyperparameter for controlling the trade-off between the relevance of recommended items to users and supplier fairness. I set the value for this hyperparameter to $\{0.2,0.4,0.6,0.8,1\}$.
\end{enumerate}

I also used two simple methods to show the extreme case in bias mitigation for comparison purposes.
\begin{enumerate}

  \item \textbf{Reverse.} Given a recommendation list of size $t$ for each user generated by base recommendation algorithm, in this method, instead of picking the $n$ items from the top (most relevant items), we pick them from the bottom of the list (least relevant items). In this approach, it is expected to see an increase in aggregate diversity as we are giving higher priority to the items with lower relevance to be picked first. However, the accuracy of the recommendations will decrease as we give higher priority to less relevant items.
  \item \textbf{Random.} Given a recommendation list of size $t$ for each user generated by base recommendation algorithm, we randomly choose $n$ items from that list and create a final recommendation list for that user. Note that this is different from randomly choosing items from all catalog to recommend to users. The reason we randomly choose the items from the original recommended list of items (size $t$) is to compare other post-processing and re-ranking techniques with a simple random re-ranking. 
\end{enumerate}

\textit{Random} and \textit{Reverse} are mainly included to demonstrate the extreme version of a re-ranking algorithm where the sole focus is on improving aggregate diversity and exposure and we ignore the relevance of the recommended items as can be seen by the low precision for these two algorithms.\\
\indent Extensive experiments are performed using each re-ranking algorithm with multiple hyperparameter values. For the purpose of fair comparison, from each of those re-ranking algorithms (FA*IR, xQuAD, DM, ProbPolicy, and the both variations of FairMatch algorithm) the configuration which yields, more of less, the same precision loss is reported. These results enable us to better compare the performance of each technique on improving exposure fairness and other non-accuracy metrics while maintaining the same level of accuracy. The precision of each re-ranking algorithm on both datasets is reported in Tables~\ref{tlb_lf_50} and \ref{tbl_ml_50}. 

\begin{figure*}[t]
    \centering
    \begin{subfigure}[b]{0.99\textwidth}
        \includegraphics[width=\textwidth]{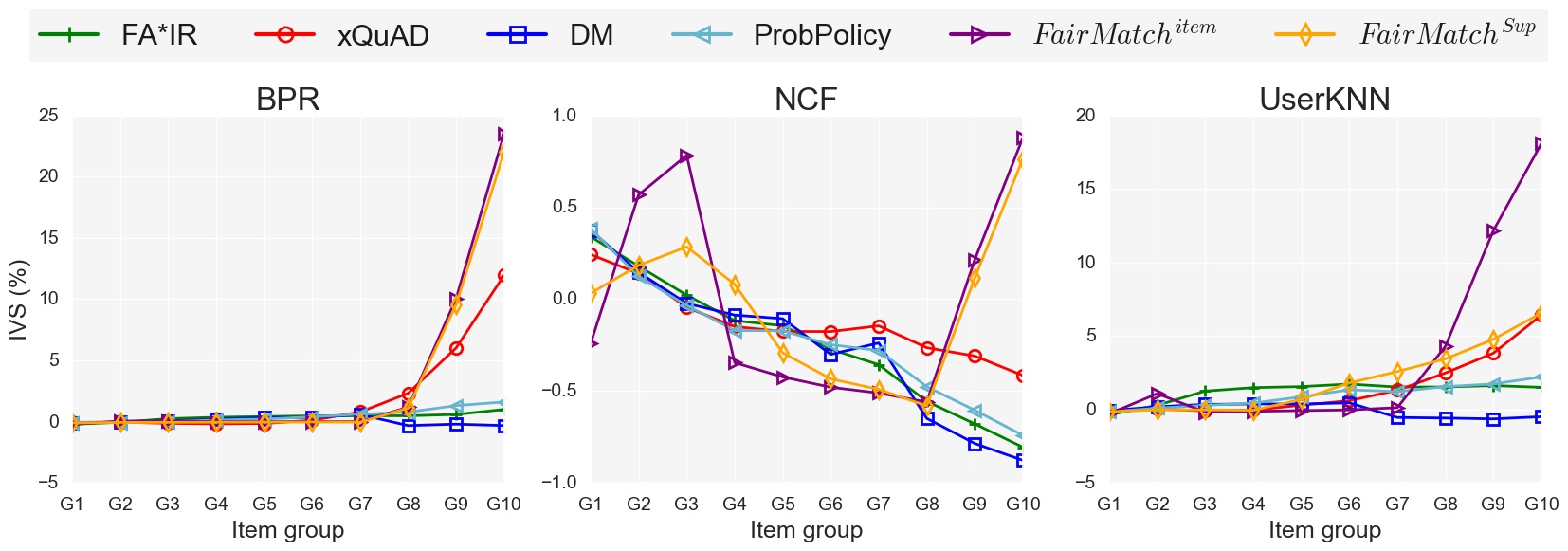}
        \caption{Last.fm} \label{vis_lf_50_item}
    \end{subfigure}
    \begin{subfigure}[b]{0.99\textwidth}
        \includegraphics[width=\textwidth]{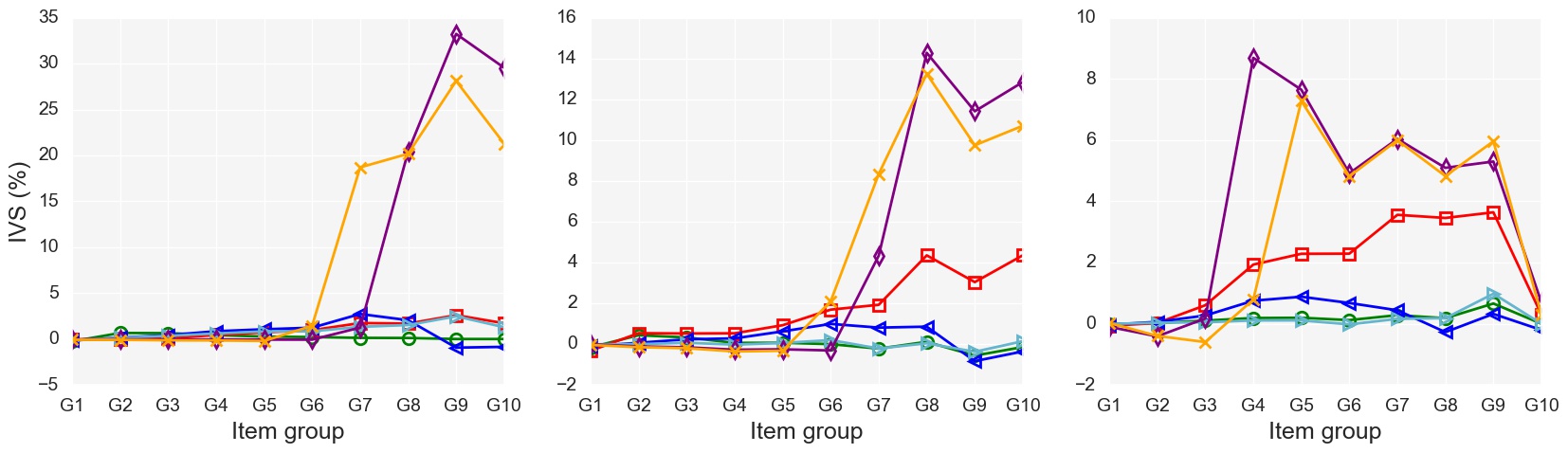}
        \caption{ML1M} \label{vis_ml_50_item}
    \end{subfigure}%
\caption{Percentage increase/decrease ($IVS$) in visibility of item groups for different reranking algorithms.} \label{vis_50_item}
\end{figure*}

\begin{figure*}[t]
    \centering
    \begin{subfigure}[b]{0.99\textwidth}
        \includegraphics[width=\textwidth]{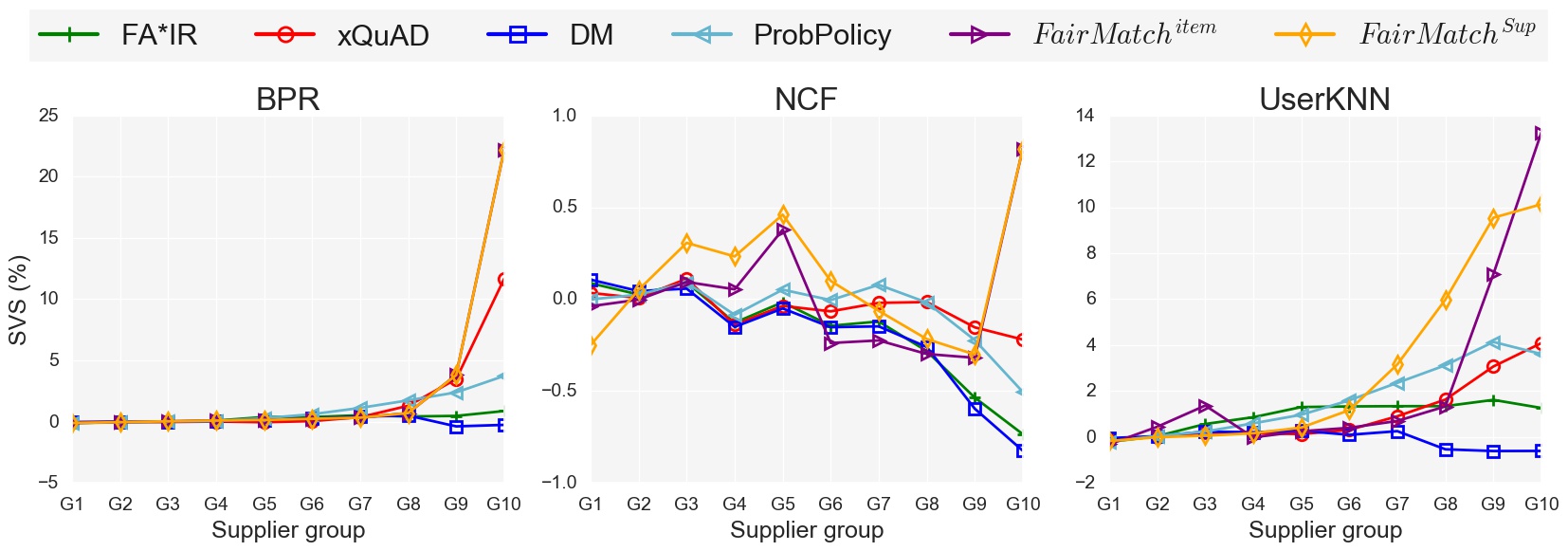}
        \caption{Last.fm} \label{vis_lf_50_sup}
    \end{subfigure}
    \begin{subfigure}[b]{0.99\textwidth}
        \includegraphics[width=\textwidth]{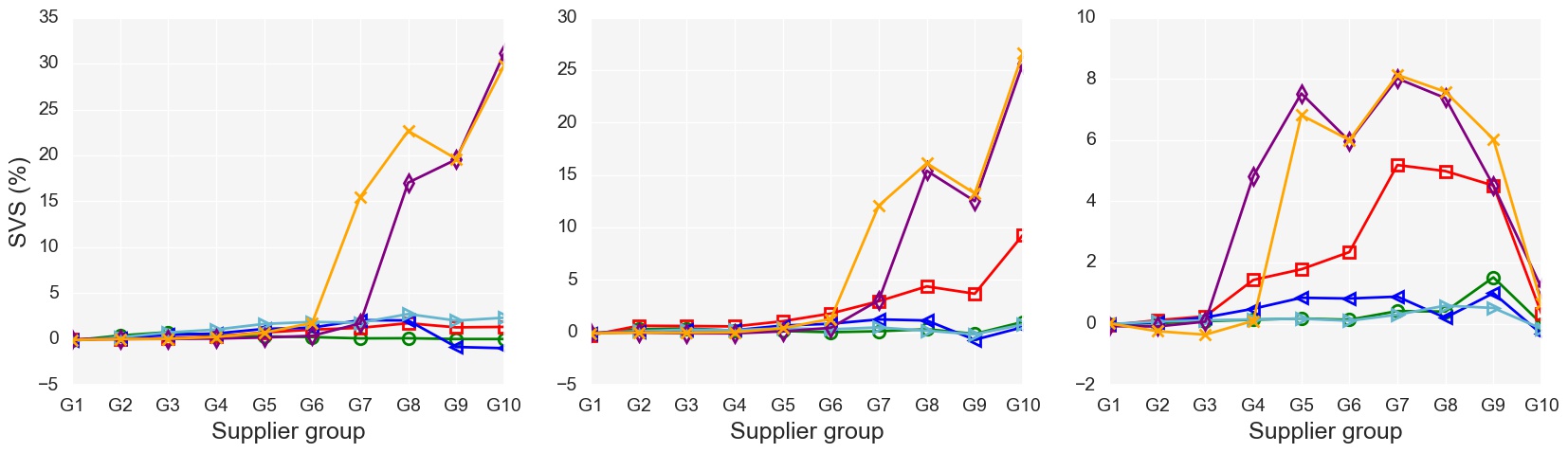}
        \caption{ML1M} \label{vis_ml_50_sup}
    \end{subfigure}%
\caption{Percentage increase/decrease ($SVS$) in visibility of supplier groups for different reranking algorithms.} \label{vis_50_sup}
\end{figure*}

\subsection{Visibility analysis}

Since the proposed FairMatch algorithm aims at improving the visibility of different items in the recommendations, I start my analysis with comparing different algorithms in terms of the visibility change ($IVS$) of the recommended items. Figure~\ref{vis_50_item} shows the percentage change in the visibility of the recommended item groups in recommendation lists generated by each re-ranking algorithm compared to their visibility in the recommendation lists generated by three base recommenders. In these plots, horizontal axis is the recommended item groups (created as explained in Section~\ref{metrics}) and vertical axis is $IVS$ metric. Item groups are sorted from the highest visibility (i.e.\ $G_1$) to the lowest visibility (i.e.\ $G_{10}$). It can be seen that both versions of FairMatch algorithm on both datasets have significantly increased the visibility of item groups with lower visibility while slightly taking away from the visibility of items with originally extreme visibility. $FairMatch^{item}$ performs slightly better than $FairMatch^{Sup}$ especially for item groups for very low visibility ($G_9$ and $G_{10}$) as it was expected since $FairMatch^{item}$ directly optimizes for improving the exposure of the low visibility items. 

Looking at \algname{NCF} on ML1M, it seems different re-ranking algorithms do not have a predictable behavior in terms of improving visibility of different item groups and, in some cases, even decreasing the visibility of item groups with already low visibility. However, a closer look at the scale of vertical axis reveals that these changes are very small and not significant. The reason is, on this dataset, \algname{NCF} has already done a good job in terms of fair item visibility and not much can be done via a re-ranking method. Among other reranking methods, xQuAD seems to also perform relatively well but still is outperformed by FairMatch. One interesting observation in this figure is that, using \algname{UserKNN} on ML1M, we can see that both FairMatch algorithms have significantly improved the visibility of item groups with medium visibility even more than the ones with lower visibility. Although these are items with medium visibility using our grouping strategy, they still get significantly less visibility compared to $G_1$ and $G_2$ in the base algorithm as we saw in Figure~\ref{dist_ml_50_item}. Therefore, we can still consider these item groups as items with relatively low visibility and FairMatch has increased their visibility.\\   
\indent Figure~\ref{vis_50_sup} is similar to Figure~\ref{vis_50_item} but here we show the percentage change in the visibility of the supplier groups in recommendation lists generated by each re-ranking algorithm compared to their visibility in the recommendation lists generated by three base recommenders. The first thing that can be observed from this figure is that, on both datasets, FairMatch algorithms outperform the other re-ranking methods especially for groups with lower visibility. \algname{NCF} on Last.fm has the same problem as we observed in Figure~\ref{vis_50_item} where the changes in vertical axis are not significant and all algorithms more or less perform equally. $FairMatch^{item}$ and $FairMatch^{Sup}$ are performing equally well for groups with extremely low visibility although $FairMatch^{Sup}$ tend to also improve the visibility of some other item groups such as $G_7$, $G_8$, and $G_9$ on ML1M using \algname{BPR} and on Last.fm using \algname{UserKNN}. Overall, $FairMatch^{Sup}$ has done a better job in terms of supplier visibility fairness and that was also expected since supplier visibility was directly incorporated into the objective function.\\ 
\indent In addition to measuring the improvement in visibility of different items, I also conducted an extensive analysis on other existing metrics in the literature to have a better picture of how each of these re-ranking methods help reducing the over-concentration of the recommendations around few highly visible items. Tables~\ref{tlb_lf_50} and \ref{tbl_ml_50} show the results for different re-ranking algorithms on Last.fm and ML1M datasets, respectively. I compare these algorithms in terms of item and supplier aggregate diversity and also fair distribution of recommended items and suppliers.



\input{./Tables/tbl_lf_50}
\input{./Tables/tbl_ml_50}

\subsection{Item aggregate diversity}

When it comes to increasing the number of unique recommended items (aggregate diversity), we can see that all re-ranking algorithms have improved this metric over the base algorithms on both datasets. I have only included $1\mbox{-}IA$ (each item should be recommended at least once to be counted) and $5\mbox{-}IA$ (each item should be recommended at least 5 times to be counted). I experimented with different values of $\alpha$ from 1 to 20 and the results can be seen in Figure~\ref{k_a} which I will describe afterwards. Generally speaking, all re-ranking methods have lost a certain degree of precision in order to improve aggregate diversity and other metrics related to fair distribution of recommended items and suppliers as can be seen from Tables~\ref{tlb_lf_50} and \ref{tbl_ml_50}. The reason is that the base algorithms are mainly optimized for relevance and therefore it is more likely for the items on top of the recommended list to be relevant to the users. As a result, when we re-rank the recommended lists and push some items in the bottom to go up to the top-n, we might swipe some relevant items with items that may not be as relevant.\\
\indent Regarding $1\mbox{-}IA$, $FairMatch^{item}$ seems to perform relatively better than the other re-rankers using all three base algorithms (\algname{BPR}, \algname{NCF} and \algname{UserKNN}) on both datasets indicating it recommends a larger number of items across all users. The same pattern can be seen for $LT$ which measures only the unique recommended items that fall into the long-tail category. This is however, not the case for $5\mbox{-}IA$ where in some cases $FairMatch^{item}$ is outperformed by other re-rankers. That shows, the improvement in recommending more unique items using $FairMatch^{item}$ is not achieved by recommending them frequent enough. On ML1M, however, $FairMatch^{item}$ performs very well on $5\mbox{-}IA$ metric. This difference in behavior across the datasets can be explained by the characteristics of the data as we saw in Figure~\ref{train_dist}. Overall, $FairMatch^{Sup}$ seems to also perform relatively well on item aggregate diversity and in some cases even better than $FairMatch^{item}$ such as on $5\mbox{-}IA$ using \algname{NCF} and \algname{UserKNN} on Last.fm and \algname{BPR} and \algname{NCF} on ML1M.

\subsection{Supplier aggregate diversity}

Suppliers of the recommended items are also important to be fairly represented in the recommendations. First and foremost, looking at the Tables~\ref{tlb_lf_50} and \ref{tbl_ml_50}, we can see that there is an overall positive connection between improving item aggregate diversity and supplier aggregate diversity indicating optimizing for either item or supplier visibility, can benefit the other side as well. However, when we directly incorporate the supplier visibility into our recommendation process as I did in $FairMatch^{Sup}$, we can see that the supplier aggregate diversity can be significantly improved. For example, we can see that $FairMatch^{Sup}$ has the best $1\mbox{-}SA$ on both datasets except for when the base algorithm is \algname{UserKNN} on ML1M where it was outperformed by $FairMatch^{item}$. So, overall, we can say that FairMatch (either $FairMatch^{Sup}$ or $FairMatch^{item}$) has the best $1\mbox{-}SA$ on both datasets using all three base algorithms.\\
\indent Regarding $5\mbox{-}SA$, FairMatch algorithms tend to also perform better than other re-rankers. Between the two variations of FairMatch, we can see that $FairMatch^{Sup}$ gives a better supplier aggregate diversity in most cases which is something that we expected. Similar to item aggregate diversity, we only included $1\mbox{-}SA$ and $5\mbox{-}SA$ for supplier aggregate diversity in the Tables. A more comprehensive analysis of the effect of $\alpha$ on this metric is illustrated in Section~\ref{alphaanalysis} which I will describe later.

\subsection{Fair distribution of recommended items}

I also wanted to evaluate different re-rankers in terms of fair distribution of recommendations across different items. I used Gini ($IG$) and Entropy ($IE$) as ways to measure how equally the recommendations are distributed across different recommended items. Even though I have not directly optimized for equal representation of different items, these two metrics show that the proposed FairMatch algorithm has given a much fairer chance to different items to be recommended compared to the base algorithms and some of the other re-rankers by having a low Gini and high Entropy.\\
\indent Between $FairMatch^{Sup}$ and $FairMatch^{item}$ there is no clear winner in terms of Gini and entropy for items as in some cases $FairMatch^{Sup}$ has a better Gini while in other cases $FairMatch^{item}$ performs better. Among other re-rankers, DM and FA*IR seem to also perform well on these two metrics indicating they also give a fair chance to different items to be recommended. 

\begin{figure*}[t]
    \centering
    \begin{subfigure}[b]{0.99\textwidth}
        \includegraphics[width=\textwidth]{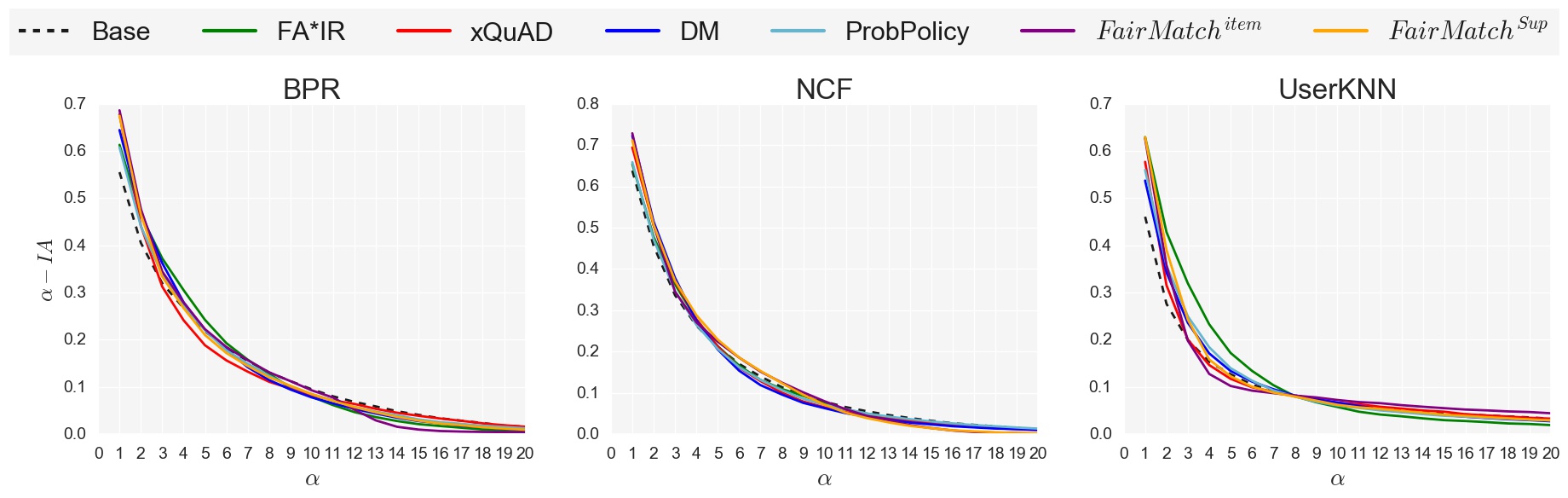}
        \caption{Last.fm} \label{k_a_lf}
    \end{subfigure}
    \begin{subfigure}[b]{0.98\textwidth}
        \includegraphics[width=\textwidth]{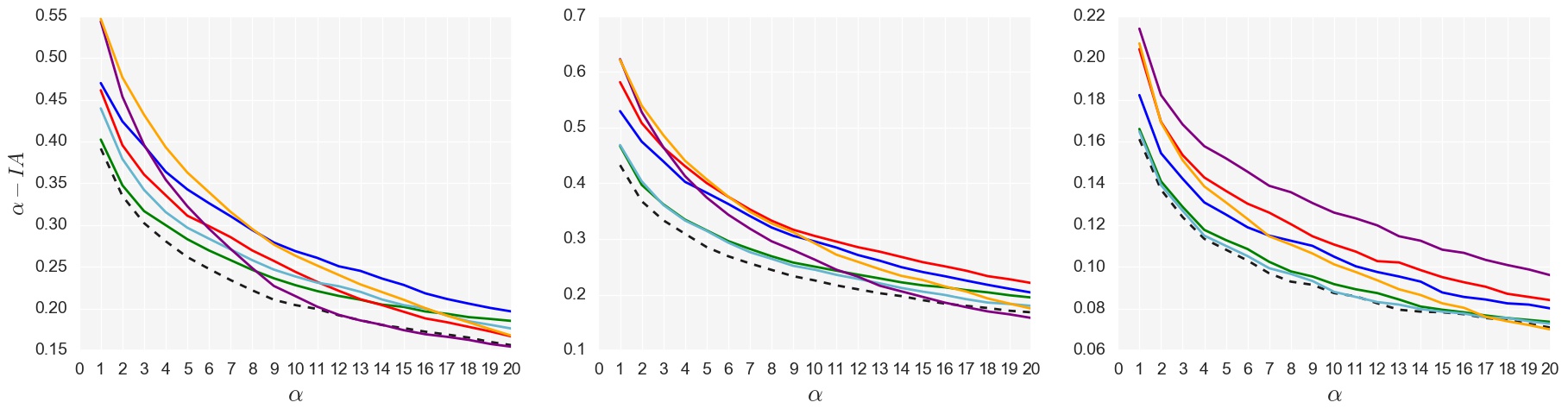}
        \caption{ML1M} \label{}
    \end{subfigure}%
\caption{Comparison of reranking algorithms in terms of item aggregate diversity ($\alpha\mbox{-}IA$) with different $\alpha$ values.} \label{k_a}
\end{figure*}

\subsection{Fair distribution of suppliers in recommendation lists}

In addition to standard Gini (i.e.\ $IG$) and Entropy (i.e.\ $IE$) which are generally calculated in an item level, I also measured the same metric but from the suppliers perspective and it can be seen in the Tables~\ref{tlb_lf_50} and \ref{tbl_ml_50} as $SG$ and $SE$ which measure the extent to which different suppliers are fairly recommended across different users. Overall, $FairMatch^{Sup}$ has the best $SG$ and $SE$ on both datasets in all situations except for \algname{NCF} and \algname{UserKNN} on ML1M. Also, using \algname{UserKNN} on ML1M, $FairMatch^{item}$ has the second best $SG$ and $SE$. This shows that incorporating the supplier visibility directly into the recommendation process can positively affect the fairness of representation across different suppliers and it is indeed supporting my initial hypothesis about the importance of incorporating suppliers in the recommendation process. The Probpolicy algorithm which also incorporates the supplier fairness in its recommendation generation, has also performed better than other re-rankers in terms of $SG$ and $SE$.

\begin{figure*}[t]
    \centering
    \begin{subfigure}[b]{0.99\textwidth}
        \includegraphics[width=\textwidth]{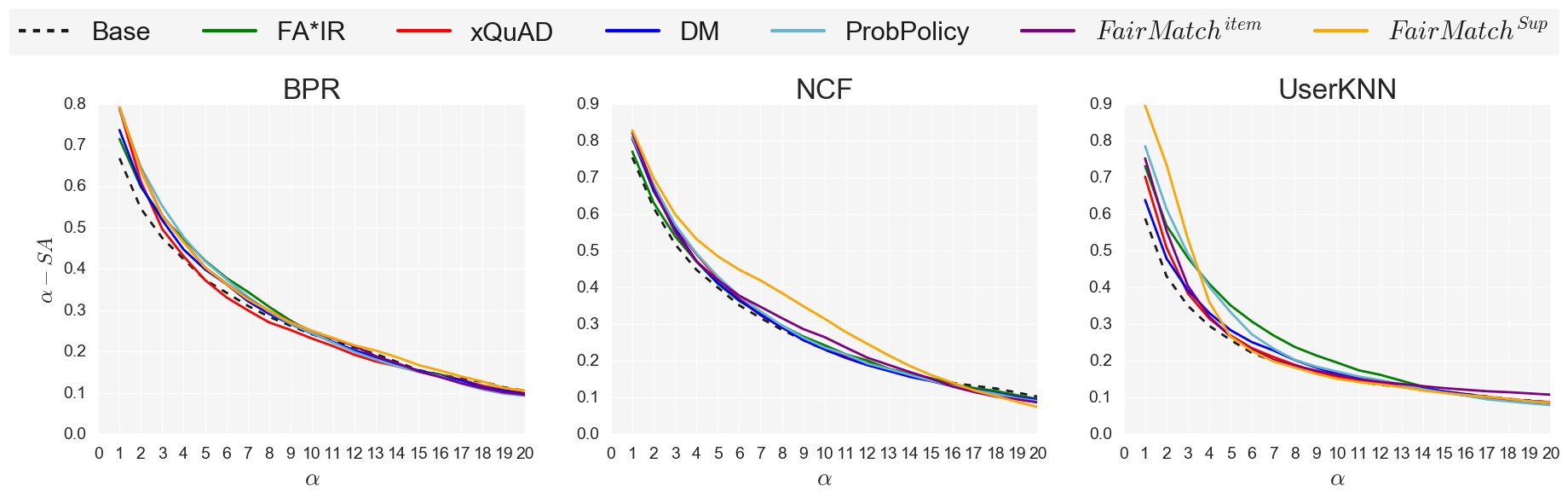}
        \caption{Last.fm} \label{k_sa_lf}
    \end{subfigure}
    \begin{subfigure}[b]{0.98\textwidth}
        \includegraphics[width=\textwidth]{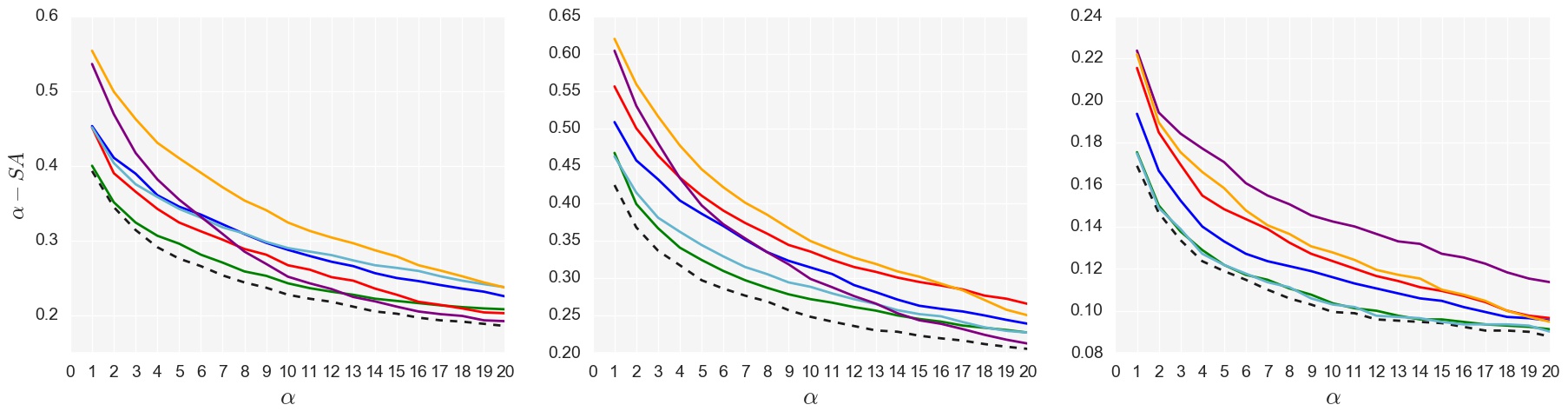}
        \caption{ML1M} \label{k_sa_ml}
    \end{subfigure}%
\caption{Comparison of reranking algorithms in terms of supplier aggregate diversity ($\alpha\mbox{-}SA$) with different $\alpha$ values.} \label{k_sa}
\end{figure*}

\begin{figure*}[t]
    \centering
    \begin{subfigure}[b]{0.8\textwidth}
        \includegraphics[width=\textwidth]{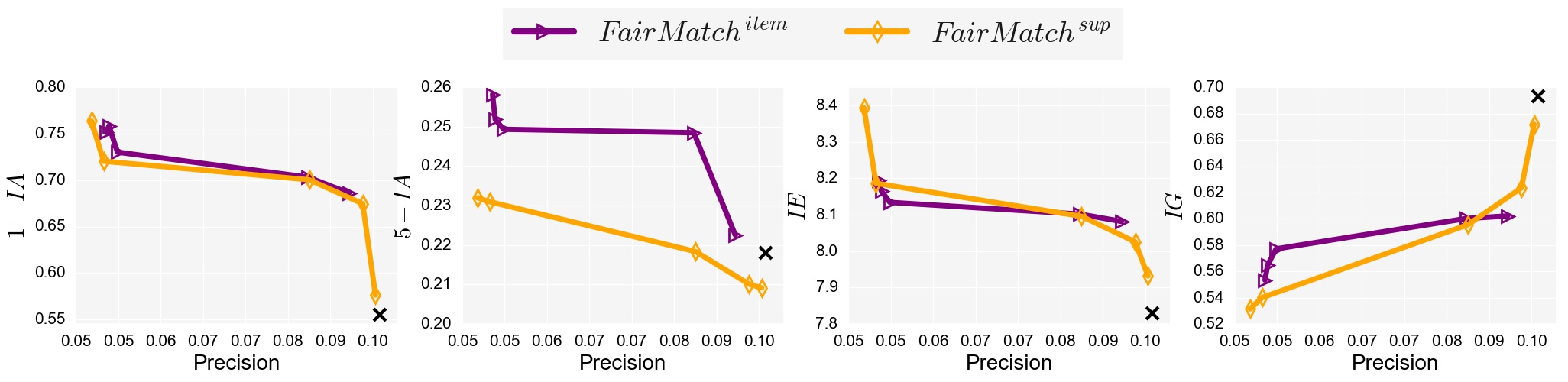}
        \caption{Last.fm, BPR} \label{sensitivity_lf_item_bpr}
    \end{subfigure}
    \begin{subfigure}[b]{0.8\textwidth}
        \includegraphics[width=\textwidth]{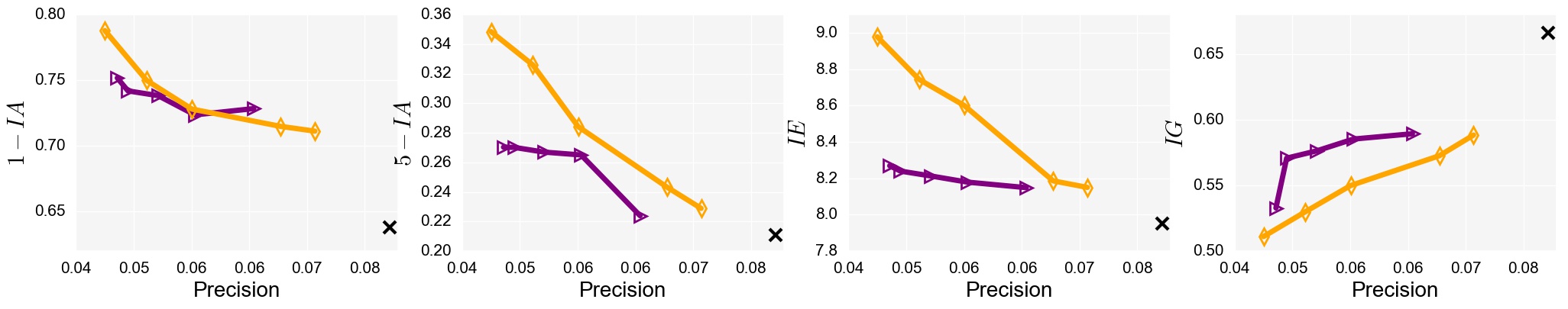}
        \caption{Last.fm, NCF} \label{sensitivity_lf_item_ncf}
    \end{subfigure}
    \begin{subfigure}[b]{0.8\textwidth}
        \includegraphics[width=\textwidth]{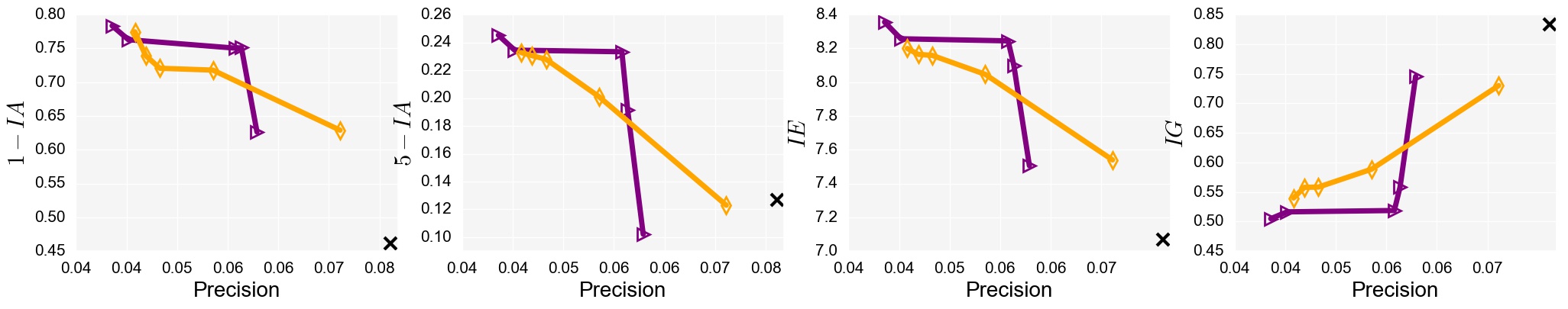}
        \caption{Last.fm, UserKNN} \label{sensitivity_lf_item_userknn}
    \end{subfigure}
    \begin{subfigure}[b]{0.8\textwidth}
        \includegraphics[width=\textwidth]{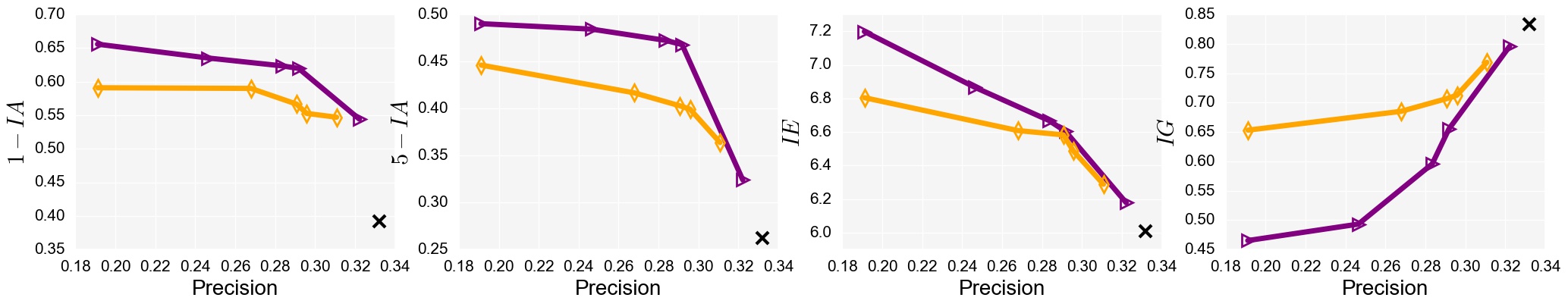}
        \caption{ML1M, BPR} \label{sensitivity_ml_item_bpr}
    \end{subfigure}
    \begin{subfigure}[b]{0.8\textwidth}
        \includegraphics[width=\textwidth]{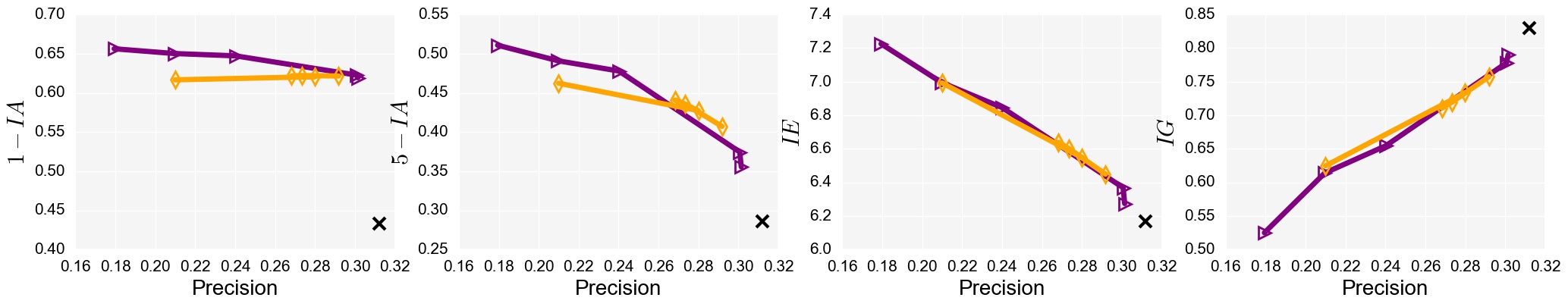}
        \caption{ML1M, NCF} \label{sensitivity_ml_item_ncf}
    \end{subfigure}
    \begin{subfigure}[b]{0.8\textwidth}
        \includegraphics[width=\textwidth]{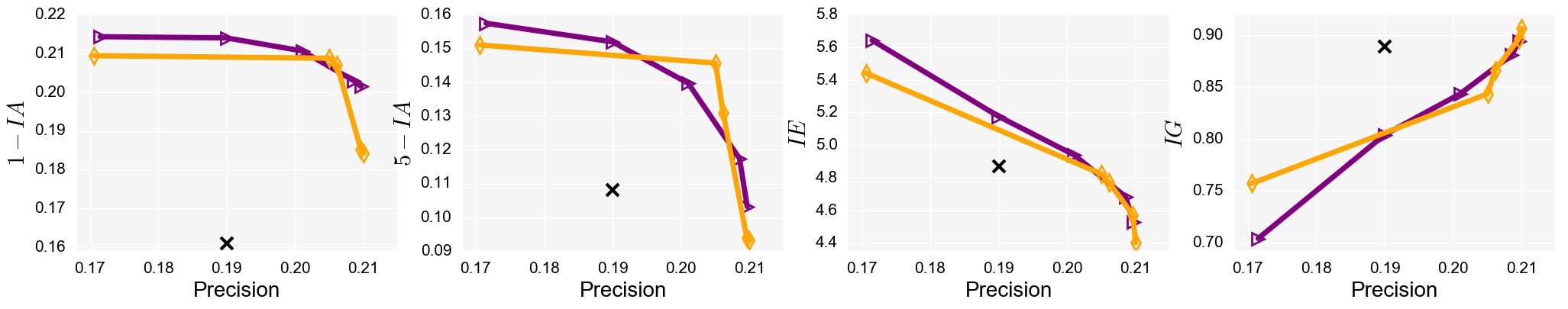}
        \caption{ML1M, UserKNN} \label{sensitivity_ml_item_userknn}
    \end{subfigure}%
\caption{Trade-off between accuracy and non-accuracy metrics for measuring the exposure fairness of items in FairMatch algorithms on \textbf{Last.fm} and \textbf{ML1M} datasets using all three base recommenders. The black cross shows the performance of original recommendation lists at size 10.
} \label{sensitivity_item}
\end{figure*}

\subsection{The effect of $\alpha$ in aggregate diversity}\label{alphaanalysis}

Standard aggregate diversity metric as it is used in \cite{vargas2011rank,adomavicius2011maximizing} counts an item even if it is recommended only once. Therefore, it is possible for an algorithm to perform really well on this metric while it has not really given enough visibility to different items. For this reason, I introduced $\alpha\mbox{-}IA$ and $\alpha\mbox{-}SA$ which are the generalization of standard aggregate diversity where we only count an item or supplier if it is recommended at least $\alpha$ times.\\ 
\indent Figures~\ref{k_a} and \ref{k_sa} show the behavior of different re-ranking algorithms on aggregate diversity for different values of $\alpha$. The most important takeaway from this figure is that some algorithms perform better than others for smaller values of $\alpha$ while they are outperformed for larger values of $\alpha$. That means if we only look at standard aggregate diversity ($1\mbox{-}IA$ or $1\mbox{-}SA$), we might think a certain algorithm is giving more visibility to different items while in reality that is not the case. For example, using \algname{BPR} as base on ML1M, $FairMatch^{Sup}$ has better aggregate diversity for smaller values of $\alpha$ ($\alpha \leq8$) than DM while for lager values of $\alpha$ its curve goes under DM indicating lower aggregate diversity. That shows that if we want to make sure different items are recommended more than 8 times, DM would be a better choice but if we want more items to be recommended even if they are recommended less than 8 times, then $FairMatch^{Sup}$ can be better.\\ 
\indent On supplier aggregate diversity, however, we can see that $FairMatch^{Sup}$ performs better than DM for all values of $\alpha$ indicating no matter how frequent we want the recommended items to appear in the recommendations, $FairMatch^{Sup}$ is still superior.  


\subsection{Trade-off between accuracy and non-accuracy metrics for FairMatch}

\begin{figure*}[t]
    \centering
    \begin{subfigure}[b]{0.8\textwidth}
        \includegraphics[width=\textwidth]{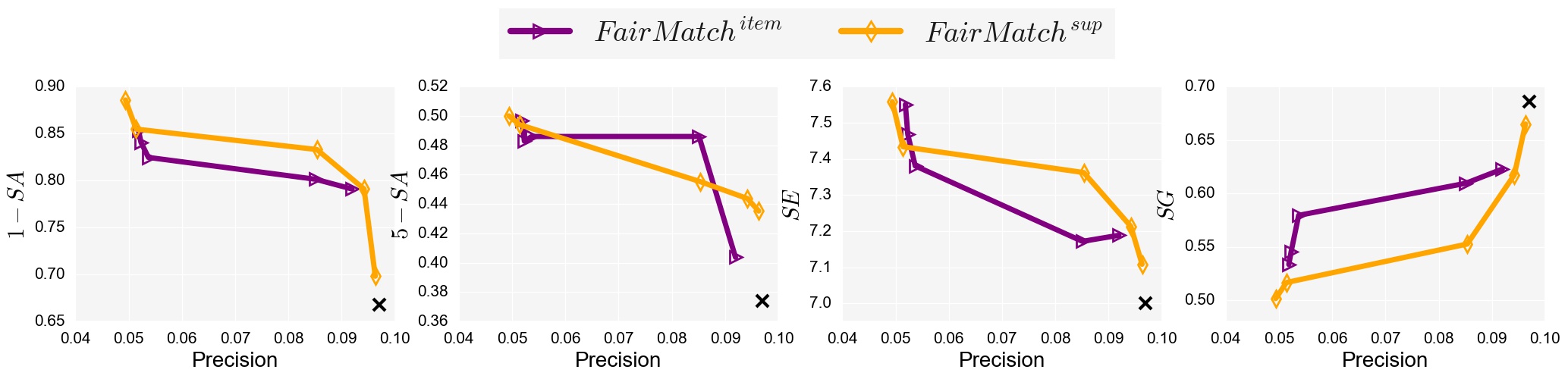}
        \caption{Last.fm, BPR} \label{sensitivity_lf_item_bpr}
    \end{subfigure}
    \begin{subfigure}[b]{0.8\textwidth}
        \includegraphics[width=\textwidth]{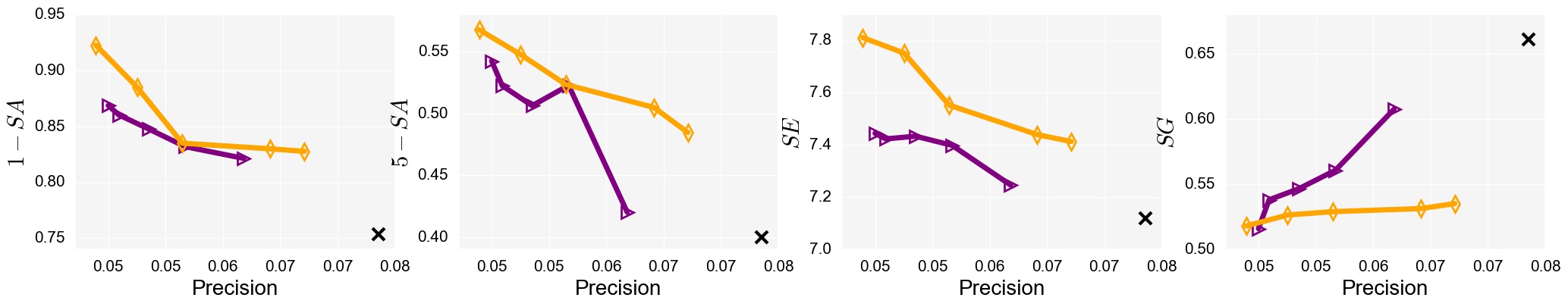}
        \caption{Last.fm, NCF} \label{sensitivity_lf_item_ncf}
    \end{subfigure}
    \begin{subfigure}[b]{0.8\textwidth}
        \includegraphics[width=\textwidth]{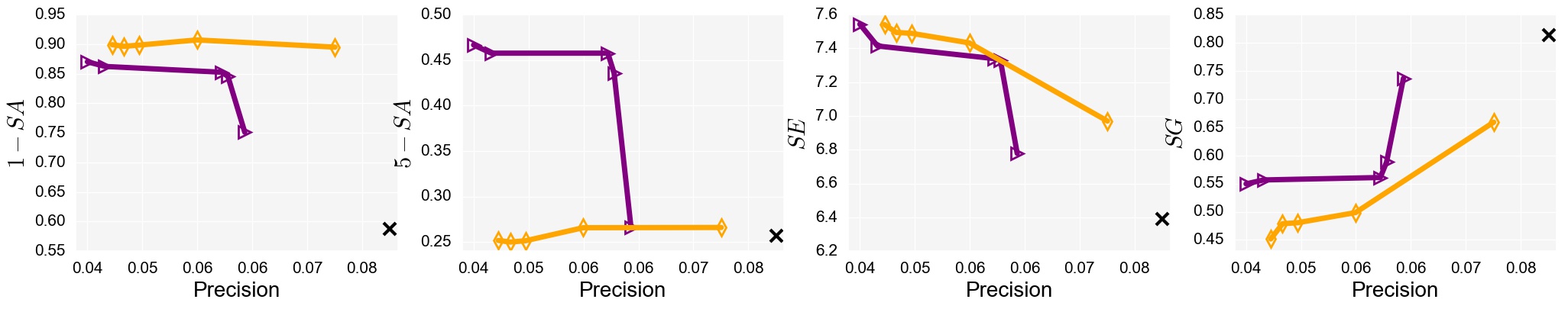}
        \caption{Last.fm, UserKNN} \label{sensitivity_lf_item_userknn}
    \end{subfigure}
    \begin{subfigure}[b]{0.8\textwidth}
        \includegraphics[width=\textwidth]{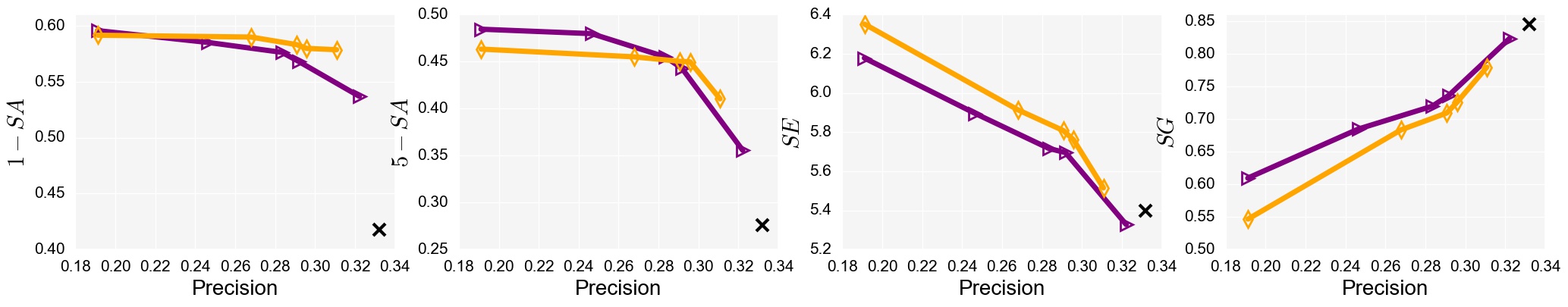}
        \caption{ML1M, BPR} \label{sensitivity_ml_item_bpr}
    \end{subfigure}
    \begin{subfigure}[b]{0.8\textwidth}
        \includegraphics[width=\textwidth]{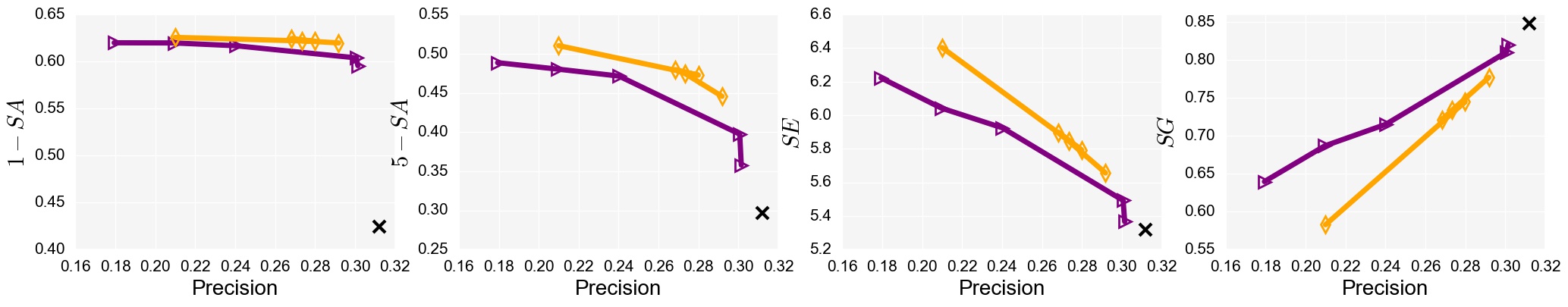}
        \caption{ML1M, NCF} \label{sensitivity_ml_item_ncf}
    \end{subfigure}
    \begin{subfigure}[b]{0.8\textwidth}
        \includegraphics[width=\textwidth]{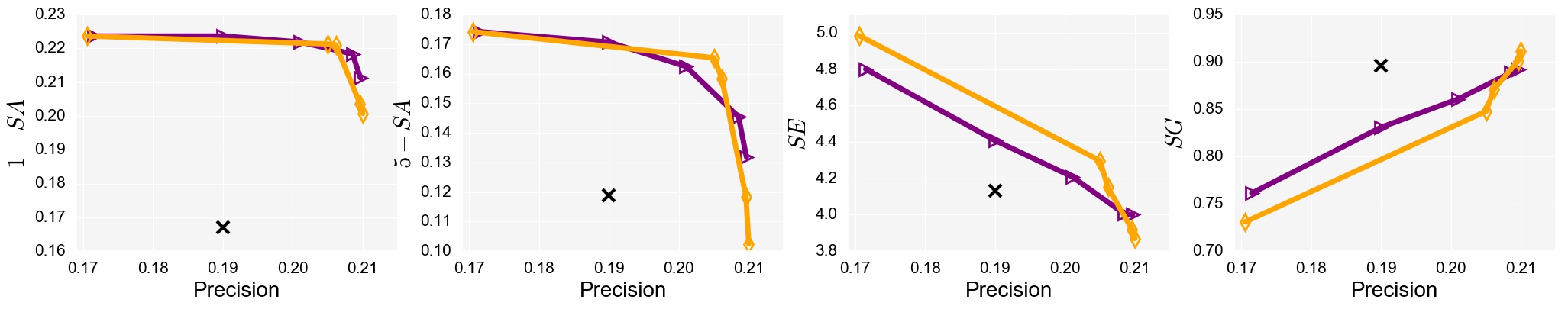}
        \caption{ML1M, UserKNN} \label{sensitivity_ml_item_userknn}
    \end{subfigure}%
\caption{Trade-off between accuracy and non-accuracy metrics for measuring the exposure fairness of suppliers in FairMatch algorithms on \textbf{Last.fm} and \textbf{ML1M} datasets using all three base recommenders. The black cross shows the performance of original recommendation lists at size 10.
} \label{sensitivity_supplier}
\end{figure*}

I investigated the trade-off between the precision and non-accuracy metrics under various settings. Figures~\ref{sensitivity_item} and \ref{sensitivity_supplier} show the experimental results for item and supplier exposure, respectively, on Last.fm and ML1M datasets using all three base recommenders. In these plots, horizontal axis shows the precision and vertical axis shows the non-accuracy metrics (i.e.\ $1\mbox{-}IA$, $5\mbox{-}IA$, $IE$, and $IG$ in Figure~\ref{sensitivity_item} for measuring item exposure and $1\mbox{-}SA$, $5\mbox{-}SA$, $SE$, and $SG$ in Figure~\ref{sensitivity_supplier} for measuring supplier exposure) of the recommendation results at size 10. 
Each point on the plot corresponds to a specific $\lambda$ value and the black cross shows the performance of original recommendation lists at size 10. 




Results in Figures~\ref{sensitivity_item} and \ref{sensitivity_supplier} show that $\lambda$ plays an important role in controlling the trade-off between the relevance of the recommended items for users (precision) and improving the utility for items and suppliers (non-accuracy metrics). As we decrease the $\lambda$ value, precision decreases, while non-accuracy metrics increase. According to Equations~\ref{wi} and \ref{ws}, for a higher $\lambda$ value, FairMatch will concentrate more on improving the accuracy of the recommendations, while for lower $\lambda$ value, it will have a higher concentration on improving the utility for items and suppliers.

I only report the results for $t=50$, but my analysis on longer initial recommendation lists (e.g.\ $t=100$) showed that by increasing the size of the initial recommendation lists we will obtain higher improvement on non-accuracy metrics especially on aggregate diversity metrics. However, we will lose accuracy as more items with lower relevance might be added to the final recommendation lists. These parameters allow system designers to better control the trade-off between the precision and non-accuracy metrics.

\subsection{Complexity analysis of FairMatch algorithm}

Solving the maximum flow problem is the core computation part of the FairMatch algorithm. I used Push-relabel algorithm as one of the efficient algorithms for solving the maximum flow problem. This algorithm has a polynomial time complexity as $O(V^{2}E)$ where $V$ is the number of nodes and $E$ is the number of edges in bipartite graph. For other parts of the FairMatch algorithm, the time complexity would be in the order of the number of edges as it mainly iterates over the edges in the bipartite graph.\\  
\indent Since FairMatch is an iterative process, unlike other maximum flow based techniques \cite{adomavicius2011maximizing, antikacioglu2017}, it requires solving maximum flow problem on the graph multiple times and this could be one limitation of this algorithm. However, except for the first iteration that FairMatch executes on the original graph, at the next iterations, the graph will be shrunk as FairMatch removes some parts of the graph at each iteration. Regardless, the upper-bound for the complexity of FairMatch will be $O(V^{3}E)$ assuming in each iteration we still have the entire graph (which is not the case). Therefore, the complexity of FairMatch is certainly less than $O(V^{3}E)$ which is still polynomial.

%% file: Tables/tbl_lf_50.tex
\captionsetup[table]{skip=4pt}
\begin{table*}[btp]
\footnotesize
\centering
\setlength{\tabcolsep}{3pt}
\captionof{table}{Comparison of different reranking algorithms on \textbf{Last.fm} dataset for long recommendation lists of size 50 ($t=50$) and final recommendation lists of size 10 ($n=10$). The bolded entries show the best values and the underlined entries show the statistically significant change from the second-best baseline with $p<0.05$ (comparison between FairMatch algorithms and other baselines ignoring Random and Reverse).} \label{tlb_lf_50}
\begin{tabular}{llrrrrrrrrrrrrr}
\toprule

 
 algorithms & baselines & Precision & $1\mbox{-}IA$ & $5\mbox{-}IA$ & $LT$ & $1\mbox{-}SA$ & $5\mbox{-}SA$ & $IG$ & $IE$ & $SG$ & $SE$ \\
 \bottomrule
 
  \multirow{9}{*}{\algname{BPR}}   
 & Base & 0.097 & 0.555 & 0.218 & 0.53 & 0.668 & 0.374 & 0.693 & 7.83 & 0.686 & 7 \\
 & Random & 0.062 & 0.695 & 0.237 & 0.678 & 0.781 & 0.424 & 0.568 & 8.16 & 0.607 & 7.23 \\
 & Reverse & 0.041 & 0.768 & 0.243 & 0.755 & 0.847 & 0.455 & 0.492 & 8.31 & 0.564 & 7.33 \\
 & FA*IR & 0.096 & 0.613 & \underline{\textbf{0.242}} & 0.591 & 0.715 & \textbf{0.421} & 0.627 & 8.01 & 0.642 & 7.13 \\ 
 & xQuAD & 0.094 & 0.677 & 0.188 & 0.659 & 0.787 & 0.373 & 0.646 & 7.95 & 0.653 & 7.1 \\
 & DM & 0.096 & 0.644 & 0.221 & 0.625 & 0.736 & 0.399 & 0.627 & 8.01 & 0.649 & 7.11 \\
 & ProbPolicy & 0.092 & 0.607 & 0.22 & 0.586 & 0.784  & 0.419 & 0.659 & 7.93 & 0.618 & 7.19 \\
 & $FairMatch^{item}$ & 0.092 & \textbf{0.686} & 0.223 & \textbf{0.669} & \textbf{0.791} & 0.404 & \underline{\textbf{0.602}} & \textbf{8.08} & 0.623 & 7.19 \\
 & $FairMatch^{Sup}$ & 0.095 & 0.675 & 0.21 & 0.657 & \textbf{0.791} & 0.404 & 0.623 & 8.03 & \textbf{0.617} & \textbf{7.22} \\
 \hline
 
 \multirow{9}{*}{\algname{NCF}}   
 & Base & 0.08 & 0.638 & 0.211 & 0.62 & 0.754 & 0.4 & 0.666 & 7.95 & 0.661 & 7.12 \\
 & Random & 0.056 & 0.74 & 0.221 & 0.726 & 0.824 & 0.441 & 0.552 & 8.22 & 0.592 & 7.28 \\
 & Reverse & 0.044 & 0.791 & 0.227 & 0.779 & 0.857 & 0.463 & 0.492 & 8.34 & 0.561 & 7.35 \\
 & FA*IR & 0.079 & 0.653 & 0.21 & 0.639 & 0.768 & 0.41 & 0.639 & 8.01 & 0.639 & 7.16 \\ 
 & xQuAD & 0.075 & 0.694 & 0.212 & 0.683 & 0.804 & 0.424 & 0.61 & 8.08 & 0.616 & 7.22 \\ 
 & DM & 0.079 & 0.723 & 0.205 & 0.707 & 0.808 & 0.412 & 0.594 & 8.11 & 0.624 & 7.2 \\ 
 & ProbPolicy & 0.072 & 0.659 & 0.207 & 0.643 & 0.809 & 0.43 &  0.647 & 7.98 & 0.611 & 7.22 \\ 
 & $FairMatch^{item}$ & 0.064 & \textbf{0.729} & 0.224 & \textbf{0.716} & 0.821 & 0.42 & 0.589 & \textbf{8.15} & 0.608 & 7.25 \\ 
 & $FairMatch^{Sup}$ & 0.071 & 0.711 & \underline{\textbf{0.229}} & 0.699 & \underline{\textbf{0.828}} & \underline{\textbf{0.485}} & \textbf{0.588} & \textbf{8.15} & \underline{\textbf{0.535}} & \underline{\textbf{7.41}} \\ 
 \hline
 
 \multirow{9}{*}{\algname{UserKNN}}   
 & Base & 0.08 & 0.461 & 0.127 & 0.431 & 0.588 & 0.257 & 0.833 & 7.07 & 0.813 & 6.39 \\
 & Random & 0.044 & 0.635 & 0.164 & 0.615 & 0.738 & 0.341 & 0.689 & 7.77 & 0.702 & 6.91 \\
 & Reverse & 0.027 & 0.712 & 0.204 & 0.696 & 0.797 & 0.394 & 0.591 & 8.09 & 0.634 & 7.14 \\
 & FA*IR & 0.074 & \textbf{0.629} & \underline{\textbf{0.172}} & \textbf{0.609} & 0.73 & \textit{0.351} & \underline{\textbf{0.687}} & \textbf{7.73} & \textit{0.706} & \textit{6.86} \\  
 & xQuAD & 0.078 & 0.577 & 0.117 & 0.554 & 0.701 & 0.269 & 0.781 & 7.31 & 0.771 & 6.58 \\ 
 & DM & 0.077 & 0.537 & 0.134 & 0.512 & 0.638 & 0.283 & 0.782 & 7.33 & 0.781 & 6.55 \\
 & ProbPolicy & 0.067 & 0.559 & 0.14 & 0.535 & 0.785 & \underline{\textbf{0.332}} & 0.765 & 7.4 & 0.7 & 6.84 \\ 
 & $FairMatch^{item}$ & 0.062 & 0.626 & 0.102 & 0.606 & 0.751 & 0.266 & 0.745 & 7.51 & 0.736 & 6.78 \\ 
 & $FairMatch^{Sup}$ & 0.073 & \textbf{0.629} & 0.123 & \textbf{0.609} & \underline{\textbf{0.895}} & 0.266 & 0.73 & 7.54 & \underline{\textbf{0.659}} & \textbf{6.97} \\ 
 
 \bottomrule
\end{tabular}
\end{table*}

%% file: Tables/tbl_ml_50.tex
\captionsetup[table]{skip=4pt}
\begin{table*}[btp]
\footnotesize
\centering
\setlength{\tabcolsep}{3pt}
\captionof{table}{Comparison of different reranking algorithms on \textbf{ML1M} dataset for long recommendation lists of size 50 ($t=50$) and final recommendation lists of size 10 ($n=10$). The bolded entries show the best values and the underlined entries show the statistically significant change from the second-best baseline with $p<0.05$ (comparison between FairMatch algorithms and other baselines ignoring Random and Reverse).} \label{tbl_ml_50}
\begin{tabular}{llrrrrrrrrrr}
\toprule

 
 algorithms & baselines & Precision & $1\mbox{-}IA$ & $5\mbox{-}IA$ & $LT$ & $1\mbox{-}SA$ & $5\mbox{-}SA$ & $IG$ & $IE$ & $SG$ & $SE$ \\
 \bottomrule
 
 \multirow{9}{*}{\algname{BPR}}   
 & Base & 0.332 & 0.392 & 0.262 & 0.351 & 0.418 & 0.276 & 0.833 & 6.01 & 0.845 & 5.4 \\
 & Random & 0.198 & 0.502 & 0.359 & 0.47 & 0.509 & 0.365 & 0.726 & 6.55 & 0.773 & 5.8 \\
 & Reverse & 0.125 & 0.547 & 0.404 & 0.518 & 0.545 & 0.394 & 0.653 & 6.81 & 0.728 & 6 \\
 & FA*IR & 0.306 & 0.402 & 0.283 & 0.362 & 0.4 & 0.296 & 0.779 & 6.33 & 0.802 & 5.5 \\ 
 & xQuAD & 0.322 & 0.461 & 0.311 & 0.426 & 0.451 & 0.324 & 0.797 & 6.19 & 0.822 & 5.34 \\
 & DM & 0.314 & 0.47 & 0.343 & 0.435 & 0.453 & 0.345 & \textbf{0.749} & \textbf{6.43} & 0.791 & \textbf{5.51} \\
 & ProbPolicy & 0.289 & 0.439 & 0.297 & 0.402 & 0.451 & 0.343 & 0.792 & 6.23 & \textit{0.782} & 5.49 \\
 & $FairMatch^{item}$ & 0.322 & 0.544 & 0.323 & 0.515 & 0.536 & 0.355 & 0.796 & 6.17 & 0.82 & 5.32 \\
 & $FairMatch^{Sup}$ & 0.311 & \underline{\textbf{0.547}} & \textbf{0.363} & \underline{\textbf{0.518}} & \underline{\textbf{0.578}} & \underline{\textbf{0.41}} & 0.768 & 6.28 & \textbf{0.779} & \textbf{5.51} \\

 \hline
 
 \multirow{9}{*}{\algname{NCF}}   
 & Base & 0.312 & 0.433 & 0.286 & 0.395 & 0.424 & 0.297 & 0.83 & 6.17 & 0.848 & 5.32 \\
 & Random & 0.198 & 0.566 & 0.396 & 0.539 & 0.571 & 0.406 & 0.728 & 6.66 & 0.776 & 5.89 \\
 & Reverse & 0.128 & 0.615 & 0.448 & 0.592 & 0.614 & 0.446 & 0.661 & 6.9 & 0.735 & 6.07 \\
 & FA*IR & 0.296 & 0.467 & 0.316 & 0.432 & 0.467 & 0.324 & 0.786 & 6.42 & 0.81 & 5.58 \\ 
 & xQuAD & 0.31 & 0.535 & 0.4 & 0.505 & 0.556 & 0.41 & 0.774 & 6.26 & \textbf{0.752} & \textbf{5.87} \\
 & DM & 0.298 & 0.53 & 0.383 & 0.499 & 0.509 & 0.386 & \textbf{0.752} & \textbf{6.54} & 0.797 & 5.6 \\
 & ProbPolicy & 0.301 & 0.469 & 0.315 & 0.434 & 0.426 & 0.344 & 0.812 & 6.24 & 0.819 & 5.45 \\
 & $FairMatch^{item}$ & 0.301 & \underline{\textbf{0.623}} & 0.375 & \underline{\textbf{0.6}} & 0.604 & 0.397 & 0.778 & 6.37 & 0.81 & 5.49 \\
 & $FairMatch^{Sup}$ & 0.292 & 0.622 & \textbf{0.407} & 0.599 & \underline{\textbf{0.62}} & \underline{\textbf{0.446}} & 0.757 & 6.45 & 0.777 & 5.66 \\
 
 \hline
 
 \multirow{9}{*}{\algname{UserKNN}}   
 & Base & 0.190 & 0.161 & 0.108 & 0.102 & 0.167 & 0.119 & 0.889 & 4.87 & 0.896 & 4.13 \\
 & Random & 0.128 & 0.219 & 0.145 & 0.164 & 0.231 & 0.157 & 0.798 & 5.51 & 0.829 & 4.79 \\
 & Reverse & 0.093 & 0.238 & 0.164 & 0.185 & 0.251 & 0.177 & 0.728 & 5.8 & 0.78 & 5.06 \\
 & FA*IR & 0.19 & 0.166 & 0.113 & 0.107 & 0.175 & 0.122 & 0.885 & 4.9 & 0.893 & 4.16 \\ 
 & xQuAD & 0.196 & 0.204 & 0.136 & 0.148 & 0.215 & 0.148 & 0.869 & 4.97 & 0.88 & 4.25 \\ 
 & DM & 0.19 & 0.183 & 0.125 & 0.125 & 0.194 & 0.133 & 0.873 & 4.98 & 0.885 & 4.22 \\ 
  & ProbPolicy & 0.19 & 0.165 & 0.11 & 0.106 & 0.175 & 0.122 & 0.886 & 4.89 & 0.891 & 4.17 \\ 
 & $FairMatch^{item}$ & 0.19 & \textbf{0.214} & \textbf{0.152} & \textbf{0.159} & \textbf{0.224} & \textbf{0.171} & \underline{\textbf{0.804}} & \textbf{5.18} & \underline{\textbf{0.831}} & \textbf{4.41} \\ 
 & $FairMatch^{Sup}$ & 0.206 & 0.207 & 0.131 & 0.152 & 0.222 & 0.158 & 0.867 & 4.77 & 0.871 & 4.15 \\ 
 
 \bottomrule
\end{tabular}
\end{table*}

%% file: Chapters/08_Conclusion.tex
\chapter{Conclusion and Future Work}
\label{chap:conclusion}
\tikzsetfigurename{conclusion_}

In this dissertation, I studied the issue of unfairness in recommender systems. I observed that recommender systems suffer from algorithmic bias and unfairness against certain groups of users, items, and suppliers. In particular, I found that unfair recommendations negatively affect different actors or sides in the system and in the long run, lead to issues like declining the aggregate diversity, shifting the representation of users' taste, and homogenization of users.\\
\indent To understand the importance of multi-sided view for addressing bias and unfairness in recommender systems, I conducted a simulation study in educational systems as a case study. In this simulation, I designed a recommender system that optimizes to match students to the supervisors under different scenarios. The scenarios were considering solely the students utilities, solely the supervisors utilities, and the utilities of both. I observed that the best performance in terms of utilities and fairness is achieved when simultaneously considering the utilities of both side, indicative of the importance of multi-sided view when optimizing a recommender system.\\
\indent I further explored the issue of popularity bias and exposure bias in recommender systems, and proposed solutions to address them. The first solution was a pre-processing approach that transforms the ratings into percentile values before recommendation generation step. Through extensive experiments, I observed that the proposed transformation technique is able to compensate the high rating values of popular items and improves the exposure fairness of the recommendations by mitigating the popularity bias in input data. The second solution was a post-processing approach that improves the exposure fairness for items and suppliers by identifying high quality items that have low visibility in long recommendation lists and promoting them to appear in the final recommendation lists. Experimental results showed the superiority of the proposed technique compared to other baselines in improving exposure fairness. In the next section, I briefly discuss the contributions and achievements introduced in the dissertation.

\section{Contributions}

\begin{itemize}
    \item \textbf{Factors leading to unfair recommendations.} I investigated potential factors that lead to unfair recommendations. The factors that I explored were characteristics of users' profile such as profile size, profile anomaly, and profile entropy. Then, I investigated the relationship between those characteristics for two groups of users and the quality of recommendations that each group receives. In the dataset that I used for this analysis, females were less in terms of number of users compared to males and also they provided much fewer ratings compared to males. Hence, I defined females as minority group and males as the majority group. Based on common definition of discrimination and unfairness, the recommender systems mainly learn the preferences of majority groups and better serve them than the minority groups. My analysis also confirmed this. Besides this definition, my analysis showed that there are also some other characteristics that may affect the fairness of recommendation results. For example, I observed that females who had larger profile size consistently received more accurate recommendations even though most of the other females received less accurate recommendations. The same results was also observed for profiles with high entropy. This shows that the characteristics of users' profile can reveal important information about their treatment by recommender systems. A user might belong to a minority group (based on a sensitive attribute), but the characteristics of her/his profile may help the recommendation algorithm to better learn her/his preferences and consequently delivers more accurate recommendations to her/him.  
    \item \textbf{The impact of algorithmic bias on the system.} In my exploration on the impact of algorithmic bias on recommender systems, I simulated the recommendation process over time in a offline setting. The interaction between users and recommender systems will form \textit{feedback loop} which recommendations generated to users would be consumed and added to the users' profile in the next iterations. I formally and empirically showed that algorithmic bias can be propagated over time and substantially amplifies the bias in input data in the next iterations. 
    Moreover, I observed that this bias amplification causes some other negative impacts on the system such as declining the aggregate diversity, shifting the representation of users' tastes, and homogenization of users. Also, another interesting observation was that this bias amplification was stronger on minority group.
    \item \textbf{Evaluating multi-sided exposure bias.} I reviewed and explored existing metrics for measuring multi-sided exposure bias in the literature. My analysis showed those metrics have limitations and are unable to properly measure various aspects of exposure bias in recommender systems. I observed that some of the existing metrics for evaluating the performance of recommendation algorithms in terms of exposure bias mitigation such as aggregate diversity hide important information about the exposure fairness of items and suppliers since those metrics do not take into account how frequent different items are recommended. Although Gini index can be used to address this issue, it also has its own limitations where an algorithm can achieve a good Gini index by equally recommending large number of items or suppliers (even if they are popular) while the rest of items or suppliers still get unfair exposure. My analysis showed that it is crucial to evaluate bias mitigation algorithms using multiple metrics each of which captures a certain aspect of the algorithm's performance. To overcome the limitations of the existing metrics, I proposed new metrics and modified existing metrics for measuring exposure bias for items and suppliers in recommender systems.
    \item \textbf{Percentile transformation as a pre-processing technique for tackling exposure bias.} To mitigate exposure bias in recommender systems, I proposed \textit{percentile transformation} which works as a pre-processing step before recommendation generation. I showed that the proposed percentile transformation is able to compensate for high rating values assigned to the popular items and as a result, it can mitigate popularity bias in input data. Extensive experiments using several recommendation algorithms on two datasets showed that the proposed percentile transformation  is able to mitigate the multi-sided exposure bias, but also significantly improves the accuracy of recommendations compared to other transformation techniques.
    \item \textbf{A graph-based approach for addressing multi-sided exposure bias.} I proposed a graph-based approach, FairMatch, for improving the aggregate diversity and exposure fairness of items and suppliers in recommender systems. FairMatch is a post-processing technique that works on the top of any recommendation algorithm. In other words, it re-ranks the output from the base recommendation algorithms such that it improves the exposure fairness of final recommendation lists with minimum loss in accuracy of recommendations. FairMatch algorithm is flexible and can be used for improving the exposure fairness of items and suppliers. Experimental results on two publicly available datasets showed that the FairMatch algorithm outperforms several state-of-the-art bias mitigation re-rankers in improving multi-sided exposure fairness. 
\end{itemize}

\section{Future work}

\begin{itemize}
    \item \textbf{Considering other definitions for exposure fairness.} The definition of exposure fairness in this dissertation was solely based on the visibility of the items or suppliers in the recommendations without taking into account their original popularity in training data. One possible future work is to take this information into account such that the fairness of exposure for items or suppliers is measured relative to their original popularity. This can be in particular important when fairness of exposure is defined as the equity of representation instead of equality of representation. In equity of representation for items and suppliers, exposure fairness is defined as the representation or visibility of items or suppliers based on their merit in input data. For example, if item $A$ is more popular than item $B$, then a fair exposure is achieved when item $A$ receives more exposure than item $B$ in recommendation results. Besides these definitions, another way for defining exposure fairness is considering the position of items in the recommendation lists. All these definitions for exposure fairness do not consider the position of recommended items in the recommendation lists. When an item is shown in the first position in the recommendation list, it has better exposure than other items in the subsequent positions. Therefore, considering item position in recommendation list for defining exposure fairness can be an interesting future work.    
    \item \textbf{Long-term fairness.} The simulation study in Chapter~\ref{chap:back}, Section~\ref{fairness_impact} showed that users and recommender system are in a process of mutual dynamic evolution where users profile get updated over time via recommendations generated by the recommender system and this way algorithmic bias will amplify inherent bias in users' interaction data. In this research, I investigated several factors including popularity amplification, decline in aggregate diversity, shift in the representation of users' taste, and homogenization of users, but other interesting investigations can be considered such as content diversity of the recommended items, analysis on supplier exposure, and grouping user based on various criteria (e.g.\ interest toward popular items) for studying homogenization of users. Also, conducting empirical research using the existing bias mitigation techniques (including the ones proposed in this thesis) in this simulation (experimentation over time) can shed more lights on the effectiveness of those techniques on achieving long-terms fairness and mitigating bias amplification in recommender systems. 
    \item \textbf{Online evaluation.} Consistent with many prior work on re-ranking methods, I observed a drop in precision for different re-rankers, including the proposed FairMatch algorithm, in my offline evaluation setting. However, how users will perceive the recommendations in an online setting can better assess the effectiveness of this type of re-rankers. The reason is, the data is skewed towards popular items and it is less likely to observe a hit when recommending less popular items using offline evaluation. Another potential future work is to investigate how users will react to the re-ranked recommendations by conducting online experiments on real users. 
\end{itemize}

%% file: FrontBackMatter/Bibliography.tex

\bibliographystyle{plain} 
\label{app:bibliography}
\bibliography{./bibliography}%


%% file: FrontBackMatter/Summaries.tex

\chapter{Summary}

\begin{center}
\Large
\myTitle
\end{center}

\thispagestyle{empty}

\emph{Summary here}

\cleardoublepage

\begin{otherlanguage}{dutch}
\chapter{Samenvatting}

\begin{center}
\Large
\myTitle
\end{center}

\emph{Summary in different language (Dutch in this example) here.}

\end{otherlanguage} 

%% file: FrontBackMatter/Acknowledgments.tex

\chapter{Acknowledgments}
\pagestyle{empty}

This thesis is the end of my PhD journey, 5 years of extremely valuable experiences. During this journey, I have been very fortunate to interact with brilliant and amazing people who helped and supported me to grow personally and professionally. I would like to take this chance to thank them.\\
\indent I had a unique opportunity to be advised by multiple advisors who helped me to gain insights and new ideas for my research. I would like to thank Robin Burke for his guidance on developing new ideas and involving me in his research projects. I would like to thank Bamshad Mobasher for his comments and insights on my academic research and also for teaching me how to conduct a scientific research. I am deeply grateful to Mykola Pechenizkiy for supporting me to pursue my research interests and to attend conferences to expand my professional network.\\
\indent I would like to thank the committee members for reviewing this dissertation and providing me with their insightful comments: Nava Tintarev  from Maastricht University, Martha Larson  from Radboud University, Martijn Willemsen from Eindhoven University of Technology, and Maryam Tavakol from Eindhoven University of Technology.\\
\indent I would like to thank my colleagues at Eindhoven University of Technology for making a friendly environment and culture at our workplace. It was my pleasure to work with you all in the same department. Especially, thanks to my friends in DAI Cluster for the pleasant moments we had together. I always enjoyed chatting and discussing with you about our PhD and social life at the coffee corner.\\
\indent I would like to express my gratitude to my friends and colleagues in Web Intelligence Lab at DePaul University. I have learnt enormously about the field of recommender systems during our weekly group meetings. The scientific discussion and presentation that we had in the meeting helped me to expand my knowledge and understanding on the field. Many thanks to my friend, Himan Abdollahpouri, for our productive collaboration on publishing papers which helped me to make significant progress on my PhD dissertation.\\
\indent I am also very fortunate to be part of a great team under supervision of Robin Burke in That Recommender Systems Lab at University of Colorado Boulder working on \textit{librec-auto} project. I am involved in this project since 2018 when the tool was developed and released for the first time. I would like to thank my teammates and Robin for numerous interesting discussions we have had in our weekly meetings on both technical and theoretical aspects of the project.\\
\indent Most importantly, I am deeply grateful to my family. Special thanks to my parents for your love and support over the years. Without you, none of this would have been possible. I hope I could make you happy by this thesis as a small gift. Many thanks to my brothers, Bijan and Hamed, for their advice and encouragement to successfully accomplish this phase of my life. Last but not least, my warm thanks to my wife, Sepideh, for constantly supporting me and being by my side.

\begin{flushright}
Masoud Mansoury\\
Eindhoven, October 2021
\end{flushright}

%% file: FrontBackMatter/curriculumVitae.tex

\chapter{Curriculum Vitae}
\pagestyle{empty}
Masoud Mansoury was born on April 15, 1987 in Noor, Iran. He obtained his B.S. degree in Information Technology Engineering from Mazandaran University of Science and Technology, Iran in 2010. Then, he received his M.S. degree in Electronic Commerce from Amirkabir University of Technology, Iran in 2012. During this period, he started researching on recommender systems and the subject of his master's thesis was on this field. From 2012 to 2016, he worked on various data mining and software engineering projects including customer clustering in banking and development of web applications.\\
\indent In 2016, Masoud started his PhD program in Computer and Information Science in Center for Web Intelligence at DePaul University, United States under supervision of prof. Robin Burke and prof. Bamshad Mobasher. In 2019, he moved to the Netherlands and continued his PhD program in Department of Mathematics and Computer Science at Eindhoven University of Technology under supervision of prof. Mykola Pechenizkiy.\\
\indent During his PhD program, Masoud has been actively involved in research and teaching activities. He has developed a comprehensive recommendation tool, \textit{librec-auto}, under supervision of prof. Robin Burke for performing flexible, automated, and efficient experiments on recommender systems. He has also assisted in various Data science courses including Machine Learning, Responsible Data Science, and Data Analysis and Regression.    

\newpage
\section*{List of Publications}
Masoud Mansoury has the following publications:

\subsection*{Journals and Conferences}
\begin{itemize}
  \item Masoud Mansoury, Himan Abdoollahpouri, Mykola Pechenizkiy, Bamshad Mobasher, Robin Burke. A Graph-based Approach for Mitigating Multi-sided Exposure Bias in Recommender Systems. In \textit{ACM Transactions on Information Systems (TOIS)}, 2021, Accepted.
  \item Nasim Sonboli, Masoud Mansoury, Ziyue Guo, Shreyas Kadekodi, Weiwen Liu, Zijun Liu, Andrew Schwartz and Robin Burke. librec-auto: A Tool for Recommender Systems Experimentation. In \textit{ACM International Conference on Information and Knowledge Management (CIKM)}, 2021, Accepted.
  \item \bibentry{mansoury2021flatter}
  \item \bibentry{masoud2021wsdm}
  \item \bibentry{elahi2021beyond}
  \item \bibentry{abdollahpouri2021user}
  \item \bibentry{mansoury2020feedback}
  \item \bibentry{mansoury2020fairmatch}
  \item \bibentry{mansoury2020investigating}
  \item \bibentry{sonboli2020fairness}
  \item \bibentry{abdollahpouri2020connection}
  \item \bibentry{abdollahpouri2020multi}
  \item \bibentry{burke2020experimentation}
  \item \bibentry{mansoury2019relationship}
  \item \bibentry{mansoury2019bias}
  \item \bibentry{himan2019c}
  \item \bibentry{mansoury2019algorithm}
  \item \bibentry{masoud2018}
  \item \bibentry{burke2017}
\end{itemize}







%% file: FrontBackMatter/SIKSdissertations.tex


\chapter{SIKS dissertations}


\newcommand{\SIKSdiss}[3]{{\bf #1}\hspace*{1ex}#2, {\it #3.}\\}

\begin{longtabu}{@{}l@{ }l@{\hspace{1em}}X}
\toprule
2011	&	 01	&	 Botond Cseke (RUN), Variational Algorithms for Bayesian Inference in Latent Gaussian Models\\
	&	 02	&	 Nick Tinnemeier (UU), Organizing Agent Organizations. Syntax and Operational Semantics of an Organization-Oriented Programming Language\\
	&	 03	&	 Jan Martijn van der Werf (TUE), Compositional Design and Verification of Component-Based Information Systems\\
	&	 04	&	 Hado van Hasselt (UU), Insights in Reinforcement Learning; Formal analysis and empirical evaluation of temporal-difference\\
	&	 05	&	 Bas van der Raadt (VU), Enterprise Architecture Coming of Age - Increasing the Performance of an Emerging Discipline.\\
	&	 06	&	 Yiwen Wang (TUE), Semantically-Enhanced Recommendations in Cultural Heritage\\
	&	 07	&	 Yujia Cao (UT), Multimodal Information Presentation for High Load Human Computer Interaction\\
	&	 08	&	 Nieske Vergunst (UU), BDI-based Generation of Robust Task-Oriented Dialogues\\
	&	 09	&	 Tim de Jong (OU), Contextualised Mobile Media for Learning\\
	&	 10	&	 Bart Bogaert (UvT), Cloud Content Contention\\
	&	 11	&	 Dhaval Vyas (UT), Designing for Awareness: An Experience-focused HCI Perspective\\
	&	 12	&	 Carmen Bratosin (TUE), Grid Architecture for Distributed Process Mining\\
	&	 13	&	 Xiaoyu Mao (UvT), Airport under Control. Multiagent Scheduling for Airport Ground Handling\\
	&	 14	&	 Milan Lovric (EUR), Behavioral Finance and Agent-Based Artificial Markets\\
	&	 15	&	 Marijn Koolen (UvA), The Meaning of Structure: the Value of Link Evidence for Information Retrieval\\
	&	 16	&	 Maarten Schadd (UM), Selective Search in Games of Different Complexity\\
	&	 17	&	 Jiyin He (UVA), Exploring Topic Structure: Coherence, Diversity and Relatedness\\
	&	 18	&	 Mark Ponsen (UM), Strategic Decision-Making in complex games\\
	&	 19	&	 Ellen Rusman (OU), The Mind's Eye on Personal Profiles\\
	&	 20	&	 Qing Gu (VU), Guiding service-oriented software engineering - A view-based approach\\
	&	 21	&	 Linda Terlouw (TUD), Modularization and Specification of Service-Oriented Systems\\
	&	 22	&	 Junte Zhang (UVA), System Evaluation of Archival Description and Access\\
	&	 23	&	 Wouter Weerkamp (UVA), Finding People and their Utterances in Social Media\\
	&	 24	&	 Herwin van Welbergen (UT), Behavior Generation for Interpersonal Coordination with Virtual Humans On Specifying, Scheduling and Realizing Multimodal Virtual Human Behavior\\
	&	 25	&	 Syed Waqar ul Qounain Jaffry (VU), Analysis and Validation of Models for Trust Dynamics\\
	&	 26	&	 Matthijs Aart Pontier (VU), Virtual Agents for Human Communication - Emotion Regulation and Involvement-Distance Trade-Offs in Embodied Conversational Agents and Robots\\
	&	 27	&	 Aniel Bhulai (VU), Dynamic website optimization through autonomous management of design patterns\\
	&	 28	&	 Rianne Kaptein (UVA), Effective Focused Retrieval by Exploiting Query Context and Document Structure\\
	&	 29	&	 Faisal Kamiran (TUE), Discrimination-aware Classification\\
	&	 30	&	 Egon van den Broek (UT), Affective Signal Processing (ASP): Unraveling the mystery of emotions\\
	&	 31	&	 Ludo Waltman (EUR), Computational and Game-Theoretic Approaches for Modeling Bounded Rationality\\
	&	 32	&	 Nees-Jan van Eck (EUR), Methodological Advances in Bibliometric Mapping of Science\\
	&	 33	&	 Tom van der Weide (UU), Arguing to Motivate Decisions\\
	&	 34	&	 Paolo Turrini (UU), Strategic Reasoning in Interdependence: Logical and Game-theoretical Investigations\\
	&	 35	&	 Maaike Harbers (UU), Explaining Agent Behavior in Virtual Training\\
	&	 36	&	 Erik van der Spek (UU), Experiments in serious game design: a cognitive approach\\
	&	 37	&	 Adriana Burlutiu (RUN), Machine Learning for Pairwise Data, Applications for Preference Learning and Supervised Network Inference\\
	&	 38	&	 Nyree Lemmens (UM), Bee-inspired Distributed Optimization\\
	&	 39	&	 Joost Westra (UU), Organizing Adaptation using Agents in Serious Games\\
	&	 40	&	 Viktor Clerc (VU), Architectural Knowledge Management in Global Software Development\\
	&	 41	&	 Luan Ibraimi (UT), Cryptographically Enforced Distributed Data Access Control\\
	&	 42	&	 Michal Sindlar (UU), Explaining Behavior through Mental State Attribution\\
	&	 43	&	 Henk van der Schuur (UU), Process Improvement through Software Operation Knowledge\\
	&	 44	&	 Boris Reuderink (UT), Robust Brain-Computer Interfaces\\
	&	 45	&	 Herman Stehouwer (UvT), Statistical Language Models for Alternative Sequence Selection\\
	&	 46	&	 Beibei Hu (TUD), Towards Contextualized Information Delivery: A Rule-based Architecture for the Domain of Mobile Police Work\\
	&	 47	&	 Azizi Bin Ab Aziz (VU), Exploring Computational Models for Intelligent Support of Persons with Depression\\
	&	 48	&	 Mark Ter Maat (UT), Response Selection and Turn-taking for a Sensitive Artificial Listening Agent\\
	&	 49	&	 Andreea Niculescu (UT), Conversational interfaces for task-oriented spoken dialogues: design aspects influencing interaction quality\\

\midrule
2012&	 01	&	 Terry Kakeeto (UvT), Relationship Marketing for SMEs in Uganda\\
	&	 02	&	 Muhammad Umair (VU), Adaptivity, emotion, and Rationality in Human and Ambient Agent Models\\
	&	 03	&	 Adam Vanya (VU), Supporting Architecture Evolution by Mining Software Repositories\\
	&	 04	&	 Jurriaan Souer (UU), Development of Content Management System-based Web Applications\\
	&	 05	&	 Marijn Plomp (UU), Maturing Interorganisational Information Systems\\
	&	 06	&	 Wolfgang Reinhardt (OU), Awareness Support for Knowledge Workers in Research Networks\\
	&	 07	&	 Rianne van Lambalgen (VU), When the Going Gets Tough: Exploring Agent-based Models of Human Performance under Demanding Conditions\\
	&	 08	&	 Gerben de Vries (UVA), Kernel Methods for Vessel Trajectories\\
	&	 09	&	 Ricardo Neisse (UT), Trust and Privacy Management Support for Context-Aware Service Platforms\\
	&	 10	&	 David Smits (TUE), Towards a Generic Distributed Adaptive Hypermedia Environment\\
	&	 11	&	 J.C.B. Rantham Prabhakara (TUE), Process Mining in the Large: Preprocessing, Discovery, and Diagnostics\\
	&	 12	&	 Kees van der Sluijs (TUE), Model Driven Design and Data Integration in Semantic Web Information Systems\\
	&	 13	&	 Suleman Shahid (UvT), Fun and Face: Exploring non-verbal expressions of emotion during playful interactions\\
	&	 14	&	 Evgeny Knutov (TUE), Generic Adaptation Framework for Unifying Adaptive Web-based Systems\\
	&	 15	&	 Natalie van der Wal (VU), Social Agents. Agent-Based Modelling of Integrated Internal and Social Dynamics of Cognitive and Affective Processes.\\
	&	 16	&	 Fiemke Both (VU), Helping people by understanding them - Ambient Agents supporting task execution and depression treatment\\
	&	 17	&	 Amal Elgammal (UvT), Towards a Comprehensive Framework for Business Process Compliance\\
	&	 18	&	 Eltjo Poort (VU), 	Improving Solution Architecting Practices\\
	&	 19	&	 Helen Schonenberg (TUE), What's Next? Operational Support for Business Process Execution\\
	&	 20	&	 Ali Bahramisharif (RUN), Covert Visual Spatial Attention, a Robust Paradigm for Brain-Computer Interfacing\\
	&	 21	&	 Roberto Cornacchia (TUD), Querying Sparse Matrices for Information Retrieval\\
	&	 22	&	 Thijs Vis (UvT), Intelligence, politie en veiligheidsdienst: verenigbare grootheden?\\
	&	 23	&	 Christian Muehl (UT), Toward Affective Brain-Computer Interfaces: Exploring the Neurophysiology of Affect during Human Media Interaction\\
	&	 24	&	 Laurens van der Werff (UT), Evaluation of Noisy Transcripts for Spoken Document Retrieval\\
	&	 25	&	 Silja Eckartz (UT), Managing the Business Case Development in Inter-Organizational IT Projects: A Methodology and its Application\\
	&	 26	&	 Emile de Maat (UVA), Making Sense of Legal Text\\
	&	 27	&	 Hayrettin Gurkok (UT), Mind the Sheep! User Experience Evaluation \& Brain-Computer Interface Games\\
	&	 28	&	 Nancy Pascall (UvT), Engendering Technology Empowering Women\\
	&	 29	&	 Almer Tigelaar (UT), Peer-to-Peer Information Retrieval\\
	&	 30	&	 Alina Pommeranz (TUD), Designing Human-Centered Systems for Reflective Decision Making\\
	&	 31	&	 Emily Bagarukayo (RUN), A Learning by Construction Approach for Higher Order Cognitive Skills Improvement, Building Capacity and Infrastructure\\
	&	 32	&	 Wietske Visser (TUD), 	Qualitative multi-criteria preference representation and reasoning\\
	&	 33	&	 Rory Sie (OUN), Coalitions in Cooperation Networks (COCOON)\\
	&	 34	&	 Pavol Jancura (RUN), Evolutionary analysis in PPI networks and applications\\
	&	 35	&	 Evert Haasdijk (VU), Never Too Old To Learn -- On-line Evolution of Controllers in Swarm- and Modular Robotics\\
	&	 36	&	 Denis Ssebugwawo (RUN), Analysis and Evaluation of Collaborative Modeling Processes\\
	&	 37	&	 Agnes Nakakawa (RUN), A Collaboration Process for Enterprise Architecture Creation\\
	&	 38	&	 Selmar Smit (VU), Parameter Tuning and Scientific Testing in Evolutionary Algorithms\\
	&	 39	&	 Hassan Fatemi (UT), Risk-aware design of value and coordination networks\\
	&	 40	&	 Agus Gunawan (UvT), Information Access for SMEs in Indonesia\\
	&	 41	&	 Sebastian Kelle (OU), Game Design Patterns for Learning\\
	&	 42	&	 Dominique Verpoorten (OU), Reflection Amplifiers in self-regulated Learning\\
	&	 43	&	 Withdrawn \\
	&	 44	&	 Anna Tordai (VU), On Combining Alignment Techniques\\
	&	 45	&	 Benedikt Kratz (UvT), A Model and Language for Business-aware Transactions\\
	&	 46	&	 Simon Carter (UVA), Exploration and Exploitation of Multilingual Data for Statistical Machine Translation\\
	&	 47	&	 Manos Tsagkias (UVA), Mining Social Media: Tracking Content and Predicting Behavior\\
	&	 48	&	 Jorn Bakker (TUE), Handling Abrupt Changes in Evolving Time-series Data\\
	&	 49	&	 Michael Kaisers (UM), Learning against Learning - Evolutionary dynamics of reinforcement learning algorithms in strategic interactions\\
	&	 50	&	 Steven van Kervel (TUD), Ontologogy driven Enterprise Information Systems Engineering\\
	&	 51	&	 Jeroen de Jong (TUD), Heuristics in Dynamic Sceduling; a practical framework with a case study in elevator dispatching\\

\midrule
2013&    01	&    Viorel Milea (EUR), News Analytics for Financial Decision Support\\
	&	 02	&	 Erietta Liarou (CWI), MonetDB/DataCell: Leveraging the Column-store Database Technology for Efficient and Scalable Stream Processing\\
	&	 03	&	 Szymon Klarman (VU), Reasoning with Contexts in Description Logics\\
	&	 04	&	 Chetan Yadati (TUD), Coordinating autonomous planning and scheduling\\
	&	 05	&	 Dulce Pumareja (UT), Groupware Requirements Evolutions Patterns\\
	&	 06	&	 Romulo Goncalves (CWI), The Data Cyclotron: Juggling Data and Queries for a Data Warehouse Audience\\
	&	 07	&	 Giel van Lankveld (UvT), Quantifying Individual Player Differences\\
	&	 08	&	 Robbert-Jan Merk (VU), Making enemies: cognitive modeling for opponent agents in fighter pilot simulators\\
	&	 09	&	 Fabio Gori (RUN), Metagenomic Data Analysis: Computational Methods and Applications\\
	&	 10	&	 Jeewanie Jayasinghe Arachchige (UvT), A Unified Modeling Framework for Service Design.\\
	&	 11	&	 Evangelos Pournaras (TUD), Multi-level Reconfigurable Self-organization in Overlay Services\\
	&	 12	&	 Marian Razavian (VU), Knowledge-driven Migration to Services\\
	&	 13	&	 Mohammad Safiri (UT), Service Tailoring: User-centric creation of integrated IT-based homecare services to support independent living of elderly\\
	&	 14	&	 Jafar Tanha (UVA), Ensemble Approaches to Semi-Supervised Learning Learning\\
	&	 15	&	 Daniel Hennes (UM), Multiagent Learning - Dynamic Games and Applications\\
	&	 16	&	 Eric Kok (UU), Exploring the practical benefits of argumentation in multi-agent deliberation\\
	&	 17	&	 Koen Kok (VU), The PowerMatcher: Smart Coordination for the Smart Electricity Grid\\
	&	 18	&	 Jeroen Janssens (UvT), Outlier Selection and One-Class Classification\\
	&	 19	&	 Renze Steenhuizen (TUD), Coordinated Multi-Agent Planning and Scheduling\\
	&	 20	&	 Katja Hofmann (UvA), Fast and Reliable Online Learning to Rank for Information Retrieval\\
	&	 21	&	 Sander Wubben (UvT), Text-to-text generation by monolingual machine translation\\
	&	 22	&	 Tom Claassen (RUN), Causal Discovery and Logic\\
	&	 23	&	 Patricio de Alencar Silva (UvT), Value Activity Monitoring\\
	&	 24	&	 Haitham Bou Ammar (UM), 	Automated Transfer in Reinforcement Learning\\
	&	 25	&	 Agnieszka Anna Latoszek-Berendsen (UM), 	Intention-based Decision Support. A new way of representing and implementing clinical guidelines in a Decision Support System\\
	&	 26	&	 Alireza Zarghami (UT), 	Architectural Support for Dynamic Homecare Service Provisioning\\
	&	 27	&	 Mohammad Huq (UT), 	Inference-based Framework Managing Data Provenance\\
	&	 28	&	 Frans van der Sluis (UT), 	When Complexity becomes Interesting: An Inquiry into the Information eXperience\\
	&	 29	&	 Iwan de Kok (UT), 	Listening Heads\\
	&	 30	&	 Joyce Nakatumba (TUE), 	Resource-Aware Business Process Management: Analysis and Support\\
	&	 31	&	 Dinh Khoa Nguyen (UvT), 	Blueprint Model and Language for Engineering Cloud Applications\\
	&	 32	&	 Kamakshi Rajagopal (OUN), 	Networking For Learning; The role of Networking in a Lifelong Learner's Professional Development\\
	&	 33	&	 Qi Gao (TUD), User Modeling and Personalization in the Microblogging Sphere\\
	&	 34	&	 Kien Tjin-Kam-Jet (UT), 	Distributed Deep Web Search\\
	&	 35	&	 Abdallah El Ali (UvA), Minimal Mobile Human Computer Interaction\\
	&	 36	&	 Than Lam Hoang (TUe), 	Pattern Mining in Data Streams\\
	&	 37	&	 Dirk B\"{o}rner (OUN), Ambient Learning Displays\\
	&	 38	&	 Eelco den Heijer (VU), 	Autonomous Evolutionary Art\\
	&	 39	&	 Joop de Jong (TUD), A Method for Enterprise Ontology based Design of Enterprise Information Systems\\
	&	 40	&	 Pim Nijssen (UM), Monte-Carlo Tree Search for Multi-Player Games\\
	&	 41	&	 Jochem Liem (UVA), Supporting the Conceptual Modelling of Dynamic Systems: A Knowledge Engineering Perspective on Qualitative Reasoning\\
	&	 42	&	 L\'{e}on Planken (TUD), Algorithms for Simple Temporal Reasoning\\
	&	 43	&	 Marc Bron (UVA), Exploration and Contextualization through Interaction and Concepts\\

\midrule
2014&	 01	&	 Nicola Barile (UU), Studies in Learning Monotone Models from Data\\
	&	 02	&	 Fiona Tuliyano (RUN), Combining System Dynamics with a Domain Modeling Method\\
	&	 03	&	 Sergio Raul Duarte Torres (UT), Information Retrieval for Children: Search Behavior and Solutions\\
	&	 04	&	 Hanna Jochmann-Mannak (UT), Websites for children: search strategies and interface design - Three studies on children's search performance and evaluation\\
	&	 05	&	 Jurriaan van Reijsen (UU), Knowledge Perspectives on Advancing Dynamic Capability\\
	&	 06	&	 Damian Tamburri (VU), Supporting Networked Software Development\\
	&	 07	&	 Arya Adriansyah (TUE), Aligning Observed and Modeled Behavior\\
	&	 08	&	 Samur Araujo (TUD), Data Integration over Distributed and Heterogeneous Data Endpoints\\
	&	 09	&	 Philip Jackson (UvT), Toward Human-Level Artificial Intelligence: Representation and Computation of Meaning in Natural Language\\
	&	 10	&	 Ivan Salvador Razo Zapata (VU), Service Value Networks\\
	&	 11	&	 Janneke van der Zwaan (TUD), An Empathic Virtual Buddy for Social Support\\
	&	 12	&	 Willem van Willigen (VU), Look Ma, No Hands: Aspects of Autonomous Vehicle Control\\
	&	 13	&	 Arlette van Wissen (VU), Agent-Based Support for Behavior Change: Models and Applications in Health and Safety Domains\\
	&	 14	&	 Yangyang Shi (TUD), Language Models With Meta-information\\
	&	 15	&	 Natalya Mogles (VU), Agent-Based Analysis and Support of Human Functioning in Complex Socio-Technical Systems: Applications in Safety and Healthcare\\
	&	 16	&	 Krystyna Milian (VU), Supporting trial recruitment and design by automatically interpreting eligibility criteria\\
	&	 17	&	 Kathrin Dentler (VU), Computing healthcare quality indicators automatically: Secondary Use of Patient Data and Semantic Interoperability\\
	&	 18	&	 Mattijs Ghijsen (UVA), Methods and Models for the Design and Study of Dynamic Agent Organizations\\
	&	 19	&	 Vinicius Ramos (TUE), 	Adaptive Hypermedia Courses: Qualitative and Quantitative Evaluation and Tool Support\\
	&	 20	&	 Mena Habib (UT), Named Entity Extraction and Disambiguation for Informal Text: The Missing Link\\
	&	 21	&	 Kassidy Clark (TUD), 	Negotiation and Monitoring in Open Environments\\
	&	 22	&	 Marieke Peeters (UU), Personalized Educational Games - Developing agent-supported scenario-based training\\
	&	 23	&	 Eleftherios Sidirourgos (UvA/CWI), 	Space Efficient Indexes for the Big Data Era\\
	&	 24	&	 Davide Ceolin (VU), Trusting Semi-structured Web Data\\
	&	 25	&	 Martijn Lappenschaar (RUN), 	New network models for the analysis of disease interaction\\
	&	 26	&	 Tim Baarslag (TUD), What to Bid and When to Stop\\
	&	 27	&	 Rui Jorge Almeida (EUR), 	Conditional Density Models Integrating Fuzzy and Probabilistic Representations of Uncertainty\\
	&	 28	&	 Anna Chmielowiec (VU), 	Decentralized k-Clique Matching\\
	&	 29	&	 Jaap Kabbedijk (UU), 	Variability in Multi-Tenant Enterprise Software\\
	&	 30	&	 Peter de Cock (UvT), 	Anticipating Criminal Behaviour\\
	&	 31	&	 Leo van Moergestel (UU), 	Agent Technology in Agile Multiparallel Manufacturing and Product Support\\
	&	 32	&	 Naser Ayat (UvA), 	On Entity Resolution in Probabilistic Data\\
	&	 33	&	 Tesfa Tegegne (RUN), Service Discovery in eHealth\\
	&	 34	&	 Christina Manteli (VU), 	The Effect of Governance in Global Software Development: Analyzing Transactive Memory Systems.\\
	&	 35	&	 Joost van Ooijen (UU), 	Cognitive Agents in Virtual Worlds: A Middleware Design Approach\\
	&	 36	&	 Joos Buijs (TUE), 	Flexible Evolutionary Algorithms for Mining Structured Process Models\\
	&	 37	&	 Maral Dadvar (UT), 	Experts and Machines United Against Cyberbullying\\
	&	 38	&	 Danny Plass-Oude Bos (UT), 	Making brain-computer interfaces better: improving usability through post-processing.\\
	&	 39	&	 Jasmina Maric (UvT), 	Web Communities, Immigration, and Social Capital\\
	&	 40	&	 Walter Omona (RUN), 	A Framework for Knowledge Management Using ICT in Higher Education\\
	&	 41	&	 Frederic Hogenboom (EUR), 	Automated Detection of Financial Events in News Text\\
	&	 42	&	 Carsten Eijckhof (CWI/TUD), 	Contextual Multidimensional Relevance Models\\
	&	 43	&	 Kevin Vlaanderen (UU), 	Supporting Process Improvement using Method Increments\\
	&	 44	&	 Paulien Meesters (UvT), 	Intelligent Blauw. Met als ondertitel: Intelligence-gestuurde politiezorg in gebiedsgebonden eenheden.\\
	&	 45	&	 Birgit Schmitz (OUN), 	Mobile Games for Learning: A Pattern-Based Approach\\
	&	 46	&	 Ke Tao (TUD), 	Social Web Data Analytics: Relevance, Redundancy, Diversity\\
	&	 47	&	 Shangsong Liang (UVA), 	Fusion and Diversification in Information Retrieval\\

\midrule
2015&	 01	&	 Niels Netten (UvA), Machine Learning for Relevance of Information in Crisis Response\\
	&	 02	&	 Faiza Bukhsh (UvT), Smart auditing: Innovative Compliance Checking in Customs Controls\\
	&	 03	&	 Twan van Laarhoven (RUN), Machine learning for network data\\
	&	 04	&	 Howard Spoelstra (OUN), Collaborations in Open Learning Environments\\
	&	 05	&	 Christoph B\"{o}sch (UT), Cryptographically Enforced Search Pattern Hiding\\
	&	 06	&	 Farideh Heidari (TUD), Business Process Quality Computation - Computing Non-Functional Requirements to Improve Business Processes\\
	&	 07	&	 Maria-Hendrike Peetz (UvA), Time-Aware Online Reputation Analysis\\
	&	 08	&	 Jie Jiang (TUD), 	Organizational Compliance: An agent-based model for designing and evaluating organizational interactions\\
	&	 09	&	 Randy Klaassen (UT), HCI Perspectives on Behavior Change Support Systems\\
	&	 10	&	 Henry Hermans (OUN), OpenU: design of an integrated system to support lifelong learning\\
	&	 11	&	 Yongming Luo (TUE), Designing algorithms for big graph datasets: A study of computing bisimulation and joins\\
	&	 12	&	 Julie M. Birkholz (VU), Modi Operandi of Social Network Dynamics: The Effect of Context on Scientific Collaboration Networks\\
	&	 13	&	 Giuseppe Procaccianti (VU), Energy-Efficient Software\\
	&	 14	&	 Bart van Straalen (UT), A cognitive approach to modeling bad news conversations\\
	&	 15	&	 Klaas Andries de Graaf (VU), Ontology-based Software Architecture Documentation\\
	&	 16	&	 Changyun Wei (UT), Cognitive Coordination for Cooperative Multi-Robot Teamwork\\
	&	 17	&	 Andr\'{e} van Cleeff (UT), Physical and Digital Security Mechanisms: Properties, Combinations and Trade-offs\\
	&	 18	&	 Holger Pirk (CWI), Waste Not, Want Not! - Managing Relational Data in Asymmetric Memories\\
	&	 19	&	 Bernardo Tabuenca (OUN), Ubiquitous Technology for Lifelong Learners\\
	&	 20	&	 Lois Vanh\'{e}e (UU), 	Using Culture and Values to Support Flexible Coordination\\
	&	 21	&	 Sibren Fetter (OUN), Using Peer-Support to Expand and Stabilize Online Learning\\
	&	 22	&	 Zhemin Zhu (UT), 	Co-occurrence Rate Networks\\
	&	 23	&	 Luit Gazendam (VU), Cataloguer Support in Cultural Heritage\\
	&	 24	&	 Richard Berendsen (UVA), 	Finding People, Papers, and Posts: Vertical Search Algorithms and Evaluation\\
	&	 25	&	 Steven Woudenberg (UU), Bayesian Tools for Early Disease Detection\\
	&	 26	&	 Alexander Hogenboom (EUR), Sentiment Analysis of Text Guided by Semantics and Structure\\
	&	 27	&	 S\'{a}ndor H\'{e}man (CWI), Updating compressed colomn stores\\
	&	 28	&	 Janet Bagorogoza (TiU), Knowledge Management and High Performance; The Uganda Financial Institutions Model for HPO\\
	&	 29	&	 Hendrik Baier (UM), Monte-Carlo Tree Search Enhancements for One-Player and Two-Player Domains\\
	&	 30	&	 Kiavash Bahreini (OU), Real-time Multimodal Emotion Recognition in E-Learning\\
	&	 31	&	 Yakup Ko\c{c} (TUD), On the robustness of Power Grids\\
	&	 32	&	 Jerome Gard (UL), Corporate Venture Management in SMEs\\
	&	 33	&	 Frederik Schadd (TUD), Ontology Mapping with Auxiliary Resources\\
	&	 34	&	 Victor de Graaf (UT), Gesocial Recommender Systems\\
	&	 35	&	 Jungxao Xu (TUD), Affective Body Language of Humanoid Robots: Perception and Effects in Human Robot Interaction\\

\midrule
2016&	 01	&	 Syed Saiden Abbas (RUN), Recognition of Shapes by Humans and Machines\\
	&	 02	&	 Michiel Christiaan Meulendijk (UU), Optimizing medication reviews through decision support: prescribing a better pill to swallow\\
	&	 03	&	 Maya Sappelli (RUN), Knowledge Work in Context: User Centered Knowledge Worker Support\\
	&	 04	&	 Laurens Rietveld (VU), Publishing and Consuming Linked Data\\
	&	 05	&	 Evgeny Sherkhonov (UVA), Expanded Acyclic Queries: Containment and an Application in Explaining Missing Answers\\
	&	 06	&	 Michel Wilson (TUD), Robust scheduling in an uncertain environment\\
	&	 07	&	 Jeroen de Man (VU), Measuring and modeling negative emotions for virtual training\\
	&	 08	&	 Matje van de Camp (TiU), A Link to the Past: Constructing Historical Social Networks from Unstructured Data\\
	&	 09	&	 Archana Nottamkandath (VU), Trusting Crowdsourced Information on Cultural Artefacts\\
	&	 10	&	 George Karafotias (VUA), Parameter Control for Evolutionary Algorithms\\
	&	 11	&	 Anne Schuth (UVA), Search Engines that Learn from Their Users\\
	&	 12	&	 Max Knobbout (UU), Logics for Modelling and Verifying Normative Multi-Agent Systems\\
	&	 13	&	 Nana Baah Gyan (VU), The Web, Speech Technologies and Rural Development in West Africa - An ICT4D Approach\\
	&	 14	&	 Ravi Khadka (UU), Revisiting Legacy Software System Modernization\\
	&	 15	&	 Steffen Michels (RUN), Hybrid Probabilistic Logics - Theoretical Aspects, Algorithms and Experiments\\
	&	 16	&	 Guangliang Li (UVA), Socially Intelligent Autonomous Agents that Learn from Human Reward\\
	&	 17	&	 Berend Weel (VU), Towards Embodied Evolution of Robot Organisms\\
	&	 18	&	 Albert Mero\~{n}o Pe\~{n}uela (VU), Refining Statistical Data on the Web\\
	&	 19	&	 Julia Efremova (Tu/e), Mining Social Structures from Genealogical Data\\
	&	 20	&	 Daan Odijk (UVA), Context \& Semantics in News \& Web Search\\
	&	 21	&	 Alejandro Moreno C\'{e}lleri (UT), From Traditional to Interactive Playspaces: Automatic Analysis of Player Behavior in the Interactive Tag Playground\\
	&	 22	&	 Grace Lewis (VU), Software Architecture Strategies for Cyber-Foraging Systems\\
	&	 23	&	 Fei Cai (UVA), Query Auto Completion in Information Retrieval\\
	&	 24	&	 Brend Wanders (UT), Repurposing and Probabilistic Integration of Data; An Iterative and data model independent approach\\
	&	 25	&	 Julia Kiseleva (TU/e), Using Contextual Information to Understand Searching and Browsing Behavior\\
	&	 26	&	 Dilhan Thilakarathne (VU), In or Out of Control: Exploring Computational Models to Study the Role of Human Awareness and Control in Behavioural Choices, with Applications in Aviation and Energy Management Domains\\
	&	 27	&	 Wen Li (TUD), Understanding Geo-spatial Information on Social Media\\
	&	 28	&	 Mingxin Zhang (TUD), Large-scale Agent-based Social Simulation - A study on epidemic prediction and control\\
	&	 29	&	 Nicolas H\"{o}ning (TUD), Peak reduction in decentralised electricity systems - Markets and prices for flexible planning\\
	&	 30	&	 Ruud Mattheij (UvT), The Eyes Have It\\
	&	 31	&	 Mohammad Khelghati (UT), Deep web content monitoring\\
	&	 32	&	 Eelco Vriezekolk (UT), Assessing Telecommunication Service Availability Risks for Crisis Organisations\\
	&	 33	&	 Peter Bloem (UVA), Single Sample Statistics, exercises in learning from just one example\\
	&	 34	&	 Dennis Schunselaar (TUE), Configurable Process Trees: Elicitation, Analysis, and Enactment\\
	&	 35	&	 Zhaochun Ren (UVA), Monitoring Social Media: Summarization, Classification and Recommendation\\
	&	 36	&	 Daphne Karreman (UT), Beyond R2D2: The design of nonverbal interaction behavior optimized for robot-specific morphologies\\
	&	 37	&	 Giovanni Sileno (UvA), Aligning Law and Action - a conceptual and computational inquiry\\
	&	 38	&	 Andrea Minuto (UT), Materials that Matter - Smart Materials meet Art \& Interaction Design\\
	&	 39	&	 Merijn Bruijnes (UT), Believable Suspect Agents; Response and Interpersonal Style Selection for an Artificial Suspect\\
	&	 40	&	 Christian Detweiler (TUD), Accounting for Values in Design\\
	&	 41	&	 Thomas King (TUD), Governing Governance: A Formal Framework for Analysing Institutional Design and Enactment Governance\\
	&	 42	&	 Spyros Martzoukos (UVA), Combinatorial and Compositional Aspects of Bilingual Aligned Corpora\\
	&	 43	&	 Saskia Koldijk (RUN), Context-Aware Support for Stress Self-Management: From Theory to Practice\\
	&	 44 &	 Thibault Sellam (UVA), Automatic Assistants for Database Exploration\\
	&	 45	&	 Bram van de Laar (UT), Experiencing Brain-Computer Interface Control\\
	&	 46	&	 Jorge Gallego Perez (UT), Robots to Make you Happy\\
	&	 47	&	 Christina Weber (UL), Real-time foresight - Preparedness for dynamic innovation networks\\
	&	 48	&	 Tanja Buttler (TUD), Collecting Lessons Learned\\
	&	 49	&	 Gleb Polevoy (TUD), Participation and Interaction in Projects. A Game-Theoretic Analysis\\
	&	 50	&	 Yan Wang (UVT), The Bridge of Dreams: Towards a Method for Operational Performance Alignment in IT-enabled Service Supply Chains\\
	
\midrule
2017&	 01	&	 Jan-Jaap Oerlemans (UL), Investigating Cybercrime\\
	&	 02	&	 Sjoerd Timmer (UU), Designing and Understanding Forensic Bayesian Networks using Argumentation\\
	&	 03	&	 Dani\"{e}l Harold Telgen (UU), Grid Manufacturing; A Cyber-Physical Approach with Autonomous Products and Reconfigurable Manufacturing Machines\\
	&	 04	&	 Mrunal Gawade (CWI), Multi-core Parallelism in a Column-store\\
	&	 05	&	 Mahdieh Shadi (UVA), Collaboration Behavior\\
	&	 06	&	 Damir Vandic (EUR), Intelligent Information Systems for Web Product Search\\
	&	 07	&	 Roel Bertens (UU), Insight in Information: from Abstract to Anomaly\\
	&	 08	& 	 Rob Konijn (VU)	, Detecting Interesting Differences:Data Mining in Health Insurance Data using Outlier Detection and Subgroup Discovery\\
	&	 09	&	 Dong Nguyen (UT), Text as Social and Cultural Data: A Computational Perspective on Variation in Text\\
	&	 10	&	 Robby van Delden (UT), (Steering) Interactive Play Behavior\\       
	&	 11	&	 Florian Kunneman (RUN), Modelling patterns of time and emotion in Twitter \#anticipointment\\	      										
	&	 12	&	 Sander Leemans (TUE), Robust Process Mining with Guarantees\\      
 	&	 13	& 	 Gijs Huisman (UT),  Social Touch Technology - Extending the reach of social touch through haptic technology\\     
 	&	 14	&	 Shoshannah Tekofsky (UvT), You Are Who You Play You Are: Modelling Player Traits from Video Game Behavior\\ 
	&	 15	&	 Peter Berck (RUN),  Memory-Based Text Correction\\  
	&	 16	&	 Aleksandr Chuklin (UVA), Understanding and Modeling Users of Modern Search Engines\\   
	&	 17	&	 Daniel Dimov (UL), Crowdsourced Online Dispute Resolution\\
	&	 18	&	 Ridho Reinanda (UVA), Entity Associations for Search\\
	&	 19	& 	 Jeroen Vuurens (UT), Proximity of Terms, Texts and Semantic Vectors in Information Retrieval\\
	&	 20	&	 Mohammadbashir Sedighi (TUD), Fostering Engagement in Knowledge Sharing: The Role of Perceived Benefits, Costs and Visibility\\
	&	 21	&	 Jeroen Linssen (UT), Meta Matters in Interactive Storytelling and Serious Gaming (A Play on Worlds)\\
	&	 22	&	 Sara Magliacane (VU), Logics for causal inference under uncertainty\\
	&	 23	&	 David Graus (UVA), Entities of Interest --- Discovery in Digital Traces\\
	&	 24	&	 Chang Wang (TUD), Use of Affordances for Efficient Robot Learning\\
	&	 25	&	 Veruska Zamborlini (VU), Knowledge Representation for Clinical Guidelines, with applications to Multimorbidity Analysis and Literature Search\\
	&	 26	&	 Merel Jung (UT), Socially intelligent robots that understand and respond to human touch\\	
	&	 27	&	 Michiel Joosse (UT), Investigating Positioning and Gaze Behaviors of Social Robots: People's Preferences, Perceptions and Behaviors\\
	&	 28	&	 John Klein (VU), Architecture Practices for Complex Contexts\\
	&	 29	&	 Adel Alhuraibi (UvT), From IT-BusinessStrategic Alignment to Performance: A Moderated Mediation Model of Social Innovation, and Enterprise Governance of    IT"\\
	&	 30	&	 Wilma Latuny (UvT), The Power of Facial Expressions\\
	&	 31	&	 Ben Ruijl (UL), Advances in computational methods for QFT calculations\\
	&	 32	& 	 Thaer Samar (RUN), Access to and Retrievability of Content in Web Archives\\
	&	 33	&	 Brigit van Loggem (OU), Towards a Design Rationale for Software Documentation: A Model of Computer-Mediated Activity\\
	&	 34	&	 Maren Scheffel (OU), The Evaluation Framework for Learning Analytics \\
	&	 35	&	 Martine de Vos (VU), Interpreting natural science spreadsheets \\
	&	 36	&	 Yuanhao Guo (UL), Shape Analysis for Phenotype Characterisation from High-throughput Imaging \\
	&	 37	&	 Alejandro Montes Garcia (TUE), WiBAF: A Within Browser Adaptation Framework that Enables Control over Privacy \\
	&	 38	&	 Alex Kayal (TUD), Normative Social Applications \\	 
	&	 39	&	 Sara Ahmadi (RUN), Exploiting properties of the human auditory system and compressive sensing methods to increase   noise robustness in ASR \\
	&	 40	&	 Altaf Hussain Abro (VUA), Steer your Mind: Computational Exploration of Human Control in Relation to Emotions, Desires and Social Support For applications in human-aware support systems \\
	&	 41	&	 Adnan Manzoor (VUA), Minding a Healthy Lifestyle: An Exploration of Mental Processes and a Smart Environment to Provide Support for a Healthy Lifestyle\\
	&	 42	&	 Elena Sokolova (RUN), Causal discovery from mixed and missing data with applications on ADHD  datasets\\
	&	 43	&	 Maaike de Boer (RUN), Semantic Mapping in Video Retrieval\\
	&	 44	&	 Garm Lucassen (UU), Understanding User Stories - Computational Linguistics in Agile Requirements Engineering\\
	&	 45	&	 Bas Testerink	(UU), Decentralized Runtime Norm Enforcement\\
	&	 46	&	 Jan Schneider	(OU), Sensor-based Learning Support\\
	&	 47	&	 Jie Yang (TUD), Crowd Knowledge Creation Acceleration\\
	&	 48	&	 Angel Suarez (OU), Collaborative inquiry-based learning\\

\midrule
2018&	 01	&	 Han van der Aa (VUA), Comparing and Aligning Process Representations \\ 
	&	 02	&	 Felix Mannhardt (TUE), Multi-perspective Process Mining \\
	&	 03	&	 Steven Bosems (UT), Causal Models For Well-Being: Knowledge Modeling, Model-Driven Development of Context-Aware Applications, and Behavior Prediction\\
	&	 04	&	 Jordan Janeiro (TUD), Flexible Coordination Support for Diagnosis Teams in Data-Centric Engineering Tasks \\
	&	 05	&	 Hugo Huurdeman (UVA), Supporting the Complex Dynamics of the Information Seeking Process \\
	&	 06	&	 Dan Ionita (UT), Model-Driven Information Security Risk Assessment of Socio-Technical Systems \\
	&	 07	&	 Jieting Luo (UU), A formal account of opportunism in multi-agent systems \\
	&	 08	&	 Rick Smetsers (RUN), Advances in Model Learning for Software Systems \\
	&	 09	&	 Xu Xie	(TUD), Data Assimilation in Discrete Event Simulations \\	
	&	 10	&	 Julienka Mollee (VUA), Moving forward: supporting physical activity behavior change through intelligent technology \\
	&	 11	&	 Mahdi Sargolzaei (UVA), Enabling Framework for Service-oriented Collaborative Networks \\
	&	 12	&	 Xixi Lu (TUE), Using behavioral context in process mining \\
	&	 13	&	 Seyed Amin Tabatabaei (VUA), Computing a Sustainable Future \\
	&	 14	&	 Bart Joosten (UVT), Detecting Social Signals with Spatiotemporal Gabor Filters \\
	&	 15	&	 Naser Davarzani (UM), Biomarker discovery in heart failure \\
	&	 16	&	 Jaebok Kim (UT), Automatic recognition of engagement and emotion in a group of children \\
	&	 17	&	 Jianpeng Zhang (TUE), On Graph Sample Clustering \\
	&	 18	& 	 Henriette Nakad (UL), De Notaris en Private Rechtspraak \\
	&	 19	&	 Minh Duc Pham (VUA), Emergent relational schemas for RDF \\
	&	 20	&	 Manxia Liu (RUN), Time and Bayesian Networks \\
	&	 21	&	 Aad Slootmaker (OUN), EMERGO: a generic platform for authoring and playing scenario-based serious games \\
	&	 22	&	 Eric Fernandes de Mello Araujo (VUA), Contagious: Modeling the Spread of Behaviours, Perceptions and Emotions in Social Networks \\
	&	 23	&	 Kim Schouten (EUR), Semantics-driven Aspect-Based Sentiment Analysis \\
	&	 24	&	 Jered Vroon (UT), Responsive Social Positioning Behaviour for Semi-Autonomous Telepresence Robots \\
	&	 25	&	 Riste Gligorov (VUA), Serious Games in Audio-Visual Collections \\ 
	&	 26	& 	 Roelof Anne Jelle de Vries (UT),Theory-Based and Tailor-Made: Motivational Messages for Behavior Change Technology \\ 
	&	 27	&	 Maikel Leemans (TUE), Hierarchical Process Mining for Scalable Software Analysis \\
	&	 28	&	 Christian Willemse (UT), Social Touch Technologies: How they feel and how they make you feel \\	
	&	 29	&	 Yu Gu (UVT), Emotion Recognition from Mandarin Speech \\
	&	 30	&	 Wouter Beek,  The "K" in "semantic web" stands for "knowledge": scaling semantics to the web \\
	
\midrule
2019
	&	 01	&	 Rob van Eijk (UL),Web privacy measurement in real-time bidding systems. A graph-based approach to RTB system classification \\
	&	 02	&	 Emmanuelle Beauxis Aussalet (CWI, UU), Statistics and Visualizations for Assessing Class Size Uncertainty \\
	&	 03	&	 Eduardo Gonzalez Lopez de Murillas (TUE), Process Mining on Databases: Extracting Event Data from Real Life Data 
				 Sources \\
	&	 04	&	 Ridho Rahmadi (RUN), Finding stable causal structures from clinical data \\				 
	& 	 05	&	 Sebastiaan van Zelst (TUE), Process Mining with Streaming Data \\
	&	 06	& 	 Chris Dijkshoorn (VU), Nichesourcing for Improving Access to Linked Cultural Heritage Datasets \\
	&	 07	&	 Soude Fazeli (TUD), Recommender Systems in Social Learning Platforms \\
	& 	 08	&	 Frits de Nijs (TUD), Resource-constrained Multi-agent Markov Decision Processes \\
	&	 09	&	 Fahimeh Alizadeh Moghaddam (UVA), Self-adaptation for energy efficiency in software systems \\
	&	 10	&	 Qing Chuan Ye (EUR), Multi-objective Optimization Methods for Allocation and Prediction \\
	&	 11	&	 Yue Zhao (TUD), Learning Analytics Technology to Understand Learner Behavioral Engagement in MOOCs \\
	&	 12	&	 Jacqueline Heinerman (VU), Better Together \\
	&	 13	&	 Guanliang Chen (TUD), MOOC Analytics: Learner Modeling and Content Generation \\
	&	 14	&	 Daniel Davis (TUD), Large-Scale Learning Analytics: Modeling Learner Behavior \& Improving Learning Outcomes in Massive Open Online Courses \\
	&	 15	&	 Erwin Walraven (TUD), Planning under Uncertainty in Constrained and Partially 
				 Observable Environments \\
	&	 16	&	 Guangming Li (TUE), Process Mining based on Object-Centric Behavioral Constraint (OCBC) Models \\ 			 
	&	 17	&	 Ali Hurriyetoglu (RUN),Extracting actionable information from microtexts \\
	&	 18	&	 Gerard Wagenaar (UU), Artefacts in Agile Team Communication \\
	&	 19	&	 Vincent Koeman (TUD), Tools for Developing Cognitive Agents \\
	&	 20	&	 Chide Groenouwe (UU), Fostering technically augmented human collective intelligence \\
	&	 21	&	 Cong Liu (TUE), Software Data Analytics: Architectural Model Discovery and Design Pattern Detection \\
	&	 22	&	 Martin van den Berg (VU),Improving IT Decisions with Enterprise Architecture \\
	&	 23	&	 Qin Liu (TUD), Intelligent Control Systems: Learning, Interpreting, Verification\\
	&	 24	&	 Anca Dumitrache (VU),  Truth in Disagreement - Crowdsourcing Labeled Data for Natural Language Processing\\
	&	 25	&	 Emiel van Miltenburg (VU), Pragmatic factors in (automatic) image description \\	
	&	 26	&	 Prince Singh (UT), An Integration Platform for Synchromodal Transport \\
	&	 27	&	 Alessandra Antonaci (OUN), The Gamification Design Process applied to (Massive) Open Online Courses\\
	&	 28	&	 Esther Kuindersma (UL), Cleared for take-off: Game-based learning to prepare airline pilots for critical situations \\
	&	 29	&	 Daniel Formolo (VU), Using virtual agents for simulation and training of social skills in safety-critical circumstances \\ 	
	&	 30	&	 Vahid Yazdanpanah (UT), Multiagent Industrial Symbiosis Systems \\
	&	 31	&	 Milan Jelisavcic (VU), Alive and Kicking: Baby Steps in Robotics \\
	&	 32	&	 Chiara Sironi (UM), Monte-Carlo Tree Search for Artificial General Intelligence in Games \\
	&	 33	&	 Anil Yaman (TUE), Evolution of Biologically Inspired Learning in Artificial Neural Networks \\
	&	 34	&	 Negar Ahmadi (TUE), EEG Microstate and Functional Brain Network Features for Classification of Epilepsy and PNES \\
	&	 35	&	 Lisa Facey-Shaw (OUN), Gamification with digital badges in learning programming \\
	&	 36	&	 Kevin Ackermans (OUN), Designing Video-Enhanced Rubrics to Master Complex Skills \\
	&	 37	&	 Jian Fang (TUD), Database Acceleration on FPGAs \\
	&	 38	&	 Akos Kadar (OUN), Learning visually grounded and multilingual representations \\

\midrule
2020

	&	 01	&	 Armon Toubman (UL), Calculated Moves: Generating Air Combat Behaviour \\
	&	 02	&	 Marcos de Paula Bueno (UL), Unraveling Temporal Processes using Probabilistic Graphical Models \\
	&	 03	&	 Mostafa Deghani (UvA), Learning with Imperfect Supervision for Language Understanding \\
	&	 04	&	 Maarten van Gompel (RUN), Context as Linguistic Bridges \\
	&	 05	&	 Yulong Pei (TUE), On local and global structure mining \\
	&	 06	&	 Preethu Rose Anish (UT), Stimulation Architectural Thinking during Requirements Elicitation - An Approach and Tool Support \\
	&	 07	&	 Wim van der Vegt (OUN), Towards a software architecture for reusable game components \\
	&	 08	&	 Ali Mirsoleimani (UL),Structured Parallel Programming for Monte Carlo Tree Search \\
	&	 09	&	 Myriam Traub (UU), Measuring Tool Bias and Improving Data Quality for Digital Humanities Research \\
	&	 10	&	 Alifah Syamsiyah (TUE), In-database Preprocessing for Process Mining \\
	&	 11	&	 Sepideh Mesbah (TUD), Semantic-Enhanced Training Data AugmentationMethods for Long-Tail Entity Recognition Models \\
	&	 12	&	 Ward van Breda (VU), Predictive Modeling in E-Mental Health: Exploring Applicability in Personalised Depression Treatment \\
	&	 13	&	 Marco Virgolin (CWI), Design and Application of Gene-pool Optimal Mixing Evolutionary Algorithms for Genetic Programming \\
	&	 14	&	 Mark Raasveldt (CWI/UL), Integrating Analytics with Relational Databases \\
	&	 15 	&	 Konstantinos Georgiadis (OUN),  Smart CAT: Machine Learning for Configurable Assessments in Serious Games \\
	&	 16	&	 Ilona Wilmont (RUN), Cognitive Aspects of Conceptual Modelling \\
	&	 17	&	 Daniele Di Mitri (OUN), The Multimodal Tutor: Adaptive Feedback from Multimodal Experiences \\
  	&	 18	&	 Georgios Methenitis (TUD), Agent Interactions \& Mechanisms in Markets with Uncertainties: Electricity Markets in Renewable Energy Systems \\
	&	 19	&	 Guido van Capelleveen (UT), Industrial Symbiosis Recommender Systems \\ 
	&	 20	&	 Albert Hankel (VU), Embedding Green ICT Maturity in Organisations \\
	&	 21	&	 Karine da Silva Miras de Araujo (VU), Where is the robot?: Life as it could be \\
	&	 22	&	 Maryam Masoud Khamis (RUN), Understanding complex systems implementation through a modeling approach: the case of e-government in Zanzibar \\
	&	 23	&	 Rianne Conijn (UT), The Keys to Writing: A writing analytics approach to studying writing processes using keystroke logging \\
	&	 24	&	 Lenin da Nobrega Medeiros (VUA/RUN), How are you feeling, human? Towards emotionally supportive chatbots \\
	&	 25	&	 Xin Du (TUE), The Uncertainty in Exceptional Model Mining \\
	&	 26	&	 Krzysztof Leszek Sadowski (UU), GAMBIT: Genetic Algorithm for Model-Based mixed-Integer opTimization \\
	&	 27	&	 Ekaterina Muravyeva (TUD), Personal data and informed consent in an educational context \\
	&	 28	&	 Bibeg Limbu (TUD), Multimodal interaction for deliberate practice: Training complex skills with augmented reality \\
	&	 29	&	 Ioan Gabriel Bucur (RUN), Being Bayesian about Causal Inference \\
	&	 30	&	 Bob Zadok Blok (UL), Creatief, Creatieve, Creatiefst \\
	&	 31 	&	 Gongjin Lan (VU), Learning better -- From Baby to Better \\
	&	 32 	& 	 Jason Rhuggenaath (TUE), Revenue management in online markets: pricing and online advertising \\
	&	 33	& 	 Rick Gilsing (TUE), Supporting service-dominant business model evaluation in the context of business model innovation \\
	&	 34 	&	 Anna Bon (MU), Intervention or Collaboration? Redesigning Information and Communication Technologies for Development \\
	&	 35	&	 Siamak Farshidi (UU), Multi-Criteria Decision-Making in Software Production \\

\midrule
2021

	&	 01	&	 Francisco Xavier Dos Santos Fonseca (TUD),Location-based Games for Social Interaction in Public Space \\
	&	 02	&	 Rijk Mercuur (TUD), Simulating Human Routines:Integrating Social Practice Theory in Agent-Based Models \\
	&	 03	&	 Seyyed Hadi Hashemi (UVA), Modeling Users Interacting with Smart Devices \\
	&	 04	&	 Ioana Jivet (OU), The Dashboard That Loved Me: Designing adaptive learning analytics for self-regulated learning \\
	&	 05	&	 Davide Dell'Anna (UU), Data-Driven Supervision of Autonomous Systems \\
	&	 06	&	 Daniel Davison (UT), "Hey robot, what do you think?" How children learn with a social robot \\
	&	 07	&	 Armel Lefebvre (UU), Research data management for open science \\
	&	 08	&	 Nardie Fanchamps (OU), The Influence of Sense-Reason-Act Programming on Computational Thinking \\
	&	 09	&	 Cristina Zaga (UT), The Design of Robothings. Non-Anthropomorphic and Non-Verbal Robots to Promote Children's Collaboration Through Play \\
	&	 10	&	 Quinten Meertens (UvA), Misclassification Bias in Statistical Learning \\
	&	 11	&	 Anne van Rossum (UL), Nonparametric Bayesian Methods in Robotic Vision \\
	&	 12	&	 Lei Pi (UL), External Knowledge Absorption in Chinese SMEs \\
	&	 13	&	 Bob R. Schadenberg (UT), Robots for Autistic Children: Understanding and Facilitating Predictability for Engagement in Learning \\
	&	 14	&	 Negin Samaeemofrad (UL), Business Incubators: The Impact of Their Support \\
	&	 15 	& 	 Onat Ege Adali (TU/e), Transformation of Value Propositions into Resource Re-Configurations through the Business Services Paradigm  \\
	&	 16 	&	 Esam A. H. Ghaleb (UM), BIMODAL EMOTION RECOGNITION FROM AUDIO-VISUAL CUES \\
	&	 17	&	 Dario Dotti (UM), Human Behavior Understanding  from motion and bodily cues using deep neural networks \\
	&	 18	&	 Remi Wieten (UU), Bridging the Gap Between Informal Sense-Making Tools and Formal Systems - Facilitating the Construction of Bayesian Networks and Argumentation Frameworks \\

\bottomrule
\end{longtabu}

%% file: main.bbl
\begin{thebibliography}{100}

\bibitem{IQ}
Intelligence quotient.
\newblock \url{https://en.wikipedia.org/wiki/Intelligence_quotient }.

\bibitem{abdollahpouri2019popularity}
Himan Abdollahpouri.
\newblock Popularity bias in ranking and recommendation.
\newblock In {\em Proceedings of the 2019 AAAI/ACM Conference on AI, Ethics,
  and Society}, pages 529--530, 2019.

\bibitem{abdollahpouri2020popularity}
Himan Abdollahpouri.
\newblock {\em Popularity bias in recommendation: A multi-stakeholder
  perspective}.
\newblock PhD thesis, University of Colorado at Boulder, 2020.

\bibitem{himan2019a}
Himan Abdollahpouri, Gediminas Adomavicius, Robin Burke, Ido Guy, Dietmar
  Jannach, Toshihiro Kamishima, Jan Krasnodebski, and Luiz~Augusto Pizzato.
\newblock Beyond personalization: Research directions in multistakeholder
  recommendation.
\newblock {\em CoRR}, abs/1905.01986, 2019.

\bibitem{abdollahpouri2020unfair}
Himan Abdollahpouri, Robin Burke, and Masoud Mansoury.
\newblock Unfair exposure of artists in music recommendation.
\newblock {\em arXiv preprint arXiv:2003.11634}, 2020.

\bibitem{abdollahpouri2017controlling}
Himan Abdollahpouri, Robin Burke, and Bamshad Mobasher.
\newblock Controlling popularity bias in learning to rank recommendation.
\newblock In {\em Proceedings of the 11th ACM conference on Recommender
  systems}, pages 42--46. ACM, 2017.

\bibitem{abdollahpouri2019managing}
Himan Abdollahpouri, Robin Burke, and Bamshad Mobasher.
\newblock Managing popularity bias in recommender systems with personalized
  re-ranking.
\newblock In {\em The Thirty-Second International Flairs Conference}, 2019.

\bibitem{abdollahpouri2020multi}
Himan Abdollahpouri and Masoud Mansoury.
\newblock Multi-sided exposure bias in recommendation.
\newblock {\em KDD workshop on Industrial Recommendation Systems}, 2020.

\bibitem{abdollahpouri2019impact}
Himan Abdollahpouri, Masoud Mansoury, Robin Burke, and Bamshad Mobasher.
\newblock The impact of popularity bias on fairness and calibration in
  recommendation.
\newblock {\em arXiv preprint arXiv:1910.05755}, 2019.

\bibitem{himan2019c}
Himan Abdollahpouri, Masoud Mansoury, Robin Burke, and Bamshad Mobasher.
\newblock The unfairness of popularity bias in recommendation.
\newblock In {\em RecSys Workshop on Recommendation in Multistakeholder
  Environments (RMSE)}, 2019.

\bibitem{abdollahpouri2020addressing}
Himan Abdollahpouri, Masoud Mansoury, Robin Burke, and Bamshad Mobasher.
\newblock Addressing the multistakeholder impact of popularity bias in
  recommendation through calibration.
\newblock {\em arXiv preprint arXiv:2007.12230}, 2020.

\bibitem{abdollahpouri2020connection}
Himan Abdollahpouri, Masoud Mansoury, Robin Burke, and Bamshad Mobasher.
\newblock The connection between popularity bias, calibration, and fairness in
  recommendation.
\newblock In {\em Fourteenth ACM Conference on Recommender Systems}, pages
  726--731, 2020.

\bibitem{abdollahpouri2021user}
Himan Abdollahpouri, Masoud Mansoury, Robin Burke, Bamshad Mobasher, and Edward
  Malthouse.
\newblock User-centered evaluation of popularity bias in recommender systems.
\newblock In {\em Proceedings of the 29th ACM Conference on User Modeling,
  Adaptation and Personalization}, pages 119--129, 2021.

\bibitem{adamopoulos2014over}
Panagiotis Adamopoulos and Alexander Tuzhilin.
\newblock On over-specialization and concentration bias of recommendations:
  Probabilistic neighborhood selection in collaborative filtering systems.
\newblock In {\em Proceedings of the 8th ACM Conference on Recommender
  systems}, pages 153--160, 2014.

\bibitem{adomavicius2011improving}
Gediminas Adomavicius and YoungOk Kwon.
\newblock Improving aggregate recommendation diversity using ranking-based
  techniques.
\newblock {\em IEEE Transactions on Knowledge and Data Engineering},
  24(5):896--911, 2011.

\bibitem{adomavicius2011maximizing}
Gediminas Adomavicius and YoungOk Kwon.
\newblock Maximizing aggregate recommendation diversity: A graph-theoretic
  approach.
\newblock In {\em Proc. of the 1st International Workshop on Novelty and
  Diversity in Recommender Systems (DiveRS 2011)}, pages 3--10. Citeseer, 2011.

\bibitem{adomavicius2011}
Gediminas Adomavicius and YoungOk Kwon.
\newblock Maximizing aggregate recommendation diversity: A graph-theoretic
  approach.
\newblock In {\em In Proceedings of the 1st International Workshop on Novelty
  and Diversity in Recommender Systems (DiveRS 2011)}, pages 3--10, 2011.

\bibitem{Adomavicius:2005}
Gediminas Adomavicius, Ramesh Sankaranarayanan, Shahana Sen, and Alexander
  Tuzhilin.
\newblock Incorporating contextual information in recommender systems using a
  multidimensional approach.
\newblock {\em ACM Transactions on Information Systems (TOIS)}, 23(1):103--145,
  2005.

\bibitem{Adomavicius:2015}
Gediminas Adomavicius and Alexander Tuzhilin.
\newblock Context-aware recommender systems.
\newblock In {\em Recommender systems handbook}, pages 191--226. Springer US,
  2015.

\bibitem{anderson2006long}
Chris Anderson.
\newblock {\em The long tail: Why the future of business is selling more for
  less}.
\newblock Hyperion, 2006.

\bibitem{antikacioglu2017}
Arda Antikacioglu and R.~Ravi.
\newblock Post processing recommender systems for diversity.
\newblock In {\em In Proceedings of the 23rd ACM SIGKDD International
  Conference on Knowledge Discovery and Data Mining}, pages 707--716, 2017.

\bibitem{balabanovic1997fab}
Marko Balabanovi{\'c} and Yoav Shoham.
\newblock Fab: content-based, collaborative recommendation.
\newblock {\em Communications of the ACM}, 40(3):66--72, 1997.

\bibitem{baluja2008video}
Shumeet Baluja, Rohan Seth, Dharshi Sivakumar, Yushi Jing, Jay Yagnik, Shankar
  Kumar, Deepak Ravichandran, and Mohamed Aly.
\newblock Video suggestion and discovery for youtube: taking random walks
  through the view graph.
\newblock In {\em Proceedings of the 17th international conference on World
  Wide Web}, pages 895--904, 2008.

\bibitem{beutel2019fairness}
Alex Beutel, Jilin Chen, Tulsee Doshi, Hai Qian, Li~Wei, Yi~Wu, Lukasz Heldt,
  Zhe Zhao, Lichan Hong, Ed~H Chi, et~al.
\newblock Fairness in recommendation ranking through pairwise comparisons.
\newblock In {\em Proceedings of the 25th ACM SIGKDD International Conference
  on Knowledge Discovery \& Data Mining}, pages 2212--2220, 2019.

\bibitem{longtailnichesriche}
Brynjolfsson, Erik, Hu, Yu~Jeffrey, Smith, and Michael D.
\newblock From niches to riches: Anatomy of the long tail.
\newblock {\em Sloan Management Review}, 47(4):67--71, 2006.

\bibitem{brynjolfsson2003}
Erik Brynjolfsson, Yu~Hu, and Michael~D. Smith.
\newblock Consumer surplus in the digital economy: Estimating the value of
  increased product variety at online booksellers.
\newblock {\em Management Science}, 49(11):1580--1596, 2003.

\bibitem{buolamwini2018gender}
Joy Buolamwini and Timnit Gebru.
\newblock Gender shades: Intersectional accuracy disparities in commercial
  gender classification.
\newblock In {\em Conference on fairness, accountability and transparency},
  pages 77--91. PMLR, 2018.

\bibitem{burke2002hybrid}
Robin Burke.
\newblock Hybrid recommender systems: Survey and experiments.
\newblock {\em User modeling and user-adapted interaction}, 12(4):331--370,
  2002.

\bibitem{burke2007hybrid}
Robin Burke.
\newblock Hybrid web recommender systems.
\newblock In {\em The adaptive web}, pages 377--408. Springer, 2007.

\bibitem{burke2017b}
Robin Burke.
\newblock Multisided fairness for recommendation.
\newblock {\em CoRR}, abs/1707.00093, 2017.

\bibitem{burke2020experimentation}
Robin Burke, Masoud Mansoury, and Nasim Sonboli.
\newblock Experimentation with fairness-aware recommendation using librec-auto:
  hands-on tutorial.
\newblock In {\em Proceedings of the 2020 Conference on Fairness,
  Accountability, and Transparency}, 2020.

\bibitem{burke2017}
Robin Burke, Nasim Sonboli, Masoud Mansoury, and Aldo Ordoñez-Gauger.
\newblock Balanced neighborhoods for fairness-aware collaborative
  recommendation.
\newblock In {\em RecSys workshop on Fairness, Accountability and Transparency
  in Recommender Systems}, 2017.

\bibitem{burke2016}
Robin~D. Burke, Himan Abdollahpouri, Bamshad Mobasher, and Trinadh Gupta.
\newblock Towards multi-stakeholder utility evaluation of recommender systems.
\newblock In {\em In UMAP (Extended Proceedings)}, 2016.

\bibitem{calders2009building}
Toon Calders, Faisal Kamiran, and Mykola Pechenizkiy.
\newblock Building classifiers with independency constraints.
\newblock In {\em 2009 IEEE International Conference on Data Mining Workshops},
  pages 13--18. IEEE, 2009.

\bibitem{calders2010three}
Toon Calders and Sicco Verwer.
\newblock Three naive bayes approaches for discrimination-free classification.
\newblock {\em Data mining and knowledge discovery}, 21(2):277--292, 2010.

\bibitem{camacho2018social}
Lesly Alejandra~Gonzalez Camacho and Solange~Nice Alves-Souza.
\newblock Social network data to alleviate cold-start in recommender system: A
  systematic review.
\newblock {\em Information Processing \& Management}, 54(4):529--544, 2018.

\bibitem{castells2015novelty}
Pablo Castells, Neil~J Hurley, and Saul Vargas.
\newblock Novelty and diversity in recommender systems.
\newblock In {\em Recommender systems handbook}, pages 881--918. Springer,
  2015.

\bibitem{chaney2018}
Allison~JB Chaney, Brandon~M. Stewart, and Barbara~E. Engelhardt.
\newblock How algorithmic confounding in recommendation systems increases
  homogeneity and decreases utility.
\newblock In {\em Proceedings of the 12th ACM Conference on Recommender
  Systems}, pages 224--232, 2018.

\bibitem{chen2020bias}
Jiawei Chen, Hande Dong, Xiang Wang, Fuli Feng, Meng Wang, and Xiangnan He.
\newblock Bias and debias in recommender system: A survey and future
  directions.
\newblock {\em arXiv preprint arXiv:2010.03240}, 2020.

\bibitem{chen2018sequential}
Xu~Chen, Hongteng Xu, Yongfeng Zhang, Jiaxi Tang, Yixin Cao, Zheng Qin, and
  Hongyuan Zha.
\newblock Sequential recommendation with user memory networks.
\newblock In {\em Proceedings of the eleventh ACM international conference on
  web search and data mining}, pages 108--116, 2018.

\bibitem{chouldechova2017fair}
Alexandra Chouldechova.
\newblock Fair prediction with disparate impact: A study of bias in recidivism
  prediction instruments.
\newblock {\em Big data}, 5(2):153--163, 2017.

\bibitem{ciampaglia2018algorithmic}
Giovanni~Luca Ciampaglia, Azadeh Nematzadeh, Filippo Menczer, and Alessandro
  Flammini.
\newblock How algorithmic popularity bias hinders or promotes quality.
\newblock {\em Scientific reports}, 8(1):1--7, 2018.

\bibitem{collins2018study}
Andrew Collins, Dominika Tkaczyk, Akiko Aizawa, and Joeran Beel.
\newblock A study of position bias in digital library recommender systems.
\newblock {\em arXiv preprint arXiv:1802.06565}, 2018.

\bibitem{Cremonesi:2010a}
Paolo Cremonesi, Yehuda Koren, and Roberto Turrin.
\newblock Performance of recommender algorithms on top-n recommendation tasks.
\newblock In {\em Proceedings of the fourth ACM conference on Recommender
  systems}, pages 39--46. ACM, September 2010.

\bibitem{damour2020}
Alexander D'Amour, Hansa Srinivasan, James Atwood, Pallavi Baljekar,
  D.~Sculley, and Yoni Halpern.
\newblock Fairness is not static: deeper understanding of long term fairness
  via simulation studies.
\newblock In {\em Proceedings of the 2020 Conference on Fairness,
  Accountability, and Transparency}, pages 525--534, 2020.

\bibitem{davidson2010youtube}
James Davidson, Benjamin Liebald, Junning Liu, Palash Nandy, Taylor Van~Vleet,
  Ullas Gargi, Sujoy Gupta, Yu~He, Mike Lambert, Blake Livingston, et~al.
\newblock The youtube video recommendation system.
\newblock In {\em Proceedings of the fourth ACM conference on Recommender
  systems}, pages 293--296, 2010.

\bibitem{mukund2004}
Mukund Deshpande and George Karypis.
\newblock Item-based top-n recommendation algorithms.
\newblock {\em ACM Transactions on Information Systems (TOIS)}, 22(1):143--177,
  2004.

\bibitem{dinic1970}
Efim~A. Dinic.
\newblock Algorithm for solution of a problem of maximum flow in networks with
  power estimation.
\newblock {\em In Soviet Math. Doklady}, 11:1277--1280, 1970.

\bibitem{dubois2009improving}
Tom DuBois, Jennifer Golbeck, John Kleint, and Aravind Srinivasan.
\newblock Improving recommendation accuracy by clustering social networks with
  trust.
\newblock {\em Recommender Systems \& the Social Web}, 532:1--8, 2009.

\bibitem{dwork2012}
Cynthia Dwork, Moritz Hardt, Toniann Pitassi, Omer Reingold, and Richard Zemel.
\newblock Fairness through awareness.
\newblock In {\em In Proceedings of the 3rd innovations in theoretical computer
  science conference}, pages 214--226, 2012.

\bibitem{edizel2020}
Bora Edizel, Francesco Bonchi, Sara Hajian, André Panisson, and Tamir Tassa.
\newblock Fairecsys: Mitigating algorithmic bias in recommender systems.
\newblock {\em International Journal of Data Science and Analytics},
  9(2):197--213, 2020.

\bibitem{ekstrand2019fairness}
Michael~D Ekstrand, Robin Burke, and Fernando Diaz.
\newblock Fairness and discrimination in retrieval and recommendation.
\newblock In {\em Proceedings of the 42nd International ACM SIGIR Conference on
  Research and Development in Information Retrieval}, pages 1403--1404, 2019.

\bibitem{ekstrand2012}
Michael~D. Ekstrand, Michael Ludwig, Jack Kolb, and John~T. Riedl.
\newblock Lenskit: a modular recommender framework.
\newblock In {\em RecSys '11 Proceedings of the fifth ACM conference on
  Recommender systems}, pages 349--350, 2011.

\bibitem{ekstrand2018}
Michael~D. Ekstrand, Mucun Tian, Ion~Madrazo Azpiazu, Jennifer~D. Ekstrand,
  Oghenemaro Anuyah, David McNeill, and Maria~Soledad Pera.
\newblock All the cool kids, how do they fit in?: Popularity and demographic
  biases in recommender evaluation and effectiveness.
\newblock In {\em In Conference on Fairness, Accountability and Transparency},
  pages 172--186, 2018.

\bibitem{elahi2021beyond}
Mehdi Elahi, Himan Abdollahpouri, Masoud Mansoury, and Helma Torkamaan.
\newblock Beyond algorithmic fairness in recommender systems.
\newblock In {\em Adjunct Proceedings of the 29th ACM Conference on User
  Modeling, Adaptation and Personalization}, pages 41--46, 2021.

\bibitem{eskandanianUMAP2020}
Farzad Eskandanian and Bamshad Mobasher.
\newblock Using stable matching to optimize the balance between accuracy and
  diversity in recommendation.
\newblock In {\em Proceedings of the 28th ACM Conference on User Modeling,
  Adaptation and Personalization}, pages 71–--79. ACM, 2020.

\bibitem{fix1951discriminatory}
Evelyn Fix.
\newblock {\em Discriminatory analysis: nonparametric discrimination,
  consistency properties}.
\newblock USAF school of Aviation Medicine, 1951.

\bibitem{ford1956}
Lester~Randolph Ford and Delbert~R. Fulkerson.
\newblock Maximal flow through a network.
\newblock {\em Canadian Journal of Mathematics}, 8:399--404, 1956.

\bibitem{gajane2017formalizing}
Pratik Gajane and Mykola Pechenizkiy.
\newblock On formalizing fairness in prediction with machine learning.
\newblock {\em arXiv preprint arXiv:1710.03184}, 2017.

\bibitem{bonchi2018}
David Garc{\'{\i}}a{-}Soriano and Francesco Bonchi.
\newblock Fair-by-design algorithms: matching problems and beyond.
\newblock {\em CoRR}, abs/1802.02562, 2018.

\bibitem{gedikli2013improving}
Fatih Gedikli and Dietmar Jannach.
\newblock Improving recommendation accuracy based on item-specific tag
  preferences.
\newblock {\em ACM Transactions on Intelligent Systems and Technology (TIST)},
  4(1):1--19, 2013.

\bibitem{ghazanfar2010scalable}
Mustansar~Ali Ghazanfar and Adam Prugel-Bennett.
\newblock A scalable, accurate hybrid recommender system.
\newblock In {\em 2010 Third International Conference on Knowledge Discovery
  and Data Mining}, pages 94--98. IEEE, 2010.

\bibitem{gogna2015comprehensive}
Anupriya Gogna and Angshul Majumdar.
\newblock A comprehensive recommender system model: Improving accuracy for both
  warm and cold start users.
\newblock {\em IEEE Access}, 3:2803--2813, 2015.

\bibitem{goldberg1988}
Andrew~V. Goldberg and Robert~E. Tarjan.
\newblock A new approach to the maximum-flow problem.
\newblock {\em Journal of the ACM (JACM)}, 35(4):921--940, 1988.

\bibitem{gravino2019towards}
Pietro Gravino, Bernardo Monechi, and Vittorio Loreto.
\newblock Towards novelty-driven recommender systems.
\newblock {\em Comptes Rendus Physique}, 20(4):371--379, 2019.

\bibitem{grvcar2005data}
Miha Gr{\v{c}}ar, Dunja Mladeni{\v{c}}, Bla{\v{z}} Fortuna, and Marko
  Grobelnik.
\newblock Data sparsity issues in the collaborative filtering framework.
\newblock In {\em International workshop on knowledge discovery on the web},
  pages 58--76. Springer, 2005.

\bibitem{Guo2015}
Guibing Guo, Jie Zhang, Zhu Sun, and Neil Yorke-Smith.
\newblock Librec: A java library for recommender systems.
\newblock In {\em UMAP Workshops}, 2015.

\bibitem{hardt2016equality}
Moritz Hardt, Eric Price, and Nati Srebro.
\newblock Equality of opportunity in supervised learning.
\newblock In {\em Advances in neural information processing systems}, pages
  3315--3323, 2016.

\bibitem{hariri2012context}
Negar Hariri, Bamshad Mobasher, and Robin Burke.
\newblock Context-aware music recommendation based on latenttopic sequential
  patterns.
\newblock In {\em Proceedings of the sixth ACM conference on Recommender
  systems}, pages 131--138, 2012.

\bibitem{Harper:2016}
F.~Maxwell Harper and Joseph~A. Konstan.
\newblock The movielens datasets: History and context.
\newblock {\em ACM Transactions on Interactive Intelligent Systems (TiiS)},
  5(4):19, January 2016.

\bibitem{he2017neural}
Xiangnan He, Lizi Liao, Hanwang Zhang, Liqiang Nie, Xia Hu, and Tat-Seng Chua.
\newblock Neural collaborative filtering.
\newblock In {\em Proceedings of the 26th international conference on world
  wide web}, pages 173--182, 2017.

\bibitem{Herlocker:1999a}
Jonathan~L. Herlocker, Joseph~A. Konstan, Al~Borchers, and John Riedl.
\newblock An algorithmic framework for performing collaborative filtering.
\newblock In {\em Proceedings of the 22nd annual international ACM SIGIR
  conference on Research and development in information retrieval}, pages
  230--237. ACM, 1999.

\bibitem{hernandez2014probabilistic}
Jos{\'e}~Miguel Hern{\'a}ndez-Lobato, Neil Houlsby, and Zoubin Ghahramani.
\newblock Probabilistic matrix factorization with non-random missing data.
\newblock In {\em International Conference on Machine Learning}, pages
  1512--1520, 2014.

\bibitem{hiranandani2020cascading}
Gaurush Hiranandani, Harvineet Singh, Prakhar Gupta, Iftikhar~Ahamath
  Burhanuddin, Zheng Wen, and Branislav Kveton.
\newblock Cascading linear submodular bandits: Accounting for position bias and
  diversity in online learning to rank.
\newblock In {\em Uncertainty in Artificial Intelligence}, pages 722--732.
  PMLR, 2020.

\bibitem{ho2008}
Daniel~E. Ho and Kevin~M. Quinn.
\newblock Improving the presentation and interpretation of online ratings data
  with model-based figures.
\newblock {\em The American Statistician}, 62(4), 2008.

\bibitem{Hyndman:1996}
Rob~J. Hyndman and Yanan Fan.
\newblock Sample quantiles in statistical packages.
\newblock {\em The American Statistician}, 50(4):361--365, November 1996.

\bibitem{hyndman2006another}
Rob~J Hyndman and Anne~B Koehler.
\newblock Another look at measures of forecast accuracy.
\newblock {\em International journal of forecasting}, 22(4):679--688, 2006.

\bibitem{isufi58accuracy}
Elvin Isufi, Matteo Pocchiari, and Alan Hanjalic.
\newblock Accuracy-diversity trade-off in recommender systems via graph
  convolutions.
\newblock {\em Information Processing \& Management}, 58(2):102459.

\bibitem{dietmar2013}
Dietmar Jannach, Lukas Lerche, Fatih Gedikli, and Geoffray Bonnin.
\newblock What recommenders recommend–an analysis of accuracy, popularity,
  and sales diversity effects.
\newblock In {\em International conference on user modeling, adaptation, and
  personalization}, pages 25--37. Springer, Berlin, Heidelberg, 2013.

\bibitem{Jannach2015}
Dietmar Jannach, Lukas Lerche, Iman Kamehkhosh, and Michael Jugovac.
\newblock What recommenders recommend: an analysis of recommendation biases and
  possible countermeasures.
\newblock {\em User Modeling and User-Adapted Interaction}, 25(5):427--491,
  2015.

\bibitem{jannach2010recommender}
Dietmar Jannach, Markus Zanker, Alexander Felfernig, and Gerhard Friedrich.
\newblock {\em Recommender systems: an introduction}.
\newblock Cambridge University Press, 2010.

\bibitem{javari2015probabilistic}
Amin Javari and Mahdi Jalili.
\newblock A probabilistic model to resolve diversity--accuracy challenge of
  recommendation systems.
\newblock {\em Knowledge and Information Systems}, 44(3):609--627, 2015.

\bibitem{jeh2003scaling}
Glen Jeh and Jennifer Widom.
\newblock Scaling personalized web search.
\newblock In {\em Proceedings of the 12th international conference on World
  Wide Web}, pages 271--279, 2003.

\bibitem{kaminskas2016}
Marius Kaminskas and Derek Bridge.
\newblock Diversity, serendipity, novelty, and coverage: a survey and empirical
  analysis of beyond-accuracy objectives in recommender systems.
\newblock {\em ACM Transactions on Interactive Intelligent Systems (TiiS)},
  7(1):1--42, 2016.

\bibitem{kamiran2013techniques}
Faisal Kamiran, Toon Calders, and Mykola Pechenizkiy.
\newblock Techniques for discrimination-free predictive models.
\newblock In {\em Discrimination and privacy in the information society}, pages
  223--239. Springer, 2013.

\bibitem{DBLP:conf/recsys/KamishimaAAS14}
Toshihiro Kamishima, Shotaro Akaho, Hideki Asoh, and Jun Sakuma.
\newblock Correcting popularity bias by enhancing recommendation neutrality.
\newblock In {\em Poster Proceedings of the 8th {ACM} Conference on Recommender
  Systems, RecSys 2014, Foster City, Silicon Valley, CA, USA, October 6-10,
  2014}, 2014.

\bibitem{kamishima2011}
Toshihiro Kamishima, Shotaro Akaho, and Jun Sakuma.
\newblock Fairness-aware learning through regularization approach.
\newblock In {\em In 11th International Conference on Data Mining Workshops},
  pages 643--650, 2011.

\bibitem{kang2018self}
Wang-Cheng Kang and Julian McAuley.
\newblock Self-attentive sequential recommendation.
\newblock In {\em 2018 IEEE International Conference on Data Mining (ICDM)},
  pages 197--206. IEEE, 2018.

\bibitem{karatzoglou2013learning}
Alexandros Karatzoglou, Linas Baltrunas, and Yue Shi.
\newblock Learning to rank for recommender systems.
\newblock In {\em Proceedings of the 7th ACM conference on Recommender
  systems}, pages 493--494, 2013.

\bibitem{khenissi2020modeling}
Sami Khenissi.
\newblock Modeling and counteracting exposure bias in recommender systems.
\newblock Master's thesis, University of Louisville, 2019.

\bibitem{kleinberg2016inherent}
Jon Kleinberg, Sendhil Mullainathan, and Manish Raghavan.
\newblock Inherent trade-offs in the fair determination of risk scores.
\newblock {\em arXiv preprint arXiv:1609.05807}, 2016.

\bibitem{klockner2004depth}
Kerstin Kl{\"o}ckner, Nadine Wirschum, and Anthony Jameson.
\newblock Depth-and breadth-first processing of search result lists.
\newblock In {\em CHI'04 extended abstracts on Human factors in computing
  systems}, pages 1539--1539, 2004.

\bibitem{Koren:2008a}
Yehuda Koren.
\newblock Factorization meets the neighborhood: a multifaceted collaborative
  filtering model.
\newblock In {\em Proceedings of the 14th ACM SIGKDD international conference
  on Knowledge discovery and data mining}, pages 426--434. ACM, 2008.

\bibitem{Koren:2009a}
Yehuda Koren, Robert Bell, and Chris Volinsky.
\newblock Matrix factorization techniques for recommender systems.
\newblock {\em Computer}, 42(8), 2009.

\bibitem{kowald2020unfairness}
Dominik Kowald, Markus Schedl, and Elisabeth Lex.
\newblock The unfairness of popularity bias in music recommendation: A
  reproducibility study.
\newblock In {\em European Conference on Information Retrieval}, pages 35--42.
  Springer, 2020.

\bibitem{krishnan2014methodology}
Sanjay Krishnan, Jay Patel, Michael~J Franklin, and Ken Goldberg.
\newblock A methodology for learning, analyzing, and mitigating social
  influence bias in recommender systems.
\newblock In {\em Proceedings of the 8th ACM Conference on Recommender
  systems}, pages 137--144, 2014.

\bibitem{kullback1997information}
Solomon Kullback.
\newblock {\em Information theory and statistics}.
\newblock Courier Corporation, 1997.

\bibitem{kyropoulou2020almost}
Maria Kyropoulou, Warut Suksompong, and Alexandros~A Voudouris.
\newblock Almost envy-freeness in group resource allocation.
\newblock {\em Theoretical Computer Science}, 841:110--123, 2020.

\bibitem{lam2008addressing}
Xuan~Nhat Lam, Thuc Vu, Trong~Duc Le, and Anh~Duc Duong.
\newblock Addressing cold-start problem in recommendation systems.
\newblock In {\em Proceedings of the 2nd international conference on Ubiquitous
  information management and communication}, pages 208--211, 2008.

\bibitem{Langford:2006}
Eric Langford.
\newblock Quartiles in elementary statistics.
\newblock {\em Journal of Statistics Education}, 14(3):1--27, November 2006.

\bibitem{larsen2005introduction}
Richard~J Larsen and Morris~L Marx.
\newblock {\em An introduction to mathematical statistics}.
\newblock Prentice Hall, 2005.

\bibitem{lee2016improving}
Jongwuk Lee, Dongwon Lee, Yeon-Chang Lee, Won-Seok Hwang, and Sang-Wook Kim.
\newblock Improving the accuracy of top-n recommendation using a preference
  model.
\newblock {\em Information Sciences}, 348:290--304, 2016.

\bibitem{lemire2005slope}
Daniel Lemire and Anna Maclachlan.
\newblock Slope one predictors for online rating-based collaborative filtering.
\newblock In {\em Proceedings of the 2005 SIAM International Conference on Data
  Mining}, pages 471--475. SIAM, 2005.

\bibitem{li2020cascading}
Chang Li, Haoyun Feng, and Maarten~de Rijke.
\newblock Cascading hybrid bandits: Online learning to rank for relevance and
  diversity.
\newblock In {\em Fourteenth ACM Conference on Recommender Systems}, pages
  33--42, 2020.

\bibitem{li2017}
Jingjing Li, Ke~Lu, Zi~Huang, and Heng~Tao Shen.
\newblock Two birds one stone: On both cold-start and long-tail recommendation.
\newblock In {\em MM '17 Proceedings of the 2017 ACM on Multimedia Conference},
  pages 898--906, 2017.

\bibitem{lika2014facing}
Blerina Lika, Kostas Kolomvatsos, and Stathes Hadjiefthymiades.
\newblock Facing the cold start problem in recommender systems.
\newblock {\em Expert Systems with Applications}, 41(4):2065--2073, 2014.

\bibitem{linden2003amazon}
Greg Linden, Brent Smith, and Jeremy York.
\newblock Amazon. com recommendations: Item-to-item collaborative filtering.
\newblock {\em IEEE Internet computing}, 7(1):76--80, 2003.

\bibitem{liu2004personalized}
Fang Liu, Clement Yu, and Weiyi Meng.
\newblock Personalized web search for improving retrieval effectiveness.
\newblock {\em IEEE Transactions on knowledge and data engineering},
  16(1):28--40, 2004.

\bibitem{liu2015trust}
Haifeng Liu, Xiaomei Bai, Zhuo Yang, Amr Tolba, and Feng Xia.
\newblock Trust-aware recommendation for improving aggregate diversity.
\newblock {\em New Review of Hypermedia and Multimedia}, 21(3-4):242--258,
  2015.

\bibitem{liu2016context}
Qiang Liu, Shu Wu, Diyi Wang, Zhaokang Li, and Liang Wang.
\newblock Context-aware sequential recommendation.
\newblock In {\em 2016 IEEE 16th International Conference on Data Mining
  (ICDM)}, pages 1053--1058. IEEE, 2016.

\bibitem{liu2018}
Weiwen Liu and Robin Burke.
\newblock Personalizing fairness-aware re-ranking.
\newblock {\em CoRR}, abs/1809.02921, 2018.

\bibitem{liu2016you}
Yiming Liu, Xuezhi Cao, and Yong Yu.
\newblock Are you influenced by others when rating? improve rating prediction
  by conformity modeling.
\newblock In {\em Proceedings of the 10th ACM Conference on Recommender
  Systems}, pages 269--272, 2016.

\bibitem{lops2011content}
Pasquale Lops, Marco De~Gemmis, and Giovanni Semeraro.
\newblock Content-based recommender systems: State of the art and trends.
\newblock {\em Recommender systems handbook}, pages 73--105, 2011.

\bibitem{lu2012recommender}
Linyuan L{\"u}, Mat{\'u}{\v{s}} Medo, Chi~Ho Yeung, Yi-Cheng Zhang, Zi-Ke
  Zhang, and Tao Zhou.
\newblock Recommender systems.
\newblock {\em Physics reports}, 519(1):1--49, 2012.

\bibitem{ma2014improving}
Xiao Ma, Hongwei Lu, and Zaobin Gan.
\newblock Improving recommendation accuracy by combining trust communities and
  collaborative filtering.
\newblock In {\em Proceedings of the 23rd ACM international conference on
  conference on information and knowledge management}, pages 1951--1954, 2014.

\bibitem{masoud2021wsdm}
Masoud Mansoury.
\newblock Fairness-aware recommendation in multi-sided platforms.
\newblock In {\em Proceedings of the 14th ACM International Conference on Web
  Search and Data Mining}, WSDM '21, page 1117–1118, New York, NY, USA, 2021.

\bibitem{mansoury2021unbiased}
Masoud Mansoury, Himan Abdollahpouri, Bamshad Mobasher, Mykola Pechenizkiy,
  Robin Burke, and Milad Sabouri.
\newblock Unbiased cascade bandits: Mitigating exposure bias in online learning
  to rank recommendation.
\newblock {\em arXiv preprint arXiv:2108.03440}, 2021.

\bibitem{mansoury2020fairmatch}
Masoud Mansoury, Himan Abdollahpouri, Mykola Pechenizkiy, Bamshad Mobasher, and
  Robin Burke.
\newblock Fairmatch: A graph-based approach for improving aggregate diversity
  in recommender systems.
\newblock In {\em Proceedings of the 28th ACM Conference on User Modeling,
  Adaptation and Personalization}, pages 154--162, 2020.

\bibitem{mansoury2020feedback}
Masoud Mansoury, Himan Abdollahpouri, Mykola Pechenizkiy, Bamshad Mobasher, and
  Robin Burke.
\newblock Feedback loop and bias amplification in recommender systems.
\newblock In {\em Proceedings of the 29th ACM International Conference on
  Information \& Knowledge Management}, pages 2145--2148, 2020.

\bibitem{mansoury2021tois}
Masoud Mansoury, Himan Abdollahpouri, Mykola Pechenizkiy, Bamshad Mobasher, and
  Robin Burke.
\newblock A graph-based approach for mitigating multi-sided exposure bias in
  recommender systems, 2021.

\bibitem{mansoury2019relationship}
Masoud Mansoury, Himan Abdollahpouri, Joris Rombouts, and Mykola Pechenizkiy.
\newblock The relationship between the consistency of users' ratings and
  recommendation calibration.
\newblock {\em Workshop on Designing Human-Centric MIR Systems}, 2019.

\bibitem{mansoury2020investigating}
Masoud Mansoury, Himan Abdollahpouri, Jessie Smith, Arman Dehpanah, Mykola
  Pechenizkiy, and Bamshad Mobasher.
\newblock Investigating potential factors associated with gender discrimination
  in collaborative recommender systems.
\newblock {\em The 32nd International FLAIRS Conference in Cooperation with
  AAAI}, 2020.

\bibitem{mansoury2019algorithm}
Masoud Mansoury and Robin Burke.
\newblock Algorithm selection with librec-auto.
\newblock In {\em AMIR@ECIR}, pages 11--17, 2019.

\bibitem{mansoury2021flatter}
Masoud Mansoury, Robin Burke, and Bamshad Mobasher.
\newblock Flatter is better: percentile transformations for recommender
  systems.
\newblock {\em ACM Transactions on Intelligent Systems and Technology (TIST)},
  12(2):1--16, 2021.

\bibitem{masoud2018}
Masoud Mansoury, Robin Burke, Aldo Ordonez-Gauger, and Xavier Sepulveda.
\newblock Automating recommender systems experimentation with librec-auto.
\newblock In {\em Proceedings of the 12th ACM Conference on Recommender
  Systems}, pages 500--501, 2018.

\bibitem{mansoury2019bias}
Masoud Mansoury, Bamshad Mobasher, Robin Burke, and Mykola Pechenizkiy.
\newblock Bias disparity in collaborative recommendation: Algorithmic
  evaluation and comparison.
\newblock {\em RMSE Workshop at RecSys'19}, 2019.

\bibitem{mansoury2016improving}
Masoud Mansoury and Mehdi Shajari.
\newblock Improving recommender systems’ performance on cold-start users and
  controversial items by a new similarity model.
\newblock {\em International Journal of Web Information Systems}, 2016.

\bibitem{marlin2012collaborative}
Benjamin Marlin, Richard~S Zemel, Sam Roweis, and Malcolm Slaney.
\newblock Collaborative filtering and the missing at random assumption.
\newblock {\em arXiv preprint arXiv:1206.5267}, 2012.

\bibitem{marlin2007collaborative}
Benjamin~M Marlin, Richard~S Zemel, Sam Roweis, and Malcolm Slaney.
\newblock Collaborative filtering and the missing at random assumption.
\newblock In {\em Proceedings of the Twenty-Third Conference on Uncertainty in
  Artificial Intelligence}, pages 267--275. AUAI Press, 2007.

\bibitem{massa2004trust}
Paolo Massa and Paolo Avesani.
\newblock Trust-aware collaborative filtering for recommender systems.
\newblock In {\em OTM Confederated International Conferences" On the Move to
  Meaningful Internet Systems"}, pages 492--508. Springer, 2004.

\bibitem{Massa:2007a}
Paolo Massa and Paolo Avesani.
\newblock Trust-aware recommender systems.
\newblock In {\em Proceedings of the 2007 ACM conference on Recommender
  systems}, pages 17--24. ACM, October 2007.

\bibitem{mehrotra2018towards}
Rishabh Mehrotra, James McInerney, Hugues Bouchard, Mounia Lalmas, and Fernando
  Diaz.
\newblock Towards a fair marketplace: Counterfactual evaluation of the
  trade-off between relevance, fairness \& satisfaction in recommendation
  systems.
\newblock In {\em Proceedings of the 27th acm international conference on
  information and knowledge management}, pages 2243--2251, 2018.

\bibitem{melville2010recommender}
Prem Melville and Vikas Sindhwani.
\newblock Recommender systems.
\newblock {\em Encyclopedia of machine learning}, 1:829--838, 2010.

\bibitem{mendoza2020evaluating}
Marcelo Mendoza and Nicol{\'a}s Torres.
\newblock Evaluating content novelty in recommender systems.
\newblock {\em Journal of Intelligent Information Systems}, 54(2):297--316,
  2020.

\bibitem{moller2018not}
Judith M{\"o}ller, Damian Trilling, Natali Helberger, and Bram van Es.
\newblock Do not blame it on the algorithm: an empirical assessment of multiple
  recommender systems and their impact on content diversity.
\newblock {\em Information, Communication \& Society}, 21(7):959--977, 2018.

\bibitem{natarajan2020resolving}
Senthilselvan Natarajan, Subramaniyaswamy Vairavasundaram, Sivaramakrishnan
  Natarajan, and Amir~H Gandomi.
\newblock Resolving data sparsity and cold start problem in collaborative
  filtering recommender system using linked open data.
\newblock {\em Expert Systems with Applications}, 149:113248, 2020.

\bibitem{Ning2011}
Xia Ning and George Karypis.
\newblock Slim: Sparse linear methods for top-n recommender systems.
\newblock In {\em Data Mining (ICDM), 2011 IEEE 11th International Conference
  on}, pages 497--506. IEEE, 2011.

\bibitem{opitz1999popular}
David Opitz and Richard Maclin.
\newblock Popular ensemble methods: An empirical study.
\newblock {\em Journal of artificial intelligence research}, 11:169--198, 1999.

\bibitem{o2006modeling}
Maeve O’Brien and Mark~T Keane.
\newblock Modeling result-list searching in the world wide web: The role of
  relevance topologies and trust bias.
\newblock In {\em Proceedings of the 28th annual conference of the cognitive
  science society}, volume~28, pages 1881--1886. Citeseer, 2006.

\bibitem{park2008long}
Yoon-Joo Park and Alexander Tuzhilin.
\newblock The long tail of recommender systems and how to leverage it.
\newblock In {\em Proceedings of the 2008 ACM conference on Recommender
  systems}, pages 11--18, 2008.

\bibitem{patro2020fairrec}
Gourab~K Patro, Arpita Biswas, Niloy Ganguly, Krishna~P Gummadi, and Abhijnan
  Chakraborty.
\newblock Fairrec: Two-sided fairness for personalized recommendations in
  two-sided platforms.
\newblock In {\em Proceedings of The Web Conference 2020}, pages 1194--1204,
  2020.

\bibitem{pazzani2007content}
Michael~J Pazzani and Daniel Billsus.
\newblock Content-based recommendation systems.
\newblock In {\em The adaptive web}, pages 325--341. Springer, 2007.

\bibitem{pedreshi2008discrimination}
Dino Pedreshi, Salvatore Ruggieri, and Franco Turini.
\newblock Discrimination-aware data mining.
\newblock In {\em Proceedings of the 14th ACM SIGKDD international conference
  on Knowledge discovery and data mining}, pages 560--568, 2008.

\bibitem{powell2017love}
Derek Powell, Jingqi Yu, Melissa DeWolf, and Keith~J Holyoak.
\newblock The love of large numbers: A popularity bias in consumer choice.
\newblock {\em Psychological science}, 28(10):1432--1442, 2017.

\bibitem{quadrana2018sequence}
Massimo Quadrana, Paolo Cremonesi, and Dietmar Jannach.
\newblock Sequence-aware recommender systems.
\newblock {\em ACM Computing Surveys (CSUR)}, 51(4):1--36, 2018.

\bibitem{rendle2009bpr}
Steffen Rendle, Christoph Freudenthaler, Zeno Gantner, and Lars Schmidt-Thieme.
\newblock Bpr: Bayesian personalized ranking from implicit feedback.
\newblock In {\em Proceedings of the twenty-fifth conference on uncertainty in
  artificial intelligence}, pages 452--461. AUAI Press, 2009.

\bibitem{Resnick:1994a}
Paul Resnick, Neophytos Iacovou, Mitesh Suchak, Peter Bergstrom, and John
  Riedl.
\newblock Grouplens: an open architecture for collaborative filtering of
  netnews.
\newblock In {\em Proceedings of the 1994 ACM conference on Computer supported
  cooperative work}, pages 175--186. ACM, October 1994.

\bibitem{resnick1997recommender}
Paul Resnick and Hal~R Varian.
\newblock Recommender systems.
\newblock {\em Communications of the ACM}, 40(3):56--58, 1997.

\bibitem{safoury2013exploiting}
Laila Safoury and Akram Salah.
\newblock Exploiting user demographic attributes for solving cold-start problem
  in recommender system.
\newblock {\em Lecture Notes on Software Engineering}, 1(3):303--307, 2013.

\bibitem{sanders1987pareto}
Robert Sanders.
\newblock The pareto principle: its use and abuse.
\newblock {\em Journal of Services Marketing}, 1987.

\bibitem{santos2010exploiting}
Rodrygo~LT Santos, Craig Macdonald, and Iadh Ounis.
\newblock Exploiting query reformulations for web search result
  diversification.
\newblock In {\em Proceedings of the 19th international conference on World
  wide web}, pages 881--890, 2010.

\bibitem{sarwar2000application}
Badrul Sarwar, George Karypis, Joseph Konstan, and John Riedl.
\newblock Application of dimensionality reduction in recommender system-a case
  study.
\newblock Technical report, Minnesota Univ Minneapolis Dept of Computer
  Science, 2000.

\bibitem{sarwar2001}
Badrul Sarwar, George Karypis, Joseph Konstan, and John Riedl.
\newblock Item-based collaborative filtering recommendation algorithms.
\newblock In {\em WWW'01 Proceedings of the 10th international conference on
  World Wide Web}, pages 285--295, May 2001.

\bibitem{sarwar2001sparsity}
Badrul~Munir Sarwar.
\newblock Sparsity, scalability, and distribution in recommender systems.
\newblock 2001.

\bibitem{sarwat2013lars}
Mohamed Sarwat, Justin~J Levandoski, Ahmed Eldawy, and Mohamed~F Mokbel.
\newblock Lars*: An efficient and scalable location-aware recommender system.
\newblock {\em IEEE Transactions on Knowledge and Data Engineering},
  26(6):1384--1399, 2013.

\bibitem{schedl2016lfm}
Markus Schedl.
\newblock The lfm-1b dataset for music retrieval and recommendation.
\newblock In {\em Proceedings of the 2016 ACM on International Conference on
  Multimedia Retrieval}, pages 103--110, 2016.

\bibitem{shani2011evaluating}
Guy Shani and Asela Gunawardana.
\newblock Evaluating recommendation systems.
\newblock In {\em Recommender systems handbook}, pages 257--297. Springer,
  2011.

\bibitem{shi2013trading}
Lei Shi.
\newblock Trading-off among accuracy, similarity, diversity, and long-tail: a
  graph-based recommendation approach.
\newblock In {\em Proceedings of the 7th ACM conference on Recommender
  systems}, pages 57--64, 2013.

\bibitem{shi2010}
Yue Shi, Martha Larson, and Alan Hanjalic.
\newblock List-wise learning to rank with matrix factorization for
  collaborative filtering.
\newblock In {\em Proceedings of the fourth ACM conference on Recommender
  systems}, pages 269--272. ACM, September 2010.

\bibitem{shi2010list}
Yue Shi, Martha Larson, and Alan Hanjalic.
\newblock List-wise learning to rank with matrix factorization for
  collaborative filtering.
\newblock In {\em Proceedings of the fourth ACM conference on Recommender
  systems}, pages 269--272. ACM, 2010.

\bibitem{singh2018fairness}
Ashudeep Singh and Thorsten Joachims.
\newblock Fairness of exposure in rankings.
\newblock In {\em Proceedings of the 24th ACM SIGKDD International Conference
  on Knowledge Discovery \& Data Mining}, pages 2219--2228, 2018.

\bibitem{sinha2016}
Ayan Sinha, David~F. Gleich, and Karthik Ramani.
\newblock Deconvolving feedback loops in recommender systems.
\newblock In {\em Advances in neural information processing systems}, pages
  3243--3251, 2016.

\bibitem{smith2017two}
Brent Smith and Greg Linden.
\newblock Two decades of recommender systems at amazon. com.
\newblock {\em Ieee internet computing}, 21(3):12--18, 2017.

\bibitem{sonboli2020fairness}
Nasim Sonboli, Robin Burke, Zijun Liu, and Masoud Mansoury.
\newblock Fairness-aware recommendation with librec-auto.
\newblock In {\em Fourteenth ACM Conference on Recommender Systems}, pages
  594--596, 2020.

\bibitem{harald2011}
Harald Steck.
\newblock Item popularity and recommendation accuracy.
\newblock In {\em RecSys '11 Proceedings of the fifth ACM Conference on
  Recommender Systems}, pages 125--132, 2011.

\bibitem{steck2013evaluation}
Harald Steck.
\newblock Evaluation of recommendations: rating-prediction and ranking.
\newblock In {\em Proceedings of the 7th ACM conference on Recommender
  systems}, pages 213--220, 2013.

\bibitem{steck2018calibrated}
Harald Steck.
\newblock Calibrated recommendations.
\newblock In {\em Proceedings of the 12th ACM conference on recommender
  systems}, pages 154--162, 2018.

\bibitem{suhr2019two}
Tom S{\"u}hr, Asia~J Biega, Meike Zehlike, Krishna~P Gummadi, and Abhijnan
  Chakraborty.
\newblock Two-sided fairness for repeated matchings in two-sided markets: A
  case study of a ride-hailing platform.
\newblock In {\em Proceedings of the 25th ACM SIGKDD International Conference
  on Knowledge Discovery \& Data Mining}, pages 3082--3092, 2019.

\bibitem{sun2005cubesvd}
Jian-Tao Sun, Hua-Jun Zeng, Huan Liu, Yuchang Lu, and Zheng Chen.
\newblock Cubesvd: a novel approach to personalized web search.
\newblock In {\em Proceedings of the 14th international conference on World
  Wide Web}, pages 382--390, 2005.

\bibitem{sun2019debiasing}
Wenlong Sun, Sami Khenissi, Olfa Nasraoui, and Patrick Shafto.
\newblock Debiasing the human-recommender system feedback loop in collaborative
  filtering.
\newblock In {\em Companion Proceedings of The 2019 World Wide Web Conference},
  pages 645--651, 2019.

\bibitem{takacs2012alternating}
G{\'a}bor Tak{\'a}cs and Domonkos Tikk.
\newblock Alternating least squares for personalized ranking.
\newblock In {\em Proceedings of the sixth ACM conference on Recommender
  systems}, pages 83--90, 2012.

\bibitem{tang2018personalized}
Jiaxi Tang and Ke~Wang.
\newblock Personalized top-n sequential recommendation via convolutional
  sequence embedding.
\newblock In {\em Proceedings of the Eleventh ACM International Conference on
  Web Search and Data Mining}, pages 565--573, 2018.

\bibitem{virginia2018}
Virginia Tsintzou, Evaggelia Pitoura, and Panayiotis Tsaparas.
\newblock Bias disparity in recommendation systems.
\newblock {\em CoRR}, abs/1811.01461, 2018.

\bibitem{vargas2011rank}
Sa{\'u}l Vargas and Pablo Castells.
\newblock Rank and relevance in novelty and diversity metrics for recommender
  systems.
\newblock In {\em Proceedings of the fifth ACM conference on Recommender
  systems}, pages 109--116, 2011.

\bibitem{vargas2014improving}
Sa{\'u}l Vargas and Pablo Castells.
\newblock Improving sales diversity by recommending users to items.
\newblock In {\em Proceedings of the 8th ACM Conference on Recommender
  systems}, pages 145--152, 2014.

\bibitem{vargas2011}
Saúl Vargas and Pablo Castells.
\newblock Rank and relevance in novelty and diversity metrics for recommender
  systems.
\newblock In {\em In Proceedings of the fifth ACM conference on Recommender
  systems}, pages 109--116, 2011.

\bibitem{vargas2014}
Saúl Vargas and Pablo Castells.
\newblock Improving sales diversity by recommending users to items.
\newblock In {\em In Proceedings of the 8th ACM Conference on Recommender
  systems}, pages 145--152, 2014.

\bibitem{wang2013opportunity}
Jian Wang and Yi~Zhang.
\newblock Opportunity model for e-commerce recommendation: right product; right
  time.
\newblock In {\em Proceedings of the 36th international ACM SIGIR conference on
  Research and development in information retrieval}, pages 303--312, 2013.

\bibitem{wang2016multi}
Shanfeng Wang, Maoguo Gong, Haoliang Li, and Junwei Yang.
\newblock Multi-objective optimization for long tail recommendation.
\newblock {\em Knowledge-Based Systems}, 104:145--155, 2016.

\bibitem{wang2014amazon}
Ting Wang and Dashun Wang.
\newblock Why amazon's ratings might mislead you: The story of herding effects.
\newblock {\em Big data}, 2(4):196--204, 2014.

\bibitem{willmott2005advantages}
Cort~J Willmott and Kenji Matsuura.
\newblock Advantages of the mean absolute error (mae) over the root mean square
  error (rmse) in assessing average model performance.
\newblock {\em Climate research}, 30(1):79--82, 2005.

\bibitem{xiang2010temporal}
Liang Xiang, Quan Yuan, Shiwan Zhao, Li~Chen, Xiatian Zhang, Qing Yang, and
  Jimeng Sun.
\newblock Temporal recommendation on graphs via long-and short-term preference
  fusion.
\newblock In {\em Proceedings of the 16th ACM SIGKDD international conference
  on Knowledge discovery and data mining}, pages 723--732, 2010.

\bibitem{xiao2017fairness}
Lin Xiao, Zhang Min, Zhang Yongfeng, Gu~Zhaoquan, Liu Yiqun, and Ma~Shaoping.
\newblock Fairness-aware group recommendation with pareto-efficiency.
\newblock In {\em Proceedings of the Eleventh ACM Conference on Recommender
  Systems}, pages 107--115, 2017.

\bibitem{yao2017}
Sirui Yao and Bert Huang.
\newblock Beyond parity: Fairness objectives for collaborative filtering.
\newblock In {\em In Advances in Neural Information Processing Systems}, pages
  2921--2930, 2017.

\bibitem{yi2014beyond}
Xing Yi, Liangjie Hong, Erheng Zhong, Nanthan~Nan Liu, and Suju Rajan.
\newblock Beyond clicks: dwell time for personalization.
\newblock In {\em Proceedings of the 8th ACM Conference on Recommender
  systems}, pages 113--120, 2014.

\bibitem{yin2012challenging}
Hongzhi Yin, Bin Cui, Jing Li, Junjie Yao, and Chen Chen.
\newblock Challenging the long tail recommendation.
\newblock {\em arXiv preprint arXiv:1205.6700}, 2012.

\bibitem{yu2019adaptive}
Zeping Yu, Jianxun Lian, Ahmad Mahmoody, Gongshen Liu, and Xing Xie.
\newblock Adaptive user modeling with long and short-term preferences for
  personalized recommendation.
\newblock In {\em IJCAI}, pages 4213--4219, 2019.

\bibitem{zafar2017fairness}
Muhammad~Bilal Zafar, Isabel Valera, Manuel Gomez~Rodriguez, and Krishna~P
  Gummadi.
\newblock Fairness beyond disparate treatment \& disparate impact: Learning
  classification without disparate mistreatment.
\newblock In {\em Proceedings of the 26th international conference on world
  wide web}, pages 1171--1180, 2017.

\bibitem{zehlike2017fa}
Meike Zehlike, Francesco Bonchi, Carlos Castillo, Sara Hajian, Mohamed Megahed,
  and Ricardo Baeza-Yates.
\newblock Fa* ir: A fair top-k ranking algorithm.
\newblock In {\em Proceedings of the 2017 ACM on Conference on Information and
  Knowledge Management}, pages 1569--1578, 2017.

\bibitem{zembel2013}
Rich Zemel, Yu~Wu, Kevin Swersky, Toni Pitassi, and Cynthia Dwork.
\newblock Learning fair representations.
\newblock In {\em In International Conference on Machine Learning}, pages
  325--333, 2013.

\bibitem{zhang2013definition}
Liang Zhang.
\newblock The definition of novelty in recommendation system.
\newblock {\em Journal of Engineering Science \& Technology Review}, 6(3),
  2013.

\bibitem{zhao2017improving}
Tong Zhao, Julian McAuley, Mengya Li, and Irwin King.
\newblock Improving recommendation accuracy using networks of substitutable and
  complementary products.
\newblock In {\em 2017 International Joint Conference on Neural Networks
  (IJCNN)}, pages 3649--3655. IEEE, 2017.

\bibitem{zhao2019recommending}
Zhe Zhao, Lichan Hong, Li~Wei, Jilin Chen, Aniruddh Nath, Shawn Andrews, Aditee
  Kumthekar, Maheswaran Sathiamoorthy, Xinyang Yi, and Ed~Chi.
\newblock Recommending what video to watch next: a multitask ranking system.
\newblock In {\em Proceedings of the 13th ACM Conference on Recommender
  Systems}, pages 43--51, 2019.

\bibitem{zheng2010photo}
Nan Zheng, Qiudan Li, Shengcai Liao, and Leiming Zhang.
\newblock Which photo groups should i choose? a comparative study of
  recommendation algorithms in flickr.
\newblock {\em Journal of Information Science}, 36(6):733--750, 2010.

\bibitem{zheng2017multi}
Yong Zheng.
\newblock Multi-stakeholder recommendation: Applications and challenges.
\newblock {\em arXiv preprint arXiv:1707.08913}, 2017.

\bibitem{zheng2018utility}
Yong Zheng and Aviana Pu.
\newblock Utility-based multi-stakeholder recommendations by multi-objective
  optimization.
\newblock In {\em 2018 IEEE/WIC/ACM International Conference on Web
  Intelligence (WI)}, pages 128--135. IEEE, 2018.

\bibitem{zheng2020disentangling}
Yu~Zheng, Chen Gao, Xiang Li, Xiangnan He, Yong Li, and Depeng Jin.
\newblock Disentangling user interest and popularity bias for recommendation
  with causal embedding.
\newblock {\em arXiv preprint arXiv:2006.11011}, 2020.

\bibitem{zhou2010impact}
Renjie Zhou, Samamon Khemmarat, and Lixin Gao.
\newblock The impact of youtube recommendation system on video views.
\newblock In {\em Proceedings of the 10th ACM SIGCOMM conference on Internet
  measurement}, pages 404--410, 2010.

\bibitem{zhu2018}
Ziwei Zhu, Xia Hu, and James Caverlee.
\newblock Fairness-aware tensor-based recommendation.
\newblock In {\em In Proceedings of the 27th ACM International Conference on
  Information and Knowledge Management}, pages 1153--1162, 2018.

\bibitem{zong2016cascading}
Shi Zong, Hao Ni, Kenny Sung, Nan~Rosemary Ke, Zheng Wen, and Branislav Kveton.
\newblock Cascading bandits for large-scale recommendation problems.
\newblock In {\em Proceedings of the Thirty-Second Conference on Uncertainty in
  Artificial Intelligence}, pages 835--844, 2016.

\bibitem{surer2018}
Özge Sürer, Robin Burke, and Edward~C. Malthouse.
\newblock Multistakeholder recommendation with provider constraints.
\newblock In {\em In Proceedings of the 12th ACM Conference on Recommender
  Systems}, pages 54--62, 2018.

\end{thebibliography}
